\documentclass[doublespacing]{elsart}
\usepackage[english]{babel}

\usepackage{epsfig}
\usepackage{graphics}

\usepackage{amssymb}

\begin{document}

\begin{frontmatter}

\title{Spontaneous and Induced Radiation by Relativistic Particles in Natural and Photonic Crystals. Crystal X-ray Lasers and Volume Free Electron Lasers (VFEL)}

\author{V.G. Baryshevsky}


%
\begin{abstract}
The  mechanisms of spontaneous and induced radiation produced by relativistic particles passing through natural and photonic crystals are reviewed. The theory of Volume Free Electron  Lasers based on spontaneous radiation in natural and photonic crystals is presented.
\end{abstract}

\tableofcontents

\end{frontmatter}

\section*{Introduction}
Current development of Free Electron Lasers (FELs) operating in
various ranges of wavelengths -- from microwave to optical and
even X-ray --  which is carried out in some research centers gives
reason to return to the problem of the possibility to design the
FEL on the basis of the interaction between relativistic electron
(positron) beams and either natural or artificial (photonic)
crystals. In the X-ray range of wavelengths, operation of this
type of crystal X-ray FELs is based on several spontaneous
radiation mechanisms: parametric X-ray radiation
\cite{berk_2,vfel_PXRbook,para_1,para_2,lanl_1}, diffracted
radiation from relativistic channeled particles
\cite{berk_2,berk_13,berk_28,lanl_1,lanl_4,lanl_5,lanl_6,lanl_7,lanl_14},
generation of X-ray radiation by particles moving in a crystal
undulator
\cite{chan_67,berk_2,berk_129,lanl_1,lanl_15,lanl_16,lanl_17,lanl_18}.
It was shown that despite serious difficulties, the law of
radiative instability of relativistic beams passing through the
crystal, which was discovered in \cite{berk_133}, enables
achieving the generation threshold for induced X-ray radiation
even at the beam current density $j\sim 10^8$ ~A/cm$^2$, while the
current density  required for X-ray laser generation in crystals
by means of channeling X-ray radiation or radiation in a crystal
undulator is $j\sim10^{12}\div 10^{13}$  \cite{lanl_1}.

Below, a brief review of the theory of the
 crystal FEL (based on either a natural or a photonic crystal) is presented.  The review emphasizes the general significance of
the law of radiative instability discovered in \cite{berk_133},
which finally led to formulating the idea of the possibility to
develop a new type of FELs, called Volume Free Electron Lasers
\cite{berk_104}.

\section{Diffraction phenomena accompanying spontaneous and stimulated
radiation from relativistic particles in crystals (\cite{lanl_1})}
\label{sec:berk1}

The emission of photons by relativistic particles in media has
been calling attention for a long time. The main reason why this
phenomenon arises such interest is because a wide variety of tasks
can  be solved by using different radiation mechanisms, such as
bremsstrahlung, transition and Cherenkov radiations and so on. The
last decades have seen a growing interest in the study of
radiation by relativistic particles in both natural and artificial
(photonic) crystals. A number of new radiation mechanisms
{associated with} a periodic structure of crystals have been
considered theoretically and confirmed experimentally
\cite{171,rins_66,lanl_1,lanl_2,lanl_3}.

All characteristic properties of radiation are
in this case
determined by the periodic  structure of a crystal.
 The medium can
influence radiation processes under passing of relativistic
charged particles through crystals in two ways.
First of all, it
is well known
that the trajectory of a charged particle incident on a crystal at
a small angle relative to crystallographic planes or axes
is formed by a series of grazing collisions with the atoms of a
crystal.
As a result, the particle moves in an averaged potential of
crystallographic planes or axes.
In this case we speak about the channeling phenomenon and can
consider the motion of channeled particles as the motion inside a
potential well -- one-dimensional (for plane channeling) or
two-dimensional (for axial channeling).
According to quantum mechanics a particle moving inside a
potential well has a discrete energy spectrum, which is the
spectrum of the particle transverse motion.
Consequently, such particles can be considered as one-dimensional
or two-dimensional atoms (oscillators) characterized by a spectrum
of bound states (zones) of transverse energy $\varepsilon_{n}$,
$\varepsilon_{f}$. The number of bound states and their
characteristics depend on the longitudinal particle energy.
One can conclude that many phenomena observed for ordinary atoms
will manifest themselves when channeled particles pass through
crystals. It is obvious that thus excited atoms should emit
photons with the energy equal to the difference between atomic
state energies $\varepsilon_{n}$ and $\varepsilon_{f}$.
The frequency of transition
$\Omega_{nf}=\varepsilon_n\varepsilon_f$ and depends, in a
laboratory frame, on the total particle energy.
By analogy with an ordinary moving oscillator, the frequency of
the emitted photons is evaluated by the Doppler effect and
determined by the following expression:
\begin{equation}
\label{berk_1.0} \omega=\frac{\Omega_{nf}}{1-\beta
n(\omega)\cos\theta},
\end{equation}
where $\theta$ is the radiation angle, $\beta=u/c$,  $u$ is the
longitudinal particle velocity, $n(\omega)$ is the refraction
index of the photon
with a frequency $\omega$ in the medium.

The relativistic oscillator can be formed not only by an
unperturbed crystal channel,
but also by an external ultrasonic or
laser wave which propagates in the crystal, creating a bent
crystal channel
\cite{chan_67,berk_2,para_1,lanl_1,lanl_15,lanl_16,lanl_17,lanl_18}

On the other hand, when the wavelength of the emitted photons is
of the order of the interplanar spacing of atoms in the crystal,
radiation diffraction can essentially modify the photon state
\cite{para_1}. In this case the radiation process is characterized
by several  indices of refraction $n_i(\omega)$ dependent on the
direction of the photon momentum, which, in turn, leads to the
modification of all mechanisms of radiation formation by
relativistic particles in the X-ray range of spectrum. For
example, radiation at a large angle relative to the direction of
particle motion  becomes possible. As a result, the diffraction
pattern characterizing a given crystal is formed. The analysis of
dielectric properties of a crystal under diffraction conditions
shows that at least one of several  indices of refraction
$n_i(\omega)$, characterizing the crystal under this condition,
becomes greater than unity within a frequency interval. As a
consequence, the Vavilov-Cherenkov condition can be fulfilled. In
this case spontaneous and induced X-ray radiation analogous to
optical Cherenkov radiation appears as well as diffracted
transition radiation \cite{para_1}. Spontaneous quasi-Cherenkov
radiation, currently referred to as parametric (quasi-Cherenkov)
X-ray radiation (PXR), is thoroughly investigated both
theoretically and experimentally
(see \cite{vfel_PXRbook}). 

Under diffraction conditions, the radiation of a relativistic
oscillator also modifies essentially. Now the periodic structure
of a crystal affects both the particle motion and the particle
state. This  leads to the formation of diffracted radiation of
oscillator (DRO), which cannot be reduced to the sequence  of two
independent processes: radiation by oscillator and diffraction of
radiated photons. In this case the process of photon emission and
its diffraction are developing simultaneously and coherently and
result in the radiation with new properties. Radiation produced by
channeled particles is often called the  diffracted channeling
radiation (DCR).

 These types of X-radiation (DRO, DCR), also associated with a change in the  indices of refraction under diffraction conditions, was considered in \cite{berk_13} and then in \cite{chan_44,chan_45,berk_19}. Currently, there is an increased interest in DCR (see \cite{lanl_4,lanl_5,lanl_6,lanl_7,lanl_14}). If, in the absence of
 diffraction, the X-ray spectrum of the oscillator is determined by the complex Doppler effect ($n(\omega)< 1$), then, under diffraction, the  index of refraction can become greater than unity and, consequently, the anomalous Doppler effect is possible. In this case the photon emitted by the oscillator is accompanied by the excitation of the oscillator itself. This is one of the important features of diffracted radiation of the oscillator (DRO).

So, the modification of refractive properties in periodic media
under diffraction leads to the appearance of two types of
radiation with angular distribution forming a diffraction pattern
determined by the parameters of the periodic medium. Depending on
the excited reflex, the spectrum of PXR and DRO for crystals with
lattice parameters of the order of {\AA} lies in the range of
X-ray and even higher frequencies.

It is well known that in an amorphous medium, the ordinary
Cherenkov radiation can be considered as a specific case of
radiation of the oscillator with  zero eigenfrequency
\cite{berk_10}. Similarly, in a periodic medium,  the frequency of
PXR can be expressed by (\ref{berk_1.0}) in the specific case of
$\Omega_{nf}=0$, i.e., $1-\beta n_i(\omega) \cos\theta=0$, where
$n_i(\omega)$ is the refraction index of  the crystal under
diffraction conditions.

\section{Dispersion characteristics of parametric X-ray radiation (PXR) and diffracted radiation of oscillator (DRO)}
\label{sec:berk_a}

Under diffraction the index of refraction depends on the direction
of particle motion and the frequency of radiated photon, therefore
equation (\ref{berk_1.0}) determining the radiated photon spectrum
is the equation with several solutions
\cite{chan_44,berk_13,berk_19,berk_27,berk_28}.

For example, let us consider the case of two-wave diffraction,
when the diffraction condition is fulfilled only for a reciprocal
lattice vector $\vec{\tau}$. It means that two strong waves with
wave vectors $\vec{k}$ and
$\vec{k}_{\vec{\tau}}=\vec{k}+\vec{\tau}$ are excited under
diffraction. For the simplicity of analysis of photon frequencies,
let us represent (\ref{berk_1.0}) in the form:
\begin{equation}
\label{berk_1.1} n(\omega)=\frac{\omega-\Omega_{nf}}{\omega\beta
\cos\theta},
\end{equation}

In this case the index of refraction in a crystal under diffraction conditions is characterized by two dispersion branches $n_{1,2}$. Using a well-known expression for $n_{1,2}$, one can rewrite (\ref{berk_1.1}) as follows:
\begin{eqnarray}
\label{berk_1.2}
& &\frac{1}{\beta\cos\theta}-\frac{\Omega_{nf}(1-\delta)}{\omega_B \beta\cos\theta}\\
& &=1-\frac{\omega^2_L}{4/\omega^2}\left\{(1+\beta_1)-
A\beta_1\frac{\delta}{|g_0|}\mp\sqrt{\left[(\beta_1-1)-A\beta_1\frac{\delta}{|g_0|}\right]^2+4\beta_1\frac{|g_{\tau}|^2}{g_0^2}}\right\}\nonumber
\end{eqnarray}
for the diffracted radiation of the oscillator (DRO) and
\begin{eqnarray}
\label{berk_1.3}
& &\frac{1}{\beta\cos \theta}\\
& &=1-\frac{\omega^2_L}{4/\omega^2}\left\{(1+\beta_1)-
A\beta_1\frac{\delta}{|g_0|}\mp\sqrt{\left[(\beta_1-1)^2-
A\beta_1\frac{\delta}{|g_0|}\right]^2+4\beta_1\frac{|g_{\tau}|^2}{|g_0|^2}}\right\}\nonumber
\end{eqnarray}
for parametric quasi-Cherenkov radiation (PXR), generated by a particle passing through a crystal at constant velocity.

We have introduced the following notations:
\[
\alpha=\frac{2(\vec{k}\vec{\tau})+\tau^2}{\omega^2/ c^2},
\]
is the deviation from the exact Bragg conditions,
\[
\alpha\cong\alpha_{\omega_B}+\left(\frac{\partial \alpha}{\partial \omega}\right)_{\omega_B}(\omega-\omega_B)=\frac{\tau^2c^2}{\omega_B^3}(\omega-\omega_B)=-A\delta,
\]
where
\[
A=\frac{\tau^2c^2}{\omega_B^2}>0
\]
$\delta=\omega-\omega_B/\omega_B$,  $\omega_B=\tau^2c^2/2|\vec{\tau}\vec{\beta}|$ is the Bragg frequency, corresponding to $\alpha=0$, $\beta_1=k_z/k_z+\tau_z$ is the geometry factor of diffraction asymmetry, the $z$-axis is chosen as a normal to the target surface directed inward the crystal. Let us assume that the particle with a mean velocity $\vec{u}$ moves along the $z$-axis, $\beta=u/c$, $g_0$, $g_{\tau}$ are the coefficients in a series expansion in terms of the reciprocal lattice vectors of the crystal susceptibility. For simplicity, we shall assume that a crystal is center-symmetric and absorption is neglected, $\omega_L^2=4\pi e n_0/m_c$ is the Langmuir frequency of the medium.

Let us consider in detail the conditions of the existence of
parametric X-ray radiation. Indeed, for the existence of this
radiation it is sufficient that relation (\ref{berk_1.3}) be
fulfilled at least for one of the two  indices of refraction
characterizing the crystal at a given frequency. On the left-hand
side of (\ref{berk_1.3}), there is a term greater than unity
($\beta<1$ and $|\cos \theta|< 1$). Consequently, for the
existence of the solution of equation (\ref{berk_1.3}), the
expression between the braces on the right-hand side of
(\ref{berk_1.3}) should be less than zero. Far from Bragg
conditions  $\delta\rightarrow \infty$, and we transit to a
well-known case of amorphous medium., i.e., to the  index of
refraction $n(\omega)=1-\omega^2_L/2\omega^2 <1$  for any
frequencies within the X-ray range. As a result, in this range
Cherenkov radiation is impossible. The analysis of
(\ref{berk_1.3}) near Bragg condition $|\alpha|\leq |g_0|$ shows
that the expression between the braces is always positive for one
dispersion branch corresponding to the sign $(-)$; consequently,
the fulfilment of (\ref{berk_1.3}) is impossible. For the second
branch corresponding to the sign $(+)$, this expression can be
negative at $|\alpha|\geq |g_0|$. For example, in the case of
\[
\left|(\beta_1-1)-A_{\beta_1}\frac{\delta}{|g_0|}\right|\gg 4\beta_1\left|\frac{g_{\tau}}{g_0}\right|^2,
\]
we can approximately write
\[
\frac{1}{\beta \cos\theta}\cong 1-\frac{\omega^2_L}{2\omega^2}\beta_1\left[1-A\frac{\delta}{|g_0|}\right]
\]
and, obviously, the fulfilment of the Cherenkov condition is possible for Laue diffraction case ($\beta_1> 0$) at frequencies for which
\[
\left|A\frac{\delta}{|g_0|}\right|>1.
\]
In Bragg diffraction case ($\beta_1<0$), the Cherenkov condition can be fulfilled not only for one dispersion branch but even for two branches at the degeneration point.

Comparison of (\ref{berk_1.3}) and (\ref{berk_1.2}) shows that the relation (\ref{berk_1.2}) can be satisfied for both dispersion branches at $\Omega_{nf}>0$, i.e., for radiation accompanied by the transition of the channeled particle to a lower energy level ($\varepsilon_n >\varepsilon_f$). This means that photons with two different frequencies are radiated at a given angle. In this situation we observe a normal complex Doppler effect. In this case, equation (\ref{berk_1.2}) is satisfied for radiation angles larger than the angle of parametric (quasi-Cherenkov) radiation. At the same time, the fulfilment of (\ref{berk_1.2}) leads to the strong limitation of the particle energy, the radiation angle and the value of deviation from the exact Bragg condition $\alpha$ for the radiation accompanied by the oscillator excitation $\Omega_{nf}< 0$ ($\varepsilon_n<\varepsilon_f$). In this case, according to (\ref{berk_1.2}), radiation of a photon (anomalous Doppler effect) with the wave vector directed at a smaller  angle relative to the  direction of the particle motion  than the angle of parametric (quasi-Cherenkov) radiation  is possible for one of the dispersion branches.

\begin{figure}[htp]
\centering
\epsfxsize = 8 cm \centerline{\epsfbox{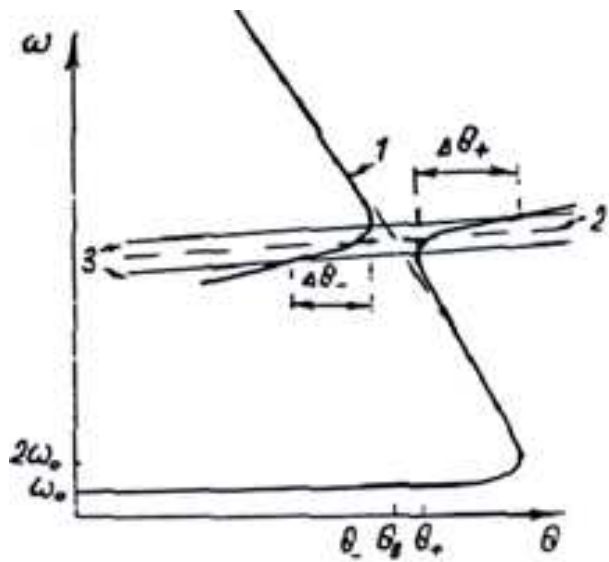}}
\caption{The dependence $\omega=\omega(\theta)$ for the case of
symmetric diffraction $\beta_1=1$} \label{berkley Figure 1}
\end{figure}

Equation (\ref{berk_1.2}) can be analytically treated for a specific case when
\begin{equation}
\label{berk_1.4}
2\frac{\omega^2_L}{\Omega^2}\phi(\omega,\theta)(1-\beta\cos\theta)\ll 1,
\end{equation}
where
\begin{equation}
\label{berk_1.5}
\phi(\omega,\theta)=\frac{1}{2}\left\{1+\beta_1-
A\beta_1\frac{\delta}{|g_0|}\mp\sqrt{\left(\beta_1-1-A\beta_1\frac{\delta}{|g_0|}\right)^2
+4\beta_1\left|\frac{g_{\tau}}{g_0}\right|^2}\right\}
\end{equation}
Let us represent (\ref{berk_1.2}) in the form:
\begin{equation}
\label{berk_1.6}
\omega=\frac{\Omega_{nf}}{2(1-\beta \cos\theta)}\left(1\pm\sqrt{1-2\frac{\omega^2_L}{\Omega^2}\phi(\omega,\theta)(1-\beta\cos\theta)}\right)
\end{equation}
In view of (\ref{berk_1.4}), (\ref{berk_1.6}) splits into two independent equations corresponding to the upper and the lower radiation branches in the absence of diffraction.
\begin{equation}
\label{berk_1.7}
\omega_I=\frac{\Omega_{nf}}{1-\beta\cos\theta}-\frac{\omega^2_L}{2\Omega_{nf}}\phi(\omega,\theta),
\end{equation}
\begin{equation}
\label{berk_1.8}
\omega_{II}=\frac{\omega^2_L}{2\Omega_{nf}}\phi(\omega,\theta).
\end{equation}
Neglecting the dependence of $\beta_1$ on $\omega$, one can obtain the frequency solutions of  (\ref{berk_1.7}) and (\ref{berk_1.8}) as a function of a radiation angle:
\begin{eqnarray}
\label{berk_1.9}
\omega_{I\pm}&=&\omega_m-\omega_0\left[\frac{1+\beta_1}{2}-\frac{x}{2}\left\{1+\frac{\omega_m-\omega_B}{\omega_0}\right.\right.\\
& & \left.\left.\pm\sqrt{\left(\frac{1-\beta_1}{x}+1-\frac{\omega_m-\omega_B}{\omega_0}\right)^2-4\left|\frac{g_{\tau}}{g_0}\right|^2\frac{1}{x}}\right]
(1-x)^{-1}\right.\nonumber,
\end{eqnarray}

\begin{eqnarray}
\label{berk_1.10}
\omega_{II\pm}&=&\omega_0\left[\frac{1+\beta_1}{2}+\frac{x}{2}\left\{1+\frac{\omega_B}{\omega_0}\right.\right.\\
& & \left.\left.\pm\sqrt{\left(\frac{1-\beta_1}{x}+1-\frac{\omega_B}{\omega_0}\right)^2+4\left|\frac{g_{\tau}}{g_0}\right|^2\frac{1}{x}}\right]
(1+x)^{-1}\right.\nonumber,
\end{eqnarray}
where

\[
\omega_0=\frac{\omega^2_L}{2\Omega},\quad \omega_m=\frac{\Omega}{1-\beta\cos\theta};\quad _B=\frac{\tau^2}{2}\left\{|\vec{\tau}_{\parallel}|\cos\theta-|\vec{\tau}_{\perp}|\sin\theta\cos\varphi\right\}^{-1}
\]
\[
x=\frac{\beta_1}{\Omega}\left(|\vec{\tau}_{\parallel}|\cos\theta-|\vec{\tau}_{\perp}|\sin\theta\cos\varphi\right),
\]
$\vec{\tau}_{\parallel}$ and $\vec{\tau}_{\perp}$ are the
projections of the reciprocal lattice vector onto the direction of
the particle mean velocity and onto the plane perpendicular to
particle velocity, respectively; $\varphi$ is the angle between
$\vec{\tau}_{\perp}$ and $\vec{k}_{\perp}$. The dependence
$\omega=\omega(\theta)$ for the case of symmetric diffraction
$\beta_1=1$ is shown in
Figure \ref{berkley Figure 1}

\begin{figure}[htp]
\centering
\epsfxsize = 8 cm \centerline{\epsfbox{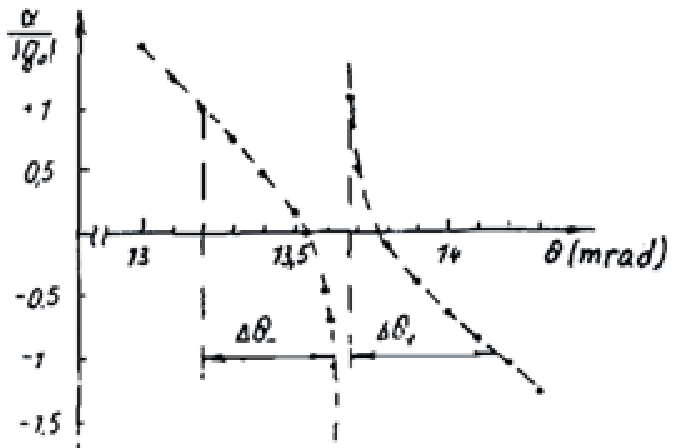}}
\caption{} \label{berkley Figure 2}
\end{figure}

According to Figure \ref{berkley Figure 1} and  (\ref{berk_1.9})
and (\ref{berk_1.10}), the spectrum of radiated photons
essentially modifies under diffraction. In the absence of
diffraction, each radiation branch splits  into two subbranches.
Thus, diffraction results in the excitation of an additional
branch in the complex Doppler effect with the frequency close to
the Bragg frequency $\omega_B$, and in the formation of a
radiation non-transparency range in the angular distribution
$\Delta\theta=\theta_+-\theta_-=10^{-5}\div 10^{-6}$ rad. When the
angle $\theta$ changes, the solution is first realized for one
dispersion branch and then for the other. Figure  \ref{berkley
Figure 1} shows the angular range in which $|\alpha|\leq|g_0|$. It
should be noted that the radiation frequency of the additional
diffraction branch changes a little with the change in the
radiation angle. As a result, the angular range, in which
$|\alpha|\leq|g_0|$, may considerably exceed the ordinary angular
interval, characterizing diffraction of an X-ray external
monochromatic wave, when $\Delta\theta$ is of the order of several
angular minutes (in Figure \ref{berkley Figure 1}, this is the
angular interval $\Delta\theta=\Delta\theta_++\Delta\theta_-$). As
it is rather complicated to obtain the analytical solution,  the
numerical calculation of the dependence of $|\alpha|/|g_0|$ on the
radiation angle $\theta$ near the Bragg angle $\theta_B$ was made
for the oscillator moving along the crystal direction
$\langle110\rangle$ and photon diffraction by crystallographic
planes $(400)$ in $Si$. According to these calculations, the
magnitude of $\Delta\theta$ weakly depends on the energy and
eigenfrequency of the oscillator and retains within the interval
$10^{-4}-10^{-3}$. In Figure \ref{berkley Figure 2}, the magnitude
of $|\alpha|/|g_0|$ is shown as a function of the angle $\theta$
at the following parameters of the oscillator: $\Omega=1$~ eV,
${\gamma=2\cdot 10^3}$. As one can see, $\Delta\theta_-\approx
4\cdot 10^{-3}$ rad, $\Delta\theta_+ =5.2 \cdot 10^{-3}$ rad and
the total interval $|\alpha|\leq|g_0|$, $\Delta\theta=9.2 \cdot
10^{-3}$ rad.

The values of the angles $\theta_{\pm}$ at $\alpha=0$ are determined from the equation
\begin{equation}
\label{berk_1.11}
\omega_{B}=\omega_m-\omega_0\left(1\pm\frac{g_{\tau}}{g_0}\right).
\end{equation}
A similar picture is also observed for the second branch (see (\ref{berk_1.10})). So, if $\omega_0\neq \omega_B$, and  the second term in the radicand of (\ref{berk_1.10}) is negligible, then, for  example, $\omega_{II+}=\omega_B$ and $\omega_{II-}=\omega_0$ at $\omega_0<\omega_B$. This means that, as in the previous case, we have the excitation of the wave with the frequency close to the Bragg frequency in addition to the solution far from diffraction at a given radiation angle $\theta$. As the calculation shows, the magnitude of $\alpha$ does not depend on the frequency of the diffracted wave and keeps practically constant, being determined by the oscillator eigenfrequency within the whole interval of radiation angles corresponding to the fulfilment of the condition (\ref{berk_1.4}). Although the parameter $\alpha$ depends on the eigenfrequency of the oscillator $\Omega$ and equals zero at
\begin{equation}
\label{berk_1.12}
\Omega_{\pm}=\frac{\omega^2_L}{2\omega_B}\left(1\pm\frac{g_{\tau}}{g_0}\right),
\end{equation}
it remains small ($|\alpha|\leq|g_0|$) within rather a wide interval of  eigenfrequencies. For example, in the case considered above $\Omega_{\pm}=0.092$ eV and $|\alpha|\leq|g_0|$ for the interval $\Omega_{+\mathrm{min}}=0.08$ eV
and $\Omega_{+\mathrm{max}}=0.15$ eV.

It should be noted that the equality (\ref{berk_1.10}) also has the solution for the negative eigenfrequency of the oscillator (at the frequency close to $\omega_B$) that corresponds to the anomalous Doppler effect, i.e., the radiation of the oscillator is accompanied by its excitation. Such a process is possible because under diffraction condition the  index of refraction can be greater than unity.

The analysis of the dispersion expression for radiation propagating at a large angle and for arbitrary geometry was made in \cite{berk_29}. Here we discuss the spectra of DRO and PXR only in the two-wave diffraction case. However, due to crystal symmetry, the diffraction condition can be satisfied for many waves, that is,  the case of multi-wave diffraction can be realized. In this case several indices of refraction $n_i(\omega)$ corresponding to different dispersion branches can be greater than unity. It appears that the possibility of new effects in radiation, such as the effect of excitation of radiation in a roundabout way takes place. This means that the PXR intensity in a diffraction peak may differ from zero even in the case when a given reflection is forbidden because of  the lattice symmetry. The particular properties in the angular distribution of radiation are observed in the vicinity of the point of degeneration of dispersion branches.

\subsection{General expression for spectral-angular distribution of radiation generated by a particle in a photonic crystal}
\label{berk_b1}
Both the spectral-angular density of radiation energy per unit solid angle $W_{\vec{n}\omega}$ and the differential number of emitted photons
  $dN_{\vec{n}\omega}\omega=1/\hbar\omega\cdot W_{\vec{n}\omega}$ can be easily obtained if the field $\vec{E}(\vec{r},\omega)$ produced by a particle at a large distance $\vec{r}$ from the crystal is known \cite{berk_2}
\begin{equation}
\label{berk_2.1}
W_{\vec{n}\omega}=\frac{er^2}{4\pi^2}\overline{\left|\vec{E}(\vec{r},\omega)\right|^2},
\end{equation}
The vinculum here means averaging over all possible states of the radiating system. In order to obtain $\vec{E}(\vec{r},\omega)$, Maxwell's equation describing the interaction of particles with the medium should be solved. The transverse solution can be found with the help of Green's function of this equation, which satisfies the expression:
\begin{equation}
\label{berk_2.2}
G=G_0+G_0\frac{\omega^2}{4\pi c^2}(\hat{\varepsilon}-1)G,
\end{equation}
$G_0$ is the transverse Green's function of Maxwell's equation at $\hat{\varepsilon}=1$. It is given, for example, in \cite{66}.

Using $G$, we can find the field we are concerned with
\begin{equation}
\label{berk_2.3}
E_n(\vec{r},\omega)=\int G_{ne}(\vec{r},\vec{r}^{\prime},\omega)\frac{i\omega}{c^2}j_{0e}(\vec{r},\omega)d^3 r^{\prime},
\end{equation}
where $n, e=x, y, z, j_{0e}(\vec{r},\omega)$ is the Fourier transformation of the e-th component of the current produced by a moving beam of charged particles (in the linear field approximation, the current is determined by the velocity and the trajectory of a particle, which are obtained from the equation of particle motion in the external field, by neglecting the influence of the radiation field on the particle motion). Under the quantum-mechanical consideration the current $j_0$ should be considered as the current of transition of the particle-medium system  from one state to another.

According to \cite{berk_2,vfel_PXRbook}, Green's function is expressed at $r\rightarrow \infty$ through the solution of homogeneous Maxwell's equations $E_n^{(-)}(\vec{r},\omega)$ containing incoming spherical waves:
\begin{eqnarray}
\label{berk_2.4}
& &\lim G_{ne}(\vec{r},\vec{r}^{\prime},\omega)=\frac{e^{ikr}}{r}\sum\limits_S e^s_n E^{(-)s*}_{\vec{k}e}(\vec{r}^{\prime},\omega),\\
& &r\rightarrow \infty\nonumber
\end{eqnarray}
where $\vec{e}^s$ is the unit polarization vector, $s-1,2$, $\vec{e}^1\perp\vec{e}^2\perp\vec{k}$.

If the electromagnetic wave is incident on a crystal of finite size, then at $r\rightarrow \infty$
\[
\vec{E}_k^{s(-)}(\vec{r}, \omega)=\vec{e}^s e^{i\vec{k}\vec{r}}+\mbox{const}\frac{e^{ikr}}{r},
\]
and one can show that the relation between the solution $\vec{E}_k^{s(-)}$ and the solution of Maxwell's equation $\vec{E}^{(+)}(\vec{k}, \omega)$ describing scattering of  a plane wave by the target (crystal), is given by:
\begin{equation}
\label{berk_2.5}
\vec{E}^{s(-)*}_{\vec{k}}=\vec{E}^{s(+)}_{-\vec{k}}
\end{equation}

Using (\ref{berk_2.3}), we obtain

\begin{equation}
\label{berk_2.6}
E_n(\vec{r}, \omega)=\frac{e^{ikr}}{r}\frac{i\omega}{c^2}
\sum\limits_S e^s_n\int E^{s(-)*}_{\vec{k}}(\vec{r}, \omega)\vec{j}_0(\vec{r}^{\prime}, \omega)d^3 r^{\prime}.
\end{equation}
As a result, the spectral energy density of photons with polarization $s$ can be written in the form:
\begin{equation}
\label{berk_2.7}
W_{\vec{n},\omega}^s=\frac{\omega^2}{4\pi^2c^2}\overline{\left|\int\vec{E}^{s(-)*}_{\vec{k}}(\vec{r}, \omega)
\vec{j}_0(\vec{r}, \omega)d^3 r\right|^2},
\end{equation}
\begin{equation}
\label{berk_2.8}
\vec{j}_0(\vec{r}, \omega)=\int e^{i\omega t}\vec{j}_0(\vec{r}, \omega) dt=eQ \int e^{i\omega t}\vec{v}(t)\delta(\vec{r}-\vec{r}(t))dt,
\end{equation}
where $eQ$ is the charge of the particle, $\vec{v}(t)$ and $\vec{r}(t)$ are the velocity and the trajectory of the particle at moment $t$. By introducing (\ref{berk_2.8}) into (\ref{berk_2.7}) we get

\begin{equation}
\label{berk_2.9}
dN^s_{\vec{n}, \omega}=\frac{e^2 Q^2\omega}{4\pi^2 \hbar c^3}\overline{\left|\int \vec{E}^{(-)s*}_{\vec{k}}(\vec{r}(t), \omega)\vec{v}(t) e^{i\omega t}d\right|^2} t.
\end{equation}
Integration in (\ref{berk_2.9}) is carried out over the whole interval of the particle motion. It should be noted that the application of the solution of a homogeneous Maxwell's equation  instead of the inhomogeneous one essentially simplifies the analysis of the radiation problem and enables one to consider various cases of radiation emission taking into account multiple scattering.


\section{Parametric X-ray radiation (PXR)}
\label{sec:para_1}

Using equations (\ref{berk_2.7})--(\ref{berk_2.9}), one can easily obtain the explicit expression for the radiation intensity and that for the effect of multiple scattering on the process under study \cite{Nuclear_optics,berk_2,vfel_PXRbook}.

Consider, for example, the PXR radiation. Let a particle moving
with a uniform velocity be incident on a crystal plate with the
thickness $L$ being $L\ll L_c$, where $L_c=(\omega q)^{-1/2}$ is
the coherent length of bremsstrahlung $q=\overline{\theta}^2/4$
and $\overline{\theta}^2$ is the mean square angle of multiple
scattering. The latter requirement allows neglecting the multiple
scattering of particles by atoms. A theoretical method describing
multiple scattering affect on the {radiation process} is given in
\cite{para_4}.

According to (\ref{berk_2.9}), in order to determine the number of
quanta emitted by a particle passing through the crystal plate,
one should first find the explicit expressions for the solutions
$\vec{E}^{(-)s}_{\vec{k}}$. As was mentioned above, the field
$\vec{E}^{(-)s}_{\vec{k}}$ can be found from the relation
$\vec{E}^{(-)s}_{\vec{k}}=(\vec{E}^{(+)s}_{-\vec{k}})^*$ if one
knows the solution $\vec{E}^{(+)s}_{\vec{k}}$ describing the
photon scattering by the crystal.

In the case of two strong waves excited under diffraction (the
so-called two-beam diffraction case \cite{132}), one can
obtain the following set of equations for determining the wave
amplitudes (see \cite{lanl_7a}):
\begin{eqnarray}
\label{para_1.15}
\left(\frac{k^2}{\omega^2}-1-\chi^*_0\right)\vec{E}^{(-)s}_{\vec{k}}c_s\chi^*_{-\vec{\tau}}\vec{E}^{(-)s}_{\vec{k}_{\tau}}=0\nonumber\\
\left(\frac{k^2}{\omega^2}-1-\chi^*_0\right)\vec{E}^{(-)s}_{\vec{k}_{\tau}}c_s\chi^*_{\vec{\tau}}\vec{E}^{(-)s}_{\vec{k}}=0.
\end{eqnarray}
Here $\vec{k}_{\vec{\tau}}=\vec{k}+\vec{\tau}$, $\vec{\tau}$ is
the reciprocal lattice vector, $\chi_0$, $\chi_{\vec{\tau}}$ are
the Fourier components of the crystal susceptibility. It is well
known that the crystal is described by a periodic susceptibility
(see, for example, \cite{132}:
\begin{equation}
\label{para_1.16}
\chi(\vec{r})=\sum_{\vec{\tau}}\chi_{\vec{\tau}}\exp(i\vec{\tau}\vec{r}).
\end{equation}
$c_s=\vec{e}^s\vec{e}^s_{\vec{\tau}}$,
$\vec{e}^s(\vec{e}^s_{\vec{\tau}})$ are the unit polarization
vectors of the incident and diffracted waves, respectively.

The condition for the linear system (\ref{para_1.15}) to be
solvable leads to a dispersion equation that determines the
possible wave vectors $\vec{k}$ in a crystal. These wave vectors
are convenient to present in the form:
\[
\vec{k}_{\mu s}=\vec{k}+\vec{\kappa}^*_{\mu s}\vec{N},\qquad
\kappa_{\mu s}^*=\frac{\omega}{c\gamma_0}\varepsilon^*_{\mu s},
\]
where $\mu=1,2$; $\vec{N}$ is the unit vector of a normal to the
entrance crystal surface which is directed into the crystal,
\begin{eqnarray}
\label{para_1.17} &
&\varepsilon_{1(2)s}=\frac{1}{4}\left[(1+\beta_1)\chi_0-\beta_1\alpha_B\right]
\pm\frac{1}{4}\left\{\left[(1-\beta_1)\chi_0+\beta_1\alpha_B\right]^2\right.\nonumber\\
&
&\left.+4\beta_1C_s^2\chi_{\vec{\tau}}\chi_{\vec{-\tau}}\right\}^{-1/2}.
\end{eqnarray}
$\alpha_B=(2\vec{k}\vec{\tau}+\tau^2)k^{-2} $ is the off-Bragg
parameter ($\alpha_B=0$ if the exact Bragg condition of
diffraction is fulfilled),
\[
\gamma_0=\vec{n}_{\gamma}\cdot\vec{N},\quad
\vec{n}_{\gamma}=\frac{\vec{k}}{k},\quad
\beta_1=\frac{\gamma_0}{\gamma_1}, \quad
\gamma_1=\vec{n}_{\gamma\tau}\cdot\vec{N},\quad
\vec{n}_{\gamma\tau}=\frac{\vec{k}+\vec{\tau}}{|\vec{k}+\vec{\tau}|}.
\]
The general solution of (\ref{para_1.15}) inside
a crystal is:
\begin{equation}
\label{para_1.18}
\vec{E}^{(-)s}_{\vec{k}}(\vec{r})=\sum\limits^2_{\mu=1}\left[\vec{e}^s
A_{\mu}\exp(i\vec{k}_{\mu s}\vec{r})+
\vec{e}^s_{\tau}A_{\tau\mu}\exp(i\vec{k}_{\mu
s\tau}\vec{r})\right].
\end{equation}
By matching these solutions with the solutions of Maxwell's
equations for the vacuum area, one can find the explicit form of
$\vec{E}^{(-)s}_{\vec{k}}(\vec{r})$ throughout the space. It is
possible to discriminate several types of diffraction geometries,
namely, the Laue (a) and the Bragg (b) schemes are most well
known.

(a) Let us consider the PXR in the Laue case.

In this case, the electromagnetic waves emitted by a
particle in both the forward and the diffracted directions leave the
crystal through the same surface ($k_z>0, k_z+\tau_z>0$), the
$z$-axis is parallel to the normal $N$ (where $N$ is the normal to
the crystal surface being directed inside a crystal). By matching
the solutions of Maxwell's equations on the crystal surfaces with
the help of (\ref{para_1.15}), (\ref{para_1.17}), (\ref{para_1.18}),
one can obtain the following expressions for the Laue case:
\begin{eqnarray}
\label{para_1.19}
\vec{E}^{(-)s}_{\vec{k}}=\left\{\vec{e}^s\left[-\sum_{\mu=1}^2\xi_{\mu s}^{0*}e^{-i\frac{\omega}{\gamma_0}\varepsilon^*_{\mu s}L}\right]e^{i\vec{k}\vec{r}}\right.\nonumber\\
\left.+e^s_{\vec{\tau}}\beta_1\left[\sum_{\mu=1}^2\xi_{\mu s}^{\tau*}e^{-i\frac{\omega}{\gamma_0}\varepsilon^*_{\mu s}L}\right]e^{i\vec{k}_{\tau}\vec{r}}\right\}\theta(-z)\nonumber\\
+\left\{\vec{e}^s\left[-\sum_{\mu=1}^2\xi_{\mu s}^{0*}e^{-i\frac{\omega}{\gamma_0}\varepsilon^*_{\mu s}(L-z)}\right]e^{i\vec{k}\vec{r}}\right.\nonumber\\
\left.+e^s_{\vec{\tau}}\beta_1\left[\sum_{\mu=1}^2\xi_{\mu s}^{\tau*}e^{-i\frac{\omega}{\gamma_0}\varepsilon^*_{\mu s}(L-z)}\right]e^{i\vec{k}_{\tau}\vec{r}}\right\}\nonumber\\
\times \theta(L-z)\theta(z)+\vec{e}^s
e^{i\vec{k}\vec{r}}\theta(z-L),
\end{eqnarray}
where
\begin{eqnarray}
\label{para_1.20}
\xi^0_{1,2 s}=\mp\frac{2\varepsilon_{2,1 s}-\chi_0}{2(\varepsilon_{2s}-\varepsilon_{1s})};\nonumber\\
\xi^{\tau}_{1,2
s}=\mp\frac{c_s\chi_{-\tau}}{2(\varepsilon_{2s}-\varepsilon_{1s})}.
\end{eqnarray}
$\theta(z)=1$ if $z\geq 0$ and $\theta(z)=0$ if $z<0$.

Substitution of (\ref{para_1.19}) into (\ref{berk_2.9}) gives for
the Laue case the differential number of quanta of the forward
directed parametric X-rays with the polarization vector
$\vec{e}_s$:
\begin{eqnarray}
\label{para_1.21} & &\frac{d^2N^L_{0s}}{d\omega
d\Omega}=\frac{e^2Q^2\omega}{4\pi^2\hbar c^3}(\vec{e}^s\vec{v})^2
\left|\sum_{\mu=1,2}\xi_{\mu s}^0
e^{i\frac{\omega}{c\gamma_0}\varepsilon_{\mu
s}L}\left[\frac{1}{\omega-\vec{k}\vec{v}}
-\frac{1}{\omega-\vec{k}^*_{\mu s}\vec{v}}\right]\right.\nonumber\\
& &\left.\times[e^{i(\omega-\vec{k}^*_{\mu
s}\vec{v})T}-1]\right|^2,
\end{eqnarray}
where $T=L/c\gamma_0$ is the particle time of flight;
$\vec{e}_1\parallel [\vec{k}\vec{\tau}]$;
$\vec{e}_2\parallel[\vec{k}\vec{e}_1]$.

 One can see that formula (\ref{para_1.21}) looks  like the formula which describes the spectral and angular distribution of the Cherenkov and transition radiations in the matter with the  index of refraction $n_{\mu s}=k_{z\mu s}/k_z=1+\kappa_{\mu s}/k_z$.

The spectral angular distribution for photons in the diffraction
direction $\vec{k}_{\tau}=\vec{k}+\vec{\tau}$ can be obtained from
(\ref{para_1.21}) by a simple substitution
\begin{eqnarray*}
& &\vec{e}_s\rightarrow\vec{e}_{s\tau},\qquad \xi^0_{\mu s}\rightarrow \beta_1\xi_{\mu s}^{\tau},\\
& &\xi^{\tau}_{1(2)s}=\pm\frac{\chi_{\tau}c_s}{2(\varepsilon_{1s}-\varepsilon_{2s})}\\
& &\vec{k}\rightarrow\vec{k}_{\tau}, \quad\vec{k}_{\mu
s}\rightarrow\vec{k}_{\tau\mu s}=\vec{k}_{\mu s}+{\tau}.
\end{eqnarray*}

(b) Now let us consider PXR in the Bragg case. In this case, side
by side with the electromagnetic wave emitted in the forward
direction, the electromagnetic wave emitted by a charged particle
in the diffracted direction and leaving the crystal through the
surface of the particle entrance can be observed. By matching the
solutions of Maxwell's equations on the crystal surface with the
help of (\ref{para_1.15}), (\ref{para_1.17}), (\ref{para_1.18}), one
can get the formulas for the Bragg diffraction schemes.

It is interesting that the spectral angular distribution for
photons emitted in the forward direction can be obtained from
(\ref{para_1.21}) by the following substitution, $\xi^0_{\mu
s}\rightarrow \gamma_{\mu s}$,
\begin{eqnarray}
\label{para_1.22} &
&\gamma^0_{1(2)s}=\left[2\varepsilon_{2(1)s}-\chi_0\right]\left[(2\varepsilon_{2(1)s}-\chi_0)
-(2\varepsilon_{1(2)s}-\chi_0)\right.\nonumber\\
&
&\left.\times\exp\left[i\frac{\omega}{\gamma_0}(\varepsilon_{2(1)s}-\varepsilon_{1(2)s})L\right]\right]^{-1}
\end{eqnarray}

The spectral angular distribution of photons emitted in the
diffracted direction can be obtained from (\ref{para_1.21}) by
substitution
\begin{eqnarray*}
& &\vec{e}_s\rightarrow\vec{e}_{s\tau},\quad \vec{k}\rightarrow\vec{k}_{\tau},\quad k_{\mu s}\rightarrow\vec{k}_{\mu\tau s},\\
& & \xi^0_{\mu
s}\exp\left[i\frac{\omega}{\gamma_0}\varepsilon_{\mu
s}L\right]\rightarrow \gamma^{\tau}_{\mu s},
\end{eqnarray*}
where
\begin{eqnarray*}
& &\gamma^{\tau}_{1(2)s}=-\beta_1[c_s\chi_{\tau}]\left[(2\varepsilon_{2(1)s}-\chi_0)-(2\varepsilon_{1(2)s}-\chi_0)\right.\\
& &
\left.\times\exp\left[i\frac{\omega}{\gamma_0}(\varepsilon_{2(1)s}-\varepsilon_{1(2)s})L\right]\right]^{-1}.
\end{eqnarray*}


 The angular
distribution for the photons emitted at large angles in the the Bragg case was
derived in \cite{para_4}. From (\ref{para_1.21}), (\ref{para_1.22})
we can obtain the angular distribution for the photons emitted in
the forward direction:
\begin{eqnarray}
\label{para_1.27}
& &dN_{0s}^B=\frac{e^2Q^2}{4\hbar c}|\beta_1||r_s|^2\nonumber\\
&
&\times\left|\left\{(\gamma^{-2}+\vartheta^2-\chi_0)^2-|\beta_1|r_s\right.\right.
\left.\left.\exp\left[-i\frac{\omega_B}{2\gamma_0c}\frac{(\gamma^{-2}+\vartheta^2-\chi_0)^2
-|\beta_1|r_s}{\gamma^{-2}+\vartheta^2-\chi_0}L\right]\right\}\right|^{-2}\nonumber\\
&
&\times\left|\frac{(\gamma^{-2}+\vartheta^2-\chi_0)^2-|\beta_1|r_s}{(\gamma^{-2}+\vartheta^2-\chi_0)^2}\right|\frac{\omega_B
T}{\sin^2\vartheta_B}\vartheta^3 d\vartheta.
\end{eqnarray}

According to (\ref{para_1.27}), the PXR angular distribution for
this case oscillates as a function of $\vartheta$, $L$,
$\omega_B$. If $\vartheta^2\gg \gamma^{-2}$, $\chi_0$, the
oscillation period is $\vartheta_{0s}=\sqrt{c/\omega_B L_0}$. For
$k_B=(\omega_B/c)=10^9$ ~cm$^{-1}$, $L_0=L/\gamma_0=10^{-2}$ ~cm,
we have $\vartheta_{0s}=3\times 10^{-4}$.

For low energy electrons, oscillations in $N_{0s}^B$ disappear.  Let us note
that the PXR photon number is proportional to $Q^2$. As a result,
the PXR intensity is very high for heavy nuclei.

For example, for $Pb$ the photon number may be 1 per nucleus for
$L=1$~ cm. It can be used for the detection of particles and for
the precise measurement of their energy.


\section{Surface parametric X-ray (quasi-Cherenkov) radiation (SPXR) and DRO}
\label{sec:berk_c}

When a particle travels in a vacuum near the surface of a
spatially periodic medium, new kinds of radiation arise
\cite{berk_104,berk_105} -- surface parametric (quasi-Cherenkov)
X-ray radiation (SPXR) and surface DRO (see Figure \ref{berkley
Figure 30}). This phenomenon takes place under the condition of
uncoplanar  surface diffraction, first considered in \cite{147}.

\begin{figure}[htp]
\centering
\epsfxsize = 8 cm \centerline{\epsfbox{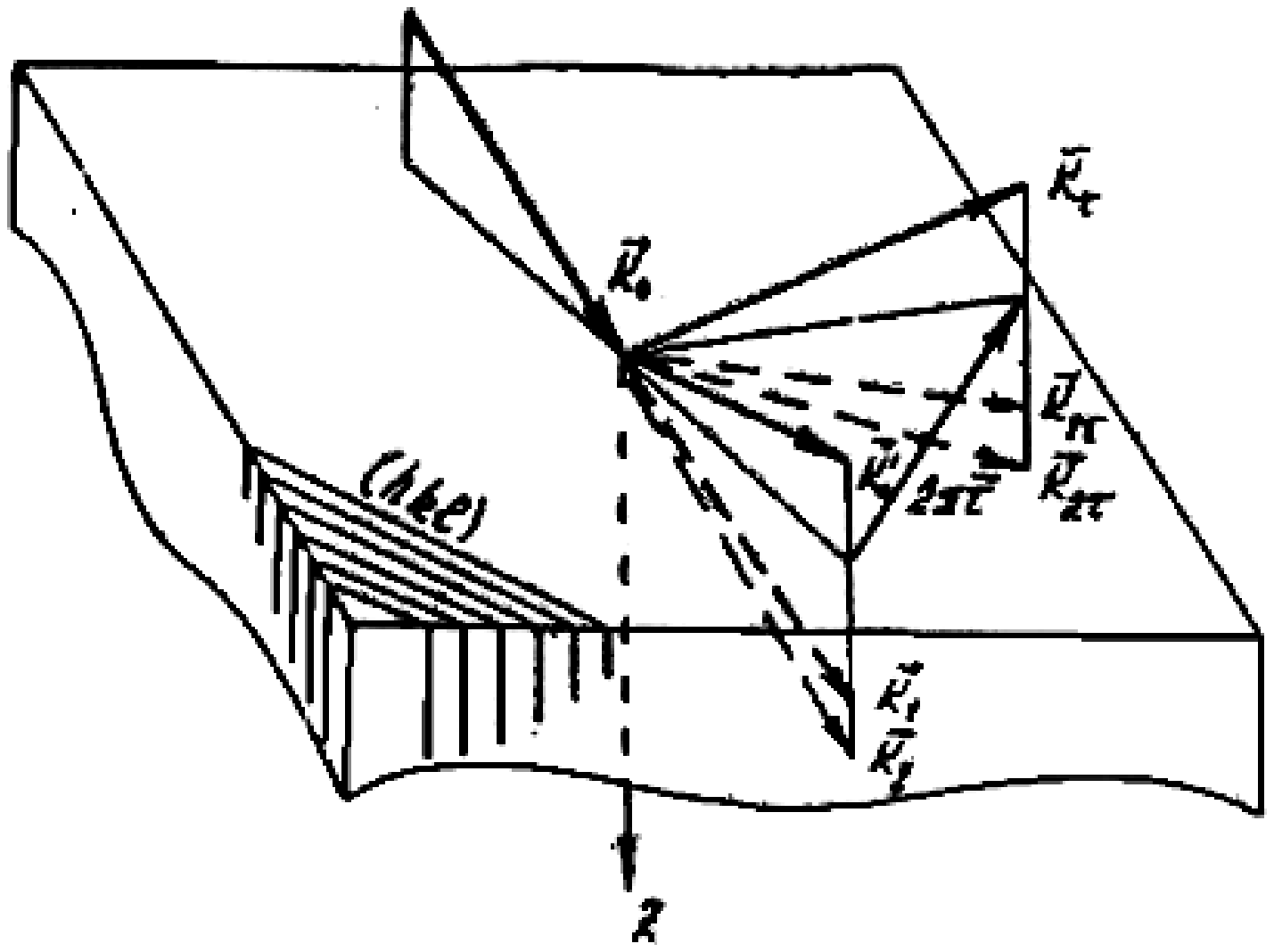}}
\caption{} \label{berkley Figure 30}
\end{figure}

The solution of Maxwell's equation $\vec{E}_{\vec{k}}^{(+)}(\vec{r})$ in this case of uncoplanar  surface diffraction was obtained in \cite{147}. It was shown that the surface diffraction in the two-wave case is characterized by two angles of total reflection (several angles in the case  of multi-wave diffraction \cite{berk_107}). The solution obtained in \cite{berk_107} contains the component, which describes the state that damps with growing distance from the surface of the medium, both within the material and in the vacuum, and which describes a surface wave, i.e., a wave in which the energy flux is directed along the boundary of the surface of a spatially periodic target (see review \cite{berk_108}). According to \cite{147}, this solution, which describes scattering of a plane wave by the target under the surface diffraction geometry, can be written in the form:
\begin{equation}
\label{berk_2.63}
\vec{E}^{(+)s}_{\vec{k}}=e_s e^{i\vec{k}\vec{r}}+A_s(\vec{k}, \omega)e^{i\vec{k}_1\vec{r}}+B_s(\vec{k},\omega)e^{i\vec{k}_2\vec{r}},
\end{equation}
where the wave vector in a vacuum $\vec{k}=(\vec{k}_t, \vec{k}_{\perp})$, $\vec{k}_1=(\vec{k}_t- \vec{k}_{\perp})|\vec{k}_{2\perp}|=\sqrt{k^2-k^2_{2t}}$, $\vec{k}_2=(\vec{k}_{2t}- \vec{k}_{2\perp})$, $\vec{k}_{2t}=\vec{k}_t+ 2\pi\vec{\tau}$, $\vec{k}_t$ is the component of the wave vector that is parallel  to the surface, $\vec{\tau}$ is the reciprocal lattice vector, $\omega$ is the photon frequency. The amplitudes $A_s$ and $B_s$ are  given in \cite{201,berk_105}. Substituting the solution $\vec{E}^{(-)s}_{\vec{k}}=(\vec{E}^{(+)s}_{-\vec{k}})^*$  into (\ref{berk_2.3}), we can find the spectral-angular distribution of SPXR and DRO. For example, in the case of DRO, the differential number of emitted photons for a particle moving parallel to the crystal surface is  \cite{berk_104}
\begin{equation}
\label{berk_2.64}
\frac{d^2 N_s}{d\omega d\Omega}=\frac{e^2\omega T}{2\pi\hbar c^3}\left|\vec{u}\vec{B}_s(\vec{k}, \omega)\right|^2\delta(\vec{k}_t\vec{u}+2\pi\vec{\tau}\vec{u}-\omega)e^{-2\texttt{Im}k_{2\perp}|z_0|}.
\end{equation}
Here we assume that a particle  moves parallel to the target surface at a distance $Z_0$ at a constant velocity $\vec{u}$; $T$  is the flight time. The argument of $\delta$-function in (\ref{berk_2.64}) is equal to zero for the frequencies
\[
\omega_u=\frac{|2\pi\vec{\tau}\vec{u}|}{1-\vec{n}_t\vec{u}/c},
\]
where $\vec{n}_t$ is the component of the unit vector in the direction of $\vec{k}$, which is parallel to the surface.

After integrating (\ref{berk_2.64}) over the frequencies, the angular distribution of radiation takes the form:
\begin{equation}
\label{berk_2.65}
\frac{d^2 N_s}{d\Omega}=\frac{e^2\omega_u T}{2\pi\hbar c^3}\left|\vec{u}\vec{B}_s(\vec{k}, \omega_u)\right|^2 e^{-2\texttt{Im}k_{2\perp}|z_0|}.
\end{equation}
The spectral-angular distribution of SPXR generated by a particle incident on the crystal at a small angle relative to the crystal surface was obtained in \cite{berk_109,andr,andr1,berk_112}.

It should be noted that, according to the analysis made in \cite{berk_104}, the formation of circularly polarized quanta  under the surface diffraction is possible and, as a consequence, such circularly polarized quanta  can be produced in the SPXR process \cite{berk_112}.


Let now an oscillator with the vibration frequency $\Omega$ in the laboratory system move along the surface. In this case we also obtain the expression similar to equation (\ref{berk_2.64}) with the following replacements made in the exponent: $\omega$ by $\omega\pm\Omega$ and $v$ by the velocity amplitude at frequency $\Omega$. As a result, the integrals appearing in equation (\ref{berk_2.64}) will give $\delta$-functions of the form $\delta(\vec{k}_t\vec{v}-\omega\pm\Omega)$, $\delta(\vec{k}_{2t}\vec{v}-\omega\pm\Omega)$. Now all three integrals are non-zero. The first one describes the normal Doppler effect, the second one, the same effect including the influence of a mirror reflected wave  on radiation. Of greatest interest is the third integral. In this case the $\delta$-function leads to the equality $\omega=(2\pi\tau_t\vec{v}\pm\Omega)(1-\vec{n}_t\vec{v}/c)^{-1}$. If $2\pi\tau_T\vec{v}>\Omega$, then both signs of the frequency $\Omega$ are allowed for radiation. From the quantum viewpoint, one sign corresponds to the emission of quantum by the oscillator (or atom) when it drops to a lower energy level, and the opposite sign corresponds to the inverse process: of the quantum  emission when the oscillator (or atom) rises to a higher energy level.
In other words, the phenomenon of surface diffraction results in the appearance of the vacuum anomalous Doppler effect  \cite{berk_104}.

Let now a beam of charged particles (or oscillators) move along the surface  of a natural or a photonic crystal in a vacuum. The phenomenon of spontaneous radiation causes the beam's instability relative to the photon emission and the formation of the charge density wave in the beam. The processes considered above also leads to the appearance of such instability. According to \cite{berk_104},  multi-wave  diffraction of emitted photons in this case also leads to a different reduction of the generation threshold similar to that appearing when a beam passes through a crystal \cite{berk_131,berk_132,berk_133,berk_134,berk_135,dan_8}.

Due to such instability the beam radiates photons collectively and
its longitudinal energy decreases. The presence of the external
field (in our case - excitation of surface diffraction by the
external field) can accelerate the beam. Note that the instability
studied in
\cite{berk_131,berk_132,berk_133,berk_134,berk_135,dan_8} is a
particular case of instabilities caused by the processes of the
emission of waves (instability may be caused by, for example, the
parametric process of the pump wave splitting into two waves, the
Mandelstam-Brillouin effect, four-wave processes)
\cite{lanl_19}. In all these cases, the
power of the root dependence of the instability increment  changes
in the periodic medium if at least for one of the waves, the
diffraction conditions are chosen according to the requirements of
\cite{dan_8} of the coincidence of the roots of the dispersion
equation characterizing the periodic medium.

\subsection{Parametric X-ray radiation in crystals under action of high frequency ultrasonic waves}
\label{berk_c1}

According to  \cite{201,berk_2}, in the presence of an external variable field (for example, an ultarsonic field), a crystal is characterized by an effective  index of refraction, depending on external field parameters. By varying these parameters, one can change the properties of parametric radiation.

As the characteristics and yield of PXR depend on the solution $\vec{E}^{(-)}(\vec{r},\omega)$ of the homogeneous Maxwell's equation
describing the diffraction process in crystals, the investigation of the influence of an external ultrasonic (US) field
on diffraction of X-ray points to the strong influence of a US external field on the PXR process.  In \cite{berk_114} it was pointed to the essential modification of scattering process
 and X-ray radiation process under the diffraction condition in crystals. Due to a dynamical character of PXR formation, according to \cite{berk_115},
 the influence of a US wave on this process will be maximum when  the US wavelength  coincides with the period of extinction beatings.

The theory of PXR under the action of an external US wave on a
crystal target was derived in
\cite{berk_116,berk_117,berk_118,berk_119,berk_120}. The boundary
problem of diffraction of X-rays by a crystal target subjected to
an  external US wave was solved for the case of two-wave
diffraction in \cite{berk_121,berk_122,berk_123}. Here we do not
give the expressions for photon wave functions
 $\vec{E}^{(-)}(\vec{k}, \vec{r})$ because they are very clumsy.
 The spectral-angular PXR distribution in the presence of a US wave
 was obtained in \cite{berk_117,berk_118} (for detail see \cite{lanl_1}). Experimental observation of the effect of the US wave on parametric radiation was performed in \cite{lanl_20,lanl_21}. Experimentally observed features of the PXR in the US wave were explained in \cite{lanl_21a}.

\section{Diffracted X-ray radiation from channeling particle (DCR)}

As we discussed above, the X-ray radiation of a relativistic
oscillator in a crystal essentially modifies under diffraction
conditions of emitted photons. A new diffracted radiation of
oscillator (DRO) appears as a result of coherent summation of two
processes -- photon radiation and photon diffraction, but it
cannot be reduced to a sequence of these two processes. The
relativistic oscillator itself can be a relativistic atom or a
relativistic charged particle channeled in the potential well of
averaged crystallographic potential of axes (planes), or an
oscillator formed by an external electromagnetic field
(ultrasonic, laser). It was shown in
\cite{berk_2,berk_13,berk_19,berk_27,berk_28} that the DRO
spectrum is rather complex and is determined by the complex and
anomalous Doppler effect (see Sections \ref{sec:berk1},
\ref{sec:berk_a}).

It is known that the transverse energy of channeled electrons (positrons) is discrete and state-to-state transitions result in radiation, i.e., in this case a channeled particle is like a one-dimensional or two-dimensional oscillator with the eigenfrequency in the laboratory frame $\Omega_{nf}=\varepsilon_n-\varepsilon_f$, where $\varepsilon_n$ and $\varepsilon_f$ are the eigenvalues of corresponding one- or two-dimensional Schr\"{o}dinger equation, in which the particle rest mass is replaced by the total energy $m\gamma$.

For the analysis of DRO characteristics, it is necessary to obtain the spectral-angular distribution. The description of the channeled particle motion with the help of one- or two-dimensional Bloch functions was given in \cite{chan_17}. The expressions for spectral-angular DRO distribution for different cases of photon dynamical diffraction were obtained in \cite{chan_44,chan_45,berk_19}. For example, in the case of two-wave Laue diffraction the spectral-angular distribution of DRO can be written in the following way \cite{chan_44,chan_45,berk_19}:
\begin{equation}
\label{berk_3.1}
\frac{d^2N_s^{\tau}}{d\omega d\Omega}=\frac{e^2 \beta_1\omega}{\pi^2\hbar c^3}
\sum\limits_{nf}Q_{nn}|\vec{e}_{0s}\vec{g}_{nf}|^2\left|\sum\limits_{\mu=1,2}\zeta^{\tau}_{\mu s}\frac{1-e^{-iq_{znf}^{\mu s}L}}{q_{znf}^{\mu s}}\right|^2,
\end{equation}
where
\begin{equation}
\label{berk_3.2} q_{znf}^{\mu
s}=\omega(1-\beta_{\parallel}\cos\theta)-\Omega_{nf}-\frac{\omega}{\gamma_0}\delta_{\mu
s},
\end{equation}
$\theta$ is the angle between the photon wave vector
$\vec{k}_{\tau}$ directed at a small angle relative to the
particle velocity and the $z$-axis. In the dipole approximation,
which is true for the X-ray radiation, we have
\[
\vec{g}_{nf}= -i\left[\beta_{\parallel}\vec{n}_z(\vec{k}_{\perp}\vec{\rho}_{nf})+\Omega_{nf}\vec{\rho}_{nf}\right]
\]
in an arbitrary nondipole case $\vec{g}_{nf}$ is defined in \cite{chan_44,chan_45},
\[
\vec{\rho}_{nf}=\int_{\Delta}=\varphi_{n\vec{k}}(\vec{\rho})\varphi^*_{f\vec{k}}(\vec{\rho})d^2\rho,
\]
$\varphi_{n\vec{k}}$ and $\varphi_{f\vec{k}}$ are the two-dimensional Bloch functions satisfying the equation similar to the Schr\"{o}dinger equation (see \cite{chan_44,chan_45,berk_19}), $L$ is the crystal target length, $Q_{nn}$ is the population probability of the particle transverse energy state $n$.

According to (\ref{berk_3.1}), the maximum intensity should be observed at the angles and frequencies that satisfy the equation \begin{equation}
\label{berk_3.3}
\omega(1-\beta_{\parallel}\cos\theta)-\Omega_{nf}-\frac{\omega}{\gamma_0}\delta_{\mu s}=0
\end{equation}
The solutions of this equation were obtained above.

In the case of rather thick crystals, the angular distribution of DRO was obtained in \cite{berk_2,chan_45,berk_19}.  For example, the angular distribution of radiation generated by plane-channeled particles can be written as \cite{absan_3}:
\begin{eqnarray}
\label{berk_3.4}
\frac{d N_{\tau}^s}{d\Omega}&=&\frac{e^2L_{\mathrm{eff}}\beta_1^2}{2\pi}\sum\limits_{nf}Q_{nn}|x_{nf}|^2\sum\limits_{\mu}
\frac{(\omega_{nf}^{\mu s})^2}{\Omega_{nf}}\left|\zeta_{\tau}^{\mu s}(\omega_{nf}^{\mu s})\right|^2\nonumber\\
&\times&\left[1-\frac{(\omega_{nf}^{\mu s})^2}{\gamma_1\Omega_{nf}}\texttt{Re}\left(\frac{\partial\delta_{\mu s}}{\partial\omega}\right)\right]^{-1}_{\omega=\omega^{\mu s}_{nf}}F_s(\theta,\varphi)
\end{eqnarray}
for $r$-polarization:
\[
F_r(\theta,\varphi)=\left\{\beta_1\omega_{nf}^{\mu\sigma}\sin^2\theta\cos\varphi
\frac{\tau_y\cos\varphi-\tau_x\sin\varphi}{|\vec{\tau}_{\perp}|}+
\Omega_{nf}\frac{\tau_z\sin\theta\sin\varphi-\tau_y\cos\theta}{|\vec{\tau}_{\perp}|}\right\}^2
\]
and for $\pi$-polarization
\[
F_{\pi}(\theta,\varphi)=\left[\frac{\beta_1\omega_{nf}^{\mu\sigma}\sin\theta\cos\varphi
[\cos\theta(\vec{n},\vec{\tau})-\tau_2]}{|\vec{\tau}_{\perp}|}+
\Omega_{nf}\frac{[\sin^2\theta\cos\varphi(\vec{n}_1\vec{\tau})-\tau_x]}{|\vec{\tau}_{\perp}|}\right]^2
\]
where
\[
\omega_{nf}^{\mu s}=\Omega_{nf}(1-\beta\cos\theta-\gamma_1^{-1}\texttt{Re}\delta_{\mu s}(\omega_{nf}^{\mu s}))^{-1},
\]
$x_{nf}$ is the matrix element obtained {from }one-dimensional
Bloch functions, $L_{\mathrm{eff}}$ is the effective length (at
$L< L_{\mathrm{abs}}$, $L_{\mathrm{eff}}=L$, $L\gg
L_{\mathrm{abs}}$, $L_{\mathrm{eff}}=L_{\mathrm{abs}}$, where
$L_{\mathrm{abs}}$ is the absorption length). The term in square
brackets takes account of the influence of  the dispersion of the
medium on the angular distribution. As the frequency, satisfying
dispersion equation (\ref{berk_3.3}), goes over from one
dispersion branch to another with changing the radiation angle
$\theta$, the summation
over  $\mu$ means that we select the corresponding root of the
dispersion equation for each definite radiation angle $\theta$;
$\vec{n}_1$ is the unit vector directed along the wave vector of
the photon propagating at a small angle relative to the mean
velocity of a channeled particle.

\begin{figure}[htp]
\centering
\epsfxsize = 8 cm \centerline{\epsfbox{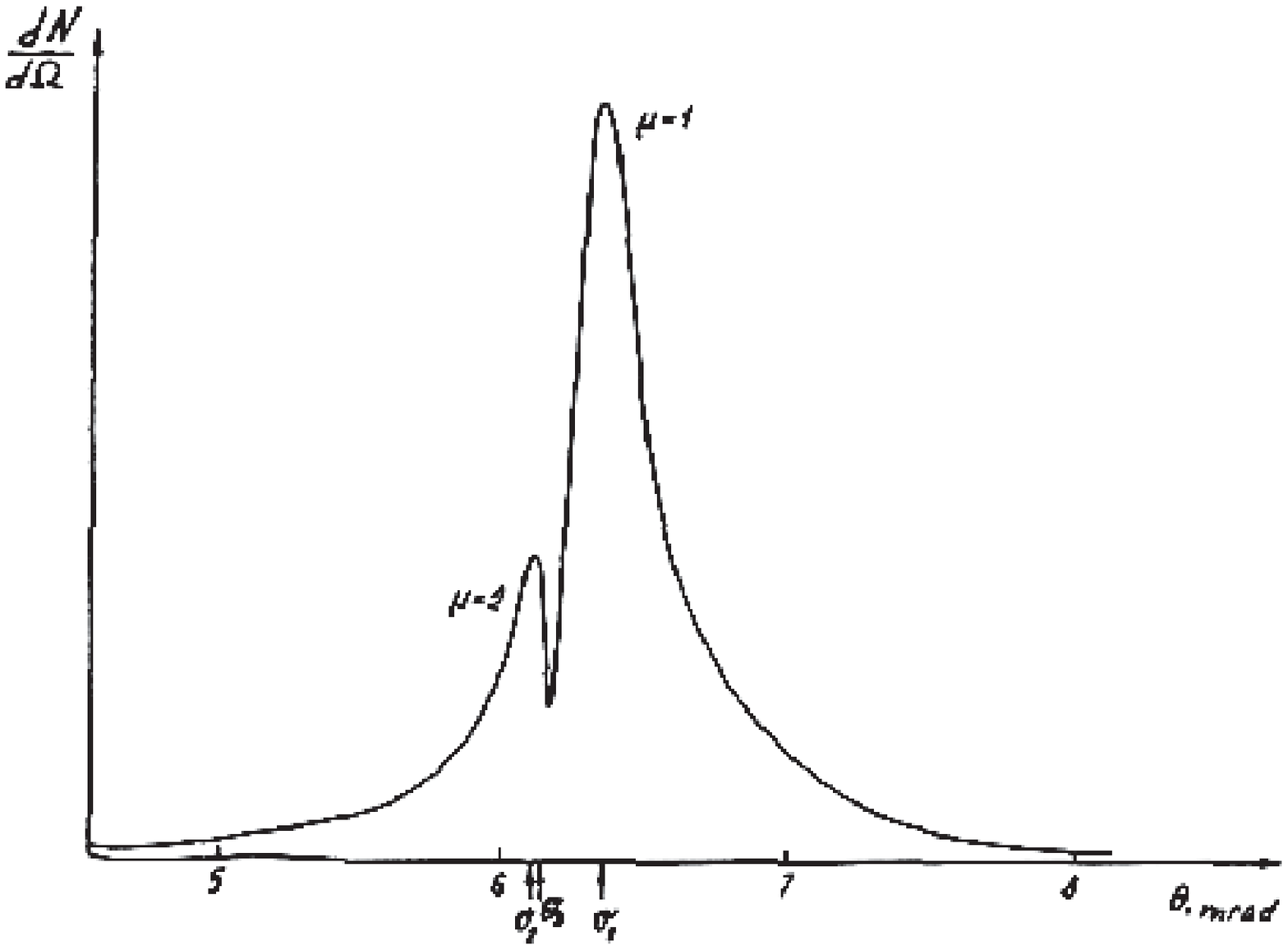}}
\caption{} \label{berkley Figure 41}
\end{figure}

Numerical calculation of the angular DCR distribution taking into
account the dispersion characteristics of the medium under
diffraction  conditions, was made in \cite{berk_29}. According to
Figure \ref{berkley Figure 41}, the angular distribution has a
fine structure which corresponds to the region of transition from
one dispersion branch to another.

One can see that the DCR distribution looks like two narrow rings -- one of which corresponds to the solution $\mu=2$, the other, to $\mu=1$ (in the case of PXR generation, only one solution satisfies the Cherenkov condition). The angular position of the DCR distribution maxima can be estimated in the first approximation as:
\begin{equation}
\label{berk_3.5}
\theta_{1,2}=\sqrt{\theta_D^2\pm\frac{\sqrt{\beta r^{\prime}_s}}{\sqrt{1+2\sin^2\theta_B}/\frac{\Omega}{\omega_B}}},
\end{equation}
where $\theta_D$ is the angle that satisfies the following equation
\[
\theta^2+\gamma^2+\frac{\omega^2_L}{\omega_B^2}-\frac{2\Omega_{nf}}{\omega_B}=0
\]
$\theta_B$ and $\omega_B$ are the angle and the frequency
satisfying the exact Bragg condition. For example, according to
Figure \ref{berkley Figure 41}, the values of these angles are
$\theta_1=6.1\cdot 10^{-3}$ rad and $\theta_2=6.2 \cdot 10^{-3}$
rad. The ratio of the angular width to the value of the angle
$\theta$ is about $\Delta\theta/\theta_D\cong 0.1$. The
expressions for the angular distribution of radiation are
simplified essentially if the particle energy is rather small
($1-\beta\gg 1/\gamma_{0,1} \texttt{Re}\delta_{\mu s}$). In this
case we can consider that the frequency corresponding to the
maximum intensity does not depend on the dielectric properties of
the crystal, being determined by the radiation angle alone. The
DRO characteristics for this case were considered in
\cite{berk_126}.

In \cite{berk_127} the possibility of experimental  observation of
the DRO  by measuring the angular distribution was analyzed. It
was shown that for such experiments particle beams of high quality
are required because the radiation characteristics are very
sensitive to the parameters of a particle beam. Indeed, the DRO
angular distribution shown in Figure \ref{berkley Figure 41} takes
place only for a particle beam whose characteristics satisfy the
following inequality:
\begin{equation}
\label{berk_3.6}
\frac{\Delta\gamma}{\gamma}+\frac{(\Delta\theta_{\mathrm{eff}}\gamma)^2}{2}+
\frac{2\pi\gamma^2}{k_BL_0}<\frac{\gamma^2\sqrt{\beta r^{\prime}_s}}{2},
\end{equation}
where $\gamma$ is the Lorentz factor, $\Delta\theta_{\mathrm{eff}}$ is the angular spread, $\Delta\gamma/\gamma$ is the energy spread, $\Delta\Omega=2\pi c/L$ is the divergence of the oscillator eigenfrequencies.

\begin{figure}[htp]
\centering
\epsfxsize = 8 cm \centerline{\epsfbox{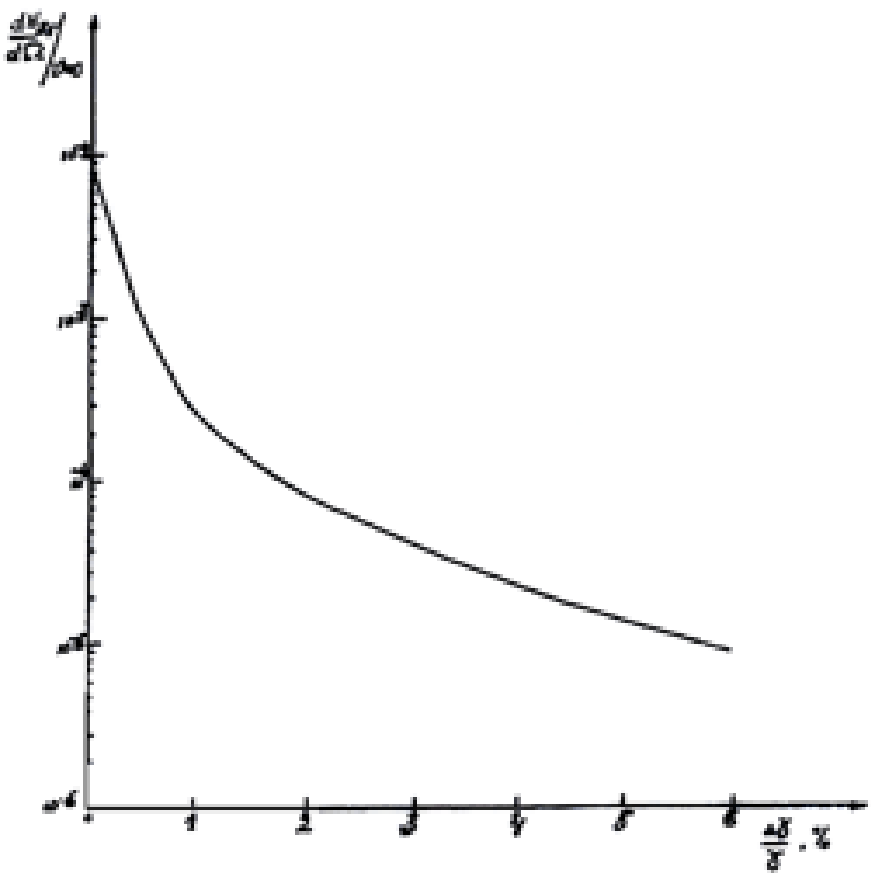}}
\caption{} \label{berkley Figure 42}
\end{figure}

In the opposite case, the angular width of the maxima is equal to
\begin{equation}
\label{berk_3.7}
\Delta\theta=+\frac{1}{\gamma}\left(\frac{2\Delta\gamma}{\gamma}+(\theta_{\mathrm{eff}}\gamma)^2
+\frac{4\pi\gamma^2}{k_BL_0}\right)^{1/2}
\end{equation}
As an example, the dependence of DRO angular distribution
characteristics on the energy divergence of the particle beam is
shown in Figure \ref{berkley Figure 42} one can see that this
dependence is rather sharp indeed.

\begin{figure}[htp]
\centering
\epsfxsize = 8 cm \centerline{\epsfbox{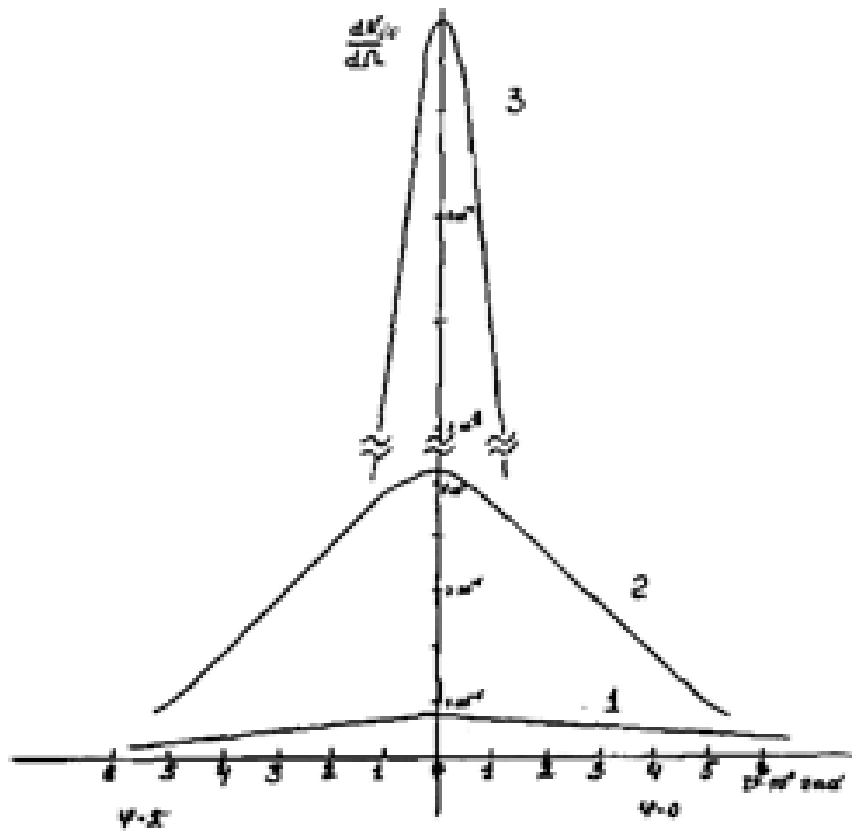}}
\caption{} \label{berkley Figure 43}
\end{figure}

The dependence of the angular density of radiation on the energy
of a relativistic oscillator, which has a "resonance" character at
a
given angle of radiation observation, was also considered in
\cite{berk_127}. If the observation angle is equal to zero, the
maximum of the angular distribution sharply increases with
$\gamma\rightarrow \gamma^R=(\omega_B/2\Omega)^{1/2}$. When the
frequency $\omega_B$ equals
$\omega_B=\omega_m\theta_x=2\Omega\gamma^2$, the maximum value of
the radiation density is observed at $\gamma$ being a little
larger than $\gamma_R$. The angular distribution in this case
looks like a bell and its width decreases sharply with
$\gamma\rightarrow \gamma_R$ (see Figure \ref{berkley Figure 43}).

In the range of $\gamma>\gamma_R$
($\omega_B<\omega_{\mathrm{max}}$), the single narrow maximum
splits into two peaks ($\varphi$ is fixed), which shift to the
range of larger radiation angles $\theta$ with increasing particle
energy $E$. In  \cite{berk_127}, the relative estimation was given
for the  contributions from different radiation mechanisms to the
total radiation angular distribution which can be observed in a
definite reflex. It was shown that at $\Delta\gamma/\gamma\sim
1\%$,  the ratio of the DRO angular density of diffracted
bremsstrahlung at $\theta=0$ is
\begin{equation}
\label{berk_3.8}
R_1=\frac{R_{\mathrm{DRO}}}{I_{DB}}=\frac{Q_{nn}\theta^2_L(\sin^2\psi+\cos^2 2\theta_B\cos^2\psi)}{(1+\cos^22\theta_B)4\overline{\theta^2_s}L(\Delta\gamma/\gamma)^2},
\end{equation}
where the estimation is given for a channeled electron (positron),
$\theta_L$ is the Lindhard angle, $\psi$ is the angle between the
particle oscillation plane and the diffraction plane,
$\overline{\theta^2_s}$ is the mean square angle of multiple
scattering per unit length, $1/4 \,\theta^2_L$ is the classical
estimation of the magnitude of $|x_{nf}|^2\Omega^2c^{-2}$.
{For a $Si$ crystal} and the channeled electron with the energy
$E=23.6$ MeV ($\gamma\approx\gamma_R$; planes of channeling (100),
$\theta=0$, diffraction plane (220)) the value of the ratio is
$R_1=25$, that is, the DRO intensity is 25 times larger than the
intensity of the diffracted bremsstrahlung at
$\Delta\gamma/\gamma\cong 1\%$ and
$\Delta\psi^2\gamma^2<\Delta\gamma/\gamma$. If the diffracted
radiation is observed at the angle $\theta \neq 0$, we should
compare it with the contribution from the parametric
(quasi-Cherenkov) radiation. In this case the analogous ratio is
estimated as  \cite{berk_127,lanl_1}
\begin{eqnarray}
\label{berk_3.9}
R_2&=&\frac{R_{\mathrm{DRO}}}{I_{RxR}} \cong Q_{nn}\left(\frac{\theta_L}{4\frac{\Delta\gamma}{\gamma}\theta_D}\right)^2\nonumber\\
 &\times&\frac{(1+\theta^2_D\gamma^2+\gamma^2\gamma_n^{-2})^2
(\sin^2\psi+\cos^2\psi\cos^2 2\theta_B)}{(\sin^2\varphi+\cos^2\varphi\cos^2 2\theta_B)},
\end{eqnarray}
where $\varphi$ is the angle between the wave vector $\vec{k}$ and
the diffraction plane, $\gamma_n=\omega_B/\omega_L$ is the Lorentz
factor corresponding to the threshold magnitude of the energy
$E=mc^2(g_0^{\prime})^{-1/2}$.
One can see that this ratio essentially depends on the value of
the azimuthal angle $\varphi$.
For example, for a $Si$ crystal this ratio is estimated as
$R_2\cong 5$ when the electron with the energy $E=34$~ MeV is
channeled between the planes (100) and the diffraction plane
 {is} (220).

Thus, the experimental observation of the diffracted radiation of
oscillator is possible with the help of the particle beams of high
quality.

A relativistic oscillator can be formed not only by an unperturbed
crystal channel but also by an external ultrasonic or laser field
\cite{chan_67,berk_104,para_1}.  A more detailed treatment of
particle radiation in crystals under the laser wave was given in
\cite{berk_128}. In \cite{berk_128} the radiation of electrons
(positrons) in a crystal subjected to a laser wave, which forms an
oscillator, was considered. The intensity of such radiation was
estimated. A relativistic oscillator can be a channeled particle,
which moves in a plane channel bent by a variable external field
(ultrasonic or laser wave), i.e., in some crystal undulator
\cite{chan_67}. In this case, the oscillator frequency in the
laboratory frame is $\Omega^{\prime}=\kappa_z u-\Omega$, where
$\vec{\kappa}$ is the wave vector of an external wave in a
crystal, $\Omega$ is its frequency (the $z$-axis is chosen along
the direction of the average particle velocity $\vec{u}$).
Diffracted radiation of the oscillator formed by an external
ultrasonic wave was considered in \cite{berk_129}.

\begin{figure}[htp]
\centering
\epsfxsize = 8 cm \centerline{\epsfbox{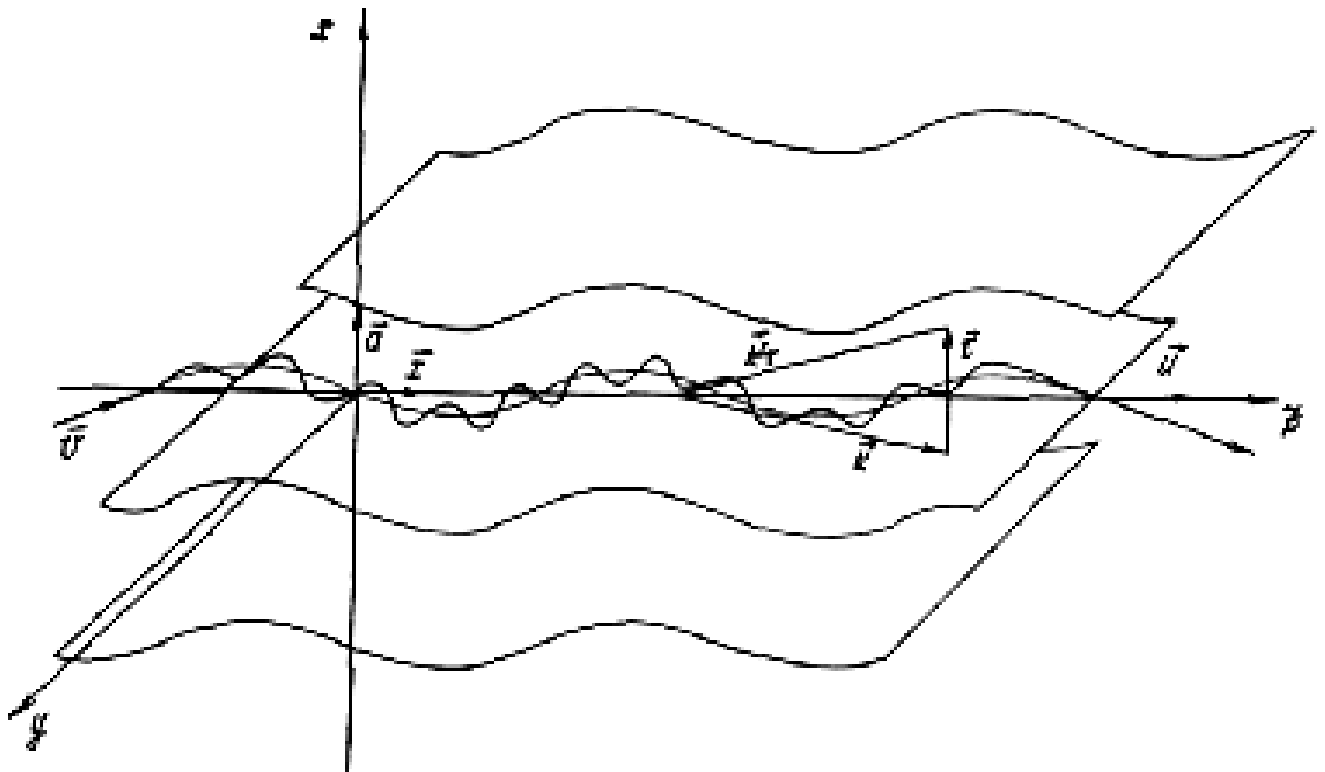}}
\caption{} \label{berkley Figure 44}
\end{figure}

According to \cite{chan_67}, the trajectory of a particle moving
in the dynamic ultrasonic undulator is written in the form (see
Figure \ref{berkley Figure 44})
\begin{equation}
\label{berk_3.10}
\vec{r}(t)=\vec{r}_{\mathrm{ch}}(t)+\vec{r}^s(t)=\vec{r}_{\mathrm{ch}}(t)+\vec{a}\cos(\Omega^{\prime}t+\delta),
\end{equation}
where $\vec{r}_{\mathrm{ch}}(t)$ is the radius vector describing
the motion of an ordinary high-frequency channeled particle, and
$\vec{r}^s(t)$ is the radius vector describing the motion of a
particle in the dynamic undulator.

Assuming that the frequency of an ultrasonic wave is much smaller
than the frequency of particle oscillations in a crystal channel,
we can consider these two kinds of particle motion independently:
the motion of an ordinary channeled particle  and the motion of
the equilibrium trajectory center of particle gravity inside the
bent channel formed under the action of the external variable
field. $\vec{a}$ and $\delta$ are the amplitude and the initial
phase of particle oscillation in the ultrasonic channel. It should
be noted that if the amplitude of the ultrasonic wave  satisfies
the condition $a\ll uU(Ed\kappa^2)^{-1}$ ($E$ is the particle
energy, $d$ is the  width of the crystal channel, and $V$ is the
depth of a potential well for a crystal channel), then the radius
of the crystal curvature due to the action of the ultrasonic wave
is much larger than the radius of the trajectory curvature for the
channeled particle incident on the crystal at the Lindhard angle.
In this case, the equilibrium trajectory of a positively charged
particle gravity center corresponds to the trajectory of a stable
channeling regime, and the curvature of the crystal channel caused
by the action of the ultrasonic wave leads only to the
displacement $\Delta$ of the equilibrium trajectory center of
gravity during the particle passage through the crystal. That is
why for positively charged particles, for which $a_f+\Delta\leq
d/2$, one can take into account the dechanneling effect, which is
due to  the channel curvature, by considering the mean square
angle of multiple scattering in this bent channel in the same way
as in an amorphous medium \cite{berk_129} ($a_f$ is the amplitude
of particle oscillation for the ordinary channeling regime).

In the case under consideration, an essential difference arises in
comparison with the case of diffracted radiation from the
oscillator caused by a channeled particle. This is that the atomic
(nuclear) oscillations, resulting in the formation of the
ultrasonic undulator, will simultaneously lead to the dielectric
constant modulation in a crystal and, consequently, can change the
diffraction process itself. As a result, the photon wave function
changes. Maxwell's equations describing this situation are given
in \cite{berk_129}. The case when the influence of an ultrasonic
wave on the X-ray diffraction process can be reduced  to the
change of the magnitude of the Fourier components of the crystal
dielectric susceptibility alone was considered in detail. The
spectral-angular distribution was obtained and  the contributions
of parametric (quasi-Cherenkov) radiation and of DRO itself were
separated. The spectral and angular characteristics were analyzed
and the total number of photons in a diffraction peak was
estimated.

It was shown that if the following inequality
$
(a\Omega^{\prime})^2>(a_f\Omega_f)^2
$
is fulfilled, the diffracted radiation from a particle in the
external field will be more intensive than the DRO from an
ordinary channeled particle ($a_f$ and $\Omega_f$ are the
amplitude and the frequency of particle oscillations in an
unperturbed channel). According to \cite{chan_67}, this inequality
can be realized for a standard ultrasonic field source and, as
shown by the estimations, the influence of this wave on
dechanneling process can be ignored in this situation.


In conclusion it should be noted that the diffracted radiation can
also be forced under the motion of the oscillator over the crystal
surface, by analogy with the surface parametric (quasi-Cherenkov)
radiation \cite{berk_104}. Radiation from particles moving in
crystal undulators is now being actively studied both
theoretically and experimentally
(see \cite{lanl_14a,lanl_15,lanl_16,lanl_17,lanl_18}).%

\section{X-ray radiation from a spatially modulated relativistic beam in a crystal (\cite{lanl_8})}
\label{sec:vesti92_1}

As was mentioned in the Introduction, in our works we 
suggested a new type of the free electron laser based on volume
multi-wave distributed feedback. It was shown that multi-wave
distributed feedback enables reducing the generation threshold
from the values of the  current densities of a relativistic beam,
such as $10^{13}$ A/cm$^{2}$, which are practically unavailable,
to acceptable values, such as $10^7-10^8$ ~A/cm$^{2}$. This
reduction can be achieved even in the X-ray range (10-100 keV).

One of the most important ways of reducing the generation
threshold in free electron lasers consists in prior modulation of
the density of a beam of relativistic particles. In particular,
modulation gives rise to radiation coherent to the beam, which
appears alongside with spontaneous radiation of electrons as the
beam gets into the undulator.

According to
\cite{berk_104,berk_133,berk_135,berk_143,vesti_9,vesti92_5}, one
of the schemes of the X-ray laser is based on using parametric
quasi-Cherenkov radiation, diffracted radiation of a relativistic
oscillator in the crystal (or radiation in a crystal undulator).
 In this case the
relativistic electrons (positrons) move not in the vacuum, but in
the crystal. Interaction of particles with the atoms of matter
result in multiple scattering, which will apparently diminish the
beam's modulation amplitude and, hence, the intensity of coherent
radiation.

In \cite{lanl_8}, the spectral angular distribution of the
intensity of parametric X-ray (quasi-Cherenkov) radiation was
obtained as well as the radiation of a relativistic oscillator
formed by a spatially modulated beam under multiple scattering.
The conditions were determined under which multiple scattering
does not affect the intensity of coherent radiation. It was shown
that the number of quanta coherently emitted by the beam in the
X-ray range with reasonable requirements to the value of the
current density of the beam of relativistic particles, such as
$\approx 10^8$ A/cm$^2$, is too small in comparison with the
number of spontaneously emitted quanta. For this reason, the
experimental observation of the effect of coherent radiation of
quanta in the X-ray range should be carried out in a rather narrow
spectral angular range. In this case, in the wavelength range of
$50$ -- $100$ ~{\AA}, it is possible to observe the effect when
the degree of beam modulation is as small as $\mu\simeq  10^{-5}$.

In our case of relativistic particles, their energy is much
greater than the energy of emitted $\gamma$-quanta. So one can
consider the radiation process within the framework of classical
electrodynamics. Let a beam of charged particles (electrons or
positrons) traverse the area occupied by matter. The spectral
density $W^s_{\vec{n}\omega}$ of the radiation energy per unit
solid angle of photons characterized by the polarization vector
$\vec{e}^s$ ($\vec{n}=\vec{k}/k$, $\vec{k}$ is the photon wave
vector) can be found when one knows the Fourier transform in time
of the electric field strength in the produced electromagnetic
wave $\vec{E}(\vec{r},\omega)$. According to
\cite{67,201}, at long distances from the target
\begin{equation}
\label{vesti92_1.1}
E_i(\vec{r},\omega)\frac{e^{ikr}}{r}\frac{i\omega}{c^2}\sum_i
e^s_i\int\vec{E}^{(-)s^*}_{\vec{k}}(\vec{r}^{\prime},\omega)\vec{\jmath}(r^{\prime},\omega)d^3r^{\prime},
\end{equation}
where $\vec{E}^{(-)s}_{\vec{k}}(\vec{r},\omega)$ is the solution
of the homogeneous Maxwell's equations, which  describes
scattering by this wave of the photon with wave vector $\vec{k}$
and polarization $s$. The scattering asymptotics has a form of an
incident plane wave plus a converging spherical wave. In this case
$\vec{E}^{s(-)*}_{\vec{k}}=\vec{E}^{s(+)}_{-\vec{k}}$, where
$\vec{E}^{s(+)}_{-\vec{k}}$ is the ordinary solution of the
homogeneous Maxwell's equations with asymptotic behavior
containing a diverging spherical wave;
\[
\vec{\jmath}(\vec{r},\omega)=\int e^{i\omega
t}\vec{\jmath}(\vec{r},\omega) dt
\]
is the Fourier transform of the beam's current, which equals
\[
\vec{\jmath}(\vec{r},\omega)=e\sum_i\vec{v}_i(t)\delta(\vec{r}-\vec{r}^{\prime}(t)),
\]
$\vec{r}_i(t)$ is the coordinate of the $i$-th particle of the
beam at moment $t$, $\vec{v}_i(t)$ is the particle velocity; $e$
is the particle's electric charge. Using (\ref{vesti92_1.1}), we
obtain the following expression for  spectral-angular energy
distribution of radiation generated by the particle beam in the
target:
\begin{equation}
\label{vesti92_1.2}
W_{\vec{n}\omega}^s=\frac{\omega^2}{4\pi^2c^3}\overline{\left|\int
\vec{E}^{(-)s^*}_{\vec{k}}(\vec{r},\omega)\vec{\jmath}(\vec{r},\omega)d^3r\right|^2}.
\end{equation}
The vinculum in (\ref{vesti92_1.2}) denotes averaging over the
distribution of the coordinates and velocities of the particles in
the beam (allowing for multiple scattering in the target). With
the help of the distribution function $w(\vec{r}_l, \vec{v}_l, t;
\, \vec{r}_m^{\prime}, \vec{v}_m^{\prime}, t^{\prime})$ defining
the joint probability density of finding in the $l$-th particle
the coordinate $\vec{r}_l$ and velocity $\vec{v}_l$ at moment $t$,
and finding in the $m$-th particle  the coordinate
$\vec{r}_m^{\prime}$ and velocity $\vec{v}_m^{\prime}$ at moment
$t^{\prime}$, (\ref{vesti92_1.2}) can be written as follows:
\begin{eqnarray}
\label{vesti92_1.3}
W_{\vec{n}\omega}^s=\frac{e^2\omega^2}{4\pi^2c^3}\sum_l\sum_m e^{i\omega t}e^{-i\omega t^{\prime}}\vec{v}_l\vec{E}^{s(-)*}_{\vec{k}}(\vec{r}_l,\omega)\nonumber\\
\times\vec{v}_m^{\prime}\vec{E}^{s(-)}_{\vec{k}}(\vec{r}_m^{\prime},\omega)w(\vec{r}_l,
\vec{v}_l, t; \, \vec{r}_m^{\prime}, \vec{v}_m^{\prime},
t^{\prime})d^3r_ld^3v_ld^3r_m^{\prime}d^3v_m^{\prime}dt
dt^{\prime}.
\end{eqnarray}
According to (\ref{vesti92_1.3}), the spectral-angular distribution
of energy $W_{\vec{n}\omega}^s$ can be represented as follows:
\begin{equation}
\label{vesti92_1.4}
W_{\vec{n}\omega}^s=\sum_l\overline{|\mu_l|^2}+\sum_{l\neq m}\overline{\mu_l\mu_m^*},
\end{equation}
where
\[
\mu_l=\frac{e\omega}{2\pi
c}\int\vec{E}^{s(-)*}_{\vec{k}}(\vec{r}_l(t),\omega)\vec{\beta}_l(t)e^{i\omega
t}dt
\]
has a meaning of the amplitude of photon emission by the
$l$-electron; $\vec{\beta}_l=\vec{v}/c$.

In the case when correlations in the positions of different
particles can be neglected, $w(\vec{r}_l, \vec{v}_l, t; \,
\vec{r}_m, \vec{v}_m, t^{\prime})=w_1(\vec{r}_l, \vec{v}_i,
t)w_1(\vec{r}_m, \vec{v}_m, t^{\prime})$ ($l\neq m$). As a result,
(\ref{vesti92_1.4}) can be recast as
\begin{equation}
\label{vesti92_1.5}
W_{\vec{n}\omega}^s=\sum_l(\overline{|\mu_l|^2}-\overline{|\mu_l|^2})+\left|\sum_{l}\overline{\mu^2}\right|^2.
\end{equation}
The first sum in (\ref{vesti92_1.5}) describes spontaneous
incoherent radiation of photons, the second one -- which is
proportional to the squared modulus of the sum of averaged
emission amplitudes -- describes coherent across the beam photon
emission. To define the conditions under which the distribution of
coordinates and velocities in the beam has no effect on the
process of coherent radiation of a photon, let us take into
account that when a photon is emitted in a homogeneous medium
(e.g., due to the Cherenkov effect), $\vec{E}^{s(-)}_{\vec{k}}$
has a form of a plane wave, while in the case when parametric
X-ray radiation or diffracted radiation of the oscillator is
generated in the crystal (in a medium with spatially-periodic
dielectric permittivity), the expression for
$\vec{E}^{s(-)}_{\vec{k}}$ has a form of a superposition of plane
waves \cite{201,berk_2,berk_19,vfel_PXRbook}. In particular, in
the space region occupied by the target,
$\vec{E}^{s(-)}_{\vec{k}}$ can be represented in the form:
\begin{eqnarray}
\label{vesti92_1.6}
\vec{E}^{s(-)}_{\vec{k}}(\vec{r}_l(t), \omega)=\sum_n A_n e^{i\vec{k}_n\vec{r}_l(t)},\nonumber\\
\vec{r}_l(t)=\vec{r}_{l 0}+\vec{u}_l t+\delta\vec{r}_l(t).
\end{eqnarray}
Here $\vec{r}_{l 0}$ is the electron coordinate at time $t=0$;
$\vec{u}_l$ is the electron velocity in a vacuum;
$\delta\vec{r}_l(t)$ describes the variation of the electron
trajectory  under the action of the forces (in particular, those
leading to multiple scattering) that affect the particle in the
area occupied by the target. Let us assume that
$\vec{u}_l=\vec{u}+\Delta\vec{u}_l$, where
$\Delta\vec{u}_l\ll\vec{u}$. To be more specific, take the
$z$-axis as directed along $\vec{u}$.

Now, let us determine the conditions under which the distribution
of the coordinates and velocities in the beam has no effect on the
intensity of coherent radiation.

Let  the crystal surface through which the electrons enter the
target be located at point $z=0$. At the moment of time $t=0$,
their coordinates were $\vec{r}_{0l}$ and were located in the area
$z< 0$. Consequently, the $l$-electron reaches the surface $z=0$
at time $t_l=|z_{l0}|/u_{zl}$. During this time, the electron will
move over the distance
$\delta\vec{r}_{l\perp}=\vec{u}_{l\perp}|\vec{z}_{l0}|/u_{zl}$ in
the transverse direction. As a result, the electron's transverse
coordinate at entering the target will be
\[
\vec{r}_{l\perp}=\vec{r}_{l0\perp}+\vec{u}_{l\perp}\frac{|z_{l0}|}{u_{zl}}=\vec{r}_{l0\perp}
-\vec{u}_{l\perp}\frac{z_{l0}}{u_{zl}}.
\]
The major contribution to the amplitude $\mu_l$ comes from
integration over the interval corresponding to particle motion
inside the target \cite{63}. We shall consider this very
contribution. In the integral over time involved in the amplitude
$\mu_l$, we shall shift the zero-time position corresponding to
the particle motion in the target area to point $t=0$.

As a result, the amplitudes $\mu_l$ can be represented in the
form:
\begin{equation}
\label{vesti92_1.7} \mu_l=A_l
l^{-i\vec{k}_{\perp}\vec{r}_{l0\perp}}e^{-i\frac{\omega}{u_{zl}}z_{l0}}e^{i\vec{k}\vec{u}_{l\perp}\frac{z}{u_{zl}}},
\end{equation}
where $A_l$ is the amplitude of radiation in the crystal of the
$l$-electron,  which entered the target at point $r=0$ at time
$t=0$. If the additional phase shift occurring during the time $T$
of the electron motion in the crystal is small due to velocity
distribution $\Delta u_l$, i.e., $\vec{k}\vec{\Delta}\vec{u}_l
T<1$, then multiple scattering has no effect on the amplitude of
radiation. Here, in the case of parametric (quasi-Cherenkov)
radiation,  all the amplitudes $A_l$ equal one another, while in
the case of diffracted radiation of the oscillator (radiation of
channeled particles), the amplitudes $A_l$ depend on the the
particle entry point into the channel, which should be taken into
account in averaging the amplitudes. It is known that the root
mean square angle of multiple scattering
$
\langle\vartheta^2\rangle=g\frac{E^2_s}{E^2}\frac{l}{L},
$
where $E_s=$21 ~MeV, $E$ is the particle energy, $l$ is the
traveled path length, $L$  is the radiation length, $g$ is the
coefficient for the difference between $\langle\vartheta^2\rangle$
and $\langle\vartheta^2\rangle_{\mathrm{am}}$ in an amorphous
medium. For positrons moving in the regime of planar channeling,
the magnitude of $g$ can be much less than unity. When $e^{\pm}$
move at small angles relative to the axes, the magnitude of $g$
can become much greater than unity. Two requirements follows from
inequality $\vec{k}\Delta\vec{u}_l T<1$:
$k_{\perp}\sqrt{\langle\vartheta^2\rangle}$, $l< 1$ and
$kl\langle\vartheta^2\rangle <1$. The former can always be chosen
with the help of the observation angle. The latter yields the
requirement
$
l<\frac{1}{\sqrt{k\langle\vartheta^2\rangle_1}},
$
where $\langle\vartheta^2\rangle_1$ is the root mean square angle
of multiple scattering over the unit length ($l\sim \gamma$,
$\gamma$ is the particle Lorentz factor). This leads to the fact
that even for quite hard radiation with the quantum energy of the
order of $100$~ keV, the target thickness required to avoid the
effect of multiple scattering on the radiation process, appears
rather large ($l\leq 10^{-3}$ ~cm) for electrons with the energy
of $1$ GeV. We shall further assume that this condition is
fulfilled. So from (\ref{vesti92_1.5}) we have the following
expression for the spectral angular  distribution of the number of
emitted quanta $d^2 N/d\omega d\Omega=W^s_{\vec{n}\omega}/\hbar
\omega$:
\begin{equation}
\label{vesti92_1.8} \frac{d^2 N}{d\omega
d\Omega}=\frac{d^2N_1}{d\omega d \Omega}N_l+\frac{d^2N_1}{d\omega
d\Omega}\sum_{l\neq
m}\overline{e^{-i\vec{K}\vec{r}_l}e^{i\vec{K}^*\vec{r}_m}},
\end{equation}
where
\[\vec{r}_l=\left(\vec{r}_{0\perp l}-\vec{u}_{l\perp}\frac{z_{0l}}{u_{zl}}, \frac{u}{u_{zl}}z_{0l}\right),
\quad \vec{K}=\left(\vec{k}_{\perp},\frac{\omega}{u}\right)
\]
in the case of parametric radiation; if radiation is produced by a
relativistic oscillator (radiation from channeled particles,
diffracted radiation of the oscillator) with the vibration
frequency $\Omega$ in the laboratory system of coordinates, vector
$
\vec{K}=\left(\vec{k}_{\perp},\,\frac{\omega\pm\Omega}{u}\right).
$
From (\ref{vesti92_1.8})  follows that to achieve the conditions
under which the velocity distribution $\Delta u_l$ in a beam has
no effect on (\ref{vesti92_1.8}), the longitudinal dimensions $L_b$
of the bunch of particles  should satisfy the relations:
\[
\frac{\omega}{u}\frac{\Delta u_{zl}}{u}L_b< 1 \quad\mbox{and}\quad
k_{\perp}\frac{\Delta u_{\perp}}{u}L_b< 1,
\]
i.e.,
\[
L_b<\frac{1}{k\nu^2}\quad\mbox{and}\quad
L_b<\frac{1}{k\nu_{\gamma}\nu},
\]
where $\nu$ is the characteristic angular distribution of
particles' velocities in the beam; $\nu_{\gamma}$ is the quantum
emission angle. When the longitudinal dimensions of the bunch
satisfy these conditions, the magnitude of $\vec{r}_l$ in
(\ref{vesti92_1.8}) can be taken equal to $\vec{r}_{0l}$. Now
averaging in (\ref{vesti92_1.8}) is easy to perform. Since the
double sum in (\ref{vesti92_1.8}) does not depend on velocities of
particles, averaging is reduced to averaging over the initial
distribution of the coordinates of particles in the beam. Let us
introduce the beam density $\rho(\vec{r})$;
$\int_{V_b}\rho(\vec{r})d^3 r=N_e$ ($V_b$ is the bunch volume,
$N_e$ is the number of particles in the bunch). Upon averaging
(\ref{vesti92_1.8}) with the distribution $\rho(r)$, we obtain
\begin{equation}
\label{vesti92_1.9} \frac{d^2 N}{d\omega
d\Omega}=\frac{d^2N_1}{d\omega d \Omega}N_e+\frac{d^2N_1}{d\omega
d\Omega}\left|\int e^{-i\vec{K}\vec{r}}\rho(\vec{r})d^3r\right|^2,
\end{equation}
where $d^2N_1/d\omega d\Omega$ is the spectral angular
distribution of quanta formed as a result of spontaneous radiation
by a single electron. Let us consider a beam modulated as
$\rho(\vec{r})=\rho_0+\rho_1\cos(\vec{\tau}\vec{r})$. As a result,
(\ref{vesti92_1.10}) will include rapidly oscillating integrals,
which can be replaced by $\delta$-functions with good accuracy.
This enables writing (\ref{vesti92_1.9}) in the form
\begin{equation}
\label{vesti92_1.10} \frac{d^2 N}{d\omega
d\Omega}=\frac{d^2N_1}{d\omega d \Omega}N_e+\frac{d^2N_1}{d\omega
d \Omega}\frac{\pi}{2}N_e\mu^2\rho_0\delta(\vec{K}-\vec{\tau}),
\end{equation}
where $\mu=\rho_1/\rho_0$. From (\ref{vesti92_1.10}) follows that
the total number of quanta emitted by the bunch equals
\begin{equation}
\label{vesti92_1.11} N=N_e\left(N_1+c\left.\frac{d^2N_1}{d\omega
d\Omega}\right|_{\vec{K}=\vec{\tau}}\frac{\pi}{2}\mu^2\frac{\rho_0}{k_0^2}\right).
\end{equation}
Here $N_1$ is the number of quanta of incoherent spontaneous
radiation, which are produced by a single electron traversing the
target  under study;
$
k_0=|\vec{\tau}|=\frac{2\pi}{\lambda_0},
$
$\lambda_0$ is the radiation wavelength equal to the  spatial
modulation period $d$ of the beam.

Note that (\ref{vesti92_1.11}) is easy to obtain from the following
quantitative considerations: In performing integration  over
$d\omega d\Omega$ in (\ref{vesti92_1.9}), the characteristic range
of values, where the second term in (\ref{vesti92_1.9}) is
nonzero is: $\Delta k_{\perp}\sim 1/L_{\perp b}$ for the
transverse dimension  and $\Delta\omega/c\sim1/L_b$, for the
longitudinal dimension. As a result, we have
\begin{eqnarray}
\label{vesti92_1.12}
& &\int\frac{d^2N_1}{d\omega d\Omega}\left|\int e^{-i\vec{K}\vec{r}}\rho(r)d^3r\right|^2d\omega d\Omega\simeq\left.\frac{d^2N_1}{d\omega d\Omega}\right|_{K=\tau}N^2_e\mu^2 c\frac{\Delta^2k_{\perp}\Delta k_{\parallel}}{k_0^2}\nonumber\\
& &\simeq c\left.\frac{d^2N_1}{d\omega
d\Omega}\right|_{K=\tau}N^2_e\frac{1}{L^2_{\perp
b}}\frac{1}{L_b}\frac{\mu^2}{k_0^2}\simeq N_e
c\left.\frac{d^2N_1}{d\omega
d\Omega}\right|_{K=\tau}\frac{\rho_0\mu^2}{k_0^2},
\end{eqnarray}
i.e., the expression appearing in (\ref{vesti92_1.11}). The ratio
of the total number of coherently emitted quanta to that of
incoherently emitted quanta is
\begin{equation}
\label{vesti92_1.13}
\frac{N_{\mathrm{coh}}}{N_{\mathrm{incoh}}}\simeq\frac{\pi}{2}\frac{\mu^2\rho_0}{k_0^2\Delta
k\Delta\Omega}.
\end{equation}
Here $\Delta k(\Delta\Omega)$ is the characteristic range $k$ (of
solid angles), where the quantum is emitted through incoherent
spontaneous radiation; $\Delta k/k\sim 1/\gamma$,
$\Delta\Omega\sim 1/\gamma^2$; $\gamma$ is the particle Lorentz
factor. In obtaining (\ref{vesti92_1.13}), the estimate
$
N_1\simeq\frac{d^2N_1}{dkd\Omega}\Delta k\Delta\Omega
$
was used; (\ref{vesti92_1.13}) can also be recast as
\begin{equation}
\label{vesti92_1.14}
\frac{N_{\mathrm{coh}}}{N_{\mathrm{incoh}}}\simeq\frac{\mu^2\rho_0}{k_0^3}\gamma^3.
\end{equation}
Note that according  to \cite{63},  in  the case  of parametric
radiation when $1/\gamma^2< n_{\mathrm{eff}}-1$
($n_{\mathrm{eff}}$ is the effective  index of refraction in the
radiation area), the characteristic range of radiation angles (the
frequency distribution) $\Delta\vartheta$,
$\Delta\omega/\omega\sim\sqrt{n_{\mathrm{eff}}-1}$. As a result,
with growing $\gamma$, the ratio (\ref{vesti92_1.14}) tends to the
limit
$N_{\mathrm{coh}}/N_{\mathrm{incoh}}\simeq(\mu^2\rho_0/k_0^3)(n_{\mathrm{eff}}-1)^{-3/2}$.
Recall that in the X-ray range of $10\div100$ ~keV,
$n_{\mathrm{eff}}-1\simeq 10^{-5}\div 10^{-7}$. From
(\ref{vesti92_1.13}), (\ref{vesti92_1.14}) follows that with other
conditions being equal, the ratio
$N_{\mathrm{coh}}/N_{\mathrm{incoh}}$ decreases rapidly with
growing $k$ (with the decrease in the radiation  wavelength). Thus
for example, even at current densities in a beam as high as
$10^{8}$ ~A/cm$^2$, which corresponds to $\rho_0\simeq 10^{17}$
and at $\mu=1$, the ratio $N_{\mathrm{coh}}/N_{\mathrm{incoh}}\leq
10^{-1}$  for quantum energy of 10 ~keV and $\gamma=10^3$. At the
same time, the ratio of the spectral densities of radiation within
the range of the emission angles $\Delta\vartheta\sim1/kL_{\perp
b}\ll1/\gamma$ of coherent radiation is $\mu^2N_e$.

Thus, within this range of emission angles, the intensity of
coherent spontaneous radiation exceeds that of incoherent
spontaneous radiation when the modulation depth is $\mu\geq
1\sqrt{N_e}$. For example, when the number of electron in the
bunch is $N_e\simeq 10^{12}$, it is sufficient that $\mu\geq
10^{-6}$, which enables one to appreciably simplify the problem of
experimental observation  of coherent radiation in the X-ray range
of 10-100 keV. Note  that the formulas derived here are also
applicable when the target is irradiated by the laser pulse of
length greater than $L_b$. Indeed, let us divide the whole pulse
length into the bunches, each of length $L_b$, i.e., we have
(\ref{vesti92_1.10}), (\ref{vesti92_1.11}), where $N_e$ stands for
the number of electrons in the entire pulse.

Now let us estimate the number of quanta which can be produced
coherently, e.g., in the case of parametric (quasi-Cherenkov)
radiation mechanism. From general formulas (\ref{vesti92_1.8}) for
parametric radiation in the Laue case follows the below expression
for spectral-angular distribution of quanta emitted in the
direction of diffraction:
\begin{equation}
\label{vesti92_1.15} \left.\frac{d^2N_1}{dk
d\Omega}\right|_{\vec{K}=\vec{\tau}}\simeq\frac{e^2}{\hbar
c\pi^2}\left|\frac{\chi_{\tau_1}}{\vartheta_B}\right|^2k_0N^2,
\end{equation}
where $L$ is the crystal thickness; $\chi_{\tau_1}$ is the Fourier
component of the crystal susceptibility for quantum diffraction by
the system of planes defined by the crystal reciprocal lattice
vector $r_1$, $\vartheta_B$ is the Bragg diffraction angle.

According to (\ref{vesti92_1.15}), (\ref{vesti92_1.11}), for the
number of coherently emitted quanta per electron we have
\begin{equation}
\label{vesti92_1.16} N_{1\mathrm{coh}}\simeq
10^{-3}\left|\frac{\chi_{\tau_1}}{\vartheta_B}\right|^2\mu^2\frac{\rho_0L^2}{k_0}.
\end{equation}
For radiation generated in $Si$, $\chi_{\tau}\sim 10^{-5}$ for
$(400)$ plane, $k_0=10^9$, $\vartheta_B\simeq 45^{\circ}$,
$\rho_0\simeq 10^{17}$, $L=10^{-1}$ ~cm, we have
$N_{1\mathrm{coh}}\simeq 10^{-7}\mu^2$.

In the case of PXR generation in a layered medium (i.e. in $NiC)$
in the range of wavelengths $\simeq 50\div 100$ ~{\AA},
$\chi_{\tau}\sim 10^{-2}$, $k_0=10^7$, $L\simeq10^{-4}$ ~cm, we
have $N_{1\mathrm{coh}}\simeq 10^{-5}\mu^2$. At the same time, in
the case of generation of surface PXR, when the quantum absorption
length  in the medium does not restrain the radiation intensity,
which is proportional to the path length traveled over the grating
\cite{berk_104}, $N_{1\mathrm{coh}}\simeq 10^{3}\mu^2$ for $L=1$
~cm. When PXR is generated in the optical range $k_0=10^5$,
$\chi_{\tau}=10^{-1}$, $L\simeq10^{-1}$ ~cm, we have
$N_{1\mathrm{coh}}\simeq 10^{5}\mu^2$. Note here that the obtained
estimates increase by two-three orders of magnitude in the case of
Bragg diffraction because of the increase in $d^2N_1$ in narrow
spectral ranges \cite{vesti92_11}.

For the oscillator mechanism of radiation \cite{berk_19}
\begin{equation}
\label{vesti92_1.17}
\frac{dN_{1\mathrm{osc}}}{dkd\Omega}\simeq\frac{e^2}{\hbar
c\pi}\left(\frac{v_{\perp}}{c}\right)^2k_0L^2,
\end{equation}
$v_{\perp}$ is the velocity amplitude of the transverse vibrations
of the particle. For this reason, to obtain the estimates,
$|\chi_{\tau}/\vartheta_B|^2$  in (\ref{vesti92_1.16}) is replaced
by $(v_{\perp}/c)$. In particular, for positrons channeled in
$Si$, the characteristic values of $v_{\perp}/c\simeq 10^{-5}$ and
of $dN_{1\mathrm{osc}}\simeq dN_{1\mathrm{PXR}}$.

Thus, it follows from the above analysis that using a spatially
modulated beam, one can observe coherent parametric X-ray
radiation. Moreover, in a soft X-ray range, under the condition
when the surface parametric radiation is generated
\cite{berk_104}, coherent radiation can be observed even when
the degree of beam modulation is not very high (e.g., for
$\mu\simeq 10^{-5}$, we have $N_{1\mathrm{coh}}\simeq 10^{-7}$).
As a result, when the transmitted pulse contains
$N_e\simeq 10^{12}$, we obtain $N_{\gamma}\simeq 10^5$, which is
quite acceptable.
The possibility to use modulated beams for generating coherent radiation in crystal undulators has recently been considered in
\cite{lanl_9}.

\section{Crystal X-ray Free Electron Lasers on the basis of PXR and DRO (DCR) }

High spectral and angular densities of parametric
(quasi-Cherenkov) and diffracted radiation of the oscillator as
well as  narrow spectral and angular widths of radiation reflex
give a basis for application of considered spontaneous mechanisms
of X-ray radiation for the construction of an X-ray coherent
radiation source by using beams of relativistic particles  in
crystals. Such a system can be considered as a crystal X-ray free
electron laser (FEL). The idea of X-ray FELs based on  spontaneous
parametric and DRO (DCR) radiation in crystals was first expressed
in \cite{berk_130,berk_131,berk_132}. In
\cite{berk_130,berk_131,berk_132,berk_133,berk_134},  the
dispersion equation for the eigenstates of the system consisting
of electromagnetic radiation, a beam of relativistic oscillators
and a crystal was obtained. The increment of the beam instability
was also analyzed. The possibility in principle of obtaining X-ray
coherent radiation with the help of a beam of relativistic
oscillators in crystals was shown. In \cite{berk_135},  the
parametric (PXR) relativistic beam instability in a crystal was
considered and the corresponding increment was obtained. Radiative
instability caused by spontaneous radiation in crystal undulators
and by a laser wave propagating through a crystal was studied in
\cite{berk_144}. 
Thus, in
\cite{berk_130,berk_131,berk_132,berk_133,berk_134,berk_135,berk_144},
a new kind of the X-ray FEL -- the solid X-ray free electron laser
(SXFEL) was suggested. As we have pointed out above, several
mechanisms of spontaneous X-ray radiation generated by a
relativistic electron beam in crystals can constitute the basis
for such SXFELs: parametric (quasi-Cherenkov) X-ray radiation and
diffracted radiation of oscillators formed in crystals, for
example, by channeling
\cite{berk_130,berk_131,berk_132,berk_133,berk_134,berk_140,berk_141,berk_142,berk_143}
or under the action of an external field \cite{berk_142,berk_144}.

The main feature of such an X-ray generator is that the crystal target, in this case, not only forms the mechanism of spontaneous radiation, but also acts as a three-dimensional resonator for X-ray radiation which produces a distributed feedback (DFB). The construction of the X-ray generator by using channeled electron beams in crystals was also considered in \cite{berk_145,berk_146,berk_147,berk_148,berk_149} and the construction of the X-ray generator on the basis of resonance transition radiation was discussed in \cite{berk_150,berk_151}. The possibility of using the crystal as a resonator that produces a one-dimensional distributed feedback for an X-ray coherent generator was first expressed in \cite{berk_152}. This idea was used for the  formation of a one-dimensional DFB in a solid X-ray FEL on the basis of channeled particles in \cite{berk_148,berk_149}. However, in all these works, the DFB was traditionally considered in a one-dimensional geometry when the radiated and diffracted waves propagate along one line in opposite directions. The authors of \cite{berk_148,berk_149} obtained a low generation threshold for such a system with a one-dimensional DFB only due to ignoring the radiation self-absorption inside the crystal. The correct consideration of absorption, as it was shown in \cite{berk_143}, leads to the threshold beam density of the order of $j^{th}\sim 10^{12}$~A/cm$^2$ for this DFB geometry.

In the solid X-ray free electron laser, suggested \\ in
\cite{berk_130,berk_131,berk_132,berk_133,berk_134,berk_135,berk_136,berk_137,berk_138,berk_139,berk_140,berk_141,berk_142,berk_143,berk_144},
the crystal resonator produces a three-dimensional DFB that allows
one to optimize the system and to essentially decrease the
generation threshold. The analysis showed that the process of
amplification and generation in such a crystal (natural or
photonic) solid resonator essentially modifies and, under definite
conditions, develops more intensively. It was shown that the
interaction between the particle beam and the electromagnetic
field is the strongest near the region of degeneration of roots of
diffraction dispersion equation, particularly, in the case of
multi-wave diffraction.

Let us consider in detail PXR and DRO (DCR) crystal X-ray FELs

\section{Parametric (Quasi-Cherenkov) X-ray FEL}
\label{sec:8FEL} The quasi-Cherenkov instability of a relativistic
electron (positron) beam in a three-dimensional periodic medium in
the X-ray range was first considered in \cite{berk_135,lanl_1}.
The authors formulated the problem of X-ray parametric radiation
amplification in an infinite medium caused by quasi-Cherenkov
instability of a beam of relativistic particles.  The dispersion
equation for the case of two-beam diffraction and the increment of
instability was obtained. It was shown that the strongest
interaction between the particle beam and radiation was close to
the region of degeneration of the dispersion equation roots. The
boundary problem of amplification of radiation in a finite
parallel-plane crystal target was solved, and the generation
threshold for the particle density was obtained. It was assumed
that a relativistic particle beam with a mean velocity $\vec{u}$
was incident at a definite angle $\psi_0$ on the parallel-plane
crystal target with the length $L$. The orientation of the
particle beam relative to crystallographic planes was made in such
a way that spontaneous photons radiated by a particle beam were
under diffraction conditions for planes with low indices. The
fulfillment of diffraction condition not only brings about the
possibility of quasi-Cherenkov radiation in the X-ray range itself
but also produces a three-dimensional distributed feedback.

The closed set of equations describing the interaction of a radiating beam with a crystal, in the general case, consists of Maxwell's equations for the electromagnetic field and the equation for particle motion in the field (for a "cold" particle beam, $\theta\psi <(kL)^{-1}$, where $\theta$ is the radiation angle, $\psi$ is the angular spread of particles in a beam, $\vec{k}$ is the photon vector) or the equation for the distribution function (in the case of a "hot" particle beam). For example, in the case of a "cold" beam we have:
\begin{eqnarray}
\label{berk_4.1}
& &\vec{\nabla}\times \vec{\nabla}\times \vec{E}(\vec{r},\omega)=\frac{4\pi i\omega}{c^2}\vec{j}(\vec{r};\omega)+\frac{\omega^2}{c^2}\vec{D}(\vec{r},\omega),\nonumber\\
& &\frac{d\vec{v}_{\alpha}(t)}{dt}=\frac{e}{m\gamma}\left\{\vec{E}(\vec{r}_{\alpha}(t),t)+
\left[\frac{\vec{v}_{\alpha}(t)}{c}\vec{H}(\vec{r}_{\alpha}(t),t)\right]\right.\\
& &\left.-\frac{\vec{v}_{\alpha}(t)}{c}\left(\frac{\vec{v}_{\alpha}(t)}{c}\vec{E}(\vec{r}_{\alpha}(t),t)\right)\right\},\nonumber
\end{eqnarray}
where $\vec{j}(\vec{r},t)=e\sum\limits_\alpha\vec{v}_\alpha(t)\delta(\vec{r}-\vec{r}_{\alpha}(t))$ is the microscopic current density of particles in a beam, $n(\vec{r},t)=e\sum\limits_{\alpha}\delta(\vec{r}-\vec{r}_{\alpha}(t))$ is the corresponding charge density, $\vec{r}_{\alpha}(t)$ and $\vec{v}_{\alpha}(t)$ are the trajectory and the velocity of the $\alpha$-th particle in a beam, $ \vec{E}(\vec{r},t)$ and $ \vec{H}(\vec{r},t)$ are the electric and magnetic strengths of the field, $\vec{D}(\vec{r},\omega)=\varepsilon(\vec{r},\omega)\vec{E}(\vec{r},\omega)$, $\varepsilon(\vec{\pi},\omega)=\sum\limits_{\tau}\varepsilon_{\tau}(\omega)e^{-i\vec{\tau}\vec{r}}$ is the crystal dielectric constant, $\varepsilon_0=1+g_0\cong1-\omega^2_L/\omega^2$, $\omega^2_L=4\pi e^2n_0/m_e$, $n_0$ is the electron density in a crystal, $g_{\tau}\equiv\varepsilon_{\tau}$ is the Fourier component of the dielectric constant, $\vec{\tau}$ is the reciprocal lattice vector.

 For the case of two-wave generation,  with the trajectory and velocity of a particle represented as $\vec{r}_{\alpha}(t)=\vec{r}_{0\alpha}+\vec{u}t+\delta\vec{r}_{\alpha}(t)$ and $\vec{v}_{\alpha}(t)=\vec{u}+\delta\vec{v}_{\alpha}(t)$, where $\vec{\tau}_{0\alpha}$ is the position of the $\alpha$-th particle in a beam at the moment of intersection of the crystal boundary, the system (\ref{berk_4.1}) can be written
as a system of Maxwell's equations for electromagnetic fields $\vec{E}(\vec{k},\omega)$ and $\vec{E}_{\tau}(\vec{k},\omega)$ in the following way \cite{berk_136,berk_137,berk_138,berk_139}
\begin{eqnarray}
\label{berk_4.2}
\left(k^2c^2-\omega^2\varepsilon_0+\frac{\tilde{\omega}^2_L}{\gamma}-
\omega^2\varepsilon_b(\vec{k})\right)E_{\sigma}-\omega^2g_{\tau}E_{\sigma}^{\tau}=0\nonumber\\
-\omega^2g_{-\tau}E_{\sigma}+\left(k_{\tau}^2 c^2-\omega^2\varepsilon_0+\frac{\tilde{\omega}^2_L}{\gamma}-
\omega^2\varepsilon_b(\vec{k}_{\tau})\right)E_{\sigma}^{\tau}=0,
\end{eqnarray}
where $\vec{k}_{\tau}=\vec{k}+\vec{\tau}$,
$E_{\sigma}=\vec{E}(\vec{k},\omega)\vec{e}_{\sigma}$,
$E_{\sigma}^{\tau}=\vec{E}(\vec{k}_{\tau},\omega)\vec{e}_{\sigma}$,
$\vec{e}_{\sigma}\parallel[\vec{k}\vec{\tau}]$. The set of
equations (\ref{berk_4.2}) is written for $\sigma$-polarization of
radiation because it is excited with maximum probability at
parametric (quasi-Cherenkov) radiation,
$\tilde{\omega}^2_L=4\pi e^2\tilde{n}_0/m_e$, $\tilde{n}_0$ is the
mean density of the unperturbed particle beam.

Comparison of (\ref{berk_4.2}) and the ordinary set of Maxwell's
equations describing X-ray dynamical diffraction in a crystal
allows one to conclude that the boundary problem of X-ray
amplification (generation) under penetration of a particle beam
through a periodic medium can be reduced to the problem of X-ray
diffraction by an "active" periodic medium, which consists of the
crystal + radiating particle beam and is characterized by the
following dielectric constant:
\begin{eqnarray}
\label{berk_4.3}
\tilde{\varepsilon}_0(\vec{k}_{\tau},\omega)=\varepsilon_0-\frac{\tilde{\omega}^2_L}{\gamma\omega^2}-
\frac{\tilde{\omega}^2_L}{\gamma\omega^2}
\frac{(\vec{u}\vec{e}_{\sigma})^2}{c^2}\frac{k^2c^2-\omega^2}{(\omega-\vec{k}\vec{u})^2},\nonumber\\
\tilde{\varepsilon}_0(\vec{k},\omega)=\varepsilon_0-\frac{\tilde{\omega}^2_L}{\gamma\omega^2}-
\frac{\tilde{\omega}^2_L}{\gamma\omega^2}
\frac{(\vec{u}\vec{e}_{\sigma})^2}{c^2}\frac{k_{\tau}^2c^2-\omega^2}{(\omega-\vec{k}_{\tau}\vec{u})^2}.
\end{eqnarray}
As the electron density in a beam is much smaller than that in a
crystal, the second term on the right-hand side of
(\ref{berk_4.3}) can be neglected. The last term has a resonance
behavior under the fulfillment of synchronism condition between
the particle beam and the electromagnetic field
$\omega-\vec{k}\vec{u}\cong 0$. In the X-ray range, the
fulfillment of this condition for a diffracted wave is impossible,
that is why it is possible to consider
$\vec{\varepsilon}_0(\vec{k}_{\tau},\omega)\cong \varepsilon_0$.

In the case of a "hot" beam, the dielectric constant of such an "active" medium is represented as:
\[
\tilde{\varepsilon}_0(\vec{k}_{\tau},\omega)=\varepsilon_0-\frac{\tilde{\omega}^2_L}{\gamma\omega^2}\frac{xe^{-x^2}\theta^2}
{(\psi_1\cos\varphi+\psi_2\sin\varphi+\psi_{\parallel}/\gamma^2\theta)^2},
\]
where $x=\omega-\vec{k}\vec{u}$, $\psi_1$, $\psi_2$, $\psi_{\parallel}$ are the transverse and longitudinal divergences of the particle velocities in a beam, $\varphi$ is the azimuthal angle of a photon, $\theta$ is the angle between the photon wave vector and the $z$-axis, directed inside the crystal as a normal to the crystal surface.

Thus, the reduction of the analysis of amplification and
generation processes in an X-ray FEL to the solution of a boundary
problem of X-ray dynamical
diffraction by an "active" crystal
target of length $L$ enables one to find generation thresholds of
such a system for different regimes of FEL operation and to
perform the optimization of parameters \cite{berk_138,berk_139}.

The dispersion relation determining the solutions for the electromagnetic wave vector inside the "active" medium in the case of two-wave generation is as follows:
\begin{eqnarray}
\label{berk_4.4}
& &(\omega-\vec{k}\vec{u})^2\left\{(k^2c^2-\omega^2\varepsilon_0)(k^2_{\tau}c^2-\omega^2\varepsilon_0)-
\omega^4g_{\tau}g_{-\tau}\right\}\nonumber\\
& & =-\frac{\tilde{\omega}^2_L}{\gamma}\frac{(\vec{u}\vec{e}_{\sigma})^2}{c^2}(k^2c^2-\omega^2)(k^2_{\tau}c^2-\omega^2\varepsilon_0).
\end{eqnarray}
and the general solution of the set of equations (\ref{berk_4.2}) can be represented in the form
\begin{equation}
\label{berk_4.5}
\vec{E}=\sum\limits^4_{\mu=1}\vec{e}_{\sigma}C_{\mu}e^{i\vec{k}_0\vec{r}}
(1+S_{\mu}e^{i\vec{\tau}\vec{r}})e^{i\delta_{\mu}z},
\end{equation}
where
\[
S_{\mu}=\frac{\omega^2 g_{-\tau}}{k^2_{\mu}c^2-\omega^2\varepsilon_0},\quad \vec{k}_{\mu}=\vec{k}_0\delta_{\mu}\vec{n}_z
\]
are the roots of dispersion equation (\ref{berk_4.4}). The wave amplitudes $C_{\mu}$ are found from the matching conditions for the electromagnetic field (\ref{berk_4.5}) on the boundaries of the crystal target and are given in \cite{berk_139}.

As a result, the generation conditions for the case of a two-wave
distributed feedback and  different cases of "cold" and "hot"
beams as well as for different regimes (weak and high gain
regimes) were obtained.
It was shown that in all cases the generation threshold has a
simple meaning: on the left-hand side of equality there is always
the term describing the "production" of radiation inside the
crystal; on the right-hand side, there are two terms describing
the radiation losses.
Particularly, the first of these {terms refers to the radiation
losses through the crystal boundaries}, while the second term
corresponds to self-absorption of radiation inside the crystal.

The analysis showed that the conditions of generation threshold
are optimal near the region of degeneration of roots of dispersion
equation (\ref{berk_4.4}). This region corresponds to the edge of
a nontransparency region in the dynamical
diffraction theory, and
the interaction of the electromagnetic field with a particle beam
and with a crystal is most effective here.

It should be noted that the condition of the degeneration of
dispersion equation roots leads to the requirement for the photon
radiation angle
$\theta^2=(-\beta_1)^{-1/2}|g_{\tau}|-|g^{\prime}_0|-\gamma^2$,
and this, in turn, gives the restriction  for the possible
geometry of Bragg diffraction in the X-ray range
$
-\frac{r^{\prime}}{|g_0^{\prime}|+\gamma^2}<\beta_1<0,
$
where $r=g_{\tau}g_{-\tau}$.

The estimations of the threshold magnitude of the beam current
density showed that the case of a two-wave solid distributed
feedback is not an optimal case for achieving the generation
regime in the X-ray range. If for a "cold" particle beam in $LiH$
and $\psi_{\perp}<10^{-6}$~rad, $\psi_{\parallel}< 10^{-8}$~ rad,
the threshold current is of the order of $10^9$ ~A/cm$^2$ at
$l\sim0.1$~ cm, then  multiple scattering of electrons makes the
beam "hot" and leads to the increase in the threshold density of a
particle beam up to $j_{th}\geq 10^{10}$ ~A/cm$^2$.

Because of the destructive influence of multiple scattering
process on the quality of a beam of relativistic particles  and,
consequently, on the threshold conditions, the crystal target
length  should be to reduced as much as possible. As is  known
from the optical laser theory, the mirror resonator, similar to a
Fabry-Perot resonator, is used for this purpose. In the X-ray
range, mirrors can be replaced by crystal plates that are oriented
in such a way that the radiation wave vector is under the Bragg
condition. Due to radiation, the generation process takes place in
a narrow angular and spectral interval
$\Delta\omega/\omega\sim\Delta\theta \leq 10^{-5}$; high
effectiveness of the radiation reflection under diffraction
conditions by an external crystal resonator can be obtained under
the corresponding coordination of the resonator and the crystal
target ("active" medium). This allows one to essentially reduce
the radiation losses through target boundaries. In
\cite{berk_138}, the generation threshold for the system with the
external Braggmirrors was derived, and it was shown that the term
connected with losses through the boundaries could be reduced by a
factor of $(1-|R|)$, where $R$ is the reflection coefficient of a
Bragg mirror. As a result, we can shorten the crystal length to
the size necessary for achieving the generation threshold.
However, the estimation showed that the threshold magnitude holds
rather high.

As the analysis in \cite{berk_138} showed, the transition to the
distributed feedback under the surface uncoplanar diffraction when
the radiating particle beam is incident on a crystal at a small
angle $\psi\sim \sqrt{g_0^{\prime}}$ relative to the crystal
surface (see Figure \ref{berkley Figure 30}), allows one to step
down the generation threshold. First of all, the destructive
influence of multiple scattering on the particle beam is
suppressed. Besides, the behavior of dispersion equation roots
changes, which modifies the process of radiation amplification.

 The disadvantage of the case of two-wave diffraction distributed feedback is that the coordination between the
 degeneration condition of dispersion equation roots and the requirement of Cherenkov synchronism hardly fixes
 the geometry of distributed feedback and leads to small magnitudes of the diffraction asymmetry factor $\beta_1$.
 This, in turn, leads to the enhancement of self-absorption of radiation inside the crystal target. In \cite{berk_132} was pointed out that the transition to the multi-wave
 diffraction allows one to modify the functional dependence of the increment of the particle beam instability and, consequently, to step down the threshold density of a beam as well. The dispersion equation for the three-wave coplanar
 diffraction geometry of distributed feedback was obtained, and the rule for writing the dispersion equation for an arbitrary multi-wave diffraction distributed feedback was formulated. In \cite{berk_138}, the expression for the generation threshold in the case of three-wave coplanar diffraction was also derived. It was shown that in this case the Cherenkov condition was fulfilled for two dispersion branches, which  provided  the possibility of the coincidence of diffraction roots with Cherenkov synchronism condition near the exact Bragg condition and, consequently, the possibility of optimization of the threshold magnitude. In the case of the Laue-Bragg diffraction geometry, the threshold density of the beam can be reduced to $j_{th}\sim 10^8$ ~A/cm$^2$,
{at} $\psi \sim 5\cdot 10^{-5}$ ~rad  in the vicinity of the
double degeneration of dispersion equation roots. It should be
noted that even in the case of three-wave generation, it becomes
possible to apply the phenomenon of anomalous X-ray penetration
under diffraction condition and, as a result, to step down
self-absorption of radiation inside the crystal.

Thus, we can conclude that the most suitable geometry for the achievement of the generation regime of quasi-Cherenkov X-ray radiation with the help of relativistic electron (positron) beams in crystals is the grazing geometry of the particle beam incidence on a target with the distributed feedback formed by multi-wave surface diffraction.

The spectral-angular distribution of the coherent PXR near the generation threshold was obtained in \cite{berk_139}.

In  \cite{berk_136}, the spectral-angular distribution of coherent
radiation far from the generation threshold was derived within the
framework of the perturbation theory, and the possibility of
experimental observation of the coherent PXR in existing
accelerators was analyzed. It was shown that observation of
coherent parametric (quasi-Cherenkov) radiation far from the
generation threshold was a very complicated problem for the X-ray
range, but it was possible to observe the coherent radiation in an
optical range  nowadays.

\section{The X-ray generator on the basis of diffracted channeling radiation (DCR)}
\label{berk_s9}

The second type of a crystal generator is based on the application
of diffracted radiation by the oscillator (DRO) as a spontaneous
radiation mechanism
\cite{berk_140,berk_141,berk_142,berk_143,berk_144}. As stated
above, the radiating oscillator can be formed in  different ways.
This can be electrons channeled in an averaged crystallographic
potential of planes or axes, or electrons moving in an
electrostatic wiggler \cite{berk_142} or, for example, an
oscillator formed by an external ultrasonic (optical) wave in a
crystal  \cite{berk_144}. It is obvious that the general approach
to the consideration of the generation problem with the help of a
relativistic oscillator beam does not depend on the formation
mechanism of the oscillator itself. Since the oscillator is a
quantum system, it is more accurate to perform the calculation of
the polarizability tensor of a particle beam  within the framework
of quantum electrodynamics. Reduction of the problem of radiation
application (generation) by a particle beam in a finite crystal
target to the problem of diffraction of X-rays by an "active"
medium, consisting of a crystal and a beam of radiating
oscillator, holds true in this case as well.

The expression for the polarizability of such an "active" medium
in the case of channeled particles in unperturbed averaged crystal
potential was obtained in  \cite{berk_141}:
\begin{equation}
\label{berk_4.7}
\tilde{\varepsilon}_0(\vec{k},\omega)=\varepsilon_0-\frac{\tilde{\omega}^2_L}{\gamma\omega^2}-\frac{4\pi e^2 n_0}{\omega^2}(W_2-W_1)\frac{\left|\vec{a}(\vec{k})_{21}\vec{e}_{\sigma}\right|^2}{\omega-\vec{u}\vec{k}-\Omega_{21}+i\Gamma};
\end{equation}
where $\vec{\alpha}_{21}(\vec{k})$ is the matrix element of the operator $\hat{\alpha}\exp(i\vec{k}\vec{r})$, which in the dipole approximation takes the form
\[
\vec{\alpha}(\vec{k})_{21}=-ix_{21}(\Omega_{21}\vec{n}_x+k_x\vec{u}_z).
\]
The axis $\vec{n}_x$ is chosen so that it lies along the
transverse particle oscillations in a channel,
$\vec{u}_{\parallel}$ is the longitudinal velocity parallel to the
channeling planes, ($k_x=\vec{k}\vec{n}_x$,
$\vec{u}_{\parallel}\vec{n}_x=0$), $\Omega_{21}$ is the frequency
of the transition, $\vec{e}_{\sigma}\parallel[\vec{k}\vec{\tau}]$,
$W_1$ and $W_2$ are the population of the states  1 and 2,
$\Gamma$ is the phenomenological constant, taking into account
inelastic oscillations; its order of  magnitude estimate gives
$(L_d)^{-1}$, where $L_d$ is the dechanneling length. In obtaining
(\ref{berk_4.7}), it was taken into account that the synchronism
condition could be fulfilled only for the wave propagating at a
small angle relative to the longitudinal velocity of the particle.
The fulfillment of the synchronism condition for the diffracted
wave is impossible in the X-ray range. As the analysis showed
\cite{berk_141}, although there were a lot of zones (states) of
transverse energy of a channeled particle, the main contribution
to the polarizability tensor was made by a certain transition with
the frequency $\Omega_{21}$. This means that the consideration is
reduced to the two-level problem.

Indeed, the contribution to the beam polarizability from the transition between the levels $m$ and $n$ is determined by the deviation from the exact synchronism condition of the radiation field with the oscillator, i.e., $\texttt{Re}(\omega-\vec{u}\vec{k}-\Omega_{mn})=0$. This contribution should be taken into account only if
\begin{equation}
\label{berk_4.8}
\left|\texttt{Re}(\omega-\vec{u}\vec{k}-\Omega_{mn})\leq\left| \texttt{Im}(\vec{k}\vec{u}-\omega)+\Gamma_{mn}\right|\right|
\end{equation}
If the magnitudes of $\texttt{Re}\omega$ and the angle between $\vec{u}$ and $\vec{k}$ are fixed, the number of transitions contributing to the polarizability depends on the relationship between $\Delta\Omega$ and $\left| \texttt{Im}(\vec{k}\vec{u}-\omega)+\Gamma_{mn}\right|$, where $\Delta\Omega$ determines the typical value of the difference  $\Omega_{n+1,n}-\Omega_{n,n-1}$ which characterizes the unharmonism  of the averaged potential (for a harmonic potential $\Delta\Omega=0$). The analysis of the magnitude of $\Delta\Omega$ for different kinds of averaged crystallographic plane potentials shows that
\[
\Delta\Omega\gg\left|\texttt{Im}(\vec{k}\vec{u}-\omega)+\Gamma_{mn}\right|,
\]
and, consequently, the synchronism condition can be fulfilled only
for a certain transition $\Omega_{21}$. Other terms in a
polarizability tensor can be neglected as nonresonant. It was
shown that the most effective interaction between the oscillator
beam and the radiated wave takes place near the degeneration
region of roots of the dispersion equation determining the
eigenstates of the field in an "active" medium. But, as contrasted to the parametric
(quasi-Cherenkov) generator, for which the radiation condition is
realized only at large deviation from the exact Bragg condition,
now there is a possibility to overlap the synchronism condition
with the exact diffraction condition. As a result, in the case
under consideration the manifestation of the effect of anomalous
X-ray penetration through the resonator under dynamic diffraction
(Borman effect) is possible. This circumstance is very important
because of strong absorption of X-rays inside a crystal target. In
\cite{berk_141} the boundary problem of X-ray diffraction by an
"active" medium of a finite size was solved, and the generation
condition was obtained. It was shown that the beam can be in
synchronism with one of the modes of the "active" medium.
These modes correspond to the waves with the wave vectors {being
the roots} ($\delta_1$ and $\delta_2$) of the dispersion equation.
According to \cite{berk_141}, the generation condition can be
realized in two cases: for the wave corresponding to the root
$\delta_2$ at the positive magnitude of $\alpha=\alpha_+$ and for
the wave corresponding to the root $\delta_1$ for the negative
deviation from the exact Bragg condition $\alpha=\alpha_-$. It was
shown that the solutions of the generation equation for different
modes are identical in structure. All of them lead to the phase
condition
\begin{equation}
\label{berk_4.9}
\delta_1^{(0)}-\delta_2^{(0)}=\frac{2\pi n}{\omega L}
\end{equation}
where $\delta_1^{(0)}$ and $\delta_2^{(0)}$ are the solutions of
the diffraction dispersion equation.
The conditions of
generation are written for the case of a channeled particle
in \cite{berk_141} and for the case of electrostatic and
magnetostatic {wiggler} in \cite{berk_142}.

If the condition (\ref{berk_4.9}) is fulfilled, the longitudinal
structure of the modes turns out to be close to the structure of a
standing wave. That is, $|E|^2$ and $|E_{\tau}|^2$ are
proportional to
$
\sim \cos^2\frac{2\pi n}{\omega L}(z-L).
 $
This condition is similar to a well known phase condition of the
stand wave appearance in a mirror resonator of an ordinary laser
\cite{berk_144}. The meaning of amplitude conditions is the same
as  in the case of the quasi-Cherenkov X-ray generator. The field amplification, due to the
radiation process, should be equal to the radiation losses caused
by absorption inside the crystal and the output of radiation
through the boundaries of the crystal target. Because the gain in
the weak-gain regime is proportional to the current density of a
beam, the formula for the threshold gain sets requirements for the
current density.  The invariant characteristics of a particle beam
are often used instead of the current density, that is, the
current  $I$, the normalized emittance $\varepsilon_n=\gamma r
\langle\psi\rangle$  and the normalized brightness
$B_n=I/\pi^2\varepsilon_n^2$, where $r$ is the beam radius,
$\langle\psi\rangle$ is the angular spread. The angular  spread
corresponds  to the divergence of the longitudinal velocity
$\sigma_{\varepsilon}\cong u\langle\psi^2\rangle/2$, and the
corresponding divergence of the particle energy is
$
\left(\frac{\Delta\gamma}{\gamma}\right)_e=\gamma^2_{\parallel}\frac{\langle\psi^2\rangle}{2}.
$
For a $LiH$ crystal, the diffraction plane (220) the values of the threshold normalized brightness of the beam, which correspond to the generation threshold for magnetic, optical undulators and channeled particles, are given in Table \ref{berk_tbl1.1}. According to Table \ref{berk_tbl1.1}, the value of brightness in the case of two-wave distributed feedback is rather high. But, as it was shown for a parametric quasi-Cherenkov generator, using the surface multi-wave
diffraction for the formation of disturbed feedback, one can decrease the threshold characteristics of a particle beam and provide the achievement of the generation regime.

\begin{table}[htp]
\caption{}
\begin{center}
\begin{tabular} {|c|c|c|c|}\hline
Parameters &  Magnetostatic  & Optical & Channeled\\
&Wiggler& Wiggler& Particle\\
\hline
\textbf{Accelerator} &  & &      \\
Energy & $=5$ GeV & $=290$ MeV & $=500$ MeV \\
Normalized brightness & $=3.5 \cdot 10^9$ & $=1.7\cdot 10^{10}$ & $=5\cdot 10^9$\\
Energy spread &$ =2.4\cdot 10^{-3}$ & $1.2\cdot 10^{-5}$ & \\
Density of current & $=5.3\cdot 10^7$ & $=1.3\cdot 10^6$ & $=3.3\cdot 10^8$\\
\textbf{Wiggler} &  & & \\
Wavelength& $=1$ mm & $=5\, \mu m$&\\
Magnetic field strength &$ =17.5$ kG & &\\
Laser energy & & $=0.75$ gW & \\
\textbf{Crystal} &  & &\\
Wavelength of radiation & $=0.05$ {\AA} & $=0.15$ {\AA} & $=1$ {\AA}\\
Asymmetry parameter & $ =9$ &$ =1$ & \\
Diffraction plane & (220) & (220) & (100)\\
\hline
\end{tabular}
\end{center}
\label{berk_tbl1.1}
\end{table}

In \cite{berk_141} the underthreshold spectral-angular
distribution of radiation was analyzed and it was shown that the
observation of collective radiation by relativistic oscillators
was a very complicated problem in the X-ray spectral range.

Thus, the parallel consideration of two kinds of crystal
three-dimensional X-ray generators, which are distinguished by the
mechanisms of spontaneous radiation, shows that three-dimensional
distributed feedback allows one to decrease the current density of
the particle beam by several orders of magnitude in comparison
with the results obtained in
\cite{berk_145,berk_146,berk_147,berk_148,berk_149}. This enables
one to consider the construction of the FEL in the hard X-ray
range as a scientific problem of nowadays, which can be analyzed
both theoretically and experimentally.

\section{Crystal X-ray FEL based on a natural crystal or an electromagnetic undulator}
\label{vesti_sec:1}

Crystal X-ray FELs based on a natural crystal and on an
electromagnetic undulator were first considered in
\cite{berk_144}. The English version of this paper is given below.

Let a relativistic electron (or positron) beam of velocity $\vec{u}_0$ and the beam velocity distribution $\Delta\vec{u}$ move in a spatially periodic medium (e.g., in a crystal). Let a linearly polarized laser pump wave be incident onto the beam along the direction $\vec{n}_{p}$. Let the wave have the wave vector $\vec{k}_{p}=\vec{k}_{p}n$, frequency $\omega_{p}$, and the field strength $\vec{E}_{p}=\vec{E}_{p}^0\cos(\vec{k}_{p}\vec{r}-\omega_{p}t+\delta$), where $\delta$ is the initial vibration phase. The $z$-axis is chosen directed along the beam's average velocity $\vec{u}_0$. The presence of the electromagnetic pump wave induces radiation leading to various sorts of instability. The most known of them is the so-called the three-wave parametric instability, which emerges due to the conversion of the pump wave into the Doppler-shifted electromagnetic wave and the charge density wave: $\omega_{p}=\omega+\omega_{ch}$, $\vec{k}_{p}=\vec{k}+\vec{k}_{ch}$ ($\omega$, $\vec{k}$ are the frequency and the wave vector of the high-frequency wave, respectively, $\omega_{ch}$, $\vec{k}_{ch}$ are the frequency and the wave vector of a charge density wave). This process is scrutinized for the case of  homogeneous matter \cite{vesti_12}.

In the case of a 3 D-periodic medium that we are considering here, the situation is basically different from that studied earlier because for $\omega$ and $\vec{k}$ satisfying the Bragg condition, the wave scattered by the beam is diffracted.

Maxwell's equations and the equations of particle motion, which describe the process under study, have the form:
\begin{eqnarray}
\label{vesti_1.1}
& &\mbox{rot}\vec{H}=\frac{1}{c}\frac{\partial\vec{D}}{\partial t}+\frac{4\pi}{c}\vec{j},\qquad \vec{D}(\vec{r},t)=\int\limits_{-\infty}^{\infty}\varepsilon(\vec{r},t-t^{\prime})
\vec{E}(\vec{r},t^{\prime})dt^{\prime},\nonumber\\
& &\mbox{rot}\vec{E}=-\frac{1}{c}\frac{\partial\vec{H}}{\partial t},\quad\qquad \qquad \mbox{div}\vec{E}=4\pi\rho,\qquad \frac{\partial\rho}{\partial t}+\mbox{div}\vec{j}=0,\nonumber\\
& &\vec{j}(\vec{r},t) = e\sum\limits_i\vec{v}_i(t)\delta(\vec{r}-\vec{r}_i(t)),\\
& &\rho(\vec{r},t)=e\sum\limits_i\delta(\vec{r}-\vec{r}_i(t)),\nonumber\\
& &\frac{d\vec{v}_i(t)}{dt}=\frac{e}{m\gamma_i}\vec{E}(\vec{r}_i(t),t)+
\frac{e}{mc\gamma_i}\left[\vec{v}_i(t)\vec{H}(\vec{r}_i(t),t)\right]\nonumber\\
& &-\frac{e}{mc^2\gamma_i}(\vec{v}_i(t)(\vec{v}_i(t)\vec{E}(\vec{r}_i(t),t))),\nonumber
\end{eqnarray}
where $\vec{E}(\vec{r}_i(t)),t)$, $\vec{H}(\vec{r}_i(t),t)$ are the electric and magnetic field strengths at moment $t$ at the location point of the $i$-th electron $\vec{r}_i(t)$, $\vec{D}$ is the electric induction, $\varepsilon(\vec{r},t)$ is the dielectric permittivity of the crystal at point $\vec{r}$, $\varepsilon(\vec{r},t)=0$ at $t<0$, $\vec{v}_i(t)$ is the velocity of the $i$-th electron of the beam at time $t$, $\vec{j}(r,t)$ is the beam's current density at point $\vec{r}$ at time $t$, $\rho(\vec{r},t)$ is the density of the beam's electron charge, $e$ is the electron charge of the particle ($e=\pm|e|$), $\gamma_i=(1-v^2/c^2)^{-1/2}$ is the particle Lorentz factor, $c$ is the velocity of light.

For the Fourier transforms $\vec{D}(\vec{r},\omega)$ and $\vec{E}(\vec{r},\omega)$, we have the relation $\vec{D}(\vec{r},\omega)=\varepsilon(\vec{r},\omega)\vec{E}(\vec{r},\omega)$, where the crystal's spatially periodic dielectric permittivity can be represented as the Fourier series: $\varepsilon(\vec{r},\omega)=\sum\varepsilon_{\tau}(\omega)e^{i\vec{\tau}r}$, $\vec{\tau}$ is the reciprocal lattice vector. Studying the possibility of occurrence of induced radiation, one should first of all determine the gain coefficient or, which is the same, the beam instability increment. To find them, in solving (\ref{vesti_1.1}) we may confine ourselves to linear approximation, finding the dielectric permittivity of the beam+crystal system in the presence of the given pump wave. Take the Fourier transform in time of (\ref{vesti_1.1}). To be more specific, let us further consider the case when a single diffracted wave $\vec{k}_{\tau}=\vec{k}+\vec{\tau}$ appears through diffraction. Let us also assume that the scattering plane of the wave, i.e., the plane defined by vectors $\vec{k}_{p}$ and $\vec{k}$ coincides with the diffraction plane defined by $\vec{k}$, $\vec{k}_{\tau}$. In a similar way as in the case of dynamical diffraction of X-rays, the generation of $\pi$- and $\sigma$-polarized waves in the crystal in the absence of the beam can be considered separately. Recall that the polarization vector of $\sigma$-polarized waves is orthogonal to the diffraction plane
\[
\vec{e}_{\sigma}\parallel\vec{e}_0^{\vec{\tau}}\parallel\frac{[\vec{k}\vec{\tau}]}{|[\vec{k}\vec{\tau}]|}.
\]
We also assume that the polarization vector $\vec{e}_{p}$ of the pump wave is parallel to $\vec{e}_{\sigma}$. In this case, upon taking the Fourier transform, (\ref{vesti_1.1}) takes the form:
\begin{eqnarray}
\label{vesti_1.2}
\left\{\begin{array}{l}
(k^2c^2-\varepsilon_0\omega^2)E_{\sigma}(\vec{k},\omega)-
\varepsilon_{\tau}\omega^2E_{\sigma}(\vec{K}_{\tau},\omega)=4\pi i\omega j_{\sigma}(\vec{k},\omega), \\
(k_{\tau}^2c^2-\varepsilon_0\omega^2)E_{\sigma}(\vec{k}_{\tau},\omega)-
\varepsilon_{-\tau}\omega^2E_{\sigma}(\vec{k},\omega)=0,\\
\omega E_{\parallel}(\vec{k},\omega)=-4\pi ij_{\parallel}(\vec{k},\omega),
\end{array}\right.
\end{eqnarray}
where $\parallel$ denotes the component of the field and current
strength parallel to the wave vector $\vec{k}$,
$\vec{E}_{\sigma}=(\vec{E}\vec{e}^{\sigma})$. In writing
(\ref{vesti_1.2}), the components
$j_{\sigma}(\vec{k}_{\tau},\omega)$ and
$j_{\parallel}(\vec{k}_{\tau},\omega)$ of the current, for which
the synchronism condition does not hold in the X-ray range  and
whose contribution is therefore small, are discarded. Without
channeling, the particle velocity in the beam can be represented
as a sum:
\[
\vec{v}_i(t)=\vec{u}_i+\vec{v}_{p i}(t)+\delta\vec{v}_{i\sigma}(t)+\delta\vec{v}_{i\parallel}(t),
\]
where $\vec{u}_i$ is the particle's longitudinal velocity in the absence of the electromagnetic wave.
\[
\vec{v}_{p i}(t)=-\frac{e\vec{E}_{p}^0}{m\gamma_i\omega_{p}}\sin(\vec{k}_{p}\vec{r}_{0i}
-\overline{\omega}_{p}t+\delta)\equiv\vec{v}_{p}^0\sin(\vec{k}_{p}\vec{r}_{0i}-\overline{\omega}_{p}t+\delta)
\]
is the velocity of the $i$-th particle under the external electromagnetic pump wave, $\overline{\omega}_{p}=\omega_{p}-(\vec{k}_{p}\vec{u}_i)$, $\delta\vec{v}_{i\sigma}(t)$ is the velocity disturbance of the $i$-th particle  in the field of a transverse scattered electromagnetic wave, $\delta\vec{v}_{i\parallel}(t)$ is the velocity disturbance induced by a longitudinal wave. The position of the $i$-th particle at time $t$ is written accordingly $\vec{r}_i(t)=\vec{r}_{0i}+\vec{u}_i(t)+\delta\vec{r}_i(t)$, where $\vec{r}_{0i}$ is the position of the $i$-th particle $(e^{\pm})$ in the beam at time $t=0$, $\delta\vec{r}_i(t)$ is the time change of the position of the $i$-th $(e^{\pm})$ in the beam.

The Fourier components of the transverse and the longitudinal current in the linear approximation with respect to perturbation  can be presented as follows:
\begin{eqnarray}
\label{vesti_1.3}
& &j_{\sigma}(\vec{k},\omega)=\frac{i(\vec{e}_{\sigma}\langle\vec{u}_i\rangle)}{4\pi}(\vec{k}\vec{E}(k,\omega))+
\left\langle\frac{i(\vec{e}_{\sigma}\vec{v}_{p}^0)}{4\pi}((\vec{k}-\vec{k}_{p})\vec{E}(\vec{k}-\vec{k}_{p}, \omega-\omega_{p}))\right.\nonumber\\
& &\left.+e\langle\sum\limits_i\delta v_{i\sigma}(\omega-\vec{k}\vec{u}_i)e^{-i\vec{k}\vec{r}_{0i}}\right>,
\end{eqnarray}
\begin{eqnarray}
\label{vesti_1.4}
j_{\parallel}(\vec{k},\omega)=\frac{i(\vec{k}\langle\vec{u}_i\rangle)}{4\pi}E_{\parallel}(\vec{k},\omega)+
e\langle\sum\limits_i\delta v_{i\parallel}(\omega-\vec{k}\vec{u}_i)e^{-i\vec{k}\vec{r}_{0i}}\rangle\nonumber\\
+\left\{\begin{array}{ll}
   e(\vec{v}_{p}^0\vec{n}_k)n_0, & \vec{k}=\vec{k}_{p}, \\
   \\
   \frac{i(\vec{v}_{p}^0\vec{n}_k)}{4\pi}((\vec{k}-\vec{k}_{p})\vec{E}(\vec{k}-\vec{k}_{p},\omega-\omega_{p})), &\vec{k}\neq\vec{k}_{p}.
 \end{array}\right.
\end{eqnarray}
$\langle\quad\rangle$ denotes averaging over the velocity distribution in the beam.

To obtain a closed system of equation, let us take the Fourier transform of the equation of particle motion (\ref{vesti_1.1})
and separate the transverse components of the velocity and the perturbation field from the longitudinal ones. The linear approximation to the disturbance of the particle velocity gives
\begin{eqnarray}
\label{vesti_1.5}
& &\langle\sum\limits_i\delta v_{i\sigma}(\overline{\omega})e^{-i\vec{k}\vec{r}_{0i}}\rangle=
\left\langle\frac{ie^2n_0}{m\omega\gamma_i}\delta E_{\sigma}(\vec{k},\omega)\right\rangle+\left\langle\frac{ie^2n_0}{4m\overline{\omega}\gamma_i}\right.\nonumber\\
& &\times\left\{\left[\frac{(\vec{e}_{\sigma}\vec{k}_{p})}{\omega_{p}-\omega}+(\vec{v}_{p}^0\vec{e}_{\sigma})
-\frac{2(\vec{v}_{p}^0(\vec{k}-\vec{k}_{p}))}{\omega-\omega_{p}}\right]\delta E_{\sigma}(\vec{k}-\vec{k}_{p},\omega-\omega_{p})\right.\\
& &\left.+2(\vec{v}_{p}^0\vec{e}_{\sigma})\langle(\vec{n}_k\vec{v}_i)\rangle\delta E_{\parallel}(\vec{k}-\vec{k}_{p},\omega-\omega_{p})\right\}-\frac{ie^2n_0}{mc^2\overline{\omega}\gamma_i}
\left[(\vec{u}_i\vec{e}_{\sigma})^2\delta E_{\sigma}(\vec{k},\omega)\right.\nonumber\\
& &\left.\left.+(\vec{u}_i+\vec{e}_{\sigma})(\vec{u}_i\vec{n}_K)\delta E_{\parallel}(\vec{k},\omega)\right]\right\rangle,\nonumber
\end{eqnarray}
\begin{eqnarray}
\label{vesti_1.6}
& &\left\langle\sum\limits_i\delta v_{i\parallel}(\overline{\omega})e^{-i\vec{k}\vec{r}_{0i}}\right\rangle=
\left\langle\frac{ien_0(1-(\vec{u}_i\vec{n}_k)^2)}{m\overline{\omega}\gamma_i}\delta E_{\parallel}(\vec{k},\omega)\right.\nonumber\\
& &+\frac{ien_0}{m\overline{\omega}\gamma_i}\left[\frac{k}{\omega}-\frac{(\vec{u}_i\vec{n}_k)}{c^2}\right]
(\vec{u}_i\vec{e}_{\sigma})
\delta E_{\sigma}(\vec{k},\omega)+\frac{ien_0}{mc\overline{\omega}\gamma_i}\\
& &\times\left[\frac{\vec{k}_{p}-\vec{k}}{\omega_{p}-\omega}-\frac{(\vec{u}_i\vec{n}_{k_{p}-k})}{c}\right]
(\vec{v}_{p}^0\vec{e}_{\sigma})\delta E_{\sigma}(\vec{k}-\vec{k}_{p}, \omega-\omega_{p}),\nonumber
\end{eqnarray}
where $\overline{\omega}=\omega-\vec{k}\vec{u}_i$, $\langle\ldots\rangle$ is averaging over $\vec{u}$.

Substitution of the obtained expressions for the velocity disturbance into (\ref{vesti_1.3}), (\ref{vesti_1.4}) and then into (\ref{vesti_1.2}) gives a closed system of equations for transverse and longitudinal fields under two-wave dynamical diffraction of scattered radiation:
\begin{eqnarray}
\label{vesti_1.7}
& &(k^2c^2-\varepsilon_0\omega^2-\langle\omega^2_{p\perp}\rangle)\delta E_{\sigma}(\vec{k},\omega)-\varepsilon_{\tau}\omega^2\delta E_{\sigma}(\vec{k}_{\tau},\omega)\nonumber\\
& &=G_1\delta E_{\sigma}(\vec{k},\omega)+G_2\delta E_{\sigma}(\vec{k},\omega),\\
& &(k^2_{\tau}c^2-\varepsilon_0\omega^2-\langle\omega^2_{p\perp}\rangle)\delta E_{\sigma}(\vec{k}_{\tau},\omega)-\varepsilon_{-\tau}\omega^2\delta E_{\sigma}(\vec{k},\omega)=0,\nonumber
\end{eqnarray}
where
\[
\langle\omega^2_{p\perp}\rangle=\frac{\omega^2_{L}}{\langle\gamma\rangle}=\frac{4\pi e^2n_0}{m_e\langle\gamma\rangle},
\quad \omega^2_{p\parallel}=\frac{\omega^2_{L}}{\langle\gamma\rangle^3},
\]
$n_0$ is the unperturbed particle density in the beam, $m_e$ is the mass of $e^{\pm}$,
\begin{eqnarray}
\label{vesti_1.8}
G_1&=&\left\langle\frac{\omega^2_{p\perp}(\vec{e}_{\sigma}\vec{u})^2}{4c^2(\overline{\omega}^2-
\omega^2_{p\parallel})}(\omega^2-k^2c^2)\right\rangle,\\
G_2&=&\left\langle\frac{(\vec{v}_{p}^0\vec{e}_{\sigma})^2\omega^2_{p\perp}((\vec{k}-
\vec{k}_{p})\vec{n}_k)}{4c^2[(\overline{\omega}
-\overline{\omega}_{p})^2-\omega^2_{p\parallel}]}\left[(\vec{k}_{p}\vec{n}_{k_p-k})c-\omega_{p}(\vec{\beta}\vec{n}_{k_p-k})\right.\right.\nonumber\\
& &\left.\left.+
(kc-\omega(\vec{\beta}\vec{n}_k))\right]\right\rangle.\nonumber
\end{eqnarray}
The determinant of (\ref{vesti_1.7})  defines the dispersion equation that specifies the relation between $\vec{k}$ and $\omega$ and enables finding the instability increment of the beam. We shall describe the beam by Maxwell's velocity distribution $f(\vec{u})$ with the effective temperatures $T_1$, $T_2$, $T_3$ over the $x$, $y$ , $z$ axes, respectively or, which is the same, with thermal-induced velocity dispersions  characterized by average thermal velocities 
$
v_{\alpha T}=\sqrt{\frac{2T_{\alpha}}{m}}:
$
\begin{equation}
\label{vesti_1.9}
f(\vec{u})=\frac{1}{\pi^{3/2}v_{1T}v_{2T}v_{3T}}e^{^-\frac{(u_x-u_{0x})^2}{v^2_{1T}}}e^{^-\frac{(u_y-u_{0y})^2}{v^2_{2T}}}
e^{^-\frac{(u_z-u_{0z})^2}{v^2_{3T}}}.
\end{equation}
Upon averaging (\ref{vesti_1.8}) with the function (\ref{vesti_1.9}), we obtain
\begin{equation}
\label{vesti_1.10}
G_1=-\frac{\omega^2_{p\perp}(\vec{e}_{\sigma}\vec{u}_0)^2(\omega^2-k^2c^2)}{2c^2\Omega^2}
\left\{1+i\frac{\sqrt{\pi}\overline{\omega}}{\Omega}W\left(\frac{\overline{\omega}}{\Omega}\right)\right\},
\end{equation}
\begin{eqnarray}
\label{vesti_1.11}
& &G_2=-\frac{(\vec{v}_{p}^0\vec{e}_{\sigma})^2\omega^2_{p\perp}((\vec{k}-
\vec{k}_{p})\vec{n}_k)}{2c^2\Omega^2_1}\left[((\vec{k}_{p}\vec{n}_{k_p-k})
c-\omega_{p}(\vec{\beta}\vec{n}_{k_p-k}))\right.\nonumber\\
& &\left.+(kc-\omega(\vec{\beta}\vec{n}_k))\right]
\left\{1+i\frac{\sqrt{\pi}(\overline{\omega}-\overline{\omega}_{p})}{\Omega_1}
W\left(\frac{\overline{\omega}-\overline{\omega}_{p}}{\Omega_1}\right)\right\},
\end{eqnarray}

where
$
W(\xi)=\left(1+\frac{2i}{\sqrt{\pi}}\int\limits^{\xi}_0e^{x^2}dx\right)e^{-\xi^2}
$
is the Kramp function (it is tabulated in \cite{vesti_13}), $\overline{\omega}=\omega-\vec{k}\vec{u}_0$, $\overline{\omega}_{p}=\omega_{p}-\vec{k}_{p}\vec{u}_0$, $\Omega^2=k_x^2v^2_{1T}+k_y^2v^2_{2T}+k_z^2v^2_{3T}$, $\Omega^2_1=(k_x-k_{px})^2v^2_{1T}+(k_y-k_{py})^2v^2_{2T}+(k_z-k_{pz})^2v^2_{3T}$.

Equations (\ref{vesti_1.7}) together with functions (\ref{vesti_1.10}), (\ref{vesti_1.11}) determine the parametric Cherenkov instability and parametric decay instability of the thermal beam. From (\ref{vesti_1.10}), (\ref{vesti_1.11}) follows that in the case when the parameters $\omega$ and $\vec{k}$ are such that $\zeta\gg 1$, the velocity distribution in the beam appears insignificant, and so one can analyze an appreciably simpler case of a cold beam. Since in radiation of a relativistic particle quanta are emitted in a narrow angular range in the direction of the particle motion, then $k_x$, $k_y\ll k_z$. As a result, at the given values of $\Omega$ and $\Omega_1$, the beam's velocity distribution in transverse direction can exceed that in the longitudinal direction.  Of particular interest is the radiation pattern for parametric decay instability, in which case there is a possibility in principle to make the differences $k-x-k_{px}$ and $k_y-k_{yp}$, appearing in the expression for $\Omega_1$, go to zero. The transverse velocity distribution in this case appears to be insufficient at all.

In a real situation, due to the finite dimensions of the system and non-monochromaticity of radiation, these differences cannot vanish, though they can be reduced appreciably. Let us estimate the maximum possible magnitude of the increment. Suppose that the parameter $\zeta\gg 1$. As stated above, in this case we go over to the approximation of a cold beam and can use the expression  $f(\vec{u})=\delta(\vec{u}-\vec{u}_0)$ for the distribution function. The parametric Cherenkov instability for this case was considered in \cite{berk_135}
 (see Section above). The term proportional to $G_2$ leads to parametric decay instability of the beam in a periodic medium. In the general case, the Cherenkov and decay instabilities develop at different frequencies and can therefore be considered separately. So here we shall focus on the decay instability, omitting the term containing $G_1$ in (\ref{vesti_1.7}) for corresponding $\vec{k}$ and $\omega$. In the case of a cold beam, using (\ref{vesti_1.7}), (\ref{vesti_1.8}), we have the following equation for decay instability:
\begin{equation}
\label{vesti_1.12}
[(k^2c^2-\omega^2\varepsilon_0)(k^2_{\tau}c^2-\omega^2\varepsilon_0)-\omega^4\varepsilon_{\tau}\varepsilon_{-\tau}]
(\overline{\omega}-\overline{\omega}_{p})^2=A_2(k^2_{\tau}c^2-\omega^2\varepsilon_0),
\end{equation}
where $A_2$ is the factor of the term $[(\overline{\omega}-\overline{\omega}_{p})^2-\omega^2_{p^\prime\prime}]^{-1}$ in $G_2$; $\omega^2_{p^\prime\prime}$ is dropped as it is small for the particle densities $n_0\simeq 10^{15}\div 10^{17}$ cm$^{-1}$ in the beam that are of interest to us.

The solution of dispersion equation (\ref{vesti_1.12}) can be found in the weakly-coupled waves approximation, which is applicable in this case because  a non-linear right-hand side of (\ref{vesti_1.12}) is small. In this approximation, the solution of dispersion equation (\ref{vesti_1.12}) is sought near the intersection of solutions of the dispersion equations for coupled waves into which (\ref{vesti_1.12}) is split  when its right-hand side equals zero, i.e.,
\begin{eqnarray}
\label{vesti_1.13}
\left\{
\begin{array}{l}
  (k^2c^2-\omega^2\varepsilon_0)(k^2_{\tau}c^2-\omega^2\varepsilon_0)-
\omega^4\varepsilon_{\tau}\varepsilon_{-\tau}=D(k_z,\vec{k}_{\perp},\omega)=0, \\
((\omega-\vec{k}\vec{u}_0)-\omega_{p})^2=0.
\end{array}\right.
\end{eqnarray}

Using weakly-coupled waves approximation of the perturbation theory  and assuming that $\varepsilon_0$, $\varepsilon_{\pm\tau}$ are real values (i.e., neglecting the intrinsic absorption of the medium), the upper estimate of the increment of the beam's parametric decay instability when a pump wave is scattered in a 3D periodic medium can be obtained from (\ref{vesti_1.12}):
\begin{equation}
\label{vesti_1.14}
\eta\simeq \frac{\omega_{B}}{c}\sqrt[4]{\frac{\omega^2_{p\perp}}{4\omega^2_{B}}
\frac{|\vec{v}_{p}^0|^2}{c^2}\frac{\overline{\omega}_{p}}{\omega_{B}}|\varepsilon_{\tau}|},
\end{equation}
where
$
\frac{|v_{p}|^2}{c^2}=\frac{e^2|E^0_{p}|^2}{m_e^2c^2\langle\gamma\rangle^2\omega^2_{p}},
$
$\omega_{B}$ is the Bragg frequency. Equation (\ref{vesti_1.14}) is obtained in the assumption that the synchronism condition is fulfilled exactly and that $D(k_z,\vec{k}_{\perp},\omega)=0$ in the degeneration point of the eigensolutions of the dispersion equation of diffraction. In this case the multiplicity of the root, determining the increment, is greater by one than in the case of modified decay in the absence of diffraction of a scattered wave. This results in the decrease in the  instability threshold, depending on the beam density and the intensity of the external electromagnetic pump wave \cite{vesti_8,vesti_9}. According to (\ref{vesti_1.14}), the numerical estimate of the maximum increment of the parametric decay instability  of a cold relativistic beam is as follows: $\eta\sim 0.5$ at $E_{p}^0\sim 10^5$ in CGS, $\gamma=2\cdot 10^2$, $\omega_{\sigma}=1.6\cdot 10^{19}$ s$^{-1}$, $\omega_{p}=10^{14}$ ~s$^{-1}$ $n_0\sim 10^{15}$ ~cm$^{-3}$, $\varepsilon_0-1=5\cdot 10^{-6}$, $\varepsilon_{\tau}\simeq 10^{-6}$.

In the above theoretical analysis we did not actually use the
explicit form of $v_{p}^0$ and the direct relation between
$\omega_{p}$ and $\vec{k}_{p}$. This theory is therefore fully
applicable to the case when, instead of the electromagnetic wave,
crystal planes bent under the ultrasonic wave act as a dynamic
wiggler. A detailed treatment of the particle motion in such an
ultrasonic wiggler was given in \cite{chan_67}. The corresponding
dispersion equation is obtained from (\ref{vesti_1.7}),
(\ref{vesti_1.8}), (\ref{vesti_1.12}) by substituting for
$v_{p}^0$ the corresponding expression for the case of the
ultrasonic wiggler
$|\vec{v}^s_{p}|=(\vec{e}_{\sigma}\vec{r}_{0\perp}^s)\Omega_s^{\prime}$
(where $\Omega_s^{\prime}=\kappa_zu_0-\Omega_s$, $\vec{\kappa}$ is
the wave vector of the ultrasonic wiggler, $\Omega_s$ is its
frequency, $\vec{r}_{0\perp}^s$ is the amplitude of the particle's
transverse vibrations in the ultrasonic wiggler \cite{chan_67})
and  by substituting $\vec{\kappa}$ for $k_{p}$; $\Omega_s$, for
$\omega_{p}$;  $\Omega_s^{\prime}$, for $\overline{\omega}_{p}$.
Upon this substitution,  for the instability increment in an
ultrasonic wiggler, we obtain the expression similar to
(\ref{vesti_1.14}). This enables readily
obtaining the following
ratio of the instability increment in  an ultrasonic wiggler to
that in a light wiggler:
\begin{equation}
\label{vesti_1.15}
R=\frac{\eta_s}{\eta}=\left(\frac{|v_{p}^s|^2\Omega_s^{\prime}}{|v_{p}^0|^2\Omega_s}\right)^{1/4}=
\left(\frac{(r_{0\perp}^s)^2(\Omega_s^{\prime})^3m_e^2\gamma^3\omega_{p}^3}{e^2E^{02}_{p}\overline{\omega}_{p}}\right)^{1/4}.
\end{equation}
The analysis shows that (\ref{vesti_1.15}) can be greater than unity. In a real situation, absorption can always reduce by one or two orders of magnitude the upper limiting estimate given above, but the conclusion about a significant magnitude of the gain coefficient holds true.


\section{Volume Free Electron Laser} \label{laser_ch1}

As has been stated, the features of radiation from relativistic
particles in crystals, which were considered earlier in this
review, have a general character and are also manifested when
radiation is generated in artificial crystals (at present termed
"photonic crystals")
\cite{berk_104,berk_144,vfel_VFELreview,vfel_FEL2002,vfel_grid-t,rins_66,lanl_23a,hyb_FirstLasing,hyb_grid-FEL06-th}.

In contrast to generation in the X-ray range, generation of
radiation in microwave and optical ranges does not require such
high particle current densities. As a result, application of
photonic crystals made it possible to develop a new type of
generator, called the Volume Free Electron Laser (VFEL). Its main
features will be considered below.


\section{Generation equations  and threshold conditions in the case of two-wave diffraction}
\label{laser_sec2}

Using multi-wave diffraction in a VFEL for the formation of a volume distributed feedback enables one, on the one hand, to appreciably reduce the length of the interaction area, and, on the other hand, to employ electron beams with a large transverse cross-section for pulse generation, which improves the electrical  endurance of the generator and prevents burning of the glass in the discharge tubes during high-power lasing. Besides,  multi-wave diffraction provides the selection of radiation modes in oversized systems  \cite{laser_8}.

When radiation is generated in an FEL, the electrons interact with the electromagnetic wave in a finite spatial region, and release energy into the wave. Depending on the length of the spatial region, different generation regimes are realized. At $|\texttt{Im} k_{z} L| \gg 1$, generation occurs in a strong (exponential) amplification regime, which is mainly employed in amplifiers. The regime of weak single-passage amplification ($|Imk_{z} L| \le 1$) is chiefly used in generators. In order to determine the structure of the fields and to describe the evolution of instability in such systems, in addition to the knowledge of the dispersion equations 
and their solutions, one should match the fields at the boundaries of the regions (joining solutions). This procedure gives the field distribution in the system.

\begin{figure}[htp]
\centering
\epsfxsize = 6 cm \centerline{\epsfbox{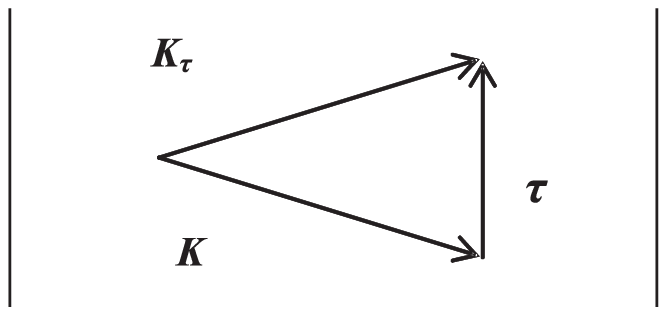}}
\caption{The geometry of two-wave Laue diffraction  $\vec {k},\vec
{k}_{\tau}  $ are the wave vectors of the incident and diffracted
waves, $\vec {\tau}  - $ is the reciprocal lattice vector of the
periodic structure. The projections of both wave vectors onto the
direction of the normal to the surface have the same sign.}
\label{fig.laser_1}
\end{figure}

Now we shall formulate the boundary problem. Let an electron beam
with  mean velocity $\vec{u}$ be incident onto a plane-parallel
spatially periodic plate of thickness $L$.  The electron beam is
oriented so that the radiation generated by the beam will be under
diffraction conditions. Under two-wave diffraction, two
fundamentally different geometries are possible. In the first case
(Laue geometry (see Figure \ref{fig.laser_1}) both waves are
emitted through one and the same boundary of the periodic
structure ($\gamma _{0} \gamma _{1} > 0$,where
$
\gamma _{0} = \frac{{\left( {\vec
{k}\vec {n}} \right)}}{{k}}\quad \mbox{and}\quad \gamma _{1} = \frac{{\left( {\vec {k}_{\tau } \vec {n}} \right)}}{{k}}
 $
 are the cosines made by the wave vectors $\vec {k}$ and $\vec {k}_{\tau}  $  with the normal vector to the surface of the periodic medium). In this geometry the amplifying regime is possible.

\begin{figure}[htp]
\centering
\epsfxsize = 6 cm \centerline{\epsfbox{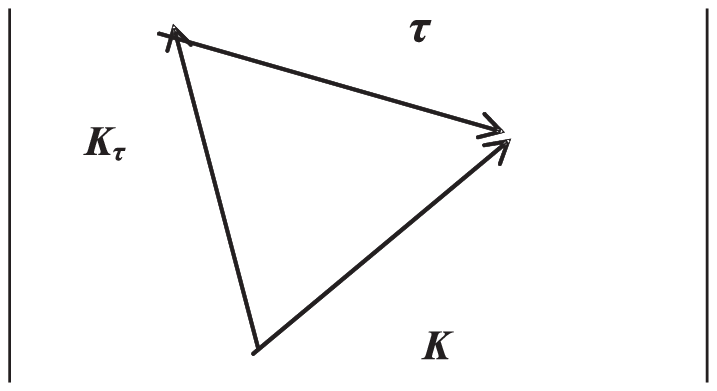}}
\caption{The geometry of  two-wave Bragg diffraction. $\vec
{k},\vec {k}_{\tau}$ are the wave vectors of the incident and
diffracted waves, respectively;  $\vec {\tau}$ is the reciprocal
lattice vector of the periodic structure. The projection of the
wave vectors onto the direction of the normal to the surface are
opposite in sign.} \label{fig.laser_2}
\end{figure}

In the latter case (Bragg geometry (see Figure
\ref{fig.laser_2})), the incident and diffracted waves leave the
plate through the opposite surfaces ($\gamma _{0} \gamma _{1} <
0$), when the electron current emits photons, positive feedback
appears, and the generation regime is available.

\bigskip


\bigskip

Let us write the expression for the fields appearing in the system described here
\begin{eqnarray}
\label{laser_eq6}
& &a\vec {e}\exp\left( i\vec {k}_{0} \vec {r} \right) +
b\vec{e}_{\tau}  \exp\left( i\vec {k}_{0\tau}  \vec {r} \right) \qquad \left( {I}
\right)\nonumber \\
& &\sum\limits_{i} {c_{i}}  \exp\left( {i\vec {k}_{i} \vec {r}} \right)\left(\vec {e} + s_{i}\vec{e} _{\tau}  \exp\left( i\vec {\tau} \vec {r} \right) \right)
\qquad \left( {II} \right)\nonumber \\
 & & f\vec {e}\exp\left( i\vec {k}_{0} \vec {r} \right)\exp\left(  - ik_{0z} L
\right) +\vec{e} _{\tau}  \exp\left( i\vec {k}_{0\tau}  \vec {r} \right)\exp\left(  -
ik_{0\tau z} L \right) \qquad
\left( {III} \right),\nonumber\\
\end{eqnarray}

\begin{eqnarray}
\label{laser_eq7}
& & a\vec {e}\exp\left( {i\vec {k}_{0} \vec {r}} \right) +
g\vec{e} _{\tau}  \exp\left( {i\vec {k}_{0\tau} ^{\left( { -}  \right)} \vec {r}}
\right)\qquad \left( {I}
\right)\nonumber\\
& & \sum\limits_{i} {c_{i}}  \exp\left( {i\vec {k}_{i} \vec {r}} \right)\left(
{\vec {e} + s_{i}\vec{e}} _{\tau}  \exp\left( {i\vec {\tau} \vec {r}} \right) \right)
\qquad \left( {II} \right) \\
& & f\vec {e}\exp\left( {i\vec {k}_{0} \vec {r}} \right)\exp\left( { - ik_{0z} L}
\right) +
b\vec{e} _{\tau}  \exp\left( {i\vec {k}_{0\tau} ^{\left( { -}  \right)} \vec {r}}
\right)\exp\left( {ik_{0\tau z} L} \right)
\qquad \left( {III} \right) \nonumber
\end{eqnarray}
 In (\ref{laser_eq6}), (\ref{laser_eq7}) $(I)$ and $(III)$ denote the fields before and after the periodic structure. In Laue geometry (\ref{laser_eq6}), in the general case two waves with specified  amplitudes are incident on the system: one along the direction with the wave vector $\vec{k}$ and amplitude $b$ and the other one along the direction with the wave vector $\vec{k}_{\tau}$ and amplitude $b$. They propagate from one side (in this case to the surface $z=0$). In Bragg diffraction geometry (\ref{laser_eq7}), the waves are incident on the system from opposite sides (the incident wave of amplitude $a$ falls onto the surface $z=0$, and the diffracted wave with amplitude $b$, onto the surface $z=L$). Wave vectors $\vec{k}_0$ and $\vec{k}_{0r}$ satisfy the standard dispersion equation describing propagation of electromagnetic waves in vacuum ($k^{2}c^{2} - \omega ^{2} = 0$). $(II)$  denotes the field in a periodic medium. The dispersion relation in the medium is defined by (\ref{laser_eq5}).
\[ s_{i} = \frac{{\omega ^{2}\chi _{ - \tau} } }{{\left[ {\left(
{\vec {k}_{i} + \vec {\tau} } \right)^{2}c^{2} - \omega
^{2}\varepsilon _{0}}  \right]}}
\]
is the coupling coefficient between the diffracted and the incident waves for the $i$-th mode determined from
equations (\ref{laser_eq6}), (\ref{laser_eq7}) and dispersion equation (\ref{laser_eq5}).

In the case of a "cold" electron beam, dispersion equation (\ref{laser_eq5}) takes the form:
\begin{eqnarray}
\label{laser_eq8}
 \left( {\omega - \vec {k}\vec {u}} \right)^{2}\left\{ {\left( {k^{2}c^{2} -
\omega ^{2}\varepsilon _{0}^{\left( {1} \right)}}  \right)\left( {k_{\tau
}^{2} c^{2} - \omega ^{2}\varepsilon _{0}^{\left( {2} \right)}}  \right) -
\omega ^{4}r} \right\}\nonumber\\
= - \frac{{\omega _{l}^{2}} }{{\gamma} }\left( {\frac{{\vec {u}\vec
{e}}}{{c}}} \right)^{2}\left( {k^{2}c^{2} - \omega ^{2}} \right)\left(
{k_{\tau} ^{2} c^{2} - \omega ^{2}\varepsilon _{0}^{\left( {2} \right)}}
\right)
\end{eqnarray}
As a result,  sum $(II)$  consists of six terms because equation (\ref{laser_eq7}) is of the sixth order with respect to $k_z$. But if the radiation geometry is not plane and the dielectric susceptibility is small, then the two waves mirror-reflected from the target surface may be ignored. In the case of a "hot" electron beam dispersion equation (\ref{laser_eq5}) reads:
\begin{eqnarray}
\label{laser_eq9}
& & \left( {k^{2}c^{2} - \omega ^{2}\varepsilon _{0}^{\left( {1} \right)}}
\right)\left( {k_{\tau} ^{2} c^{2} - \omega ^{2}\varepsilon _{0}^{\left( {2}
\right)}}  \right) - \omega ^{4}r\nonumber \\
& &= - \frac{{\sqrt {\pi}  \omega _{l}^{2}} }{{\delta _{0}^{\left( {i}
\right)} \gamma} }\left( {\frac{{\vec {u}\vec {e}}}{{c}}} \right)^{2}x_{i}
\exp\left( { - x_{i}^{2}}  \right)\quad \quad \left( {k^{2}c^{2} - \omega
^{2}} \right)\left( {k_{\tau} ^{2} c^{2} - \omega ^{2}\varepsilon
_{0}^{\left( {2} \right)}}  \right)
\end{eqnarray}
 Sum $(II)$ now contains only four terms because the corresponding dispersion equation is of the fourth order. With the mirror-reflected waves being neglected, only two terms remain, and the system of the boundary conditions for defining the unknown coefficients  takes a very simple form: the field in the periodic structure is completely determined by the two boundary conditions for the fields of the incident and diffracted waves at the boundaries of the periodic structure.
\begin{equation}
\label{laser_eq10}
\begin{array}{l}
 a = c_{1} + c_{2} \\
 b = s_{1} c_{1} + s_{2} c_{2} \\
 \end{array}
\end{equation}
\begin{equation}
\label{laser_eq11}
\begin{array}{l}
 a = c_{1} + c_{2} \\
 b = s_{1} c_{1} \exp\left( {ik_{1z} L} \right) + s_{2} c_{2} \exp\left(
{ik_{2z} L} \right). \\
 \end{array}
\end{equation}
 Equation (\ref{laser_eq10}) gives the continuity conditions for the incident and diffracted waves at the boundary $z=0$  for the Laue diffraction case. For the case of Bragg diffraction,  equation (\ref{laser_eq11}) gives the boundary conditions for the incident and diffracted waves, respectively when $z=0$ and $z=L$. In the regime of a "cold" electron beam, provided the mirror-reflected waves are neglected, dispersion equation  (\ref{laser_eq7}) has four solutions. To define the structure of the field appearing in a periodic plate, four boundary conditions should be used. Two of them have the same form as those in the case of a "hot" beam: the continuity of the incident and diffracted waves at the target boundaries.  Two additional conditions  are the continuity of the charge and current densities at the input boundary. As a result, the system of equations  defining the fields in this case has the form
\begin{equation}
\label{laser_eq12}
\left\{ \begin{array}{l}
 {\sum\limits_{i = 1}^{4} {c_{i}}  = a} \\
 {\sum\limits_{i = 1}^{4} {\frac{{c_{i}} }{{\delta _{i}} } = 0}}  \\
 {\sum\limits_{i = 1}^{4} {\frac{{c_{i}} }{{\delta _{i}^{2}} } = 0}}  \\
 {1} \left.\right)\sum\limits_{i = 1}^{4} {s_{i} c_{i}}  = b\qquad\mbox{or} \qquad 2 \left.\right)\quad \sum\limits_{i
= 1}^{4} {s_{i} c_{i}}  \exp\{ ik_{iz} L\} = b. \\
 \end{array}\right.
\end{equation}
The first equation in (\ref{laser_eq12}) expresses the continuity of the incident wave at the front boundary, the third and second ones describe the continuity of the charge and current densities at the front boundary,
$
\delta _{i} = \frac{{\vec {k}_{i} \vec
{u} - \omega} }{{ku_{z}} }.
$
 The fourth equation expresses the continuity of the diffracted wave value at the front boundary in  Laue geometry (denoted by figure 1 in line four of (\ref{laser_eq12}) or at the back boundary in the case of Bragg geometry (figure 2 in line 4 of (\ref{laser_eq12})). It is easy to see (see (\ref{laser_eq10}), (\ref{laser_eq12})) that in  Laue diffraction geometry, the electromagnetic field in the interaction area will also be absent if the amplitudes of the incident waves equal zero.

 The situation is different in Bragg geometry. From (\ref{laser_eq11}), (\ref{laser_eq12}) follows that in this case, under certain conditions it is possible that the field in the medium exists at nonzero amplitudes of the incident fields. The generation equations defining these conditions  are obtained by equating the  determinants of the systems (\ref{laser_eq11}) and (\ref{laser_eq12}) to zero.  For a "hot" electron beam, the generation equation takes the form:
 \begin{equation}
\label{laser_eq13}
s_{2} \exp\left( {ik_{2z} L} \right) - s_{1} \exp\left( {ik_{1z} L} \right) =
0.
\end{equation}
A similar equation for a "cold" electron beam reads
\begin{eqnarray}
\label{laser_eq14}
& & s_{1} \exp\{ ik_{1z} L\} \delta _{1}^{2} \left( {\delta _{2} - \delta _{3}}
\right)\left( {\delta _{2} - \delta _{4}}  \right)\left( {\delta _{3} -
\delta _{4}}  \right)\nonumber \\
& &- s_{2} \exp\{ ik_{2z} L\} \delta _{2}^{2} \left( {\delta _{1} - \delta _{3}}
\right)\left( {\delta _{1} - \delta _{4}}  \right)\left( {\delta _{3} -
\delta _{4}}  \right) \nonumber \\
& & +s_{3} \exp\{ ik_{3z} L\} \delta _{3}^{2} \left( {\delta _{1} - \delta _{2}}
\right)\left( {\delta _{1} - \delta _{4}}  \right)\left( {\delta _{2} -
\delta _{4}}  \right)  \\
& & -s_{4} \exp\{ ik_{4z} L\} \delta _{4}^{2} \left( {\delta _{1} - \delta _{2}}
\right)\left( {\delta _{1} - \delta _{3}}  \right)\left( {\delta _{2} -
\delta _{3}}  \right) = 0. \nonumber
\end{eqnarray}

 Upon solving these equations, one may determine the threshold generation conditions, i.e., the electron current and other parameters of the beam, marking the starting point from which radiation prevails over the losses. Besides, radiation generation takes place when some phase relations are fulfilled: phase shift between the two diffraction modes traversing the interaction area should be a multiple of $2\pi$ (the field at the output should be similar to that at the input):
\begin{equation}
\label{laser_eq15}
\texttt{Re}\left( {k_{1z} - k_{2z}}  \right)L = 2\pi n.
\end{equation}
In the region, where these conditions are fulfilled, the solution of (\ref{laser_eq14}), (\ref{laser_eq15}) has the form:
\begin{equation}
\label{laser_eq16}
\omega ^{\prime\prime}= \frac{{\omega} }{{2\left( {1 - \beta}  \right)}}\left\{
{G^{\left( {b} \right)} - \chi _{0}^{\prime\prime} \left( {1 - \beta \mp \frac{{\sqrt
{ - \beta}  r^{\prime\prime}}}{{|\chi _{\tau}  |\chi _{0} ^{\prime\prime}}}} \right) - \left(
{\frac{{\gamma _{0} c}}{{\vec {n}\vec {u}}}} \right)^{3}\frac{{16\pi
^{2}n^{2}}}{{ - \beta \left( {k\chi _{\tau}  L_{\ast} }  \right)^{2}kL_{\ast
}} }} \right\}.
\end{equation}
$\omega ^{\prime\prime}$ in (\ref{laser_eq16}) is the increment of absolute instability, which describes the increase of the field amplitude in time in the field linear regime,
\begin{eqnarray}
\label{laser_eq17}
G^{( {b})} = \left\{ \begin{array}{l}
  - \frac{\sqrt {\pi} } {\gamma} \frac{\omega _{l}^{2}} {\omega ^{2}}\frac{( \vec {u}\vec {e}
)^{2}}{u^{2}}\frac{k^{2}c^{2} - \omega ^{2}}{\delta _{0}^{2}
}x^{( {t})}\exp(  - x^{( t)2})\qquad\mbox{for a "hot" beam}\nonumber \\
\\
 \frac{\pi ^{2}n^{2}}{4\gamma} \left( \frac{\omega _{l}} {\omega
} \right)^{2}\frac{( \vec {u}\vec {e})^{2}}{u^{2}}(
k^{2}c^{2} - \omega ^{2})k^{2}L_{\ast} ^{2} f( y
)\qquad\mbox{for a "cold" beam}  \\
 \end{array}  \right.\nonumber
\end{eqnarray}
\begin{equation}
\label{laser_eq18}
f\left( {y} \right) = \sin y\frac{{\left( {2y + \pi n} \right)\sin y - y\left(
{y + \pi n} \right)\cos y}}{{y^{3}\left( {y + \pi n} \right)^{3}}}
\end{equation}
is the profile function depending on the detuning from  synchronism conditions
$
y =
\frac{{\delta \omega} }{{2u_{z}} }L_{\ast} .
$

\begin{figure}[h]
\centering
\epsfxsize = 12 cm \centerline{\epsfbox{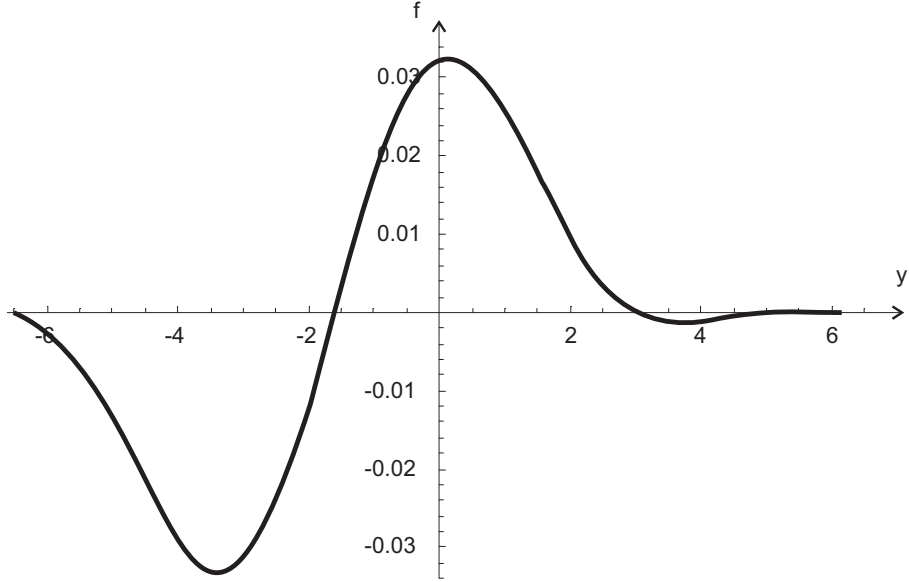}}
\caption{The
profile generation function in the root degeneration region versus
the detuning from synchronism conditions $y$}
\label{Fig:nolabel}
\end{figure}

The profile function of detuning  is plotted in Figure
\ref{Fig:nolabel}. As follows from  (\ref{laser_eq18}) and Figure
\ref{Fig:nolabel}, one of the peculiarities of generation in the
region of degeneration of the roots of the diffraction equation is
that stimulated radiation is nonzero at zero detuning. This
peculiarity appears due to the interference of contributions to
radiation coming from the two branches of the diffraction
equation. In a standard classical FEL, generation does not occur
at the exact fulfillment of the synchronism condition. This is due
to the fact that the gain  coefficient is proportional to the
difference between  spontaneous radiation and absorption. In the
case when these branches overlap each other, the gain coefficient
becomes proportional to the derivative  of the spectral function,
while at zero detuning from  synchronism conditions the spectral
intensity of spontaneous radiation reaches its maximum, and the
derivative at this point equals zero.

\begin{figure}[htbp]
\centering
\epsfxsize = 10 cm \centerline{\epsfbox{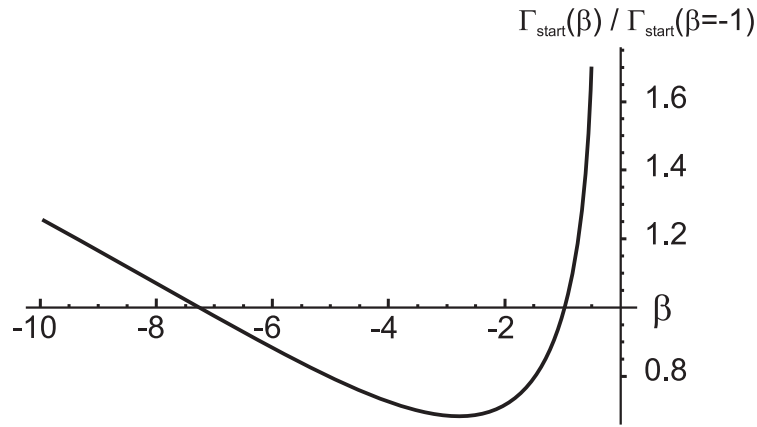}} \caption{The
relation $\Gamma _{loss} \left( {\beta} \right)/\Gamma _{loss}
\left( {\beta = - 1} \right)$ as a function of the asymmetry
factor of diffraction for $\lambda \sim 4$ mm and $L = 10$ cm}
\label{fig.laser_4}
\end{figure}

Equation (\ref{laser_eq16}) has an obvious physical meaning: the
first term between the braces is proportional to the intensity of
radiation produced by the electron beam (in the unit length). The
next two terms describe  losses of radiation appearing through
absorption and due to the fact that radiation leaves the area of
interaction with electrons. The values of the electron beam
parameters, at which the radiation generation equals radiation
losses ($\omega ^{\prime\prime}$ ) mark the starting point for the
generation process. From the obtained expression
(\ref{laser_eq16}) follows that at given values of the size of the
interaction area and absorption, there is an optimal diffraction
geometry in which the losses  $\Gamma_{\mathrm{loss}}$ are
minimum. In the general case, the optimal geometry is not at all
plane. Figure \ref{fig.laser_4} plots the relation $\Gamma
_{\mathrm{loss}} \left( {\beta}  \right)/\Gamma _{\mathrm{loss}}
\left( {\beta = - 1}\right)$ against the asymmetry factor for
diffraction in the millimeter wavelength range ($\lambda\sim 4$
mm).

\section{Generation equations and threshold conditions in the geometry of three-wave Bragg diffraction}
\label{laser_sec3}

Above we have considered the theory of a volume distributed
feedback generator in the geometry of two-wave Bragg diffraction.
It has been shown that the transition to a volume distributed
feedback opens up  wider perspectives for the optimization of the
system. Transition to a multi-wave distributed feedback provides
additional possibilities. Here we shall mention some of them. For
example, in a hard X-ray range, where the susceptibility has a
negative value $\chi_0^{\prime}<0$, the condition of fulfillment
of Cherenkov synchronism imposes restrictions on the asymmetry
factor of diffraction  $\left| {\beta}  \right| > \left( {\left|
{\chi _{0}^{/}}  \right| + \gamma ^{ - 2}} \right)/\left| {\chi
_{\tau} }  \right|^{2}$. In the X-ray spectral range, where the
absorption is large, radiation losses  under the condition of
two-wave dynamical diffraction will be rather large at such values
of the asymmetry factor. Moreover, strict requirements are set for
the parameters of the starting generation current. One of the ways
to diminish the losses is using supplementary external mirrors.
Transition to  multi-wave diffraction also enables one to reduce
losses because the parameters of the system required to initiate
the generation process change due to the rearrangement of the
field structure in the interaction area. Synchronism conditions in
this case contain additional parameters as compared to the
two-wave dynamical diffraction, which enables one to match the
Cherenkov conditions to the parameters corresponding to the
regions with smaller absorption of radiation.

The possibility of generation at the point of degeneration of several diffraction roots is another important feature of multi-wave diffraction. At this point the functional change of the threshold characteristics occurs (e.g., the coincidence of the two roots gives the  doubly degenerate case and $s=2$) and, as a result, it becomes possible to reduce the length of the generation area  (at a given operating current) or the operating current (at a given length of the generation area).

Application of multi-wave diffraction for generating in a
microwave range has one more remarkable feature -- the possibility
of selecting modes in oversized waveguides and resonators.
Production of microwave power pulses requires high electric
strength of the generator and radiation resistance of the output
window.  To reduce load on these elements, the transversal (with
respect to the direction of the electron beam velocity) dimension
of the resonator should be large (much larger than the
wavelength).  As a rule this leads to a multi-mode  generation
regime and low efficiency.
In the presence of $n$-wave diffraction, selection of modes can be
effectively carried out due to the requirement of fulfillment of
$n-1$ Bragg condition.

\
\begin{figure}[htp]
\centering
\epsfxsize = 6 cm \centerline{\epsfbox{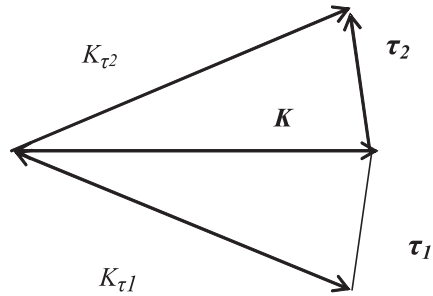}}
\caption{Figure Three-wave diffraction in Laue-Laue geometry.
Projection of wave vectors $\vec {k},\vec {k}_{\tau 1} ,\vec
{k}_{\tau 2} $  onto the normal to the surface have the same sign.
$\vec {\tau} _{1} ,\vec {\tau} _{2} $ are the reciprocal lattice
vectors of the periodic structure.} \label{fig.laser_5}
\end{figure}

  To illustrate the potential of the multi-wave distributed feedback, we shall give a more detailed consideration of three-wave diffraction. In this case, volume distributed feedback (VDF) can be realized in three different geometries:

(1) Laue-Laue diffraction, when the three  waves exit through the
same surface ($\gamma _{0} ,\gamma _{1} ,\gamma _{2} > 0$, Figure
\ref{fig.laser_5};

(2) diffraction in  Bragg-Bragg geometry, when $\gamma _{0} > 0$, while  $\gamma_{1} ,\gamma _{2} < 0$. That is, the wave vector satisfying the synchronism condition is directed along the normal to the surface of the slow-wave structure, and  the two vectors corresponding to diffracted waves point in the opposite direction;

\begin{figure}[htp]
\centering
\epsfxsize = 6 cm \centerline{\epsfbox{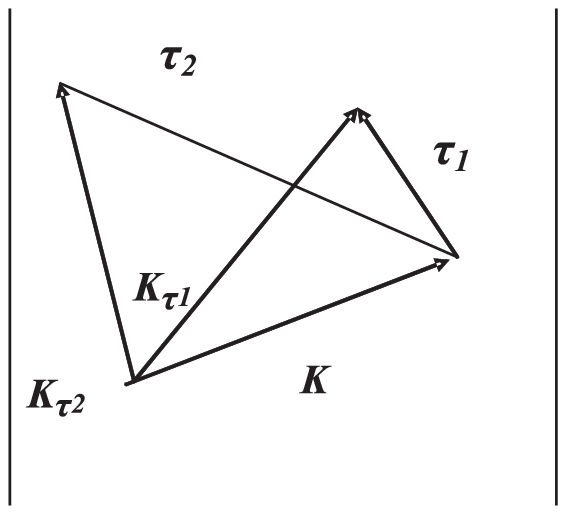}}
\caption{Figure Three-wave diffraction in  Laue-Bragg geometry.
The projection of wave vectors $\vec {k},\vec {k}_{\tau 1} $ onto
the surface normal and the projection of vector $\vec{k}_{\tau 2}$
are opposite in sign. $\vec {\tau} _{1}$, $\vec {\tau} _{2} $ are
the reciprocal lattice vectors of the periodic structure.}
\label{fig.laser_6}
\end{figure}

(3) Bragg-Laue diffraction, when  $\gamma _{0} ,\gamma _{1} > 0$, $\gamma _{2} < 0$. Vector $\vec{k}$ corresponding to the synchronous wave and one of the vectors $\vec {k}_{1} = \vec
{k} + \vec {\tau} _{1} $ corresponding to the diffracted  wave point in the same direction, while vector $\vec
{k}_{2} = \vec {k} + \vec {\tau} _{2} $ points in different direction (see Figure \ref{fig.laser_6}).

Similarly to the two-wave case, the problem of the beam interaction with the target field may be reduced to the problem of  three-wave diffraction of an  electromagnetic wave incident onto the active medium. The active medium here is the system "spatially-periodic structure + electron beam".

The system of equations describing the process of three-wave coplanar diffraction by such an active medium takes the form:
\begin{eqnarray}
\label{laser_eq19}
 D_{\sigma} ^{\left( {0} \right)} E_{\sigma} ^{\left( {0} \right)} - \omega
^{2}\chi _{1} E_{\sigma} ^{\left( {1} \right)} - \omega ^{2}\chi _{2}
E_{\sigma} ^{\left( {2} \right)} = 0 \nonumber\\
 - \omega ^{2}\chi _{ - 1} E_{\sigma} ^{\left( {0} \right)} + D_{\sigma
}^{\left( {1} \right)} E_{\sigma} ^{\left( {1} \right)} - \omega ^{2}\chi
_{2 - 1} E_{\sigma} ^{\left( {2} \right)} = 0 \\
 - \omega ^{2}\chi _{ - 2} E_{\sigma} ^{\left( {0} \right)} - \omega
^{2}\chi _{1 - 2} E_{\sigma} ^{\left( {1} \right)} + D_{\sigma} ^{\left( {2}
\right)} E_{\sigma} ^{\left( {2} \right)} = 0, \nonumber
\end{eqnarray}




where the terms
\begin{equation}
\label{laser_eq19a}
D^{\left( {i} \right)} = k_{i}^{2} c^{2} - \omega ^{2}\varepsilon
_{0}^{\left( {i} \right)} + \frac{{\omega _{l}^{2}} }{{\gamma} } + \Gamma
_{i} \left( {\left( {\vec {k}_{i} - \vec {k}_{o}}  \right)^{2}c^{2} - \left(
{\omega - \omega _{o}}  \right)^{2}} \right)
\end{equation}
 that contain the contributions due to resonance interaction of electrons with the wave,
\begin{eqnarray}
\label{laser_eq4}
\Gamma _{i} = \left\{
\begin{array}{cc}
  \frac{\omega _{l}^{2}} {\gamma} \frac{\Omega ^{2}}{( \omega -
\vec {k}_{i} \vec {u} - \omega _{o} )^{2}} & \qquad \mbox{in the case of a "cold" electron beam} \\
{}&{}\\
  \frac{\sqrt {\pi}  \omega _{l}^{2}} {\gamma} \frac{\Omega
^{2}}{\delta _{0}^{( {i} )}} x_{i} \exp( - x_{i}^{2}) & \qquad \mbox{in the case of a "hot" electron beam}
\end{array}
 \right.\nonumber\\
\end{eqnarray}
The values of $\Gamma_i$ in  equation (\ref{laser_eq4})  are given for two generation regimes: in the regime of a "cold" electron beam all the electrons have the same velocity and the velocity spread is small ($\frac{{\vec {k}\Delta \vec {u}}}{{\Delta \omega} } \ll 1$,
where $\Delta \vec {u}$ is the thermal straggling of the electron beam, $\Delta\omega$ is the width of the emission line). In this case all the electrons participate in the interaction with the electromagnetic wave. In the regime of a "hot" electron beam the relation
$
\frac{{\vec {k}\Delta \vec {u}}}{{\Delta \omega} } \ge 1
$
is fulfilled.
In this case  only a part of the electron beam participates in the interaction process. The explicit form of $\Omega$ and $\omega_0$ depends on the generation mechanism. For the Cherenkov mechanism
$
\Omega = \left( {\frac{{\vec {u}\vec {e}_{i}} }{{c}}}
\right)^{2},
$
where $\vec {e}_{i} $ is the polarization vector,
$\omega _{o} = 0$. In (\ref{laser_eq19})--(\ref{laser_eq4})

\[
\omega _{l}^{2} = \frac{{4\pi n_{e}} }{{m_{e}} } ,\quad x_{i} =
\frac{{\omega - \vec {k}_{i} \vec {u} - \omega _{o}} }{{\sqrt {2} \delta
_{0}^{\left( {i} \right)}} },\quad \delta _{0}^{\left( {i} \right)2} = k_{i1}
\psi _{x}^{2} + k_{iy}^{2} \psi _{y}^{2} + k_{iz}^{2} \psi _{z}^{2},\quad \psi_{i}
\]
are the thermal straggling for a "hot" electron beam in the distribution function:
\[
f_{\left( {0} \right)} = \frac{{n_{e}} }{{\left( {2\pi}  \right)^{3/2}\psi
_{1} \psi _{2} \psi _{3}} }exp\left\{ { - \frac{{v_{1}^{2}} }{{2u^{2}\psi
_{1}^{2}} } - \frac{{v_{2}^{2}} }{{2u^{2}\psi _{2}^{2}} } - \frac{{v_{3}^{2}
}}{{2u^{2}\psi _{3}^{2}} }} \right\}
\]
is the distribution function describing the velocity spread of the electrons in a beam before the interaction with the electromagnetic wave.

In the case of  two-wave Bragg diffraction, when only two wave vectors satisfy the Bragg condition \cite{132}
and two strong waves are excited, the dispersion equation for Cherenkov instability takes the form
\begin{eqnarray}
\label{laser_eq5}
& &\left( {k^{2}c^{2} - \omega ^{2}\varepsilon _{0}^{\left( {1} \right)} +
\frac{{\omega _{l}^{2}} }{{\gamma} } + \Gamma _{1} \left( {k^{2}c^{2} -
\omega ^{2}} \right)} \right)\nonumber\\
& &\times \left( {k_{\tau} ^{2} c^{2} - \omega
^{2}\varepsilon _{0}^{\left( {2} \right)} + \frac{{\omega _{l}^{2}
}}{{\gamma} } + \Gamma _{\tau}  \left( {k_{\tau} ^{2} c^{2} - \omega ^{2}}
\right)} \right) - \omega ^{4}r = 0.
\end{eqnarray}
Here $r = \chi _{\tau}  \chi _{ - \tau}  $

As it was noted in \cite{berk_133,berk_134}, the functional relationship between the imaginary part of the solution $k_z$ and the electron beam density may change appreciably at the points of degeneration of roots  $k_z$ of the diffraction equation
(the dispersion equation without the electron beam describing the dispersion of electromagnetic waves in a periodic medium). This occurs, in particular, at the point of $s$-fold degeneration of roots: $\texttt{Im} k_{z}
\sim n_{e}^{1/\left( {2 + s} \right)} $. In the Compton generation regime, this quantity significantly exceeds in
magnitude a similar parameter for $\texttt{Im} k_{zz} \sim n_{e}^{1/3} $ beyond the root degeneration region.
From this fact the authors of \cite{berk_133,berk_134} concluded that the instability increment in the degeneration region grows sharply.


 The dispersion equation  corresponding to the system (\ref{laser_eq19}) for the three-wave case of coplanar diffraction reads:
\begin{equation}
\label{laser_eq20}
F_{\sigma} ^{\left( {3} \right)} \left( {\vec {k},\vec {k}_{1} ,\vec {k}_{2}
} \right) = - \Gamma F_{\sigma} ^{\left( {2} \right)} \left( {\vec {k}_{1}
,\vec {k}_{2}}  \right),
\end{equation}
where
\begin{eqnarray*}
 & &F_{\sigma} ^{\left( {3} \right)} \left( {\vec {k},\vec {k}_{1} ,\vec
{k}_{2}}  \right) = \left( {k^{2}c{}^{2} - \omega ^{2}\varepsilon _{0}}
\right)\left( {k_{1}^{2} c{}^{2} - \omega ^{2}\varepsilon _{0}}
\right)\left( {k_{2}^{2} c{}^{2} - \omega ^{2}\varepsilon _{0}}  \right)\\
& &-\omega ^{4}\left( {k^{2}c{}^{2} - \omega ^{2}\varepsilon _{0}}  \right)\chi
_{1 - 2} \chi _{2 - 1} - \omega ^{4}\left( {k_{1}^{2} c{}^{2} - \omega ^{2}\varepsilon _{0}}
\right)\chi _{2} \chi _{ - 2}\\
& & - \omega ^{4}\left( {k_{2}^{2} c{}^{2} -
\omega ^{2}\varepsilon _{0}}  \right)\chi _{1} \chi _{ - 1} - \omega
^{6}\left( {\chi _{1} \chi _{ - 2} \chi _{2 - 1} + \chi _{2} \chi _{ - 1}
\chi _{1 - 2}}  \right), \\
& & F_{\sigma} ^{\left( {2} \right)} \left( {\vec {k}_{1} ,\vec
{k}_{2}}  \right) = \left( {k_{1}^{2} c{}^{2} - \omega ^{2}\varepsilon _{0}
} \right)\left( {k_{2}^{2} c{}^{2} - \omega ^{2}\varepsilon _{0}}  \right) -
\omega ^{4}\chi _{1 - 2} \chi _{2 - 1} .
\end{eqnarray*}
In deriving (\ref{laser_eq20}), we based on the assumption that the electrons are synchronous to the wave with  wave vector $\vec{k}$. The field in the three-wave system is written as follows:

(1) in  Laue-Laue geometry

\begin{eqnarray*}
& & a^{\left( {0} \right)}\vec {e}\exp\left( {i\vec {k}_{0} \vec {r}} \right) +
a^{\left( {1}
\right)}\vec
{e} _{1} \exp\left( {i\vec {k}_{01} \vec {r}} \right) + a^{\left( {2}
\right)}\vec{e} _{2} \exp\left( {i\vec {k}_{02} \vec {r}} \right) \\
& & \sum\limits_{i} {c_{i}}  \exp\left( {i\vec {k}_{i} \vec {r}} \right)\left(
{\vec {e} + s_{i}
\vec{e}} _{\tau}  \exp\left( {i\vec {\tau} \vec {r}} \right) \right) \\
& & b\vec {e}\exp\left( {i\vec {k}_{0} \vec {r}} \right)\exp\left( { - ik_{0z} L}
\right) + b^{\left( {1}
\right)}\vec
{e} _{1} \exp\left( {i\vec {k}_{01} \vec {r}} \right)\exp\left( { - ik_{01z}
L} \right)\\
& & + b^{\left( {2}
\right)}\vec
{e} _{2} \exp\left( {i\vec {k}_{02} \vec {r}} \right)\exp\left( { - ik_{02z}
L} \right), \\
\end{eqnarray*}

(2) in Bragg-Bragg geometry

\begin{eqnarray*}
& & a^{\left( {0} \right)}\vec {e}\exp\left( {i\vec {k}_{0} \vec {r}} \right) +
b^{\left( {1}
\right)}\vec{e} _{1} \exp\left( {i\vec {k}_{01}^{\left( { -}  \right)} \vec {r}} \right)
+ b^{\left( {2}\right)}\vec{e} _{2} \exp\left( {i\vec {k}_{02}^{\left( { -}  \right)} \vec {r}} \right)
\\
& & \sum\limits_{i} {c_{i}}  \exp\left( {i\vec {k}_{i} \vec {r}} \right)\left(
{\vec {e} + s_{i}
\vec{e} _{\tau}  \exp\left( {i\vec {\tau} \vec {r}} \right)} \right) \\
& & b\vec {e}\exp\left( {i\vec {k}_{0} \vec {r}} \right)\exp\left( { - ik_{0z} L}
\right) + a^{\left( {1}
\right)}\vec{e} _{1} \exp\left( {i\vec {k}_{01}^{\left( { -}  \right)} \vec {r}}
\right)\exp\left( {ik_{01z} L} \right) \\
& &+ a^{\left( {1}
\right)}\vec
{e} _{1} \exp\left( {i\vec {k}_{01}^{\left( { -}  \right)} \vec {r}}
\right)\exp\left( {ik_{01z} L} \right), \\
 \end{eqnarray*}

(3) in Bragg-Laue geometry

\begin{eqnarray*}
 & &a^{\left( {0} \right)}\vec {e}\exp\left( {i\vec {k}_{0} \vec {r}} \right) +
a^{\left( {1}
\right)}\vec
{e} _{1} \exp\left( {i\vec {k}_{01} \vec {r}} \right) + b^{\left( {2}
\right)}\vec
{e} _{2} \exp\left( {i\vec {k}_{02}^{\left( { -}  \right)} \vec {r}} \right)
\\
& & \sum\limits_{i} {c_{i}}  \exp\left( {i\vec {k}_{i} \vec {r}} \right)\left(
{\vec {e} + s_{i}
\vec
{e} _{\tau}  \exp\left( {i\vec {\tau} \vec {r}} \right)} \right) \\
 & &b\vec {e}\exp\left( {i\vec {k}_{0} \vec {r}} \right)\exp\left( { - ik_{0z} L}
\right) + b^{\left( {1}
\right)}\vec
{e} _{1} \exp\left( {i\vec {k}_{01} \vec {r}} \right)\exp\left( {ik_{01z} L}
\right)  \\
& &+a^{\left( {1}
\right)}\vec
{e} _{1} \exp\left( {i\vec {k}_{01}^{\left( { -}  \right)} \vec {r}}
\right)\exp\left( {ik_{01z} L} \right), \\
 \end{eqnarray*}
In the above expressions the waves incident on the structure have the amplitudes $a^{(i)}$, while the exit waves have the amplitudes  $b^{(i)}$. In the expression for the field in a periodic structure, summation is made over the diffraction modes.
In the case of a "cold" beam, the sum contains five terms, and in the case of a "hot" beam, three terms.
\[
 s_{i}^{\left( {1}
\right)} = \frac{{\lambda _{i} \lambda _{i2} - r_{2}} }{{\lambda _{i2} \chi
_{1} + \chi _{2} \chi _{1 - 2}} }\qquad\mbox{and}\qquad s_{i}^{\left( {2} \right)} =
\frac{{\lambda _{i} \lambda _{i1} - r_{1}} }{{\lambda _{i1} \chi _{2} + \chi
_{1} \chi _{2 - 1}} }
\]
are the coupling coefficients between the diffracted waves and the incident wave; their form is obtained by the system of equations (\ref{laser_eq19}), $\lambda _{i\alpha}  =
\{ \left( {\vec {k}_{i} + \vec {\tau} _{\alpha} }  \right)^{2}c^{2} - \omega
^{2}\varepsilon _{0} \} /\omega ^{2},r_{\alpha}  = \chi _{\alpha}  \chi _{ -
\alpha}  $.

By analogy with a two-wave case, we shall join the fields at the system boundaries and write the system of equations defining the field in the interaction area:

(1) in the case of Laue-Laue geometry:
\begin{eqnarray}
\label{laser_eq21}
 \sum\limits_{i = 1}^{5} {c_{i}}  &=& a\qquad \sum\limits_{i = 1}^{5} {s_{i}^{\left( {1} \right)} c_{i}}  = a_{1} \qquad
\sum\limits_{i = 1}^{5}{s_{i}^{\left( {2} \right)} c_{i}}  = a_{2}\nonumber \\
 \sum\limits_{i = 1}^{5} {\frac{{c_{i}} }{{\delta _{i}} }}& =& 0\qquad  \sum\limits_{i = 1}^{5} {\frac{{c_{i}
}}{{\delta _{i}^{2}} }} = 0,
\end{eqnarray}

(2) for Bragg-Bragg geometry
\begin{eqnarray}
\label{laser_eq22}
 \sum\limits_{i = 1}^{5} {c_{i}}  &=& a\qquad\sum\limits_{i = 1}^{5} {s_{i}^{\left( {1} \right)} c_{i}}  \exp\left(
{ik_{iz} L} \right) = a_{1} \qquad \sum\limits_{i = 1}^{5} {s_{i}^{\left( {2} \right)} c_{i}}  \exp\left(
{ik_{iz} L} \right) = a_{2}\nonumber \\
 \sum\limits_{i = 1}^{5} {\frac{{c_{i}} }{{\delta _{i}} }} &=& 0\qquad \sum\limits_{i = 1}^{5} {\frac{{c_{i}
}}{{\delta _{i}^{2}} }} = 0,
\end{eqnarray}

(3) for Bragg-Laue geometry
\begin{eqnarray}
\label{laser_eq23}
 \sum\limits_{i = 1}^{5} {c_{i}}  &=& a\qquad \sum\limits_{i = 1}^{5} {s_{i}^{\left( {1} \right)} c_{i}}  = a_{1} \qquad\sum\limits_{i = 1}^{5}
{s_{i}^{\left( {2} \right)} c_{i}}  \exp\left( {ik_{iz} L} \right) = a_{2}\nonumber \\
 \sum\limits_{i = 1}^{5} {\frac{{c_{i}} }{{\delta _{i}} }}& =& 0\qquad
\sum\limits_{i = 1}^{5} {\frac{{c_{i}
}}{{\delta _{i}^{2}} }} = 0.
\end{eqnarray}
In all three  of these systems (\ref{laser_eq21})--(\ref{laser_eq21}) the first three equalities are obtained from the requirement of the continuity of fields at the boundary. In the Laue-Laue case the boundary conditions are written for the boundary $z=0$. In Bragg-Bragg geometry one condition for the incident wave is written for the boundary $z=0$, two others, for diffracted waves  for the boundary $z=L$. In Laue-Bragg geometry two conditions are defined at $z=0$, and one, at $z=L$. The last two equations of the obtained systems follow from the requirement of continuity of the current and charge densities at the boundary $z=0$. In the three-wave case of Laue geometry, as well as in the case of a two-wave VDF, only  the amplification regime is possible. In Bragg-Bragg and Bragg-Laue geometry both the amplification regime and the generation regime accompanied by radiation from spontaneous noise at zero amplitude of the incident field are possible. Using the systems (\ref{laser_eq22}) and (\ref{laser_eq23}), write the generation equations for these two geometries.

Bragg-Bragg geometry
\begin{eqnarray}
\label{laser_eq24}
& & \left( {s_{1}^{\left( {1} \right)} s_{2}^{\left( {2} \right)} -
s_{2}^{\left( {1} \right)} s_{1}^{\left( {2} \right)}}  \right)\delta _{1}
\delta _{2} \left( {\delta _{3} - \delta _{4}}  \right)\exp\{ i\left( {k_{1z}
+ k_{2z}}  \right)L\} \nonumber \\
& &- \left( {s_{1}^{\left( {1} \right)} s_{3}^{\left( {2} \right)} -
s_{3}^{\left( {1} \right)} s_{1}^{\left( {2} \right)}}  \right)\delta _{1}
\delta _{3} \left( {\delta _{2} - \delta _{4}}  \right)\exp\{ i\left( {k_{1z}
+ k_{3z}}  \right)L\} \nonumber \\
& &+ \left( {s_{1}^{\left( {1} \right)} s_{4}^{\left( {2} \right)} -
s_{4}^{\left( {1} \right)} s_{1}^{\left( {2} \right)}}  \right)\delta _{1}
\delta _{4} \left( {\delta _{2} - \delta _{3}}  \right)\exp\{ i\left( {k_{1z}
+ k_{4z}}  \right)L\}  \\
& &+ \left( {s_{2}^{\left( {1} \right)} s_{3}^{\left( {2} \right)} -
s_{3}^{\left( {1} \right)} s_{2}^{\left( {2} \right)}}  \right)\delta _{2}
\delta _{3} \left( {\delta _{1} - \delta _{4}}  \right)\exp\{ i\left( {k_{2z}
+ k_{3z}}  \right)L\} \nonumber\\
& & -\left( {s_{2}^{\left( {1} \right)} s_{4}^{\left( {2} \right)} -
s_{4}^{\left( {1} \right)} s_{2}^{\left( {2} \right)}}  \right)\delta _{2}
\delta _{4} \left( {\delta _{1} - \delta _{3}}  \right)\exp\{ i\left( {k_{2z}
+ k_{4z}}  \right)L\}\nonumber \\
& & +\left( {s_{3}^{\left( {1} \right)} s_{4}^{\left( {2} \right)} -
s_{4}^{\left( {1} \right)} s_{3}^{\left( {2} \right)}}  \right)\delta _{3}
\delta _{4} \left( {\delta _{1} - \delta _{2}}  \right)\exp\{ i\left( {k_{3z}
+ k_{4z}}  \right)L\} = 0. \nonumber
 \end{eqnarray}

Bragg-Laue geometry

\begin{eqnarray}
\label{laser_eq25}
 & &s_{1}^{\left( {2} \right)} \exp\left( {ik_{1z} L} \right)\delta _{1} \{
s_{2}^{\left( {1} \right)} \delta _{2} \left( {\delta _{4} - \delta _{3}}
\right) - s_{3}^{\left( {1} \right)} \delta _{3} \left( {\delta _{4} -
\delta _{2}}  \right) + s_{4}^{\left( {1} \right)} \delta _{4} \left(
{\delta _{3} - \delta _{2}}  \right)\} \nonumber \\
& &- s_{2}^{\left( {2} \right)} \exp\left( {ik_{2z} L} \right)\delta _{2} \{
s_{1}^{\left( {1} \right)} \delta _{1} \left( {\delta _{4} - \delta _{3}}
\right) - s_{3}^{\left( {1} \right)} \delta _{3} \left( {\delta _{4} -
\delta _{1}}  \right) + s_{4}^{\left( {1} \right)} \delta _{4} \left(
{\delta _{3} - \delta _{1}}  \right)\}\nonumber  \\
 & &+s_{3}^{\left( {2} \right)} \exp\left( {ik_{3z} L} \right)\delta _{3} \{
s_{1}^{\left( {1} \right)} \delta _{1} \left( {\delta _{4} - \delta _{2}}
\right) - s_{2}^{\left( {1} \right)} \delta _{2} \left( {\delta _{4} -
\delta _{1}}  \right) + s_{4}^{\left( {1} \right)} \delta _{4} \left(
{\delta _{2} - \delta _{1}}  \right)\}\nonumber  \\
& &- s_{4}^{\left( {2} \right)} \exp\left( {ik_{4z} L} \right)\delta _{4} \{
s_{1}^{\left( {1} \right)} \delta _{1} \left( {\delta _{3} - \delta _{2}}
\right) - s_{2}^{\left( {1} \right)} \delta _{2} \left( {\delta _{3} -
\delta _{1}}  \right) + s_{3}^{\left( {1} \right)} \delta _{3} \left(
{\delta _{2} - \delta _{1}}  \right)\} = 0 \nonumber\\
 \end{eqnarray}

Similar generation equations in these geometries may be written for a "hot" electron beam \cite{laser_8}.

Before we proceed to the analysis of generation equations, let us examine the behavior of the roots of the dispersion equation for three-wave diffraction. It is known that the most effective interaction of the electron beam and the electromagnetic wave takes place at the root degeneration point when the synchronism condition is fulfilled. Thus, the parameters for the generation threshold can be determined by the weakly-coupled mode method \cite{laser_13}. According to this method, we first find the solutions of the dispersion equation without the electron beam, then the additional synchronism condition is imposed on these solutions. To complete the procedure, one should substitute the value of $\vec {k} = \vec {k}_{0} $ that satisfies the condition $\omega - \vec {k}_{0} \vec {u} = 0$ into the dispersion equation describing three-wave diffraction. Making use of the form of $F_{\sigma} ^{\left( {3} \right)} \left( {\vec {k},\vec {k}_{1} ,\vec
{k}_{2}}  \right)$ in (\ref{laser_eq20}),  we get:
\begin{eqnarray}
\label{laser_eq26}
& & ll_{1} l_{2} - lr_{12} - l_{1} r_{2} - l_{2} r_{1} - f = 0\nonumber \\
& & \beta _{1} \beta _{2} l_{1} l_{2} + l\left( {\beta _{1} l_{1} + \beta _{2}
l_{2}}  \right) - \beta _{1} \beta _{2} r_{12} - \beta _{1} r_{1} - \beta
_{2} r_{2} = 0.
 \end{eqnarray}
In (\ref{laser_eq25})
\[
l = \left( {k_{0}^{2} c^{2} - \omega ^{2}\varepsilon _{0}}
\right)/\omega ^{2} = \sin^{2}\theta + \left( {\frac{{c}}{{u\gamma} }}
\right)^{2} - \chi _{0} ,
\]
\[
l_{1,2} = l + \alpha _{1,2} ,\qquad\alpha _{1,2} =
\frac{{2\vec {k}_{0} \vec {\tau} _{1,2} + \tau _{1,2}^{2}} }{{\omega
^{2}/c^{2}}}
\]
 are the two parameters of deviation from Bragg diffraction conditions, $f =
\chi _{1} \chi _{ - 2} \chi _{2 - 1} + \chi _{2} \chi _{ - 1} \chi _{1 - 2}
$, $\beta _{1,2} = \gamma _{0} /\gamma _{1,2} $ are the asymmetry factors of diffraction,
$\gamma _{0} ,\gamma _{1} ,\gamma _{2} $ are the direction cosines of diffraction. In (\ref{laser_eq26})
the first equation is the requitement for simultaneous fulfillment of the synchronism conditions and the dispersion equation of diffraction  $F_{\sigma} ^{\left( {3} \right)} \left( {\vec {k},\vec {k}_{1}
,\vec {k}_{2}}  \right) = 0$. The second equation in  (\ref{laser_eq26}) appears due to the requirement for root degeneration of the diffraction equation. Introducing new variables
 $z_{1} = l_{1} /l$ and $z_{2} = l_{2} /l$, we obtain from (\ref{laser_eq26}):
\begin{eqnarray}
\label{laser_eq27}
& & z_{1} z_{2} - \frac{{r_{2} z_{1} + r_{1} z_{2}} }{{l^{2}}} - \frac{{r_{12}
}}{{l^{2}}} - \frac{{f}}{{l^{3}}} = 0 \\
& & \beta _{1} \beta _{2} z_{1} z_{2} + \beta _{1} z_{1} + \beta _{2} z_{2} -
\beta _{1} \beta _{2} \frac{{r_{12}} }{{l^{2}}} - \frac{{\beta _{1} r_{1} -
\beta _{2} r_{2}} }{{l^{2}}} = 0. \nonumber
 \end{eqnarray}
Solve (\ref{laser_eq27}) for $z_{i} $:
\begin{eqnarray}
\label{laser_eq28}
& & z_{1} = \frac{{r_{1} - \beta _{2} \chi _{1} \chi _{2} \frac{{\chi _{1 - 2}
}}{{l}} \pm \chi _{1 - 2} \left( {1 + \frac{{\chi _{1} \chi _{2}} }{{\chi
_{1 - 2} l}}} \right)\sqrt { - \frac{{\beta _{2}} }{{\beta _{1}} }\left(
{l^{2} + \beta _{1} r_{1} + \beta _{2} r_{2}}  \right)}} }{{l^{2} + \beta
_{2} r_{2}} } \nonumber\\
& & z_{2} = \frac{{r_{2} - \beta _{1} \chi _{1} \chi _{2} \frac{{\chi _{1 - 2}
}}{{l}} \mp \chi _{1 - 2} \left( {1 + \frac{{\chi _{1} \chi _{2}} }{{\chi
_{1 - 2} l}}} \right)\sqrt { - \frac{{\beta _{1}} }{{\beta _{2}} }\left(
{l^{2} + \beta _{1} r_{1} + \beta _{2} r_{2}}  \right)}} }{{l^{2} + \beta
_{1} r_{1}} }.\nonumber\\
\end{eqnarray}
Thus, at the point of intersection of roots, the parameters of deviation from the Bragg conditions $a_{1,2}$ appear to be expressed in terms of the angle of the radiation wave vector with respect to the velocity vector (recall here that
$l = \sin^{2}\theta + \left( c(u\gamma) \right)^{2} - \chi _{0}$), the asymmetry factors of diffraction  ($\beta _{i} $) and the polarizability  ($\chi _{i} $) of the periodic structure. From (\ref{laser_eq28}) follows that at the point of degeneration of diffraction roots and simultaneous fulfillment of Cherenkov synchronism, the following condition should hold:
\begin{equation}
\label{laser_eq29}
\beta _{1} \beta _{2} \left( {l^{2} + \beta _{1} r_{1} + \beta _{2} r_{2}}
\right) < 0.
\end{equation}
From (\ref{laser_eq29}) follows that there are no degeneration points in Laue geometry ($r_{1} ,r_{2} >
0$, while for the Laue case doth asymmetry factors $\beta _{1} $  and $\beta _{2} $  are positive values). In Bragg-Bragg geometry, a critical angle exists:
\begin{equation}
\label{laser_eq30}
sin\theta _{thr} = \sqrt {\sqrt { - \beta _{1} r_{1} - \beta _{2} r_{2}}  +
\chi _{0} - \left( {\frac{{c}}{{u\gamma} }} \right)^{2}} .
\end{equation}
The degeneration region occurs at radiation angles $\theta <
\theta _{\mathrm{thr} }$. So, in the X-ray spectral range in
Bragg-Bragg geometry at least one of the asymmetry factors should
be large. In Bragg-Laue geometry there is the opposite situation,
when the degeneration region occurs at $\theta >
\theta_{\mathrm{thr}} $. In three-wave geometry there is a
possibility of threefold degeneration of roots.  At the point of
threefold degeneration, there is a strong relationship among the
direction of the photon emission, asymmetry factors, and the
polarizabilities of the periodic structure.

The conditions for the threshold current values are obtained by
through solving equations (\ref{laser_eq24}) and
(\ref{laser_eq25}). In a weak single-stage amplification regime
($|Imk_{z} L| \ll 1$), this conditions in the region of two-fold
degeneration have the form:
\begin{equation}
\label{laser_eq31}
G^{\left( {b} \right)} = |a|\chi _{0}^{\prime\prime} + \frac{{16}}{{\left| {\beta _{1}
\beta _{2}}  \right|}}\left( {\frac{{\gamma _{0} c}}{{\vec {n}\vec {u}}}}
\right)^{3}\frac{{\pi ^{2}n^{2}}}{{\left( {klL_{\ast} }  \right)^{2}kL_{\ast
}} }|\eta _{BL\left( {BB.} \right)} |
\end{equation}
In (\ref{laser_eq31}) the following notations are used:
\[
a = \frac{{z_{1} z_{2} + z_{1} + z_{2} - \frac{{r_{1} + r_{2} + r_{12}
}}{{l^{2}}} + \frac{{r_{12} + z_{1} r_{2}^{\prime\prime} + z_{2} r_{1}^{\prime\prime}} }{{l\chi
_{0}^{\prime\prime}} } + \frac{{f"}}{{l^{2}\chi _{0}^{\prime\prime}} }}}{{z_{1} z_{2} -
\frac{{r_{12}} }{{l^{2}}}}},
\]

\[
a = \frac{{z_{1} z_{2} + z_{1} + z_{2} - \frac{{r_{1} + r_{2} + r_{12}
}}{{l^{2}}} + \frac{{r_{12} + z_{1} r_{2}^{//} + z_{2} r_{1}^{//}} }{{l\chi
_{0}^{//}} } + \frac{{f"}}{{l^{2}\chi _{0}^{//}} }}}{{z_{1} z_{2} -
\frac{{r_{12}} }{{l^{2}}}}},
\]

\[
\eta _{BL} = X\frac{{\left( {s_{3}^{\left( {1} \right)} - s_{1}^{\left( {2}
\right)}}  \right)\varsigma _{1} + s_{1}^{\left( {2} \right)} \varsigma _{2}
- s_{3}^{\left( {2} \right)} \varsigma _{2} cos\{ k\left( {\delta _{1}^{/} -
\delta _{3}^{/}}  \right)L\}} }{{s_{1}^{\left( {2} \right)} \left(
{s_{3}^{\left( {1} \right)} - s_{1}^{\left( {1} \right)}}  \right)\left(
{z_{1} z_{2} - \frac{{r_{12}} }{{l^{2}}}} \right)}},
\]

\[
\eta _{BB} = X\frac{{s_{3}^{\left( {1} \right)} \varsigma _{1} -
s_{3}^{\left( {2} \right)} \varsigma _{2} + \left( {s_{2}^{\left( {1}
\right)} \varsigma _{1} - s_{1}^{\left( {2} \right)} \varsigma _{2}}
\right)cos\{ k\left( {\delta _{1}^{/} - \delta _{3}^{/}}  \right)L\}
}}{{\left( {s_{3}^{\left( {2} \right)} s_{1}^{\left( {1} \right)} -
s_{3}^{\left( {1} \right)} s_{1}^{\left( {2} \right)}}  \right)\left( {z_{1}
z_{2} - \frac{{r_{12}} }{{l^{2}}}} \right)}},
\]

\[
\varsigma _{1} = \frac{{\left( {1 + \beta _{1} z_{1}}  \right)\left(
{\frac{{z_{1} \chi _{2}} }{{l}} + \frac{{\chi _{1} \chi _{2 - 1}
}}{{l^{2}}}} \right) - \left( {z_{1} - \frac{{r_{1}} }{{l^{2}}}}
\right)\frac{{\chi _{2}} }{{l}}}}{{\beta _{1} \left( {\frac{{z_{1} \chi _{2}
}}{{l}} + \frac{{\chi _{1} \chi _{2 - 1}} }{{l^{2}}}} \right)^{2}}},
\]
\[
\varsigma _{2} = \frac{{\left( {1 + \beta _{2} z_{2}}  \right)\left(
{\frac{{z_{2} \chi _{1}} }{{l}} + \frac{{\chi _{2} \chi _{1 - 2}
}}{{l^{2}}}} \right) - \left( {z_{2} - \frac{{r_{2}} }{{l^{2}}}}
\right)\frac{{\chi _{1}} }{{l}}}}{{\beta _{2} \left( {\frac{{z_{2} \chi _{2}
}}{{l}} + \frac{{\chi _{2} \chi _{1 - 2}} }{{l^{2}}}} \right)^{2}}}.
\]
$G^{\left( {b} \right)}$ is defined in  (\ref{laser_eq16}), (\ref{laser_eq17}).

The threshold condition in equation (\ref{laser_eq31}) holds true in the region of the two-fold degeneration of roots $k_{1z}$, $k_{2z}$, and as the modes corresponding to these roots pass through the interaction area, their relative phase shift should satisfy the condition $\left( {k_{1z} - k_{2z}}  \right)L = 2\pi n$ (here the roots are "almost" degenerated if ($2\pi n)/k|\chi _{\tau}  |L) \ll 1$). The third diffraction root is located at a distance $|k_{3z} - k_{1z} |\sim|\chi _{\tau}  |$, $|k_{3z} - k_{2z} |\sim|\chi _{\tau}  |$ from the degenerated roots.

The dependence of the threshold conditions on the length of the interaction area at the point of three-fold degeneration changes appreciably:
\begin{equation}
\label{laser_eq32}
G = A\chi _{0}^{\prime\prime} + \frac{{B}}{{kL}}\left( {\frac{{2\pi} }{{klL}}}
\right)^{4} \quad .
\end{equation}
Realization of this regime requires that the following phase
conditions be fulfilled: $\left( {k_{1z} - k_{2z}}  \right)L =
2\pi n$, $\left( {k_{2z} - k_{3z}}  \right)L = 2\pi m$, $n \ne m$.
The coefficients $A$, $B$ in (\ref{laser_eq32}) depend on the
polarizabilities, $z_{1}$, $z_{2} $ and the indices $m$,  $n$.
Since they are awkward, they are dropped here. From
(\ref{laser_eq32}) follows that in the region of three-fold
degeneration of roots, the functional dependence of the losses at
the boundaries on $L$ changes significantly. According to
(\ref{laser_eq32}), for a "cold" electron beam, when absorption is
not important,
$
j_{\mathrm{thr}} \sim\frac{{1}}{{\left( {kL} \right)^{3}}}\left(
{\frac{{2\pi} }{{klL}}} \right)^{4}
$
(it would be recalled here that in the regime of a "cold" electron beam  $G\sim jL^{2}$ (see (\ref{laser_eq17})).  Under dynamical diffraction, when inequality
$
\frac{{4\pi} }{{klL}} \ll 1
$
holds, this dependence leads to an appreciable reduction of the threshold current. Under multi-wave ($s$-wave) VDF, the threshold current depends on $L$ as
\begin{equation}
\label{laser_eq33}
j_{thr} \sim\frac{{1}}{{\left( {kL} \right)^{3}}}\left( {\frac{{2\pi
}}{{klL}}} \right)^{2\left( {s - 1} \right)},
\end{equation}
so the transition to  multi-wave diffraction enables one to significantly reduce the longitudinal dimension of the generating system.

As follows from the above results, the volume distributed feedback
(VDF) has a number of advantages that make its application
beneficial for generating stimulated radiation in a wide spectral
range (with the wavelengths from centimeters and millimeters to
angstr\"{o}ms). Moreover, in a short-wave spectral range, the size
of the radiating system reduces appreciably due to the change of
the functional dependence under multi-wave VDF. A short-wave
spectrum corresponds to greater values of the wave number $k$, so
at a given operating current, due to the presence of the factor $
\left( {\frac{{2\pi} }{{klL}}} \right)^{2\left( {s - 1} \right)},
$ $L$ can be appreciably reduced.  In optical and X-ray ranges,
where the requirements for the current density and the quality of
the beam are very strict, there appears the possibility to
noticeably reduce the threshold values of the current for a given
beam propagation area. In this case the VFEL is a unique system
providing lasing at  relatively small interaction lengths. In a
microwave range,  VFELs are beneficial for both the reduction of
the generator size and selection of modes when producing
high-power radiation pulses in oversized generators.

\section{Application of volume diffraction gratings for terahertz
lasing in Volume FELs (VFELs)} \label{nim03_sec1}

Generation of radiation in millimeter and far-infrared range
with nonrelativistic and low-relativistic electron beams is a complicated task. 
Gyrotrons and cyclotron resonance facilities are
used as sources in millimeter and sub-millimeter range, but for
their operation a magnetic field of several tens of kiloGauss
$
\left(\omega \sim \frac{eH}{mc}\gamma \right)
$
 is necessary. Slow-wave
devices (TWT, BWT, orotrons) in this range require application of
dense and thin ($<0.1$ mm) electron beams because only electrons
passing near the slowing structure at distance   $d \leq \lambda
\beta \gamma /(4\pi )$ can effectively interact with
electromagnetic waves.
It is difficult to guide thin beams near a slowing structure with
desired accuracy. And electrical endurance of resonator limits
radiation power and density of the acceptable electron beam.
Conventional waveguide systems are essentially restricted by the
requirement for transverse dimensions of a resonator, which should
not significantly exceed the radiation wavelength. Otherwise,
the generation efficiency decreases abruptly due to the excitation of
plenty of modes. Most of the above problems can be overcome in Volume Free Electron Lasers
 (VFEL) 

 In  volume FELs, the greater part of the electron beam interacts with an electromagnetic wave due to
 volume distributed interaction.
Transverse dimensions of a VFEL resonator could significantly exceed
radiation wavelength $D \gg \lambda $. In addition, the electron beam
and the radiation power are distributed over the whole volume, which is
beneficial for electrical endurance of the system. Multi-wave
Bragg dynamical diffraction provides mode discrimination in VFELs.

\subsection{Amplification and generation in a photonic crystal}
\begin{figure}[tbp]
\epsfxsize =8cm \centerline{\epsfbox{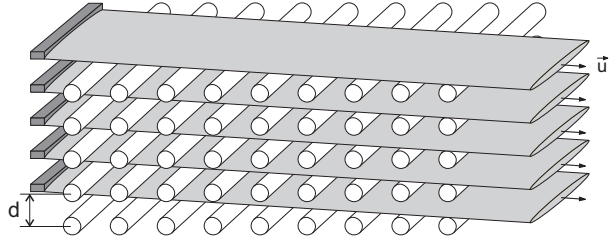}}
\caption{General view of Volume Free Electron Laser formed by
metal threads with several sheet electron beams. }
\label{nim03_f1}
\end{figure}

\begin{figure}[htbp]
\epsfxsize =8cm \centerline{\epsfbox{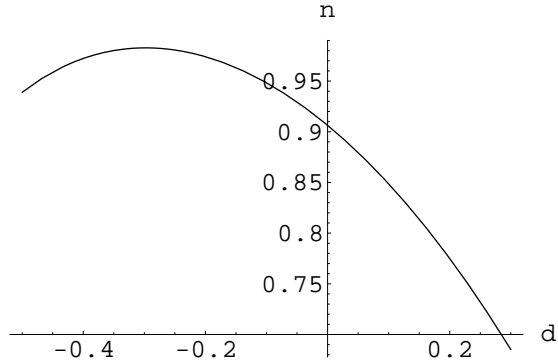}} 
\caption{Calculated dependence of {temporal increment} on tuning
out parameter} \label{nim03_Fig.1}
\end{figure}

 Let us consider an electron beam with velocity $\vec{u}$ passing
through a periodic structure composed of either dielectric or metal threads (see Figure \ref{nim03_f1})
Fields, appearing while an electron beam passes through a volume
spatially periodic medium, are described by the set of equations
given in \cite{laser_8} 
 Instability of electron beam is described by
the dispersion equation \cite{laser_8}: 
\begin{equation}
(k^{2}c^{2}-\omega ^{2}\varepsilon )(k_{\tau }^{2}c^{2}-\omega
^{2}\varepsilon +\chi _{\tau }^{(b)})-\omega ^{2}\chi _{\tau }\chi
_{-\tau }=0.  \label{nim03_disp}
\end{equation}
$\vec{k}_{\tau }=\vec{k}+\vec{\tau}$ is the wave vector of the
diffracted photon,
$
\vec{\tau}=\left\{\frac{2\pi }{a}l;\frac{2\pi
}{b}m;\frac{2\pi }{c}n\right\}
$
 are the reciprocal lattice vectors,
$a,b, c$ are the translation periods,
 $\chi _{\alpha }^{(b)}$ is the part of dielectric
susceptibility caused by the presence of the electron beam. As
synchronism conditions are incompatible with those of Bragg,  in
the instability range $k^{2}\neq k_{\tau }^{2}$. At the same time,
two different types of instability exist, depending on radiation
frequency. Amplification takes place when the electron beam is in
synchronism with the electromagnetic component $\vec{k}+\vec{\tau}$,
which has a positive projection $k_{z}$. If the projection $k_{z}$ is
negative and generation threshold is reached, then generation
evolves. In the first case, radiation propagates along the
transmitted wave which has positive projection of group velocity
$
v_{z}=\frac{c^{2}k_{z}^{(0)}%
}{\omega }\quad (k_{z}^{(0)}=\sqrt{\omega ^{2}\varepsilon -k_{\perp
}^{2}}),
$
 and beam disturbance moves along it. In the second case,
the group velocity has negative projection
$
v_{z}=-\frac{c^{2}k_{z}^{(0)}}{\omega },
$
 and radiation propagates
along the back-wave and the electromagnetic wave comes from the range of
the greatest beam disturbance to the place, where electrons come
into the interaction area. For a one-dimensional structure such a
mechanism is realized in a backward-wave tube. In amplification case,
equation (\ref{nim03_disp}) gives for the increment of instability: $\texttt{Im} k_{z}^{\prime
}=-\frac{\sqrt{3}}{2}f $, where
\[
f=\sqrt[3]{\frac{h\omega _{L}^{2}(\vec{u}\vec{e}%
^{\tau })^{2}\omega ^{4}r}{2k_{z}^{(0)}c^{4}u_{z}^{2}\left(
k_{\tau
}^{2}c^{2}-\omega ^{2}\varepsilon _{0}\right) }},
\]
 if the condition
 $ 2k_{z}^{\prime }f\gg \frac{\omega ^{2}\chi _{0}"}{c^{2}} $ is
fulfilled. Here $r=\chi_{\tau}\chi_{-\tau}$,
$h(\vec{u}\vec{e}^{\tau})^2/c^2=1/\gamma^3$ if the electron beam
propagates in a strong guiding magnetic field, otherwise,
$h=1/\gamma$. In case
$
2k_{z}^{\prime }f\ll \frac{\omega ^{2}\chi
_{0}"}{c^{2}}
$
a dissipative instability evolves. Its increment is
\[
\texttt{Im}
k_{z}=-\frac{c}{\omega}\sqrt{\frac{k_{z}^{(0)}f^{3}}{\chi_{0}}}.
\]
If inequalities $k_{z}^{`2}\gg 2k_{z}k_{z}^{`}$ and $k_{z}^{`2}\gg \frac{%
\omega ^{2}\chi _{0}"}{c^{2}}$ are fulfilled, the spatial
increment of instability can be expressed as
\[ \texttt{Im}
k_{z}^{`}=-\left( \frac{h\omega _{L}^{2}(\vec{u}\vec{e}^{\tau
})^{2}\omega ^{4}r}{c^{4}\left( k_{\tau }^{2}c^{2}-\omega
^{2}\varepsilon _{0}\right) u_{z}^{2}}\right) ^{1/4},
\]
but the parameters providing such dependence correspond to the conversion
from the amplification to the generation regime (for the Compton instability
this situation takes place at $%
k_{z}^{(0)}\approx 0$). The frequency of amplified radiation is
defined as:
\begin{equation}
\omega =\frac{\vec{\tau}\vec{u}}{1-\beta _{x}\eta _{x}-\beta
_{y}\eta _{y} - \beta _{z}\sqrt{\varepsilon -\eta _{x}^{2}-\eta
_{y}^{2}}}. \label{nim03_freqa}
\end{equation}
The instability in the generation regime is described by the
temporal increment and cannot be described by the spatial
increment. The increment of absolute instability can be found by
solving the equation $\texttt{Im} k_{z}^{(+)}(\omega )=\texttt{Im}
k_{z}^{(-)}(\omega )  $ with respect to the imaginary part of
$\omega $. Calculated dependence of temporal increment on the
parameter of detuning is presented in Figure \ref{nim03_Fig.1}.

Axes in  Figure \ref{nim03_Fig.1}
 are denoted as:
\[
d=\frac{a_{1}+k_{z}^{(0)}}{f},\quad n=\frac{\frac{\omega "}{u_{z}}+\frac{2\omega \omega "\varepsilon
_{0}+\omega ^{2}\chi _{0}"}{c^{2}k_{z}^{(0)}}}{f}.
\]
 It follows from  Figure \ref{nim03_Fig.1} that at a certain value of the parameter
 of detuning:
\begin{equation}
\frac{a_{1}+k_{z}^{(0)}}{f}\approx -0.3 \label{nim03_detuning}
\end{equation}
the increment of instability has a maximum
\begin{equation}
\frac{\frac{\omega "}{u_{z}}+\frac{2\omega \omega "\varepsilon
_{0}+\omega ^{2}\chi _{0}"}{c^{2}k_{z}^{(0)}}}{f}\approx 0.98.
\label{nim03_solution}
\end{equation}
The increment of absolute instability can be found from
(\ref{nim03_solution}).
 The absolute instability can evolve if current
exceeds start value, which is determined by dissipation.
The amplification regime has no threshold and decisive influence of
dissipation causes dissipative instability.  Frequencies,
corresponding to the generation regime, are defined by an expression
different from (\ref{nim03_freqa}):

\begin{equation}
\omega =\frac{\vec{\tau}\vec{u}}{1-\beta _{x}\eta _{x}-\beta
_{y}\eta _{y}+\beta _{z}\sqrt{\varepsilon -\eta _{x}^{2}-\eta
_{y}^{2}}}. \label{nim03_freqr}
\end{equation}

Thus, it follows from (\ref{nim03_freqa}),  (\ref{nim03_freqr}) that
the change of a radiation angle causes smooth frequency tuning. As a result,
the generation frequencies are less than those corresponding to
the amplification regime. Hence, using the system as an amplifier, one should
add dispersion elements in it to raise dissipation in the frequency
range, where generation occurs.

The use of Bragg multi-wave distributed feedback increases
generation efficiency and provides discrimination of generated
modes.  If the conditions of synchronism and Bragg conditions are
not fulfilled simultaneously, photonic crystals (diffraction
structures) with different periods can be applied
\cite{nim03_nonrel}. One of them provides the synchronism of the
electromagnetic wave with the electron beam $\omega
-\vec{k}\vec{u}=\vec{\tau}_{1}\vec{u}$. The second photonic
crystal (diffraction structure) evolves distributed Bragg
coupling $|\vec{k}%
|\approx |\vec{k}+\vec{\tau}_{j}|$, $\vec{\tau}_{j}$ ($j=2\div n$)
are the reciprocal lattice vectors of the second structure.
Threshold conditions for s-wave diffraction are converted to:
\begin{equation}
G^{(s)}=\frac{a_{s}^{3}}{(k\chi L_{\ast })^{2s}kL_{\ast }}+\chi
_{0}^{\prime \prime }b_{s}.  \label{nim03_s_wave}
\end{equation}
For dynamical diffraction, when $k|\chi |L_{\ast }\gg 1$, either
the generation start current or the length of the generation zone at certain
current value can be reduced.

 Each Bragg condition holds one of free parameters. For example, for
certain geometry and electron beam velocity, two conditions for
three-wave diffraction entirely determine transverse components of
wave vectors $k_{x}$ and $k_{y}$, and therefore generation
frequency (see (\ref{nim03_freqa}), (\ref{nim03_freqr})). Hence,
volume photonic crystal (diffraction system) provides mode
discrimination due to multi-wave diffraction.

 The above results affirm that a photonic crystal (volume diffraction structure) provides
both amplification and generation regimes even in the absence of
dynamical diffraction. In the latter case, generation evolves with
backward wave, similar to the backward-wave tube. The frequency in such
structures is changed smoothly either by  a smooth variation of
the radiation angle (variation of $k_x$ and $k_y$) or by
the rotation of the diffraction grating or the electron beam (change of $\vec{\tau}\vec{u}$)
(see (\ref{nim03_freqa})). For certain geometry and reciprocal
lattice vector,  amplification corresponds to higher frequencies than
generation. Rotation of either the diffraction grating or the electron
beam also changes the value of the  boundary frequency, which separates
generation and amplification ranges. The use of multi-wave distributed
feedback owing to Bragg diffraction allows one either to increase
the generation efficiency or to reduce the length of the interaction area
(\ref{nim03_s_wave}). In this case, generation is available with both
backward and following waves.
 In particular, the proposed volume structure can be used for
generation of sub-millimeter radiation by accelerator  LIU-3000.
The parameters of this setup are: electron beam energy $E=800$ keV, beam current $%
I=100\div 200$ A. To generate radiation with wavelength of 0.3 mm in
such a system, a volume structure composed of strained threads should have a period $%
\sim $ 2 mm, and a period of diffraction grating providing Bragg coupling $%
\sim $ 0.16 mm. Generation of radiation in the  terahertz range
can also be obtained in a photonic crystal using higher harmonics
of the Fourier expansion of the dielectric permittivity of the
crystal \cite{lanl_22}.

\section{Dependence of VFEL threshold conditions on undulator parameters}
\label{und_sec1}

A sharp increase of amplification caused by volume distributed feedback (VDFB) yields a noticeable reduction
of threshold currents necessary for the lasing start.
This fact is particularly important
for lasing in sub-millimeter and visible ranges and for shorter wavelengths.
Explicit expressions of the VFEL
threshold currents were obtained in \cite{laser_8}.
Here we shall consider the dependence of VFEL starting current on
the undulator parameters \cite{lanl_10}.

The set of equations describing the interaction of a relativistic electron beam,
which propagates in the spatially periodic structure of the undulator is \cite{laser_8}:
\begin{eqnarray}
DE-\omega ^{2}\chi _{1}E_{1}-\omega ^{2}\chi _{2}E_{2}-...=0 \nonumber \\
-\omega ^{2}\chi _{-1}E+D_{1}E_{1}-\omega ^{2}\chi _{2-1}E_{2}-...=0  \label{und_system} \\
-\omega ^{2}\chi _{-2}E-\omega ^{2}\chi _{1-2}E_{1}+D_{2}E_{2}=0-..., \nonumber \\
\end{eqnarray}
where $D_{\alpha }=k_{\alpha }^{2}c^{2}-\omega ^{2}\varepsilon +\chi
_{\alpha }^{(b)}$ , $\vec{k}_{\alpha }=\vec{k}+\vec{\tau}_{\alpha }$ are the
wave vectors of photons diffracted by the crystal planes with corresponding
reciprocal vectors $\vec{\tau}_{\alpha }$, $\varepsilon_{0} =1+\chi _{0}$ , $\chi
_{\alpha }$ are the dielectric constants of a periodic structure;
$\chi _{\alpha }^{(b)}$ is the part of dielectric susceptibility appearing from
the interaction between an electron beam propagating in the undulator
and radiation.
Setting the determinant of the linear system (\ref{und_system}) equal to zero,
one can obtain the dispersion equation for the system
"electromagnetic wave + undulator + electron beam + periodic structure".
In the case of two-wave dynamical diffraction, this equation has the following form:
\begin{eqnarray}
DD_1-\omega^4\chi _1\chi _{-1}=0 \label{und_two}
\end{eqnarray}
For the system with finite interaction length, the solution of the boundary problem
can be presented as a sum:
\begin{eqnarray}
\vec{E}+\vec{E}_{1}= \sum_{i}c_{i}\exp
\{i\vec{k}_{i}\vec{r}\}(\vec{e}+\vec{e} _{1}s_{1}^{(i)}\exp
{i\vec{\tau} \vec{r}}) \label{und_bound2}
\end{eqnarray}
In (\ref{und_bound2})  $s_{1}^{(i)}=(k_i^2 c^2-\omega^2\varepsilon
_0)/(\omega^2\chi _{1})$ and $\vec{k}_i$ are the coupling
coefficients between the diffracted and transmitted waves
 and the solutions of dispersion equation
(\ref{und_two}), respectively. The coefficients $c_i$ are defined
by boundary conditions at the system ends $z=0$ and $L$. The part
of the electron beam energy converting into radiation can be
expressed by:
\begin{eqnarray}
I&\sim &\gamma _0 |E(z=L)|^2+|\gamma _1| |E_1(z=0)|^2 \label{und_energy} \\
&=&(\gamma _0 |a|^2+|\gamma _1| |b|^2)\left(\frac{\gamma _0 c}{\vec{n}\vec{u}}\right)^3
\frac{16 \pi^2 n^2}{-\beta (k|\chi _1|L_*)^2kL_*(\Gamma _{\mathrm{start}}-\Gamma)},
\nonumber
\end{eqnarray}
where $L$ is the length of interaction in the undulator
\begin{eqnarray*}
\Gamma _{\mathrm{start}}&=&\left(\frac{\gamma _0 c}{\vec{n}\vec{u}}
\right)^3\frac{16 \pi^2 n^2}{-\beta (k|\chi _1|L_*)^2}-
\chi^{\prime\prime}\left(1-\beta \pm \frac{r^{\prime\prime}\sqrt{-\beta}}{|\chi _1|\chi^{\prime\prime}}\right) \\
\Gamma & =&\frac{\pi^2 n^2}{4}\frac{\pi \Theta _{s}^{2}j_0c^{2}}
{\gamma _{z}^{2}\gamma I_{A}\omega^2} k^2 L_*^2  f(y),
\end{eqnarray*}
$f(y)$ is the function describing the dependence of generation on
detuning from the synchronism condition, $y=(\omega
-\vec{k}\vec{v}_w-\Omega )L/(2u_z)$ is the detuning factor,
$\beta=\gamma_0/\gamma_1$ is the diffraction asymmetry factor,
 $\gamma_0$, $\gamma_1$  are diffraction cosines,
$\chi^{\prime\prime} =\texttt{Im}\chi_0$, $\Theta _{s}=eH_{w}/(mc^{2}\gamma k_{w})$, $\gamma _{z}^{-2}=\gamma
^{-2}+\Theta _{s}^{2}$, $k_{w}=2\pi/\lambda_w$, $\lambda_w$ is the undulator period, $H_w$ is the undulator field.
It follows from (\ref{und_energy}) that:

1. the starting current in the case of two-wave diffraction is proportional to
$j_{\mathrm{start}}\sim (kL)^{-1}(k\chi _1L)^{-2} $;

2. the non-one dimensional VDFB provides the possibility to decrease the starting
current of generation
by varying the angles between the waves.

3. if the electron beam current is less than the starting value
$j<j_{\mathrm{start}}$, the energy of the electromagnetic wave at
the system entrance can be written in the form:
\begin{eqnarray}
E/(\gamma _0 |a|^2+|\gamma _1| |b|^2)=1
-\beta \frac{\pi ^{2}n^{2}}{4}\frac{\pi \Theta _{s}^{2}j_{0}c^{2}}{\gamma
_{z}^{2}\gamma I_{A}\omega ^{2}}(kL)^3
\left( \frac{k\chi _{\tau }L}{4\pi }\right) ^{2} f(y)\label{und_ampl}
\end{eqnarray}

The conventional FEL gain is proportional to $(kL)^3$
\cite{laser_1}, but as follows from (\ref{und_ampl}) in the case
of two-wave diffraction, the gain gets an additional factor $
\sim~\left( \frac{k\chi _{\tau }L}{4\pi }\right) ^{2}, $ which
noticeably exceeds unity in conditions of dynamical diffraction.
Such an increase of the radiation output at the degeneration point
can be explained by the reduction of the wave group velocity,
which can be written as:
\begin{equation}
v_{g}=-\left( \frac{\partial D}{\partial k_{z}}\right) /\left( \frac{
\partial D}{\partial \omega }\right)\sim \prod\limits_{i<j}(k_{zi}-k_{zj})  \label{und_group}
\end{equation}
It follows from (\ref{und_group}) that for multi-wave dynamical diffraction at the
$s$-fold-degeneration point  $v_g\sim v_0/(kL)^s$, the starting current
$j_{\mathrm{start}}\sim (kL)^{-3}(k\chi _1L)^{-2s}$ and the amplification
are proportional to  $(kL)^3(k\chi _1L)^{2s}$.
It should be noted that  the considered effects take place in a wide spectral range
for wavelengths from centimeters to X-ray,
and the influence of the effect increases with frequency growth.

The generation threshold in the undulator VFEL in the case of VDFB  can be achieved at
a lower electron beam current and a shorter undulator length when special
conditions of  degeneration of roots are fulfilled. Changing the VDFB conditions by varying
the volume geometry parameters (for example, the angle between the wave vectors) gives the
possibility to increase the $Q$-factor and to decrease the starting current (see  Figure \ref{und_fig.ratio}).
Hence, the generation efficiency  can be increased.

\section{Use of a dynamical undulator mechanism to produce short wavelength
radiation in VFELs} \label{mop_sec1}

Here we shall consider VFEL lasing in a system with a dynamical undulator \cite{lanl_11}. In
this system radiation of long wavelengths creates the undulator for
lasing at a shorter wavelength. Two diffraction gratings with
different spatial periods form a VFEL resonator. The grating with
longer period pumps the resonator with the long wavelength radiation
to provide the necessary amplitude of the undulator field. The grating
with the shorter period is used to select the mode for the short wavelength
radiation. Lasing of such a system in the terahertz frequency range is
discussed below.

Numerous applications can benefit from the development of powerful
electromagnetic generators with frequency tuning in millimeter,
sub-millimeter and terahertz wavelength range using
low-relativistic electron beams. One of the ways to create such
generators is to use VFEL principles. The main distinction of VFEL
in comparison with the traditional FEL is the use of the non
one-dimensional distributed feedback, which allows wide range
tuning of frequency, decreases starting currents of generation,
and allows one to use a wide electron beam (or several beams)
\cite{berk_139,laser_8}. In the VFEL the generation evolves in a
large volume, which  increases the electrical strength of the
resonator (the electromagnetic power and electron beam are
distributed over a greater volume). This peculiarity of the VFEL
is  essential for generation of power and super-power
electromagnetic pulses. The mode discrimination in such an
oversized system is carried out by multi-wave dynamical
diffraction \cite{laser_8}. Low relativistic electron beams in the
undulator system can be used for radiation of short wavelength
radiation, but it requires manufacturing of undulators with a
small period. For example, to obtain radiation with the wavelength
of $ 0.3 -  1~$ mm at the beam energy $E=800$ KeV$ - 1 $ MeV an
undulator with the period $\sim 0.3-1~$cm is necessary. This is an
extremely complicated problem. The use of a two-stage FEL with the
dynamical wiggler generated by an electron beam \cite{laser_1} is
a possible solution to this problem.

The dynamical wiggler can be created with the help of any radiation
mechanism: Cherenkov, Smith-Purcell, quasi-Cherenkov
\cite{berk_139}, undulator. VFEL principles provide advantages
of  the two-stage generation scheme and, in particular, allow one  to
smoothly tune the period of dynamical wiggler by rotating the diffraction grating.
There is the possibility of smooth frequency tuning
 for both the pump wave and the signal wave either by
variation of geometric parameters of  the volume diffraction grating
or by their rotation. Moreover, the VFEL allows one to create the dynamical
wiggler in a large volume, that is a great problem for a static wiggler.

\begin{figure}[htbp]
\epsfxsize =8cm \centerline{\epsfbox{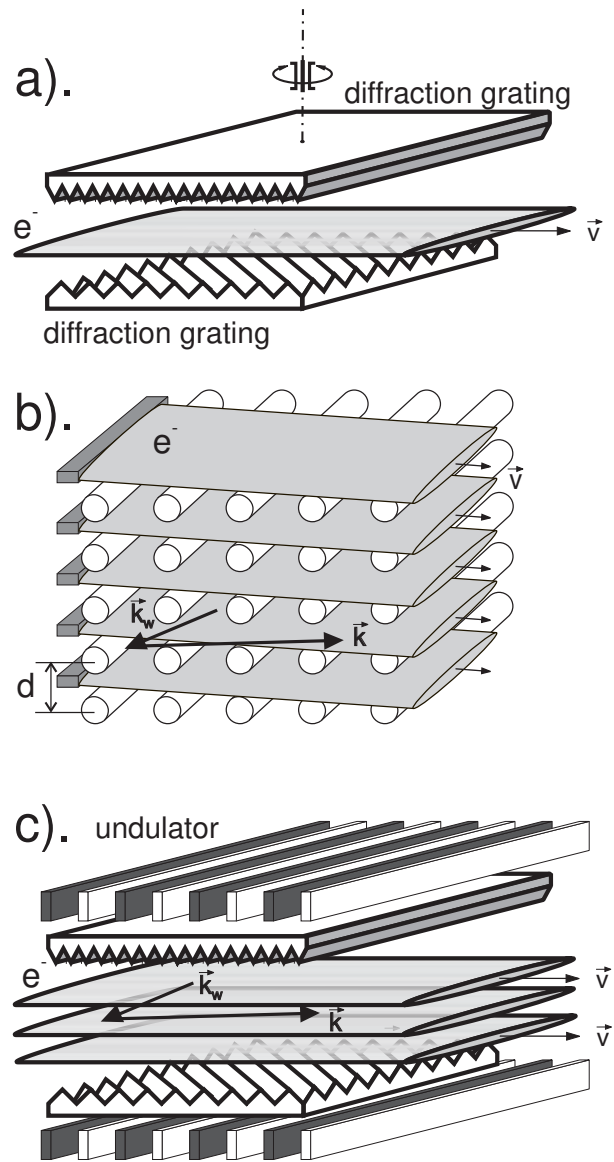}}
\caption{The schemes of dynamic wiggler. Rotation of diffraction
grating (schemes a). and c).) changes the period of dynamic
wiggler} \label{mop_wigg}
\end{figure}

  There are two stages in the generation scheme proposed above:

(a) creation of the dynamical wiggler in a system with  two-dimensional
(three-dimensional) gratings (in other words, during this stage, the electromagnetic field,
which exists inside the VEFL resonator, is used to create the dynamical wiggler).
Smooth variation of the orientation of the diffraction grating in the VEFL resonator provides  means for a smooth change of the dynamical wiggler parameters;

(b) radiation is generated by an electron beam  interacting with
the dynamical wiggler, which is created during the previous stage
(stage a). \newline Both stages evolve in the same volume.

The idea of a two-stage tunable VFEL described above can be realized in different ways. Let us consider
some examples (Figure \ref{mop_wigg}a-c). Figure \ref{mop_wigg} a displays the two-stage device in which the dynamical wiggler is realized on the basis of a VEFL generator using the Smith-Purcell radiation mechanism. This generation scheme was considered in \cite{nim03_nonrel}, when experimentally realized, first lasing of a VFEL was observed \cite{hyb_FirstLasing}

In this case only a part $\delta l_x$ of an electron beam participates in the generation process during the first stage:
$
\frac{\delta l_x}{l_x}\sim \lambda_w\beta\gamma/(4\pi l_x),
$
here $\delta l_x$ is the transverse size of the part of the
electron beam participating in the interaction, $l_x$ is the
transverse size of the electron beam, $\lambda_w$ is the
wavelength, $\beta=u/c$, $u$ is the electron velocity, $\gamma$ is
the Lorentz factor. The resonator in Figure \ref{mop_wigg}a
consists of two diffraction gratings
\cite{lanl_23a,hyb_FirstLasing}. The lower grating
provides the Smith-Purcell generation mechanism. The period $d_1$
of this grating is determined by
$d_1\sim\beta\gamma^2\lambda_w\cos\varphi$ ($d_1$ is the period of
the Smith-Purcell grating, $\varphi$ is the angle between the
direction of the electron velocity and the direction of the
grating periodicity). The upper grating provides the distributed
feedback \cite{berk_139,laser_8} by multi-wave dynamical
diffraction. The conditions of dynamical diffraction
${|\vec{k}_w|\approx |\vec{k}_w+\vec{\tau}_i|}$ are fulfilled in
this case ($\vec{\tau}_i$ are the reciprocal wave (lattice)
vectors of this grating). It should be noted that the period of
the upper grating does not coincide with that of the lower one.
Radiation accumulated in the resonator during the described first
stage creates the dynamical wiggler. Beam electrons oscillate in
this electromagnetic wiggler and radiate just as in a conventional
FEL (this is not necessarily the same electron beam that
participates in the first stage. It can be another beam of higher
energy). The field inside the resonator is a standing wave.
Traveling waves, which form this standing wave are actually the
pump waves. They are scattered by the electron beam according to
the synchronism condition $\omega-(\vec{k}\vec{u})\approx
\omega_w-(\vec{k}_w\vec{u})$. The resulting wave has the
wavelength
$
\lambda\sim\frac{1-\beta}{1+\beta}\lambda_w
$
(in the last estimation it is supposed that the wave vector $\vec{k}_w$ of the pump wave is antiparallel to the velocity, and the wave vector $\vec{k}$ is parallel to it. For a relativistic beam, this relation has the form $\lambda\sim\lambda_w/(4\gamma^2)$. It should be emphasized that in this case there is one more possible way to create the dynamical wiggler in the resonator shown in Figure \ref{mop_wigg}a. It is based on the excitation of a slow wave, which is diffracted by the lower grating (surface VFEL). The electron beam oscillates in this wiggler and radiates. In this case the upper grating forms the distributed feedback which provides VFEL lasing at shorter wavelength. The change of the radiated frequency is provided by the rotation of both the upper and lower gratings. Figure \ref{mop_wigg}b  presents the variant of the volume diffraction grating which can provide the generation mechanism and distributed feedback simultaneously \cite{berk_139,laser_8}.  Let us note that in these examples the generation mechanism during the first stage is based on the slowing of the electromagnetic wave and only a part of the electron beam participates in the creation of the dynamical wiggler. The whole electron beam participates in the generation process during the second stage. The larger the part of the beam that does not participate in the the first stage interaction, the more unperturbed electrons appear during the second stage and, therefore, they radiate more effectively. The dynamical wiggler in Figure \ref{mop_wigg}c uses the undulator radiation mechanism during the first stage. In this case the dynamical wiggler is formed during the first stage due to the interaction between the electron beam and a conventional magnetostatic undulator. The frequency of the pump wave is
$
\omega_w\sim \frac{2\pi\beta}{d_u(1-\beta\cos(\theta))},
$
where $d_u$  is the period of the magnetostatic undulator. During this stage the lower diffraction grating is used to provide  the distributed feedback for the wave with frequency $\omega_w$ and  operation of the VFEL at this frequency.
During the second stage, electrons oscillate in the field of this wave, which plays the role of the dynamical wiggler. As a result, during the second stage of the process the wave with the frequency $\omega\sim 4\gamma^2\omega_w$ is generated and the distributed feedback is provided by the upper grating, the period of which corresponds to the wave with the frequency $\omega$. The rotation of  diffraction gratings provides frequency tuning.

It is clear that the time $\tau_w$ of the dynamical  wiggler creation is a very important characteristic of the proposed system. This time is determined by $\tau_w\sim Q/\omega_w$, where $\omega_w$ is the frequency of the pump wave.

 The $Q$ factor of resonator for the frequency $ \omega_w $
should be sufficient to create a magnetic field amplitude of
about $ 100 $ G - $ 1 $ kG. It follows from the energy balance
equation in the resonator, $ (\omega_w/Q) V (H _ {m} ^
{2}/8\pi) =W _ {0} $, ($W _ {0} $ is the power of the pump wave
formed by the electron beam) that  $ Q =(\omega_w/ W _
0) V(H _ {m} ^ {2}/8\pi) $, where $V  $ is the cavity
volume, $H _ {m} $ is the amplitude of the magnetic field of the dynamical
wiggler.  It follows from the above that to create the magnetic field of about $100-1000$ G, the time ${\tau_w\sim 10^{-10}- 3\times 10^{-9}}$~ s is necessary ($V= 30$ cm$^3$, $W\sim 30$ MW). As one can see, this time is small enough for the wiggler to evolve while the electron beam passes through the system. When the pump field achieves the necessary magnitude,
stage (b) begins.

Dynamics of the signal electromagnetic wave and the electron beam  in
the system (volume diffraction grating + pump electromagnetic wave)
 in this case is described by equations
\begin{eqnarray}
\label{mop_system} D_0~&E& -\omega ^ {2} \chi _ {1} E _ {1}
-\omega ^ {2} \chi _ {2} E _
{2} - \omega ^ {2} \chi _ {3} E _ {3} -\ldots = 0 \nonumber \\
-\omega ^ {2} \chi _ {-1} &E& +D _ {1} E _ {1} -\omega ^ {2} \chi
_ {2-1} E _ {2} - \omega ^ {2} \chi _ {3-1} E _ {3} -\ldots =0,
\\
-\omega ^ {2} \chi _ {-2} &E& -\omega ^ {2} \chi _ {1-2} E _ {1}
+D _ {2} E _ {2} -\omega ^ {2} \chi _ {3-2} E _ {3} -\ldots
=0.\nonumber
\end{eqnarray}

\begin{figure}[htbp]
\epsfxsize =8cm \centerline{\epsfbox{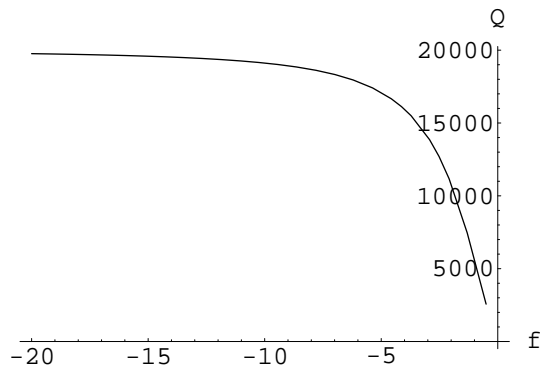}}
\caption{Calculated dependence of Q factor  on diffraction
asymmetry factor $f$ } \label{mop_Fig.2}
\end{figure}

In (\ref{mop_system}) $D _ {\alpha} =k _ {\alpha} ^ {2} c ^ {2} -\omega ^
{2} \varepsilon + \chi _ {\alpha} ^ {(b)} $, $ \vec {k} _ {\alpha}
= \vec {k} + \vec {\tau} _ {\alpha} $ is the  vector of
the diffracted wave, $ \chi _ {\alpha} ^{(b)} $ is the part of the dielectric susceptibility corresponding to
the interaction of the electron beam with radiation

\[
\chi _{\alpha }^{(b)}=\frac{q^{(w)}_{\alpha}}{\left\{ \omega
-(\vec{k}_{\alpha }\vec{v}_{w})\mp
(\omega_w-(\vec{k}_w\vec{v}))\right\} ^{2}} \nonumber
\]

\begin{eqnarray*}
& &q^{(w)}_{\alpha} =\frac{a_{w}^{2}}{4\gamma ^{3}}\frac{\omega _{L}^{2}}{(k_{w}v)^{2}}%
\left\{ \left[ \frac{(\vec{u}\vec{e}_{\alpha})}{c(k_{w}v)}(\omega _{w}\vec{u}-\vec{k}%
_{w}c^{2})(\vec{k}_{\alpha}-\vec{k}_{w})\right. \right.  \\
& &\left. \left. -(\vec{k}_{\alpha}\vec{e}_{w})c\right] \frac{(\vec{u}\vec{e}_{\alpha})}{c}-(\vec{%
k}_{w}\vec{e}_{\alpha})(\vec{u}\vec{e}_{w})-(\vec{e}_{\alpha}\vec{e}_{w})(k_{w}v)^{2}\right\}^2
\\
& &\times\left\{ (\vec{k}_{\alpha}-\vec{k}_{w})^{2}c^{2}-(\omega -\omega
_{w})^{2}\right\},
\end{eqnarray*}
$ \vec {k}, \omega, \vec {e}  $ and $ \vec {k} _ {w}, \omega _
{w}, \vec {e} _ {w} $ are the wave vectors, frequencies and
polarization vectors of both the  signal  and pump waves,
respectively, $v = (c, \vec {u}) $, $k _ {w} = (\omega _ {w}/c,
\vec {k} _ {w}) $, $a _ {w} = eH _ {w}/ mc\omega _ {w}$. The
dispersion equation corresponding to equation (\ref {mop_system})
has the following schematic form
\begin{equation}
F ^ {(n)} = -\chi _ {\alpha} ^ {(b)} F ^ {(n-1)}, \label{mop_disp}
\end{equation}
where the term $F ^ {(n)} $ on  left-hand side of (\ref{mop_disp}) corresponds
to the $n  $-wave Bragg dynamical diffraction  (equation $F ^ {(n)} =0$
is the dispersion equation defining diffraction modes in the $n $-wave
case). The continuity of the current densities, charge
densities, and transverse components of fields on the boundaries and dispersion equation
(\ref{mop_disp})  give the equation for the generation threshold \cite{laser_8}.
From  (\ref{mop_system}) we obtain that for
the $n$-fold degeneration point  of the roots of the dispersion equation
(when $n+1 $ roots of the equation $D ^ {(m)} =0 $ at $m\geq n+1 $
coincide),  the equation for the generation threshold has the following
form
\begin{equation}
\label{mop_threshold}
\frac {1} {\gamma ^ {3}} \left (\frac {\omega _ {L}} {\omega}
\right) ^ {2} a _ {w} ^ {2} k ^ {3} L _ {\ast} ^ {3} = \frac {a _
{n}} {(k |\chi |L _ {\ast}) ^ {2n}} +b _ {n} k\chi ^{\prime\prime} L _ {\ast}.
\end{equation}
In (\ref{mop_threshold}) $k =\omega /c $, $L _ {\ast} $ is the length of the interaction area of the electron beam
with electromagnetic radiation, $ \chi^{\prime\prime} $ is the
imaginary part of the dielectric susceptibility, which describes
absorption, $a _ {n} $, $b _ {n} $ are the parameters depending on
the geometric parameters of the system (except
 $L _ {\ast} $). The equality
 (\ref{mop_threshold}) has an obvious physical meaning: the left-hand
side of  (\ref{mop_threshold}) contains the term describing generation
of radiation by the electron beam, and the right-hand side includes
the terms describing losses on the boundaries (the first term) and absorption
losses (the second term) in the medium.
 One of the
peculiarities of the VFEL with multi-wave
 distributed feedback is the possibility of a sharp decrease of losses at the
boundaries (the first term on the right-hand side of  (\ref{mop_threshold})
decreases with the  increase of $s$ due to the condition $k |\chi
|L _{\ast} \gg 1 $ under dynamical diffraction).
  Let us express the synchronism condition for the above
generation mechanism $ \omega -\vec {k} \vec {v} = \Omega _ {w} $, where
$ \Omega _ {w} = \omega _ {w} - \vec {v} \vec {k} _ {w}$. Then, the frequency of the signal wave is  equal
to (if the pump wave is oncoming)
\begin{equation}
\label{mop_freq}
\omega = \frac {2\omega
_{w}(\vec{\tau}_{1},\ldots\vec{\tau}_{n},\vec{n}_{u},S)(1-\beta\cos(\Theta_w))}{1-\beta \cos
(\Theta)}.
\end{equation}
In (\ref{mop_freq}) the explicit dependence of the pump wave frequency
on the geometry of the multi-wave diffraction ($ \vec {\tau} _ {1},
\ldots\vec {\tau} _ {n} $) and the set of resonator parameters $S $  is
marked out (if a resonator is not oversized, then the dependence on $S $ disappears). Smooth change of the
VFEL geometry also varies  the $Q$ factor  and, therefore, varies the generation  efficiency. For example
the dependence of $Q $ on  the diffraction asymmetry factor $f = \gamma_0 /\gamma_1$ is shown
in Figure \ref{mop_Fig.2} ($\gamma_0 ,\gamma_1$ are diffraction cosines \cite{berk_139}).

It should be noted that the distributed feedback can be used for
both the first and the second stages. To optimize the resulting
radiation output, $Q$ factor can be controlled at both stages.
Thus, for radiation angle  $ \Theta =0 $ the frequency $ \omega
\sim 4\gamma ^ {2} \omega _ {w} $,  even the moderately
relativistic electron beam ($E\sim 1 $ MeV) gives a frequency
multiplication $ \sim 35 $ times. If during the first stage the
undulator mechanism is used (undulator period $\sim 8$ cm),
 then the wavelength of the pump
wave is $\lambda_w \sim 1$ cm. Thus the signal wave is generated
in the teraHertz range (Figure~\ref{mop_Fig.2}).

Thus,

(1) the principles of VFEL can be used for creation of a dynamical
wiggler  with variable period in a large volume,\\
(2) the two-stage scheme of generation can be used for
lasing in the teraHertz
frequency range using low-relativistic beams, \\
(3) the two-stage scheme of generation combined with the volume
distributed feedback opens up the possibility of creating powerful
generators with wide electron beams (or system of beams).

\section{Formation of distributed feedback in a FEL under
multi-wave diffraction} \label{nim95}

We shall further consider the boundary problem for a
quasi-Cherenkov FEL  for the case in which a distributed feedback
is provided by  three-wave coplanar diffraction of the emitted
photons \cite{laser_8}. As compared with a two-wave distributed
feedback,  in the three-wave case the appropriate choice of
parameters enables reducing the interaction region between a
particle beam and the electromagnetic wave necessary to reach the
oscillation threshold as well as radiation losses inside the
medium. For example, it allows a reduction in the size of an FEL
and the construction of more compact coherent radiation sources in
any spectral range.

Here we show that the two-wave DFB (in which only two waves are
strongly excited) is not optimal because the region of root
degeneration often coincides with the region of strong radiation
absorption inside the medium. This is especially important for
solid FELs. The most effective resonator is one with a multi-wave
DFB, where the distributed feedback is formed by multi-wave
dynamical diffraction. The advantages of a multi-wave DFB are
analyzed in detail for the solid quasi-Cherenkov FEL with
three-wave coplanar DFB because this analysis can be conducted
analytically. Consider, for example, the specific case of a solid
X-ray quasi-Cherenkov FEL where the crystal medium provides both a
spontaneous radiation mechanism in the X-ray spectral range
\cite{63}  and diffraction of the emitted X-rays by the crystal
forming the three-dimensional distributed feedback. For the X-ray
spectral range,  the crystal target is a radiator and a resonator
simultaneously. It should be mentioned that the analysis derived
below will be appropriate for other spectral ranges as well. For
example, a three-dimensional optical grating can be formed inside
a solid target by a laser. Moreover, a multi-wave distributed
feedback can be formed by surface dynamical diffraction if a
particle beam moves over a three-dimensional periodic structure at
a distance not larger than $\lambda_{\gamma}$ (where $\lambda$  is
the radiation wavelength,  $\gamma$ is the Lorentz factor). In
this way the multi-wave DFB can be used with an ordinary undulator
FEL.


Let a relativistic particle beam be incident on a crystal plate ($0\leq z\leq L$) at an arbitrary angle $\Psi_0$. The set of Maxwell's equations which describes the interaction of an electromagnetic wave with a crystal and with a particle beam passing through a crystal target can be written in the following form \cite{berk_139}:
 \begin{eqnarray}
 \label{nim95_1.1}
 D^{(0)}_{\sigma}E^{(0)}_{\sigma}-\omega^2\chi_1 E^{(1)}_{\sigma}-\omega^2\chi_2 E^{(2)}_{\sigma}=0,\nonumber\\
 -\omega^2\chi_{-1} E^{(0)}_{\sigma}+D_{\sigma}^{(1)}E^{(1)}_{\sigma}-\omega^2\chi_{2-1} E^{(2)}_{\sigma}=0,\\
 -\omega^2\chi_{-2} E^{(0)}_{\sigma}-\omega^2\chi_{1-2} E^{(1)}_{\sigma}+D_{\sigma}^{(2)}E^{(2)}_{\sigma}=0,\nonumber
 \end{eqnarray}
where $D_{\sigma}^{(\alpha)}=k^2_{\alpha}c^2-\omega^2\varepsilon_0^{(\alpha)}+\chi_b^{(\alpha)}$. We assume that a particle beam and a crystal plate are oriented so that the three-wave coplanar diffraction condition is satisfied for emitted photons. In this case only three strong waves with the $\sigma$-polarization are excited inside the crystal medium. (see \cite{berk_139} for two-wave diffraction geometry). The subscript $\alpha$ ($\alpha=0-2$ ) labels the transmitted wave ($\alpha=0$) and diffracted waves ($\alpha=1,2$); $E^{(\alpha)}_{\sigma}$  are the $\sigma$ components of the amplitudes of electromagnetic waves, $\vec{k}_1=\vec{k}_0+\vec{\tau}_1$, $\vec{k}_2=\vec{k}_0+\vec{\tau}_2$  are the wave vectors of photons diffracted by crystal planes with corresponding reciprocal vectors $\vec{\tau}_1$  and $\vec{\tau}_2$;  $\epsilon_0^{(\alpha)}=1+\chi_{\alpha}$ are the dielectric constants of a crystal for transmitted and diffracted waves.
\begin{equation}
\chi_b^{(\alpha)} =
\frac{1}{\gamma}(\omega_L/\omega^2)(\vec{u}\vec{e}_{\sigma}/c)^2
\frac{k_{\alpha}^2c^2-\omega^2}{(\omega-\vec{k}_{\alpha}\vec{u})^2-\frac{\hbar^2}{4m^2c^4\gamma^2}(k^2c^2-\omega^2)^2}
\end{equation}

for the "cold" beam limit and
\begin{eqnarray*}
\chi_b^{(\alpha)} =-
\frac{i\sqrt{\pi}}{\gamma}(\omega_L/\omega)^2(\vec{u}\vec{e}_{\sigma}/c)^2\frac{k_{\alpha}^2c^2-\omega^2}{\delta_{\alpha}^2}x_{\alpha}^t\exp[-(x_{\alpha}^t)^2]
\end{eqnarray*}
for the "hot" beam limit. $\chi_b$ represents the part of  the dielectric susceptibility produced by the interaction of a particle beam with radiation, $x_{\alpha}^t=(\omega-\vec{k}_{\alpha}\vec{u})/\sqrt{2}\delta_{\alpha}$, $\delta_{\alpha}^2=(k^2_{\alpha 1}\Psi^2_1+k^2_{\alpha 2}\Psi^2_2+k^2_{\alpha 3}\Psi^2_3)u^2$, and $\Psi=\Delta\overline{V}/|\overline{V}|$ is the velocity spread. As was shown in \cite{berk_139}, comparison of a standard equation of X-ray dynamical diffraction  with (\ref{nim95_1.1}) leads to the conclusion that the combination of a crystal and a particle beam may be considered as an "active" medium with dielectric susceptibility  equal to $\chi_{\alpha}+\chi^{(\alpha)}_b$. It permits the boundary problem of X-ray amplification (lasing) due to the passage of a particle beam through a periodic medium to be reduced to the problem of X-ray diffraction by an "active" periodic medium consisting of a crystal plus radiating particle beam.

\begin{figure}[htp]
\centering
\epsfxsize = 10 cm \centerline{\epsfbox{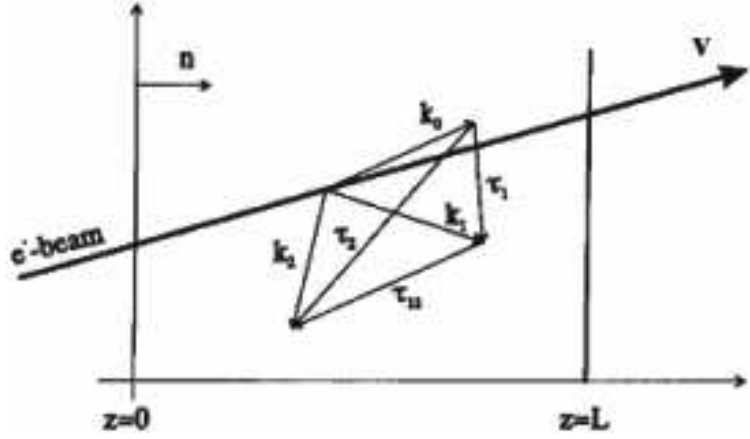}}
\caption{Figure  The geometry of three-wave diffraction. $\vec{V}$
is the mean particle beam velocity, and $z=0$ and $z=L$ are two
surfaces of a crystal plate} \label{nim95_fig1}
\end{figure}

The geometry of three-wave diffraction is shown in Figure \ref{nim95_fig1}, where $\vec{V}$ is the mean particle beam velocity, $z=0$ and $z=L$ are the two surfaces of a crystal plate.

The dispersion equation determining the electromagnetic modes inside the "active" medium can be represented in the following form:
\begin{equation}
\label{nim95_1.2}
F^{(3)}_{\sigma}(\vec{k}_0;\vec{k}_1;\vec{k}_2)=-i\chi_b^{(0)}F^{(2)}_{\sigma}(\vec{k}_1;\vec{k}_2),
\end{equation}
where
\begin{eqnarray*}
& &
F^{(3)}_{\sigma}(\vec{k}_0;\vec{k}_1;\vec{k}_2)=(k^2c^2-\omega^2\epsilon_0^{(0)})
(k_1^2c^2-\omega^2\epsilon_0^{(1)})(k_2^2c^2-\omega^2\epsilon_0^{(2)})-\\
& &-\omega^4(k^2c^2-\omega^2\epsilon_0)\chi_{1-2}\chi_{2-1}-
\omega^4(k_1^2c^2-\omega^2\epsilon_0^{(1)})\chi_{2}\chi_{-2}\\
& &-\omega^4(k_2^2c^2-\omega^2\epsilon_0^{(2)})\chi_{1}\chi_{-1}
-\omega^6(\chi_1\chi_{-2}\chi_{2-1}+\chi_2\chi_{-1}\chi_{1-2});
\end{eqnarray*}
\[
F^{(2)}_{\sigma}(\vec{k}_1;\vec{k}_2)=(k_1^2c^2-\omega^2\epsilon_0^{(1)})
(k_2^2c^2-\omega^2\epsilon_0^{(2)})-\omega^4\chi_{1-2}\chi_{2-1}
\].

From (\ref{nim95_1.2}) follows that the root degeneration (the
strongest interaction is in the root degeneration region) and the
fulfillment of the Cherenkov condition are possible simultaneously
only under the following conditions:
\begin{eqnarray}
\label{nim95_1.3}
& &l_0l_1l_2-l_0r_{12}-l_1r_2-l_2r_1-f=0,\nonumber\\
& &\beta_1\beta_2l_1l_2+l_0(\beta_1 l_1-\beta_2l_2)-\beta_1\beta_2 r_{12}-\beta_1r_1-\beta_2 r_2=0,
\end{eqnarray}
where $l_0=\theta^2-\chi_0+\gamma^{-2}$, $l_{1,2}=l_0+\alpha_{B1,2}$; $\alpha_{B1,2}=(2\vec{k}\vec{\tau}_{1,2}+\vec{\tau}^2_{1,2})/k^2$.

It is straightforward to show that for the fulfillment of (\ref{nim95_1.3}), the system parameters should satisfy the following relationship:
\begin{equation}
\label{nim95_1.4}
\beta_1\beta_2(l_0^2+\beta_1r_1+\beta_2r_2) < 0,
\end{equation}
where $f=\chi_1\chi_{-2}\chi_{2-1}+\chi_2\chi_{-1}\chi_{1-2}$, $\beta_{1,2}=\gamma_0/\gamma_{1,2}$ are the asymmetry factors of diffraction $\gamma_{\alpha}=(\vec{k}_0\vec{n})/|\vec{k}_{\alpha}|$, $r_1=\chi_1^{\prime}\chi_{-1}^{\prime}$  and  $r_2=\chi_2^{\prime}\chi_{-2}^{\prime}$. Expression (\ref{nim95_1.4}) is  more restrictive than the equivalent relation  for the two-wave case \cite{berk_139}.

In the case of Bragg-Bragg  diffraction in the X-ray spectral
range there is a restriction on the radiation angle
\begin{equation}
\label{nim95_1.5}
\theta_{B-B}^{\mathrm{max}}=(\sqrt{-\beta_1r_1-\beta_2r_2}+\chi_0^{\prime}-\gamma^{-2})^{1/2}.
\end{equation}
This relation leads to a large value of the diffraction asymmetry factor, which, in turn, leads to strong radiation absorption inside the medium in the vicinity of the root degeneration point -- just as in the two-wave DFB case  \cite{berk_139}. The situation changes in the Bragg-Laue  geometry. In this case the inequality (\ref{nim95_1.4}) can be satisfied at angles $\theta\ge \theta_{B-B}$, which makes it  possible to reduce the asymmetry factors $\beta_1$ and $\beta_2$ and, consequently, radiation absorption inside the crystal.

The solution of the corresponding boundary problem is presented as a sum:
\begin{equation}
\label{nim95_1.6}
\vec{E}=\sum\limits_i c_i\exp i\vec{k}_i\vec{r}(\vec{e}_0+\vec{e}_1s_i^{(1)}\exp i\vec{\tau}_1\vec{r}+\vec{e}_2s_i^{(2)}\exp i\vec{\tau}_2\vec{r}),
\end{equation}
where
$s^{(1)}=(\lambda\lambda_2-r_2)/(\lambda_2\chi_1+\chi_2\chi_{1-2})$,
$s^{(2)}=(\lambda\lambda_1-r_1)/(\lambda_1\chi_2+\chi_1\chi_{2-1})$,
and $\vec{k_i}$  are the solutions of the dispersion equation and
the coefficients of coupling between the transmitted and
diffracted waves, $E^{(1)}=s^{(1)}E$, $E^{(2)}=s^{(2)}E$ ,
$\lambda_{\alpha}=[(\vec{k}+\vec{\tau})^2c^2-\omega^2\epsilon_0]/\omega^2$.
To determine the unknown coefficients, it is necessary to solve
the boundary conditions for the waves on the crystal surfaces.
These can be written in the following form for the Bragg-Laue
case:
\begin{equation}
\label{nim95_1.7}
\sum\limits_i ^3c_i=1,\quad \sum\limits_i ^3 s_i^{(1)}c_i=0,\quad \sum\limits_i ^3s_i^{(2)}c_il_i=0.
\end{equation}
The condition (\ref{nim95_1.7}) is written for the  "hot" beam limit. For the "cold" beam limit the corresponding expression can be found in \cite{berk_138}.

It is well known  \cite{nim95_2} that the oscillation threshold can be determined from the condition that
$\Delta=0$, where $\Delta$  is the determinant of the system (\ref{nim95_1.7}).

Solving the equation $\Delta=0$, we  obtain the threshold in the form
\begin{equation}
\label{nim95_1.8}
G=a\chi_0^{\prime\prime}+\frac{16}{|\beta_1\beta_2|}\left[\frac{\gamma_0c}{\vec{n}\vec{u}}\right]^3
\frac{\pi^2n^2}{(kl_0L_*)^2kL_*}\eta_{B-L}
\end{equation}
with the phase condition $(k_{1z}-k_{2z})L=2\pi n$ ($n$ is an integer), $r_{12}=\chi_{1-2}\chi_{2-1}$, where
\begin{eqnarray*}
G&=&-\frac{\pi^2n^2}{4\gamma}
\left\{\frac{\omega_L}{\omega}\right\}^2k^2L^2_*(\chi^{\prime}_0\pm\sqrt{-\beta}|\chi_{\tau}|-\gamma^{-2})(\chi^{\prime}_0\pm\sqrt{-\beta}|\chi_{\tau}|)\sin\phi^2\\
&\times& \sin y[(2y+\pi n)\sin y-y(y+\pi n)\cos y]y^3(y+\pi
n)^{-3}
\end{eqnarray*}
for the "cold" beam limit and
\begin{eqnarray*}
G&=&-\sqrt{\pi}\frac{\omega^2_L}{\gamma\omega^2}
\frac{(l_0+\chi_0^{\prime}-\gamma^{-2})(l_0+\chi_0^{\prime})\sin\phi^2}{\delta_0^2/k^2}x^t
e^{-(x^t)^2}
\end{eqnarray*}
for the "hot" beam limit. $a$ and $\eta_{BL}$ are smooth functions depending on the diffraction geometry and usually are of the order unity, $y=k\delta L/2$ and $\delta$ is the deviation from the exact Cherenkov synchronism condition.

The analysis shows that in the coplanar Bragg-Laue diffraction geometry of DFB radiation, absorption inside the medium can be reduced in the region of root degeneration. For example, under dynamical diffraction by the $(111)$ and $(11\overline{1})$  planes with a symmetry factor  $\beta_1=-\beta_2=\beta=0.16$,  the current density required to achieve the threshold can be reduced by approximately one order of magnitude in comparison with a two-wave DFB FEL using $LiH$  and an electron energy of $750$  MeV with a transverse angular spread $\Psi_{\perp}=5\times 10^{-6}$ rad \cite{berk_139}.

\section{Distributed feedback under the multi-wave diffraction}
\label{nim95_sec:2}

In principle, in the general case of the $n$-wave diffraction
DFB,
it is possible to achieve the degeneration of $n$ roots. In this
case the threshold condition has the form:
\begin{equation}
\label{nim95_1.9}
G=\frac{a_n}{(k|\chi^{\prime}_{\tau}|L)^{2(n-1)}kL}+b_n\chi^{\prime\prime}_0,
\end{equation}
where $a_n$ and $b_n$ are the functions weakly dependent on the
diffraction type.

The energy loss through target surfaces can be reduced due to the
fulfillment of the inequality $k|\chi_{\tau}|L\gg 1$ under
dynamical diffraction. As the degree of dispersion root
degeneration increases, radiation remains inside the target longer
because the group velocity decreases and is proportional to $\sim
L^{n-1}$, where $n$ is the degree of dispersion root degeneration.
As a result, the region of interaction between the particle beam
and the emitted radiation can be reduced. This leads to the
possibility of reducing the FEL's dimensions, i.e., of
constructing a compact source of coherent radiation.

\section{Parametric X-ray FEL operating with external Bragg reflectors}\label{nim96}

This section is devoted to the application of external Bragg reflectors in a volume parametric X-ray FEL \cite{lanl_12}. It is shown that external Bragg reflectors permit not only a reduction of electromagnetic energy losses through target boundaries but also a reduction of the radiation self-absorption inside the target by modifying radiation modes excited in the active medium containing a particle beam plus a crystal. As a result, the starting beam current can be reduced by more than a factor of ten. It is also shown that the best conditions for lasing are realized when the diffracted wave is reflected by the external Bragg reflector.

\subsection{Generation threshold for parametric X-ray FEL with
external reflectors}\label{nim96_sec:1}

\begin{figure}[htbp]
\centering \epsfxsize = 8 cm \centerline{\epsfbox{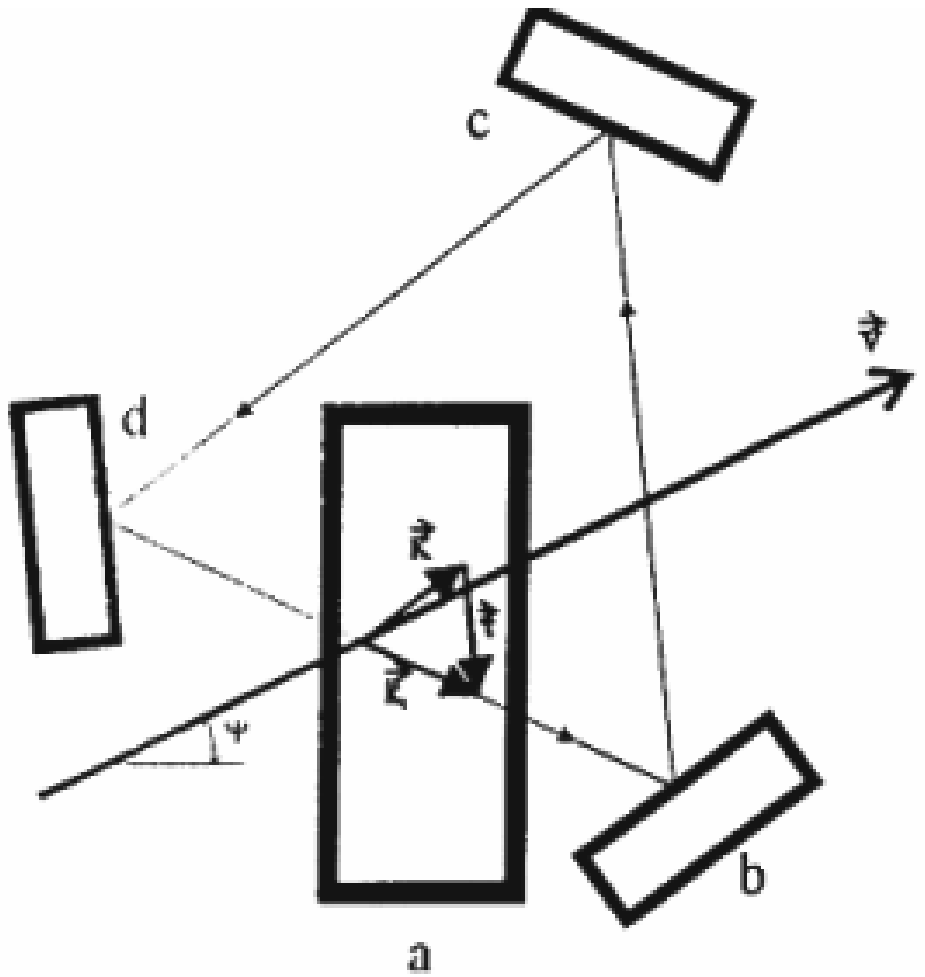}}
\caption{} \label{nim96_fig1}
\end{figure}

The scheme of an X-ray volume FEL using spontaneous X-ray
parametric radiation has been considered in \cite{berk_139}.
Following the remarks mentioned above, the scheme under
consideration differs from the case described in \cite{berk_139}
by the addition of external Bragg reflectors which can be placed
in the direction of both transmitted and diffracted waves.
For
example, in the case of a three-crystal scheme of diffracted wave
reflection, the geometry of a volume FEL can be represented in the
following way:

Here $\vec{V}$   is the mean velocity of electrons, $\vec{k}$ and
$\omega$ are the wave
vector and the frequency of the emitted
photon, $\vec{k}$  and $\omega$ are the wave vector and the
frequency of the diffracted photon, $a$, $b$, $c$, $d$ are the
crystal targets.
We assume that the crystal (a) in Fig.\ref{nim96_fig1} is oriented
relative to the particle beam so that only two strong waves are
excited inside the crystal under diffraction (two-wave
diffraction).

The closed system of Maxwell's equations, including the equation describing the motion of electrons, and the method of its solution were described in  \cite{berk_139}.

The introduction of external Bragg reflections modifies only the
boundary
conditions where the reflection coefficients for
diffracted and transmitted waves appear. Besides, the analysis
shows that there appears the possibility of realization of lasing
regime under a Laue diffraction scheme (see  Figure
\ref{nim96_fig1}) which is impossible, in principle, without
external Bragg reflectors.

Let us restrict ourselves to the analysis of Laue geometry for
distributed feedback. Solving the
boundary conditions while taking
into account the phase relation for the main crystal target (a)
and for the external reflectors we can obtain the generation
 equations for two cases.

Case (1) The "cold" beam limit and weak amplification:
\begin{eqnarray}
\label{nim96_1.1}
G=\left\{\begin{array}{c}
    (1-|\alpha|)(1-\frac{\beta l_ 1^{\prime}}{l_ 0^{\prime}})\frac{\gamma_0V}{V_z}+k[x_1^{\prime\prime}
    +\frac{\beta l_ 1^{\prime}}{l_ 0^{\prime}}x_2^{\prime\prime}]L_*=\Gamma^{(1)}_{\mathrm{th}}, \quad(1)\\
    (1-|\alpha|)(1-\frac{\beta l_ 0^{\prime}}{l_ 1^{\prime}})\frac{\gamma_0V}{V_z}+k[x_1^{\prime\prime}
    +\frac{l_ 0^{\prime}}{\beta l_ 1^{\prime}}x_2^{\prime\prime}]L_*=\Gamma^{(2)}_{\mathrm{th}}, \qquad(2)
  \end{array}\right.
\end{eqnarray}
where
\begin{eqnarray}
\label{nim96_1.2} & &
G=\frac{\beta}{4\gamma}k^2L_*^3\left(\frac{\omega_L}{\omega}\right)^2(l_
0^{\prime}+
\chi_ 0^{\prime}-\gamma^{-2})(l_ 0^{\prime}+\chi_ 0^{\prime})l_ 1^{\prime}( l_ 0^{\prime}+\beta l_ 1^{\prime})^{-1}\sin^2\Phi f(\chi), \nonumber\\
& &f(x)=\frac{\sin x}{x}\frac{x\cos  x-\sin x}{x^2},\\
& &l_ 0^{\prime}=1/2[-\alpha_B\pm\sqrt{\alpha^2_B+4r}],\quad l_ 1^{\prime}=1/2[+\alpha_B\pm\sqrt{\alpha^2_B+4r}],\nonumber\\
& &
x^{\prime\prime}_{1(2)}=\frac{1}{4}\chi_0^{\prime\prime}\left\{1+\beta\pm\frac{(l_0^{\prime}-\beta
l_1^{\prime})(\beta-1)+2\beta
r^{\prime\prime}/\chi_0^{\prime\prime}}{[(l_0^{\prime}-\beta
l_1^{\prime})^2+4\beta r]^{1/2}}\right\},\nonumber
\end{eqnarray}
the prime means the real part of the magnitude and the double prime means its imaginary part; equation (1) of the system (\ref{nim96_1.1}) describes the case of diffracted wave reflection, and  equation (2) corresponds to the case of transmitted wave reflection. The quantity $\alpha=|\alpha|\exp i\phi$  is the reflection coefficient for transmitted (2) or for diffracted (1) waves respectively.

The value of $(1-|\alpha|)$ can be very small for a narrow  interval of angles $(1-|\alpha|)=-\chi_0^{\prime\prime}|\chi_{\tau}|^{-1}(1-\exp(-W))$, where $W$ is the Debye-Waller factor. For example, for a $LiH$ crystal $(1-|\alpha|)<10^{-3}$ at $\sim 1-10$ keV photoenergies.

As a result, the radiation losses through crystal boundaries can be essentially reduced. This means that the threshold condition will be determined mainly by radiation absorption. In this case (\ref{nim96_1.1}) can be represented in the form:
\begin{eqnarray}
\label{nim96_1.3}
(1)\qquad j_{\mathrm{th}}&=&\frac{4}{\beta}\left(\Lambda^{\prime\prime}_2+
\frac{b_0}{\beta b_1}\Lambda^{\prime\prime}_1\right)\frac{b_0+\beta b_1}{(b_0-a_F)(b_0-a_F^{(1)})b_1}=\frac{m\gamma c^3}{b_0eL^2_*}\frac{\chi_0^{\prime\prime}}{|\chi\tau|^2}g_1(a,\beta),\nonumber\\
(2)\qquad j_{\mathrm{th}}&=&\frac{4}{\beta}\left(\Lambda^{\prime\prime}_1+
\frac{b_0}{\beta b_1}\Lambda^{\prime\prime}_2\right)\frac{b_0+\beta b_1}{(b_0-a_F)(b_0-a_F^{(1)})b_1}\frac{m\gamma c^3}{b_0eL^2_*}\frac{\chi_0^{\prime\prime}}{|\chi\tau|^2}g_2(a,\beta),\nonumber\\
\end{eqnarray}
here $b_{0(1)}=\mp a+[a^2+1]^{1/2}$,
$a=\frac{\alpha_B}{2|\chi\tau|}$,
$\Lambda^{\prime\prime}_{1(2)}=\frac{x^{\prime\prime}_{1(2)}}{\chi_0^{\prime\prime}}$,
$a_F=\frac{(|\chi_0^{\prime}|+\gamma^{-2})}{|\chi\tau|}$,
$a_F^{(1)}=\frac{|\chi_0^{\prime}|}{|\chi\tau|}$.

\begin{figure}[htp]
\centering
\epsfxsize = 12 cm \centerline{\epsfbox{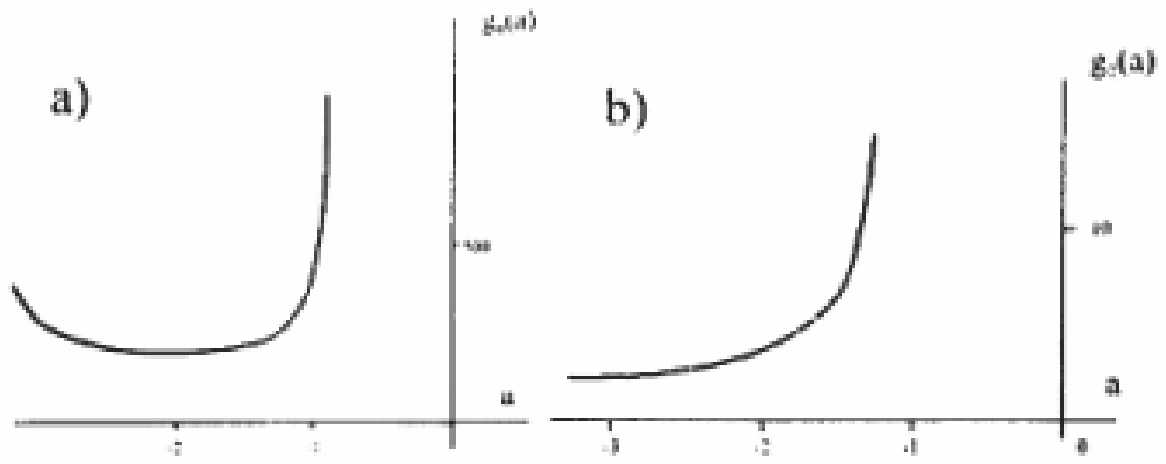}}
\caption{} \label{nim96_fig2}
\end{figure}

The dependence of  $g_1$ and $g_2$ on the parameter of deviation
from the exact Bragg condition is shown in Figure \ref{nim96_fig2}
(a) and (b) for a $LiH  $ crystal, the $(111)$ diffraction plane,
$E=700$  MeV, $\beta=1$, $\omega=3\times 10^{-18}$ c$^{-1}$ and
the angular divergence of electrons in the beam
$\Delta\Psi_{\perp}=10^{-6}$  rad.

\bigskip

From Figure \ref{nim96_fig2} (a) it can be seen that $g_2(a)$ has a minimum of $\sim 200$ at a definite deviation parameter $a=-2$. At the same time, $g_1(a)$  does not have a definite optimal value. It depends weakly on the asymmetry factor and approaches 2 as the absolute value of the deviation parameter $a$  increases.

 A detailed comparative analysis of these figures shows that under the same conditions, the threshold current in the first case is 100 times larger than in the second case. It follows from Figure
\ref{nim96_fig2} (b) and (\ref{nim96_1.3}) that
\[
j_{\mathrm{th}}^{\mathrm{opt}}\simeq 2\frac{m\gamma c^3}{b^0eL_*^2}\frac{\chi_0^{\prime\prime}}{|\chi\tau|^2}.
\]

This can be explained by the fact that the growth rate of the electromagnetic field is maximum at the deviation parameter $\alpha_B/2|\chi_{\tau}|\sim - 10$. The analysis shows that in this case the main part of radiated electromagnetic energy turns out to be accumulated in the diffracted wave. This means that for achieving the best condition for lasing, the Bragg reflector should be placed in the direction of the diffracted wave.

Case(2). The "hot" beam limit.

The boundary condition in this case is simpler, and for the diffracted wave reflection we can obtain the dispersion equation in the form:
\begin{equation}
\label{nim96_1.4}
\frac{\Gamma^t}{4}\left(1-\frac{(l^{\prime}_0-\beta
l^{\prime}_1)}{[(l^{\prime}_0-\beta l^{\prime}_1)^2+4\beta
r]^{1/2}}\right)=\Gamma_{\mathrm{th}}^{(1)},
\end{equation}
where
\begin{eqnarray}
\label{nim96_1.5}
\Gamma^t&=&-\sqrt{\pi}\frac{\omega^2_L}{\gamma}
\frac{(l_0+\chi_0^{\prime}-\gamma^{-2})(l_0+\chi_0^{\prime})\sin^2\phi}{\delta^2_0}x^t\exp(-(x_t)^2),\nonumber\\
x^t&=&\frac{(\omega-k^{\prime}U)}{\sqrt{2}\delta_0},\quad l_0=\Theta^2-\chi_0-\gamma^{-2},\quad \delta_0=\omega\Theta\Delta\Psi_{\perp},
\end{eqnarray}
$\Theta$  is the radiation angle, $\Delta\Psi_{\perp}$ is the transverse angular divergence of electron velocities in the beam.

\begin{figure}[htp]
\centering \epsfxsize = 8 cm {\epsfbox{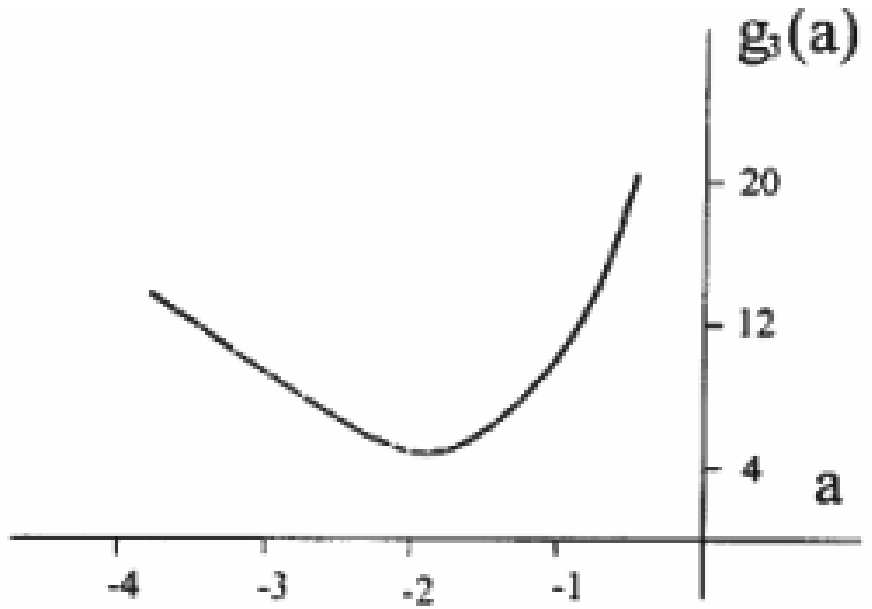}} \caption{}
\label{nim96_fig3}
\end{figure}

At a small value of $(1-|\alpha|)$, when the threshold  condition is determined by the radiation absorption, the relation  analogous to (\ref{nim96_1.3}) for the "cold" particle beam can be represented as:
$
j_{\mathrm{th}}\sim Ag_3(a,\,\beta),
$
where $a$ is the function similar to that in (\ref{nim96_1.3}). It
weakly depends on the asymmetry factor, the particle energy, and
the crystal parameters. The dependence of $g_3$  on the deviation
parameter $a$ is shown in Figure \ref{nim96_fig3}.

We can see that in this case there is an optimal value of
$\alpha_B$ at which the magnitude of $g_3$ is minimum. At the same
parameters as in the case of a "cold"  particle beam and at
$\Delta\Psi_{\perp}=5\cdot 10^{-6}$ rad $j_{\mathrm{th}}=2\cdot
10^9$ A/cm$^2$.

So, from the above analysis we can conclude that the application
of external Bragg reflectors allows one to lower the threshold
electron beam current density of the electron beam necessary for
achieving lasing in the X-ray spectral range by more than a factor
of ten. The best situation is realized when the Bragg reflector is
placed in the direction of the diffracted wave. In this case the
value of the starting beam current can be $j_{\mathrm{th}}\simeq
2\cdot 10^9$~  A/c for the "hot" particle beam and
$j_{\mathrm{th}}\simeq  10^8$~  A/c  for the "cold" particle beam
and the generation frequency $\omega\simeq 3\cdot 10^{18}$
~c$^{-1}$ in $LiH$, $E=700$~ MeV, $\beta=1$ and $L_*\leq 5\cdot
10^{-3}$~ cm.


\section{Theory of induced PXR in a crystal plate (general formulas) (\cite{berk_139})}\label{phys}

Let an ultra-relativistic  electron (positron) beam of velocity
$\vec{u}$ enter, at a certain angle, a crystal plane-parallel
plate with length $L$ (the $z$-axis is perpendicular to the
crystal surface, and the plate lies in the interval ($0<z<L$).
The set of equations describing the interaction of an
electromagnetic wave with the "crystal-beam" system consists of
Maxwell's equations and those of particle motion in the
electromagnetic field.
The dielectric susceptibility of a crystal has the form
$\varepsilon(\vec{r};\omega)=\sum_{\vec{\tau}}\varepsilon_{\tau}(\omega)\exp
(-i\vec{\tau}\vec{r})$, where $\vec{\tau}$ is the reciprocal
lattice vector. Perturbations of the current density and charge
density in the linear field approximation may be written in the
form:
\begin{eqnarray}
\label{phys_1.1}
& & \delta\vec{j}(\vec{k};\omega)=e\sum\limits_{\alpha}\exp(-i\vec{k}\vec{r}_{\alpha 0} )\left\{\delta \vec{v}_{\alpha}(\omega-\vec{k}\vec{u})\right.\nonumber\\
& & \left.-i\vec{u}[\vec{k}\delta\vec{r}_{\alpha}(\omega-\vec{k}\vec{u})]\right\}\delta n(\vec{k};\omega)\\
& & =e\sum\limits_{\alpha}\exp(-i\vec{k}\delta\vec{r}_{\alpha 0})\left\{-i[\vec{k}\vec{r}_{\alpha}(\omega-\vec{k}\vec{u})]\right\},\nonumber
\end{eqnarray}
where $\delta\vec{j}(\vec{k};\omega)$ and $\delta n(\vec{k};\omega)$ are the Fourier transformations of the expressions
\begin{eqnarray*}
\vec{j}(\vec{r};t)&=&e\sum\limits_{\alpha}\vec{v}_{\alpha}(t)\delta[\vec{r}-\vec{r}_{\alpha}(t)]\\
n(\vec{r};t)&=&\sum\limits_{\alpha}\delta[\vec{r}-\vec{r}_{\alpha}(t)].
\end{eqnarray*}
$\vec{u}$ is the unperturbed electron (positron) velocity; $\delta\vec{v}_{\alpha}$ and  $\delta\vec{r}_{\alpha}$ are perturbations of the velocity and radius vectors, respectively, due to the interaction with the radiation field:
\begin{eqnarray*}
\vec{v}_{\alpha}(t)&=&\vec{u}+\delta\vec{v}_{\alpha}(t)\\
\vec{r}_{\alpha}(t)&=&\vec{r}_{\alpha 0}+\vec{u}t+\delta\vec{r}_{\alpha}(t).
\end{eqnarray*}
The subscript $\alpha$ denotes the number of the particle.

Deriving the perturbations of velocity and radius vectors by the particle motion equation and using (\ref{phys_1.1}) for the current density, one can obtain a set of Maxwell's equations, which describes the interaction of an electromagnetic wave with a crystal, and a particle beam penetrating through it, in the following form:
\begin{eqnarray}
\label{phys_1.2}
& &k^2_{\tau}\vec{E}(\vec{k}_{\tau^{\prime}},\omega)-\vec{k}_{\tau^{\prime}}
[\vec{k}_{\tau^{\prime}}\vec{E}(\vec{k}_{\tau^{\prime}},\omega)]
-\frac{\omega^2}{c^2}\sum\limits_{\tau}
\varepsilon_{\tau}(\vec{k}_{\tau^{\prime}},\vec{\omega})\vec{E}(\vec{k}_{\tau+\tau^{\prime}},\omega)\nonumber\\
& &=-\frac{\omega^2_L}{\gamma c^2}\vec{E}(\vec{k}_{\tau^{\prime}},\omega)-\left(\frac{\omega^2_L\vec{k}_{\tau^{\prime}}}
{\gamma c^2(\omega-\vec{k}_{\tau^{\prime}} \vec{u})}+\frac{\omega^2_L(\vec{k}_{\tau^{\prime}}C^2-\omega^2)}
{\gamma C^4(\omega-\vec{k}_{\tau^{\prime}} \vec{u})^2}\right)\\
& &\times\left[(\vec{u}\vec{E}(\vec{k}_{\tau^{\prime}},\omega))-
\left(\frac{\omega^2_L\vec{u}}{\gamma c^2(\omega-\vec{k}_{\tau^{\prime}}  \vec{u})}\right)(\vec{k}_{\tau^{\prime}}\vec{E}(\vec{k}_{\tau^{\prime}},\omega))\right],\nonumber
\end{eqnarray}
$\vec{\tau}^{\prime}=0$, $\vec{\tau}_1,\vec{\tau}_2,\ldots$.

The set of equations (\ref{phys_1.2}) describes the situation when a distributed feedback is formed by many diffracted waves (it is the analogy of the case of multi-wave  X-ray diffraction in crystals \cite{132}). However, the analysis of such a general situation is very complicated. So we only consider here a two-wave distributed feedback. This allows us to obtain all the main characteristics of X-ray FELs analytically and also to show the advantages of three-dimensional geometry of distributed feedback in comparison with the one-dimensional case.

So, let us consider specifically the generation of a
$\sigma$-polarized wave, i.e., the wave polarized in the
diffraction plane. For the geometry of the so-called two-beam
diffraction \cite{L_23}, where two strong waves are excited and
diffraction occurs by the set of crystallographic planes,
determined by a reciprocal lattice vector. In this case one can
obtain a set of Maxwell's equations describing two-wave
diffraction in a crystal, having a beam penetrating through it, in
the following way:
\begin{eqnarray}
\label{phys_1.3}
& &\left(k^2c^2-\omega^2\varepsilon_0+\frac{\omega^2_L}{\gamma}+\frac{\omega^2_L}{\gamma}
\frac{(\vec{u}\vec{e}_{\sigma}^{\prime})^2}{c^2}\frac{k^2c^2-\omega^2}{(\omega-\vec{k}\vec{u})^2}\right)E_{\sigma}-
\omega^2\varepsilon_{\tau}E_{\sigma}^{\tau}=0\nonumber\\
& &-\omega^2\varepsilon_{-\tau}E_{\sigma}+\left(K^2_rc^2-\omega^2\varepsilon_0+\frac{\omega^2_L}{\gamma}+
\frac{\omega^2_L}{\gamma}\frac{(\vec{u}\vec{e}_{\sigma})^2}{e^2}
\frac{k^2_{\tau}c^2-\omega^2}{(\omega-\vec{k}_{\tau}\vec{u})^2}\right)\vec{E}_{\sigma}^{\tau}=0.\nonumber\\
\end{eqnarray}
In (\ref{phys_1.3})
$E_{\sigma}=\vec{E}(\vec{k},\omega)\cdot\vec{e}_{\sigma}$,
$E_{\sigma}^{\tau}=\vec{E}(\vec{k}+\vec{\tau},\omega)\cdot\vec{e}_{\sigma}$,
$ \vec{e}\parallel[\vec{k}\vec{\tau}]$, $\omega^2_L=4\pi
e^2n_0/m$, where $n_0$ is the average electron (positron) density
in a beam. Comparing (\ref{phys_1.3}) with the standard equation
of X-ray dynamical diffraction, one can see that the system of a
crystal and particle beam may be considered as an active medium
with dielectric susceptibility:
\begin{eqnarray*}
\tilde{\varepsilon}_0(\vec{k},\omega)-1=\varepsilon_0-1-\frac{\omega^2_L}{\gamma\omega^2}-\frac{\omega^2-L}{\gamma\omega^2}
\frac{(\vec{u}\vec{e}_{\sigma})^2}{c^2}\frac{k^2c^2-\omega^2}{(\omega-\vec{k}\vec{u})^2},~ \tilde{\varepsilon}_{\tilde{\tau}}=\varepsilon_{\tau}\\
\tilde{\varepsilon}_0(\vec{k}_{\tau},\omega)-1
=\varepsilon_0-1-\frac{\omega^2_L}{\gamma\omega^2}-\frac{\omega^2_L}{\gamma\omega^2}\frac{(\vec{u}\vec{e}_{\sigma})^2}{c^2}\frac{\vec{k}^2_{\tau}c^2-\omega^2}{(\omega-\vec{k}_{\tau}\vec{u})^2}
,~ \tilde{\varepsilon}_{-\tau}=\varepsilon_{-\tau}
\end{eqnarray*}

Further, we shall analyze the generation of the wave with a wave vector  $\vec{k}$, which makes a small angle with the particle velocity vector $\vec{u}$. In this case the wave vector $\vec{k}_{\tau}=\vec{k}+\vec{\tau}$ is directed at a large angle relative to $\vec{u}$, and consequently the magnitude of $(\omega-\vec{k}_{\tau}\vec{u})$ cannot become small. As a result, the terms containing the expression  $(\omega-\vec{k}_{\tau}\vec{u})$ in their denominators will be small and can be ignored. We shall also neglect the term $\omega^2_L/\gamma$ -- this is justified for real beam densities. It is well known that in order to provide non-zero solutions for the equation set (\ref{phys_1.3}), its determinant should be equal to zero.  This also defines the dispersion equation, and for the $\sigma$-polarized wave it can be written in the form:
\begin{eqnarray}
\label{phys_1.4}
(\omega-\vec{k}\vec{u})^2\left[(k^2c^2-\omega^2\varepsilon_0)(k^2_{\tau}c^2-\omega^2\varepsilon_0)-\omega^4\varepsilon_{\tau}\varepsilon_{-\tau}\right]
=\\
-\frac{\omega^2_L}{\gamma}\frac{(\vec{u}\vec{e}_{\sigma})^2}{c^2}
(k^2c^2-\omega^2)(k^2_{\tau}c^2-\omega^2\varepsilon_0). \nonumber
\end{eqnarray}
The dispersion equation in such a form was derived in \cite{berk_135}. To solve the boundary problem, we use field continuity, the beam density and the beam current density at the boundaries. For the last two conditions we apply the following expressions, obtained from the equations of particle motion and the expression for the particle beam current

\begin{eqnarray}
& & \delta
j_{\sigma}=\frac{ie^2n_0}{m\gamma\omega}\frac{(\vec{u}\vec{e}_{\sigma})^2}{c^2}
\frac{k^2c^2-\omega^2}{(\omega-\vec{k}\vec{u})^2}E_{\sigma}\nonumber\\
& &
j_{\sigma}=e(\vec{u}\vec{e}_{\sigma})n_0-\frac{ie^2n_0}{m\gamma}
\frac{(\vec{u}\vec{e}_{\sigma})^2}{c^2(\omega-\vec{k}\vec{u})}E_{\sigma}.
\label{phys_1.5}
\end{eqnarray}

The dispersion equation (\ref{phys_1.4}) is sixth-order and, hence
six solutions correspond to it. However, the two solutions
corresponding to mirror reflected waves can be neglected due to
the small value of $\varepsilon_{\tau}(\omega)$ in the X-ray
range, $\varepsilon_0-1\equiv g_0$; $\varepsilon_{\tau}\sim
10^{-5}$. Small values of $\varepsilon_{\tau}$ and $g_0$ will also
be utilized when performing joining (we only join electric field
strengths at boundaries).

The general solution for the field in a crystal is written as
\begin{equation}
\label{phys_1.6}
\vec{E}=\sum\limits_{i=1}^4\vec{e}_{\sigma}c_i\exp(i\vec{k_i}\vec{r})[1+s_i\exp(i\vec{\tau}\vec{r})],
\end{equation}
where $\vec{k}_i$ is the i-th solution of the dispersion equation (\ref{phys_1.4}) and $\vec{k}_{i\tau}=\vec{k}_i+\vec{\tau}$; $\vec{\tau}$ is the reciprocal lattice vector corresponding to the planes of  diffraction reflection.

Taking into account  the remarks made, the boundary conditions may be written as:
\begin{eqnarray}
\label{phys_1.7}
& &c_1+c_2+c_3+c_4=1,\nonumber\\
& &f_1c_1+f_2c_2+f_3c_3+f_4c_4=0,\nonumber\\
& &g_1c_1+g_2c_2+g_3c_3+g_4c_4=0,\\
& &s_1c_1e^{iK_{1z}L}+s_2c_2e^{iK_{2z}L}+s_3c_3e^{iK_{3z}L}+s_4c_4e^{iK_{4z}L}=0,\nonumber\\
&
&s_i=\frac{\omega^2\varepsilon_{-\tau}}{k^2_{i_{\tau}}c^2-\omega^2\varepsilon_0},\qquad
f_i=\frac{(\vec{u}\vec{e}_{\sigma})^2}{(\omega-\vec{k}_i\vec{u})},\qquad
g_i=\frac{k^2_ic^2-\omega^2}{(\omega-\vec{k}_i\vec{u})^2}\frac{(\vec{u}\vec{e}_{\sigma})^2}{c^2}.\nonumber
\end{eqnarray}
In (\ref{phys_1.7}), only those boundary conditions are written that determine the field inside a crystal.
The first equation corresponds to the continuity of an incident wave at the boundary $z=0$;
the second and the third conditions correspond to vanishing of the beam density and the beam
current density at the crystal entrance. 
The last condition  corresponds to the vanishing of the diffracted
wave at the exit boundary $z=L$ (we consider  Bragg diffraction
geometry), $\vec{k}_i$ and $k_{iz}$ ($i=1 \div 4$) are the
solutions of the dispersion equation (\ref{phys_1.4}). The linear
system (\ref{phys_1.7}), defining the coefficients $c_i$ placed
ahead of $i$-modes in (\ref{phys_1.6}), has the solution
$c_i=\Delta_i/\Delta$, where $\Delta$ is the determinant of the
system (\ref{phys_1.7}), $\Delta_i$ is the $i$-th minor, obtained
as a result of replacement of the $i$-th column by
\[
\left(
  \begin{array}{c}
    1 \\
    0 \\
    0 \\
    0 \\
  \end{array}
\right).
\]
Hence, at $\Delta\rightarrow 0$, the field amplitudes inside a crystal will increase, and thus a condition occurs  when the field is nonzero though the incident wave is equal to zero.  The condition $\Delta=0$  with $\Delta_i\neq 0$ is called the generation threshold condition \cite{nim95_2}. Substituting the expressions
\begin{eqnarray}
\label{phys_1.8}
& &\vec{k}_i=\vec{k}_0+\vec{k}\delta_i\vec{n},\qquad k_{0z}=\frac{\omega-\vec{k}_{\perp}\vec{u}_{\perp}}{u_z},\nonumber\\
& &k=\omega/c,\qquad \delta_i\ll 1,
\end{eqnarray}
(where $\vec{n}$ is the normal to the crystal surface, $\vec{k}_0=(k_{0z},\vec{k}_{\perp})$) into the determinant $\Delta$, we can represent the generation threshold condition $\Delta=0$ as
\begin{eqnarray}
\label{phys_1.9} \hspace{-10mm}& &
\frac{(\delta_1-\delta_2)(\delta_1-\delta_3)(\delta_2-\delta_3)}{\delta_1^2\delta_2^2\delta_3^2}s_4e^{ik\delta_4L}
-\frac{(\delta_1-\delta_2)(\delta_1-\delta_4)(\delta_2-\delta_4)}{\delta_1^2\delta_2^2\delta_4^2}s_3e^{ik\delta_3L}\\
\hspace{-10mm}& &
+\frac{(\delta_1-\delta_3)(\delta_1-\delta_4)(\delta_3-\delta_4)}{\delta_1^2\delta_3^2\delta_4^2}s_2e^{ik\delta_2L}
-\frac{(\delta_2-\delta_3)(\delta_2-\delta_4)(\delta_3-\delta_4)}{\delta_2^2\delta_3^2\delta_4^2}s_1e^{ik\delta_1L}=0\nonumber
\end{eqnarray}
In (\ref{phys_1.9}), the terms containing nonresonance  $f_i$ and
$g_i$ ($i=1\div 4$) were neglected.

Further investigation will be based on a common consideration of (\ref{phys_1.4}) and (\ref{phys_1.9}). We substitute (\ref{phys_1.8}) into (\ref{phys_1.4})  and transform the dispersion equation as follows:
\begin{eqnarray}
\label{phys_1.10}
\frac{(\vec{u}\vec{n})^2}{c^2}\delta^2[4\gamma_0\gamma_1\delta^2+2(\gamma_1l+\gamma_0l_{\tau})\delta+ll_{\tau}-\tau]
=\nonumber \\
-\frac{1}{\gamma}\left(\frac{\omega_L}{\omega}\right)^2\theta^2\sin^2\varphi(l+g_0)l_{\tau}.
\end{eqnarray}
In (\ref{phys_1.10}) $\theta$ is the angle between $\vec{k}$ and $\vec{u}$; $\varphi$ is the angle between $\vec{k}_{\perp_u}$ and $\vec{\tau}_{\perp_u}$, where the symbol $\perp_u$  denotes the projection onto a plane perpendicular to the velocity, $r=\varepsilon_{\tau}\varepsilon_{-\tau}$; $l=\theta^2+g_0+\gamma^{-2}$; $l_{\tau}=l+\alpha$; $\alpha=(2\vec{k}_0\vec{\tau}+\tau^2)/k^2$  is the departure from the Bragg condition.

\[
\gamma_0=\frac{(\vec{k}_0\vec{n})}{K},\qquad
\gamma_1=-\frac{(\vec{k}_{0\tau}\vec{n})}{K}
\]
are the cosines of the angles, made by the wave vectors of the transmitted and diffracted waves with the normal vector.

\section{Spontaneous and induced parametric and Smith-Purcell
radiation from electrons moving in a photonic crystal built from
the metal threads (\cite{rins_66})} \label{vfel_ch:1}

Research and development of microwave generators using radiation
from an electron beam in a periodic slow-wave circuit (traveling
wave tubes, backward wave oscillators, etc.,) has a long history
\cite{vfel_TWT1,vfel_TWT2}. First generators operated in the centimeter
wavelength range.

In 1953 Smith and Purcell \cite{vfel_SP}  made the next step and
observed generation of incoherent radiation at visible wavelengths
by using a finely-focused electron beam propagating close to the
surface of a metal diffraction grating (at the distance
$
\delta
\le \frac{{\lambda \beta \gamma} }{{4\pi} },
$
$\delta $ is the
so-called { beam impact parameter, $\lambda $ is the radiation
wavelength,} $\beta = v /c$ , $v$ is the electron beam velocity,
$\gamma $ is the electron Lorentz-factor).
Beam current densities were insufficient to produce significant
amplification of the spontaneous emission.
However, as it was shown in \cite{vfel_Salisbury1,vfel_Salisbury2}, to change this
spontaneous radiation process into a stimulated one, the generator
should be supplied with a feedback cavity formed by a pair of
reflectors.

After the discovery of the Smith-Purcell effect, it soon became
clear that it might be used as a radiation source in the
millimeter to visible range, for which tunable sources were hardly
or not available at that time \cite{vfel_Walsh1,vfel_Walsh2}.

The Smith-Purcell effect belongs to the general class of
diffracted radiation effects induced by electron interaction with
 a medium.
Diffraction of waves associated with the electromagnetic field of
the electron by an obstacle leads to the so-called diffracted
radiation \cite{vfel_Bolotovskii}.
Diffracted radiation in periodical structures is in the basis of
operation of traveling wave tubes \cite{vfel_TWT1,vfel_TWT2} and
such devices as the orotron
\cite{vfel_orotron1,vfel_orotron2,vfel_orotron3} and the ledatron
\cite{vfel_ledatron} (see also
\cite{vfel_Leavitt1,vfel_Leavitt2,vfel_Leavitt3a,vfel_Leavitt3b}).

All the above devices use feedback, which is formed  by either two
parallel mirrors placed on both sides of the working area or
a one-dimensional diffraction grating, in which incident and
diffracted (reflected) waves move along the electron beam
(one-dimensional distributed feedback).

The conception of volume distributed feedback \cite{berk_133} was
first originated in view of prediction and experimental study of
spontaneous parametric and diffracted transition X-ray radiation
from charged particles in crystals (PXR) \cite{vfel_PXRbook}.
The detailed analysis of the induced PXR and channeling radiation
revealed unique possibilities provided by volume distributed
feedback \cite{berk_133}.
It was shown that dynamical diffraction in a volume spatially
periodic medium evokes the peculiar conditions, corresponding to
degeneration of the eigenmodes that results in a new law of
electron beam instability.
Within these peculiar conditions the electron beam interacts with
the electromagnetic wave more effectively \cite{berk_133}.

{Thus, for example, even in the X-ray range the electron beam
current density necessary for running up to the generation
threshold in conditions of non-one-dimensional distributed
feedback \cite{berk_133,berk_104} appears significantly reduced
($10^8~ \textrm{A/cm}^2$ for LiH crystal against $10^{13}~
\textrm{A/cm}^2$ required in \cite{vfel_Kurizki}) that even makes
possible to reach generation threshold for the induced X-ray
radiation in crystals i.e. to create an X-ray laser.}

Moreover, all the conclusions are valid for a beam moving in
vacuum close to the surface of the periodic medium either in the presence
or absence of the undulator \cite{berk_104}.

 The
originated law is universal and valid for all wavelength ranges
(from X-ray to microwave) regardless the spontaneous radiation
mechanism\\
\cite{berk_104,berk_133,berk_139,vfel_VFELreview,vfel_FEL2002,laser_8,
nova_bar4,hyb_FirstLasing,hyb_patent}.
Radiation generators using non-one-dimensional distributed
feedback, which is created with the aid of either natural or
artificial (photonic) crystals, is called Volume Free Electron
Laser (VFEL).

 Use of the volume distributed feedback makes
available:

1. frequency tuning at fixed energy of the electron beam in the
significantly wider range than conventional systems can provide;

2. significant reduction of the threshold
current of the electron beam due to more effective interaction of the electron beam with the
electromagnetic wave allows and, as a result, miniaturization of
generators;

3. reduction of limits for the available output power by the use
of wide electron beams and diffraction gratings (photonic
crystals) of large volumes;

4. simultaneous generation at several frequencies;

5. effective modes selection in oversize systems, in which the
radiation wavelength is significantly smaller than the resonator
dimensions.

Studies of the two-dimensional distributed feedback application in
millimeter-wave FEL-oscillators began in the 90s in
\cite{vfel_G0,vfel_G4,vfel_G6,vfel_G9,vfel_G10}.

\begin{figure}[htp]
\centering
\epsfxsize = 8 cm \centerline{\epsfbox{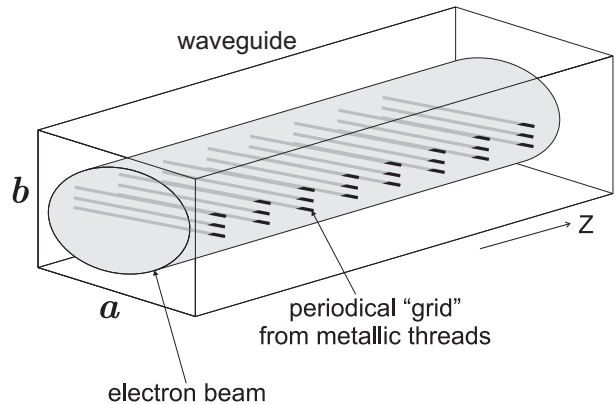}}
\caption{VFEL ''grid'' resonator (''grid'' photonic crystal)}
\label{vfel_resonator}
\end{figure}

One of the VFEL types uses a ''grid'' volume resonator (''grid''
photonic crystal) that is formed by a periodic structure built
from either dielectric or metal threads (see Figure
\ref{vfel_resonator}) \cite{vfel_grid-t,nova_bar4,tu2_grid-ex}.

The ''grid'' structure formed by periodically strained dielectric
threads was experimentally studied in \cite{nova_bar4}, where
it was shown that ''grid'' photonic crystals have sufficiently
high $Q$ factors ($10^4 - 10^6$).
Lasing from  VFELs  with the ''grid'' resonator formed by
periodically strained  metal threads was observed
in~\cite{tu2_grid-ex}.

Propagation of waves through a photonic crystal is the subject
of numerous theoretical and experimental studies
\cite{vfel_PC1,vfel_PC2,vfel_PC3,vfel_PC4}.

In \cite{rins_66} the properties of a ''grid'' photonic crystal
built from metal threads were considered along with its frequency
characteristics in view of their importance for VFEL lasing.
%
%
A challenge, which appears when considering interaction of an
electromagnetic wave with such a photonic crystal, is as follows.
In contrast to the case of wave interaction with a thin dielectric
thread, the interaction of the electromagnetic wave with a single
metal thread of the ''grid'' (i.e., the diffraction grating unit
cell) cannot be described in terms of the perturbation theory.

The approach developed in \cite{rins_66} provides the
description of diffraction in a ''grid'' grating in terms of the
amplitude of the wave scattering by a single thread.
Methods for calculation of scattering amplitudes are
 well-developed~\cite{vfel_Nikolsky,vfel_Henl}.
This approach enables developing the theory of diffraction in
a ''grid'' photonic crystal similar to the dynamical theory of
diffraction for X-rays \cite{132,nim06_James} and using the results
obtained therein.
The equations describing lasing of VFEL with such a resonator have been
obtained.

\section{Scattering by a set of metal threads in free space}
\label{vfel_sec:2}

\begin{figure}[htp]
\centering
\epsfxsize = 8 cm \centerline{\epsfbox{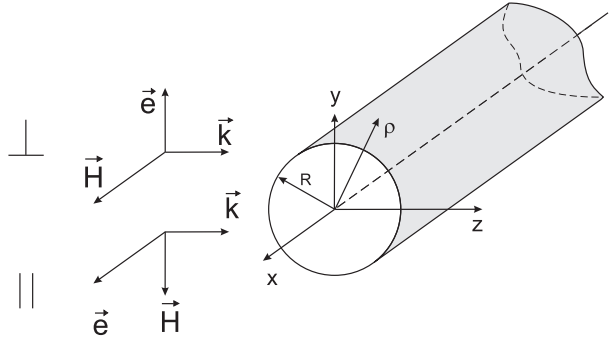}}
\caption{Coordinates and vectors} \label{vfel_cylinder}
\end{figure}

Suppose a plane electromagnetic wave $\vec{E}=\Psi_0 \vec{e}=e^{i
\vec{k} \vec{r}} \vec{e}~$ falls onto a metal thread.
Here $\vec{e}$ is the polarization vector of the wave,
$\vec{k}$ is the wavevector, $\vec{r}$ is the coordinate.
Suppose the wave falls perpendicular to the cylinder axis i.e.
along $OZ$.
The metal thread looks like a cylinder placed into the origin of
coordinates and the cylinder axis coincides with the axis $x$
Figure (\ref{vfel_cylinder}).

Two orientations of $\vec{e}$ should be considered:
$\vec{e}$ is parallel to the cylinder axis $x$ and
$\vec{e}$ is perpendicular to the cylinder axis $x$.
The solution of the problem of the  wave scattered by a cylinder is
well-known \cite{vfel_Nikolsky,vfel_Henl}.
Below the case when the radiation wavelength $\lambda$ exceeds the
thread radius $R$ ($\lambda \gg R$) is considered.

For clarity, suppose that $\vec{e} \parallel 0x$.
In this case the expression, which describes the wave appearing
due to scattering, can be expressed as a superposition of the
incident $\Psi_0$ and scattered $\Psi_{sc}$ waves as follows
\cite{vfel_Nikolsky,vfel_Henl}:
%

%
%
\begin{equation}
\Psi=\Psi_0+\Psi_{sc}=e^{ikz}+a_0 H_0^{(1)}(k \rho),
\label{vfel_1cylinder}
\end{equation}
here $\rho$ is the transverse coordinate $\rho=(y,z)$, $H_0^{(1)}$
is the Hankel function of zero order, $k$ is the wave number,
$a_0$ is the amplitude of the scattered wave.

\begin{equation}
a_{0 (\parallel)} = \frac{-J_0(k_t R)
J_0^{\prime}(kR)+\sqrt{\varepsilon_t} J_0^{\prime}(k_t R)J_0(k R)}
{J_0(k_t R) H_0^{(1)\prime}(kR)-\sqrt{\varepsilon_t}
J_0^{\prime}(k_t R)H_0^{(1)}(k R)}~,
 \label{vfel_a0_full_par}
\end{equation}
\begin{equation}
a_{0 (\perp)} = \frac{-J_0(k_t R)
J_0^{\prime}(kR)+\frac{1}{\sqrt{\varepsilon_t}} J_0^{\prime}(k_t
R)J_0(k R)}
{J_0(k_t R) H_0^{(1)\prime}(kR)-\frac{1}{\sqrt{\varepsilon_t}}
J_0^{\prime}(k_t R)H_0^{(1)}(k R)}~,
 \label{vfel_a0_full_perp}
\end{equation}
where $\varepsilon_t$ is the dielectric permittivity of the thread
material, $k_t=\sqrt{\varepsilon_t} k$.
Note that for $\lambda \gg R$ the amplitude $a_0$ does not depend
on the scattering angle.

Let us now consider a wave, which is scattered by a set of threads
placed at the points with the coordinates $\rho_n=(y_n,z_n)$.
The scattered wave can be expressed as a superposition of waves
scattered by separate threads:
\begin{equation}
\Psi=e^{ikz}+a_0 \Sigma_{n} H_0^{(1)}
(k \left\vert \vec{\rho }-\vec{\rho }_{n}\right\vert)e^{ikz_n}
\label{vfel_sum_cylinder}
\end{equation}
or, using the integral representation for the Hankel function,
$H_0^{(1)}(k \rho)$
\begin{equation}
\Psi=e^{ikz}+A_0 \Sigma_{n}
\int\limits_{-\infty }^{\infty }\frac{e^{i k\sqrt{\left\vert \vec{%
\rho }-\vec{\rho }_{n}\right\vert ^{2}-x^{2}}}}{\sqrt{\left\vert
\vec{\rho }-\vec{\rho }_{n}\right\vert ^{2}-x^{2}}}dx
e^{ikz_n},
\label{vfel_ncylinder}
\end{equation}
where $A_0=-\frac{i a_0}{\pi}$, $\left\vert
\vec{\rho }-\vec{\rho }_{n}\right\vert ^{2}=(y-y_n)^2+(z-z_n)^2$.

Note that in (\ref{vfel_ncylinder}) we neglected the contributions from
rescattered waves.

Let us consider a wave passing through a layer of cylinders whose
axes are distributed in the plane $x0y$ with the distance $d_y$
between them.
Suppose that the transversal size of the layer $L_{\perp}$ is much
larger than both $d_y$ and the radiation wavelength  ($L_{\perp}
\gg d_y$ and $L_{\perp} \gg \lambda$).
This assumption enables considering the ideal case when the layer
is supposed to have an infinite size in the plane $x0y$.
In this case summation over the coordinates $y_n$ provides the
following expression for $\Psi$:
\begin{equation}
\Psi=e^{ikz}+\frac{2 \pi i A_0}{k d_y}e^{ikz}=(1+\frac{2 \pi i
A_0}{k d_y})e^{ikz}. \label{vfel_average-n-cylinder}
\end{equation}
This expression reflects a well-known fact that a plane wave
scattered by a plane layer of the scatterers is expressed as a
plane wave with the modified amplitude.

Thus, after passing $m$ planes spaced $d_z$ apart from each other,
the scattered wave can be expressed as:
\begin{equation}
\Psi =\left( \sqrt{\left( 1-\frac{2\pi
~{\textrm{Im}}A_{0}}{k~d_{y}}\right) ^{2}+\left( \frac{2\pi
~{\textrm{Re}}A_{0}}{k~d_{y}}\right) ^{2}}\right)^m e^{i\varphi m}
e^{ikz}, \label{vfel_m_planes}
\end{equation}
where
\[
\varphi =\textrm{arctg} \left( \frac{\frac{2\pi ~\textrm{Re} A_{0}}{k~d_{y}}}{1-\frac{%
2\pi ~\textrm{Im} A_{0}}{k~d_{y}}}\right),
\]
 $m=L_z/d_z$
inside the photonic crystal formed by threads and $L_z$ is the
length of the photonic crystal. This expression can be easily
converted to the form $\Psi =e^{iknz}$, where $n$ is the
index refraction defined as
\begin{eqnarray}
\begin{array}{c}
n =  n^{\prime }+in^{\prime \prime }=
\left( 1+\frac{\lambda }{2\pi d_{z}}%
\arctan \left( \frac{\frac{\lambda }{d_{y}}{\textrm{Re}}A_{0}}{1-\frac{\lambda }{%
d_{y}}{\textrm{Im}}A_{0}}\right) \right) - \\
-i\frac{\lambda }{2\pi d_{z}}\ln \left(
\sqrt{\left( \frac{\lambda }{d_{y}}{\textrm{Re}}A_{0}\right) ^{2}+\left( 1-\frac{%
\lambda }{d_{y}}{\textrm{Im}}A_{0}\right) ^{2}}\right),
\end{array}
\label{vfel_n_exact}
\end{eqnarray}
here $\lambda=2 \pi/k$ is used, $n^{\prime }$ and
$n^{\prime \prime}$ denotes the real and imaginary parts of $n$,
respectively, ${\textrm{Re}}A_{0}$ and ${\textrm{Im}}A_{0}$ are
the real and imaginary parts of $A_0$.

Let us consider now the parameter $|\frac{\lambda}{d_y}A_0|$ and
suppose it to be small.
When $|\frac{\lambda}{d_y}A_0| \ll 1$,
(\ref{vfel_n_exact}) can be rewritten as:
\begin{equation}
n=1+\frac{2 \pi}{d_y d_z k^2}A_0.
\label{vfel_n_Bar}
\end{equation}
The same expression can also be  obtained for the  index of refraction
of the wave with polarization orthogonal to the thread axis
($\vec{e} \perp 0x$).

\begin{figure}[htp]
\centering
\epsfxsize = 8 cm \centerline{\epsfbox{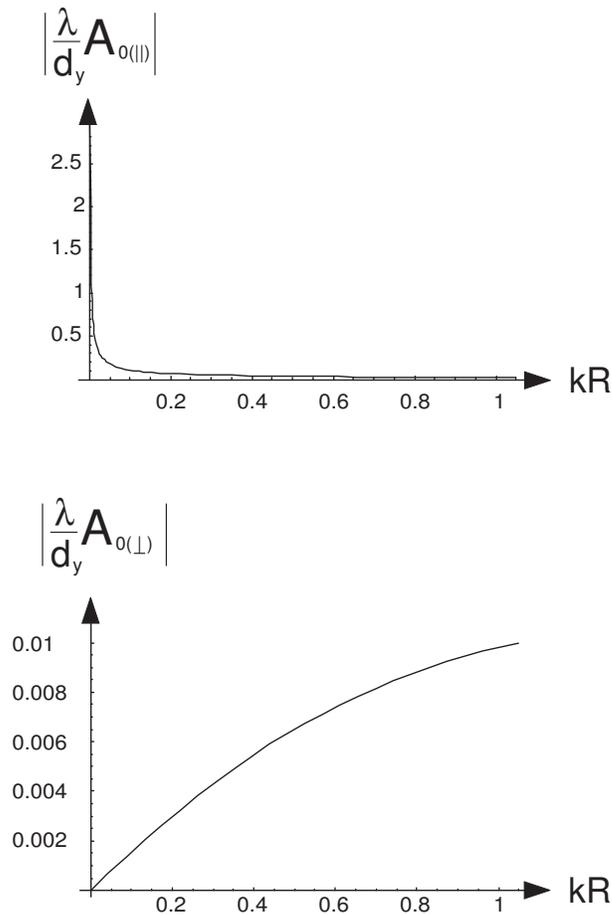}}
\caption{Dependence of parameter $|\frac{\lambda}{d_y}A_0|$ on
$kR$ for parallel and orthogonal polarizations
($d_y=0.5~\textrm{cm}$, $d_z=1.6~\textrm{cm}$)}
\label{vfel_fig:parameter}
\end{figure}

When the parameter $|\frac{\lambda}{d_y}A_0|$ grows with the
growth of the density of scatterers (i.e., with decreasing $d_y$),
the rescattered waves become important and the difference between
the mean and local fields should be considered similarly to
Clausius-Mossoti (Lorentz-Lorenz) relation \cite{vfel_Born}.
Calculations show that the parameter $|\frac{\lambda}{d_y}A_0|$
behaves differently for different polarizations (see Figure
\ref{vfel_fig:parameter}) and for the wave with polarization
parallel to the threads, this parameter can appear greater than 1.
In this case formula (\ref{vfel_n_Bar}) for parallel polarization
is not valid and the expressions considering wave rescattering
should be applied (see \cite{rins_66}).

%
It is important to remind that the conception of the refraction index is
applicable even to considering waves of different types (X-rays,
particles) passing through matter when the distance between
scatterers exceeds many times the wavelength, i.e., $kd \gg 1$
\cite{1}.
At that, if the wavelength is either comparable or smaller than
the  size of the scatterer, then the refraction index is determined by the
amplitude of the wave scattering at zero angle.
In the considered case of the wave scattering by the thread the
amplitude $A_0$ in  (\ref{vfel_n_Bar}) should be replaced
by the amplitude of forward scattering of the wave by the thread.
The explicit expression for this amplitude see in
\cite{vfel_Nikolsky,vfel_Henl}.

Radiation wavelengths of our interest are  $\lambda \le
3~\textrm{cm}$. In this wavelength range, the skin depth $\delta$
is about 1 micron for  most metals (for example,
$\delta_{Cu}=0.66$ $\mu$m, $\delta_{Al}=0.8$ $\mu$m,
$\delta_{W}=1.16$ $\mu$m, and so on). Thus, in this wavelength
range the metal threads can be considered as perfectly conducting.

From (\ref{vfel_a0_full_par}), (\ref{vfel_a0_full_perp})  follows that the
amplitude $A_0$ for a perfect conducting cylinder for polarization
of the electromagnetic wave parallel to the cylinder axis can be
expressed as:
\begin{equation}
A_{0\left( \parallel \right) }=\frac{1}{\pi }\frac{J_{0}\left(
kR\right) N_{0}\left( kR\right) }{J_{0}^{2}\left( kR\right)
+N_{0}^{2}\left( kR\right) }+i\frac{1}{\pi }\frac{J_{0}^{2}\left(
kR\right) }{J_{0}^{2}\left( kR\right) +N_{0}^{2}\left( kR\right)
}~. \label{vfel_A0_par}
\end{equation}

Using the asymptotic values for these functions for
$
kR = \frac{2 \pi R}{\lambda} \ll 1
$
\begin{eqnarray}
\begin{array}{l}
J_{0}\left( x\rightarrow 0\right) \approx 1,~
N_{0}\left( x\rightarrow 0\right) \approx -\frac{2}{\pi }\ln \frac{2}{%
1.781\cdot x},\\
J_{0}^{\prime }\left( x\rightarrow 0\right) =-J_{1}\approx
-\frac{x}{2},~
N_{0}^{\prime }\left( x\rightarrow 0\right) =-N_{1}\approx \frac{2}{\pi }%
\frac{1}{x},
\end{array}
\end{eqnarray}
one can obtain:
\begin{equation}
\textrm{Re} A_{0\left( \parallel \right)}
\approx
 -\frac{1}{\pi} ~\frac{\frac{2}{\pi} \ln (\frac{2}{1.781\cdot k R})}{1+ (\ln (\frac{2}{1.781\cdot k R}))^2}
 ~,~
 \textrm{Im} A_{0\left( \parallel \right)} \approx
 \frac{1}{\pi} ~\frac{1}{1+ (\ln (\frac{2}{1.781\cdot k R}))^2}~,\nonumber\\
\label{vfel_A0par1} \\
\end{equation}
From (\ref{vfel_A0par1}) it can be seen that for $kR \ll
1$, the values  $\textrm{Re}A_{0\left( \parallel \right)}<
0$, therefore, the real
part of the  index of refraction in this limit can be expressed as:
\begin{equation}
\label{vfel_n_Bar1}
n^{\prime}_{\left( \parallel \right)}=1-\frac{2 \pi}{d_y d_z
k^2}|\textrm{Re}A_{0 \left( \parallel \right)}|,
\end{equation}
%
%
%
%
From (\ref{vfel_n_Bar1}) follows that the growth of the wavelength (reduction
of $k$) makes $n_{\parallel}^{\prime} \rightarrow 0$, and
$n_{\parallel}^{\prime}$  can even become negative.
However, it should be reminded that (\ref{vfel_n_Bar1})
is valid only when $|n_{\parallel}^{\prime}-1| \ll 1$ (i.e
$\frac{2 \pi}{d_y d_z k^2}|\textrm{Re}A_{0(\parallel)}| \ll 1$).
It is interesting to compare $n_{\parallel}^{\prime}$ calculated
by formulas (\ref{vfel_n_exact}) and (\ref{vfel_n_Bar1}): in the range of large wavelength they yield
different results because the condition
$|n_{\parallel}^{\prime}-1| \ll 1$ is violated. In this case the
expressions given in \cite{rins_66} should be applied.
Note also that $n_{\parallel}^{\prime}=0$ corresponds to the
threshold of the wave total reflection from the medium surface.

Let us consider a particular example: suppose the radiation
wavelength is $\lambda=3~\textrm{cm}$ and the thread radius
$R=50~\mu\textrm{m}$, then
\begin{equation}
\label{vfel_A0reimpar}
ReA_{0\left( \parallel \right)} =
-0.096,~ImA_{0\left( \parallel \right)} = 0.032,
\end{equation}
Suppose $d_y=0.5~\textrm{cm}$, $d_z=1.6~\textrm{cm}$, therefore,
the parameter $|\frac{\lambda}{d_y}A_{0}| \ll 1$ for both
polarizations and (\ref{vfel_n_Bar}) gives:
\begin{equation}
\label{vfel_n_par}
n_{\parallel}=0.828+i \cdot 0.058 ,
\end{equation}



The above analysis does not include the contribution to the index of refraction
 from rescattered waves. Their contribution was investigated in \\ \cite{vfel_grid-t,vfel_new-grid-experiment,tu2_grid-ex,rins_66,lanl_13,lanl_23}. It was shown, in particular, that the the index of refraction  $n_{\perp}$ for a wave with the polarization vector orthogonal to the threads can be greater than unity, i.e., in such photonic crystals, the Cherenkov effect is possible.

\section{Complex anomalous Doppler effect in the ''grid'' photonic crystal}
\label{vfel_sec:4}

Let us now study Smith - Purcell (diffracted) radiation in a
photonic crystal when an electron beam of velocity ${\vec{v}}$
passes through the ''grid''.

The radiation condition can be expressed as
\begin{equation}
\omega - \vec{k} n({k}){\vec{v}}= \vec{\tau} \vec{v} ~,
\label{vfel_eq:3-2}
\end{equation}
where $\vec{\tau}$ is the reciprocal lattice vector and $n(k)$ is
the index of refraction.
Suppose the electron beam velocity is directed along the axis
 $OZ$, then (\ref{vfel_eq:3-2}) can be presented in the
form:
\begin{equation}
k - {\tau}_z {\beta} = k \,n(k)\, \beta \cos \theta,
\label{vfel_eq:3-3}
\end{equation}
where $\beta=\frac{v}{c}$, the angle between $\vec{k}$ and the
electron beam velocity is denoted by $\theta$ and $\tau_z=\frac{2
\pi m_h}{d_z}$, where $m_h=1,2,...$ is the harmonic number.
From (\ref{vfel_eq:3-3}) follows the equation
\begin{equation}
(\frac{k - {\tau}_z {\beta}}{\beta \cos \theta} )^2 = k^2 + \eta~,
\label{vfel_eq:3-6}
\end{equation}
which is similar to the equation for the complex and anomalous
Doppler effect \cite{berk_10}.

The roots of this equation give the spectrum of frequencies of
diffracted (Smith-Purcell) radiation, which is induced by a
particle moving in the above volume ''grid'' structure. In the
case under consideration $\frac{\eta}{\tau _{z}^2} \ll 1$.
For $\beta$ providing $\frac{\eta}{\tau _{z}^2 \beta^2} \ll 1$, the
roots of (\ref{vfel_eq:3-6}) can be easily found:
\begin{eqnarray}
& \hspace{1cm}& k _{1} = \frac{{\tau _{z} \beta}}{{1 - (\beta \cos
\theta) ^{2}}}\left( {1 - \beta \cos \theta \sqrt {1 + \frac{{
\eta} }{{\tau _{z}^{2}} }\frac{{1 - (\beta \cos
\theta) ^{2}}}{{\beta ^{2}}}}}  \right), \nonumber \\
\\
& & k _{2} = \frac{{\tau _{z} \beta}}{{1 - (\beta \cos \theta)
^{2}}}\left( {1 + \beta \cos \theta \sqrt {1 + \frac{{\eta}
}{{\tau _{z}^{2}} }\frac{{1 - (\beta \cos \theta) ^{2}}}{{\beta
^{2}}}}} \right), \nonumber \label{vfel_eq:3-07}
\end{eqnarray}
where $\eta$ is taken at $k=\frac{\tau _{z} \beta}{1 - (\beta \cos
\theta) ^{2}}$.

It should be reminded here that $\tau_z=\frac{2 \pi m_h}{d_z}$,
where $m_h=1,2,...$ is the harmonic number.
From (\ref{vfel_eq:3-07}) follows that higher harmonics provide
getting radiation with higher frequencies. For example, for the
electron beam with the energy 200 keV, considering $\theta \sim
20^{\circ}$ and $d_z=1.6$ cm, the first harmonic ($m_h=1$) gives
radiation frequencies $\sim 10$ GHz and $\sim 40$ GHz for the
roots 1 and 2 of  equation (\ref{vfel_eq:3-3}), respectively, the
30-th harmonic ($m_h=30$) provides $\sim 230$ GHz and $\sim 1$
THz.
%

Let us now study diffracted radiation in a metal waveguide  of
rectangular cross-section with the ''grid'' structure (''grid''
photonic crystal) placed inside it (see Figure
\ref{vfel_resonator}).
Suppose that the lateral dimensions $a$ and $b$ of the waveguide
noticeably exceed the distance between the next threads of the
''grid'' photonic crystal ($a,b \gg d_y, d_z$).
This makes possible to consider a waveguide with the ''grid''
photonic crystal inside as a waveguide filled with matter with the
effective dielectric permittivity $\varepsilon = n^2$.

Remember that in the waveguide with the axis parallel to $OZ$, the
$x$ and $y$ components of the wavevector are not continuous, but
quantized. And only the wavenumber $k_z$ changes continuously.
The wave field inside the waveguide represents a wave standing in
the transversal directions ($OX$ and $OY$), but traveling along
$OZ$.

In the waveguide there is the discrete set of the waves with the
eigenvalues $\kappa_{mn}$ determined by the waveguide
transverse dimensions (width $a$ and
height~$b$)~\cite{WIND_Jackson-a,vfel_Nikolsky,vfel_Henl}
%
\begin{equation}
\kappa_{mn}^2=(\frac{\pi m}{a})^2+(\frac{\pi n}{b})^2~.
\label{vfel_eq:3-03}
\end{equation}
Therefore, in the waveguide the radiation frequency $\omega$ is
related to the wavenumber $k_z$ as follows \cite{WIND_Jackson-a}:
\begin{equation}
k_z^2 (m,n)=(\frac{\omega}{c})^2 \varepsilon -\kappa_{mn}^2~.
\label{vfel_eq:3-04}
\end{equation}
Remembering that $n^2=\varepsilon=1+\chi$ ($\varepsilon$ is the
permittivity of the ''medium'' formed by the metal threads and
$\chi$ is its susceptibility) one can rewrite (\ref{vfel_eq:3-04})
as follows:
\begin{equation}
k_z^2 (m,n)=(\frac{\omega}{c})^2 - (\kappa_{mn}^2- \eta)~,
\label{vfel_eq:3-05}
\end{equation}
here $\eta=\frac{\omega^2}{c^2} \chi$. According to the previous
section $n < 1$, therefore both $\chi$ and $\eta$ are negative.

Therefore, addition of the described volume structure to the
waveguide ''replaces'' their own eigenvalues by some effective
$K_{mn}$ ($K_{mn}^2=\kappa_{mn}^2-\eta$). It is interesting to
note here that, as $\eta < 0$, the addition of the volume grid to the
waveguide increases the waveguide limiting frequency (i.e., the
effective waveguide appears as though having smaller transversal
dimensions).

Let us now consider the spontaneous Smith-Purcell radiation from a
particle moving along the waveguide axis.
The radiation condition for the Smith-Purcell radiation
(\ref{vfel_eq:3-2}) converts to
\begin{equation}
\omega - k_{z}{v}= {\tau}_z {v} ~. \label{vfel_eq:3-2-1}
\end{equation}

Combination of (\ref{vfel_eq:3-05}) and (\ref{vfel_eq:3-2-1}) enables one to
find the equation for radiation frequencies similar to
(\ref{vfel_eq:3-6}):
\begin{equation}
(\frac{\omega - {\tau}_z {v}}{v})^2 = (\frac{\omega}{c})^2 -
(\kappa_{mn}^2- \eta)~.
\end{equation}
The roots of this equation for $\frac{\eta}{\tau _{z}^2 \beta^2}
\ll 1$ can be obtained similarly (\ref{vfel_eq:3-07}):
\begin{eqnarray}
& \hspace{1cm}& \omega _{1} \left( {m,n} \right) = \frac{{\tau
_{z} v}}{{1 - \beta ^{2}}}\left( {1 - \beta \sqrt {1 -
\frac{{(\kappa_{mn}^2- \eta)} }{{\tau _{z}^{2}} }\frac{{1 -
\beta ^{2}}}{{\beta ^{2}}}}}  \right), \nonumber \\
\\
& & \omega _{2} \left( {m,n} \right) = \frac{{\tau _{z} v}}{{1 -
\beta ^{2}}}\left( {1 + \beta \sqrt {1 - \frac{{(\kappa_{mn}^2-
\eta)} }{{\tau _{z}^{2}} }\frac{{1 - \beta ^{2}}}{{\beta ^{2}}}}}
\right). \nonumber \label{vfel_eq:3-7}
\end{eqnarray}

For a detailed treatment of the theory of VFEL lasing using electron beam radiation
 in a ''grid'' photonic crystal, see \cite{rins_66}.

\section{Generation of radiation in Free Electron Lasers with photonic crystals (diffraction gratings) with the variable spatial period}
\label{vp_sec1}

In this section the equations providing the description of the
generation process in FEL with varied parameters of a photonic
crystal  (diffraction grating) are obtained \cite{lanl_13}.
It is shown that applying  photonic crystals (diffraction
gratings) with the variable period, one can significantly increase
the radiation output.
It is mentioned that photonic crystals  (diffraction gratings) can
be used for creation of the dynamical wiggler with variable period
in the system.
This makes possible to develop double-cascaded FEL with variable
parameters changing, which efficiency can be significantly higher
that of conventional system.

Generators using radiation from an electron beam in a periodic
slow-wave circuit (traveling wave tubes, backward wave
oscillators, free electron lasers) are now widespread
\cite{tu2_Granatstein}.

Diffracted radiation \cite{vfel_Bolotovskii} in periodical
structures is in the basis of operation of traveling wave tubes
(TWT) \cite{vfel_TWT1,vfel_TWT2}, backward wave oscillators (BWO)
and such devices as Smith-Purcell lasers
\cite{vfel_SP,vfel_Salisbury1,vfel_Salisbury2,vfel_Walsh1,vfel_Walsh2}
and volume FELs using two- or three-dimensional distributed
feedback
\cite{vfel_VFELreview,vfel_FEL2002,hyb_FirstLasing,hyb_patent}.

The analysis shows that during the operation of such devices electrons
lose their energy for radiation, therefore, the electron beam
slows down and gets out of synchronism with the radiating wave.
These limits the efficiency of the generator, which usually does not
exceed $\sim 10 \%$.

During the first years after the creation of the traveling wave tube, it was
demonstrated \cite{vfel_TWT2} that  synchronism between
the electron beam and the electromagnetic wave in a TWT can be  retained by changing  the
 phase velocity of the wave.
{Application} of systems with variable parameters in microwave
devices significantly increases their efficiency
\cite{vfel_TWT2,vp_Kuraev}.

The same methods  are widely used for increasing the efficiency of
undulator FELs \cite{vp_undulatorFEL}.

\subsection{Lasing equations for
 the system with a photonic crystal (diffraction grating) with changing parameters}
\label{vp_sec:2}

In the general case the equations, which describe lasing process,
follow from the Maxwell equations:
\begin{eqnarray}
& \hspace{1cm}& \textrm{rot} \vec{H}=\frac{1}{c}\frac{\partial
\vec{D}}{\partial t}+\frac{4 \pi}{c} \vec{j}, ~\textrm{rot}
\vec{E}=-\frac{1}{c}\frac{\partial
\vec{H}}{\partial t}, \nonumber \\
&  &\textrm{div} \vec{D}=4 \pi \rho,~\frac{\partial \rho}{\partial
t}+ \textrm{div} \vec{j}=0, \label{vp_sys0}
\end{eqnarray}
here $\vec{E}$ and $\vec{H}$ are the electric and magnetic fields,
$\vec{j}$ and $\rho$ are the current and charge densities, the
electromagnetic induction $D_i(\vec{r},t^{\prime})=\int
\varepsilon_{il}
(\vec{r},t-t^{\prime})E_l(\vec{r},t^{\prime})dt^{\prime}$ and,
therefore, $D_i(\vec{r},\omega)=\varepsilon_{il}
(\vec{r},\omega)E_l(\vec{r},\omega)$, the indices $i,l=1,2,3$
correspond to the axes $x,y,z$, respectively.


The current and charge densities are respectively defined as:
\begin{equation}
\vec{j}(\vec{r},t)=e
\sum_{\alpha}\vec{v}_{\alpha}(t)\delta(\vec{r}-\vec{r}_{\alpha}(t)),~
\rho(\vec{r},t)=e
\sum_{\alpha}\delta(\vec{r}-\vec{r}_{\alpha}(t)), \label{eq2}
\end{equation}
where $e$ is the electron charge, $\vec{v}_{\alpha}$ is the
velocity of the particle $\alpha$ ($\alpha$ numerates the beam
particles),
\begin{equation}
 \frac{d
\vec{v}_{\alpha}}{dt}=\frac{e}{m \gamma_{\alpha}}\left\{
 \vec{E}(\vec{r}_{\alpha}(t),t)+\frac{1}{c} [
\vec{v}_{\alpha}(t) \times \vec{H}(\vec{r}_{\alpha}(t),t)
 ]-
 \frac{\vec{v}_{\alpha}}{c^2}(\vec{v}_{\alpha}(t)\vec{E}(\vec{r}_{\alpha}(t),t))
\right\},\label{vp_sys1}
\end{equation}
here $\gamma_{\alpha}=(1-\frac{v_{\alpha}^2}{c^2})^{-\frac{1}{2}}$
is the Lorentz-factor, $\vec{E}(\vec{r}_{\alpha}(t),t)$
($\vec{H}(\vec{r}_{\alpha}(t),t)$) is the electric (magnetic)
field at the point of location $\vec{r}_{\alpha}$ of the particle
$\alpha$.

It should be reminded that  (\ref{vp_sys1}) can also be written as
\cite{L_14}:
\begin{equation}
 \frac{d
\vec{p}_{\alpha}}{dt}=m\frac{d \gamma_{\alpha}
v_{\alpha}}{dt}={e}\left\{
 \vec{E}(\vec{r}_{\alpha}(t),t)+\frac{1}{c} [
\vec{v}_{\alpha}(t) \times \vec{H}(\vec{r}_{\alpha}(t),t)
 ]
\right\},\label{vp_sys11}
\end{equation}
where $p_{\alpha}$ is the particle momentum.

Combining the equations in (\ref{vp_sys0}), we obtain:
\begin{equation}
-\Delta
\vec{E}+\vec{\nabla}(\vec{\nabla}\vec{E})+\frac{1}{c^2}\frac{\partial^2
\vec{D}}{\partial t^2}=-\frac{4 \pi}{c^2} \frac{\partial
\vec{j}}{\partial t}. \label{vp_sys01}
\end{equation}


The dielectric permittivity tensor can be expressed as
$\hat{\varepsilon}(\vec{r})=1+\hat{\chi}(\vec{r})$, where
$\hat{\chi}(\vec{r})$ is the dielectric susceptibility.
When $\hat{\chi} \ll 1$,  (\ref{vp_sys01}) can be rewritten as:

\begin{equation}
\Delta \vec{E}(\vec{r},t)-\frac{1}{c^2}\frac{\partial^2}{\partial
t^2}
\int \hat{\varepsilon}(\vec{r},t-t^{\prime})
\vec{E}(\vec{r},t^{\prime}) dt^{\prime}
=4 \pi \left( \frac{1}{c^2} \frac{\partial
\vec{j}(\vec{r},t)}{\partial t} + \vec{\nabla} \rho (\vec{r},t)
\right). \label{vp_1}
\end{equation}

When the grating is ideal $\hat{\chi}(\vec{r}) = \sum_{\tau}
\hat{\chi}_{\tau} (\vec{r}) e^{i \vec{\tau} \vec{r}}$,
%
where $\vec{\tau}$ is the reciprocal lattice vector.

Let the photonic crystal (diffraction grating) period be smoothly
varied with distance, which is much greater then the diffraction
grating {(ptotonic crystal lattice)} period.
It is convenient in this case to present the susceptibility
$\hat{\chi}(\vec{r})$ in the form, typical of the theory of X-ray
diffraction in crystals with lattice distortion \cite{tu3_Takagi}:
\begin{equation}
\hat{\chi}(\vec{r})=\sum_{\tau} e^{i \Phi_{\tau}(\vec{r})}
\hat{\chi}_{\tau} (\vec{r}), \label{vp_chi01}
\end{equation}
where $\Phi_{\tau}(\vec{r})=\int \vec{\tau} (\vec{r}^{\,\prime})
d\vec{l}^{\prime}$, $\vec{\tau} (\vec{r}^{\,\prime})$ is the
reciprocal lattice  vector in the vicinity of the point
$\vec{r}^{\,\prime}$.
In contrast to the theory of X-rays diffraction, in the case under
consideration $\hat{\chi}_{\tau}$ depends on $\vec{r}$.
Moreover, $\hat{\chi}_{\tau}$ depends on the volume
of the lattice unit cell $\Omega$, which can be significantly
varied for diffraction gratings (photonic crystals), as distinct
from natural crystals.
The volume of the unit cell $\Omega(\vec{r})$  depends on
coordinate and, for example, for a cubic lattice it is determined
as
$\Omega(\vec{r})=\frac{1}{d_1(\vec{r})d_2(\vec{r})d_3(\vec{r})}$,
where $d_i$ are the lattice periods.
If $\hat{\chi}_{\tau} (\vec{r})$ does not depend on $\vec{r}$, the
expression (\ref{vp_chi01}) converts to that usually used for
X-rays in crystals with lattice distortion \cite{tu3_Takagi}.

It should be reminded that for an ideal crystal without lattice
distortions, the  wave, which propagates in the crystal can be
presented as a superposition of  plane waves:
\begin{equation}
\vec{E}(\vec{r},t) = \sum_{\vec{\tau}=0}^{\infty}
\vec{A}_{\vec{\tau}} e^{i(\vec{k}_{\tau} \vec{r}-\omega t)},
\label{vp_field}
\end{equation}
where $\vec{k}_{\tau}=\vec{k}+\vec{\tau}$.

Let us now use the fact that in the case under consideration the
typical length for the change of the lattice parameters
significantly exceeds the lattice period. Then the field inside
the crystal with lattice distortion can be expressed similarly to
(\ref{vp_field}), but with $\vec{A}_{\vec{\tau}}$ depending on
$\vec{r}$  and $t$ and changing noticeably at the distances much
greater than the lattice period.

Similarly, the wave vector should be considered as a slowly
changing function of a coordinate.

According to the above, let us find the solution of (\ref{vp_1}) in
the form:
\begin{equation}
\vec{E}(\vec{r},t)=\textrm{Re} \left\{
\sum_{\vec{\tau}=0}^{\infty}  \vec{A}_{\vec{\tau}}
e^{i(\phi_{\tau} (\vec{r})-\omega t)} \right\}, \label{vp_*}
\end{equation}
where $\phi_{\tau} (\vec{r})= \int_0^{\vec{r}} k(\vec{r}) d
\vec{l}+ \Phi_{\tau}(\vec{r})$, where $k(\vec{r})$ can be found as
a solution of the dispersion equation in the vicinity of the point
with the coordinate vector $\vec{r}$, integration is made over the
quasiclassical trajectory, which describes motion of the
wavepacket in the crystal with lattice distortion.

Now let us consider the case when all the waves participating in
the diffraction process lie in a plane (coupled wave diffraction,
multiple-wave diffraction), i.e., all the reciprocal lattice
vectors $\vec{\tau}$ lie in one plane \cite{132,nim06_James}.
Suppose the wave polarization vector is orthogonal to the plane of
diffraction.

Let us rewrite (\ref{vp_*}) in the form
\begin{equation}
\vec{E}(\vec{r},t)=\vec{e}\,E(\vec{r},t)=\vec{e} \, \textrm{Re}
\left\{
 \vec{A}_1
e^{i(\phi_{1} (\vec{r})-\omega t)}+\vec{A}_2 e^{i(\phi_{2}
(\vec{r})-\omega t)} + ... \right\}, \label{vp_*1}
\end{equation}
where
\begin{equation}
\phi_1(\vec{r})=\int_0^{\vec{r}} \vec{k}_1(\vec{r}^{\, \prime}) d
\vec{l}, \label{vp_phi1}
\end{equation}
\begin{equation}
\phi_2(\vec{r})=\int_0^{\vec{r}} \vec{k}_1(\vec{r}^{\,\prime}) d
\vec{l} + \int_0^{\vec{r}} \vec{\tau}(\vec{r}^{\,\prime}) d
\vec{l}. \label{vp_phi2}
\end{equation}

Then multiplying (\ref{vp_1}) by $\vec{e}$, one can get:
\begin{equation}
\Delta {E}(\vec{r},t)-\frac{1}{c^2}\frac{\partial^2}{\partial t^2}
\int
\hat{\varepsilon}(\vec{r},t-t^{\prime}){E}(\vec{r},t^{\prime})
dt^{\prime}
=4 \pi \vec{e} \left( \frac{1}{c^2} \frac{\partial
\vec{j}(\vec{r},t)}{\partial t} + \vec{\nabla} \rho (\vec{r},t)
\right). \label{vp_2}
\end{equation}
Applying the equality $\Delta {E}(\vec{r},t)=\vec{\nabla}
(\vec{\nabla} E)$ and using (\ref{vp_*1}), we obtain
\begin{eqnarray}
\hspace{-1 cm} \Delta (\vec{A}_1 e^{i(\phi_{1} (\vec{r})-\omega
t)})=
e^{i(\phi_{1} (\vec{r})-\omega t)}
[2i  \vec{\nabla} \phi_1 \vec{\nabla} A_1 +i \vec{\nabla}
\vec{k}_1 (\vec{r}) A_1  - k_1^2(\vec{r})  A_1], \label{vp_3a}
\end{eqnarray}

Therefore, substitution of the above expression into (\ref{vp_2})
gives the following system:
\begin{eqnarray}
& & \frac{1}{2} e^{i(\phi_{1} (\vec{r})-\omega t)} [ 2i
\vec{k}_1(\vec{r}) \vec{\nabla} A_1 +i \vec{\nabla}
\vec{k}_1 (\vec{r}) A_1  - k_1^2(\vec{r}) A_1  \nonumber \\
& & + \frac{\omega^2}{c^2} \varepsilon_0(\omega,\vec{r}) A_1 + i
\frac{1}{c^2} \frac{\partial \omega^2
\varepsilon_0(\omega,\vec{r})}{\partial \omega} \frac{\partial
A_1}{\partial t} + \frac{\omega^2}{c^2}
\varepsilon_{-\tau}(\omega,\vec{r}) A_2 \nonumber\\
& &+ i \frac{1}{c^2} \frac{\partial \omega^2
\varepsilon_{-\tau}(\omega,\vec{r})}{\partial \omega}
\frac{\partial A_2}{\partial t}
] \nonumber \\
& & + \textrm{~conjugated~terms~}
=4 \pi \vec{e} \left( \frac{1}{c^2} \frac{\partial
\vec{j}(\vec{r},t)}{\partial t} + \vec{\nabla} \rho (\vec{r},t)
\right), \nonumber \\
& & \frac{1}{2} e^{i(\phi_{2} (\vec{r})-\omega t)} [ 2i
\vec{k}_2(\vec{r}) \vec{\nabla} A_2 +i \vec{\nabla}
\vec{k}_2 (\vec{r}) A_2  - k_2^2(\vec{r}) A_2  \nonumber \\
& & + \frac{\omega^2}{c^2} \varepsilon_0(\omega,\vec{r}) A_2 + i
\frac{1}{c^2} \frac{\partial \omega^2
\varepsilon_0(\omega,\vec{r})}{\partial \omega} \frac{\partial
A_2}{\partial t} + \frac{\omega^2}{c^2}
\varepsilon_{\tau}(\omega,\vec{r}) A_1 \nonumber\\
& &+ i \frac{1}{c^2} \frac{\partial \omega^2
\varepsilon_{\tau}(\omega,\vec{r})}{\partial \omega}
\frac{\partial A_1}{\partial t}
] \nonumber \\
& & + \textrm{~conjugated~terms~}
=4 \pi \vec{e} \left( \frac{1}{c^2} \frac{\partial
\vec{j}(\vec{r},t)}{\partial t} + \vec{\nabla} \rho (\vec{r},t)
\right),  \label{vp_3}
\end{eqnarray}
where vector $\vec{k}_2 (\vec{r})=\vec{k}_1 (\vec{r})+
\vec{\tau}$,
$\varepsilon_0(\omega,\vec{r})=1 + {\chi}_{0}(\vec{r})$, here
the notation ${\chi}_{0} (\vec{r})={\chi}_{\tau=0} (\vec{r})$ is used,
$\varepsilon_{\tau}(\omega,\vec{r})={\chi}_{\tau} (\vec{r})$.
Note here that for a numerical analysis of (\ref{vp_3}), if
${\chi}_{0} \ll 0$, it is convenient to take  vector $\vec{k}_1
(\vec{r})$ in the form $\vec{k}_1
(\vec{r})=\vec{n}\sqrt{k^2+\frac{\omega^2}{c^2} \chi_0(\vec{r})}$.

For better understanding, let us suppose that the diffraction
grating (photonic crystal lattice) period changes along one
direction and define this direction as axis $z$.

Thus, for a one-dimensional case, when
$\vec{k}(\vec(r))=(\vec{k}_{\perp},k_z(z))$, the system (\ref{vp_3})
converts to the following:
\begin{eqnarray}
& & \frac{1}{2} e^{i(\vec{k}_{\perp} \vec{r}_{\perp} +
\phi_{1z}(z)-\omega t)} [ 2i {k}_{1z}(z) \frac{\partial
A_1}{\partial z} +i \frac{\partial {k}_{1z}(z)}{\partial z}
 A_1  - (k_{\perp}^2+k_{1z}^2({z}) ) A_1  \nonumber \\
& & + \frac{\omega^2}{c^2} \varepsilon_0(\omega,z) A_1 + i
\frac{1}{c^2} \frac{\partial \omega^2
\varepsilon_0(\omega,z)}{\partial \omega} \frac{\partial
A_1}{\partial t} + \frac{\omega^2}{c^2}
\varepsilon_{-\tau}(\omega,z) A_2 \nonumber\\
& &+ i \frac{1}{c^2} \frac{\partial \omega^2
\varepsilon_{-\tau}(\omega,z)}{\partial \omega} \frac{\partial
A_2}{\partial t}
] \nonumber \\
& & + \textrm{~conjugated~terms~}
=4 \pi \vec{e} \left( \frac{1}{c^2} \frac{\partial
\vec{j}(\vec{r},t)}{\partial t} + \vec{\nabla} \rho (\vec{r},t)
\right), \nonumber \\
& & \frac{1}{2} e^{i(\vec{k}_{\perp} \vec{r}_{\perp} +
\phi_{2z}(z)-\omega t)} [ 2i {k}_{2z}(z) \frac{\partial
A_2}{\partial z} +i \frac{\partial {k}_{2z}(z)}{\partial z}
 A_2  - (k_{\perp}^2+k_{2z}^2({z}) ) A_2  \nonumber \\
& & + \frac{\omega^2}{c^2} \varepsilon_0(\omega,z) A_2 + i
\frac{1}{c^2} \frac{\partial \omega^2
\varepsilon_0(\omega,z)}{\partial \omega} \frac{\partial
A_2}{\partial t} + \frac{\omega^2}{c^2}
\varepsilon_{\tau}(\omega,z) A_1 \nonumber\\
& &+ i \frac{1}{c^2} \frac{\partial \omega^2
\varepsilon_{\tau}(\omega,z)}{\partial \omega} \frac{\partial
A_1}{\partial t}
] \nonumber \\
& & + \textrm{~conjugated~terms~}
=4 \pi \vec{e} \left( \frac{1}{c^2} \frac{\partial
\vec{j}(\vec{r},t)}{\partial t} + \vec{\nabla} \rho (\vec{r},t)
\right),   \label{vp_4}
\end{eqnarray}

Let us multiply the first equation by $e^{-i(\vec{k}_{\perp}
\vec{r}_{\perp} + \phi_{1z}(z)-\omega t)}$ and the second by
$e^{-i(\vec{k}_{\perp} \vec{r}_{\perp} + \phi_{2z}(z)-\omega t)}$.
This procedure enables neglecting the conjugated terms, which
appear fast oscillating (when averaging over the oscillation
period they become zero).

Considering the right-hand side of (\ref{vp_4}), let us take into account
that microscopic currents and densities are the sums of terms,
containing delta-functions, therefore, the right-hand side can be
rewritten as:
\begin{eqnarray}
& & e^{-i(\vec{k}_{\perp} \vec{r}_{\perp} + \phi_{1z}(z)-\omega
t)} 4 \pi \vec{e} \left( \frac{1}{c^2} \frac{\partial
\vec{j}(\vec{r},t)}{\partial t} + \vec{\nabla} \rho (\vec{r},t)
\right) \\
& &= - \frac{4 \pi i \omega e}{c^2}  \vec{e} \sum_{\alpha} \vec
{v}_{\alpha} (t) \delta(\vec(r) -\vec(r)_{\alpha}(t))
e^{-i(\vec{k}_{\perp} \vec{r}_{\perp} + \phi_{1z}(z)-\omega t)} \,
\theta (t-t_{\alpha}) \, \theta (T_{\alpha}-t) \nonumber
\label{vp_5}
\end{eqnarray}

here $t_\alpha$ is the time of entrance of particle $\alpha$  to
the resonator, $T_\alpha$ is the time of particle leaving the
resonator, $\theta-$functions in (\ref{vp_5}) indicate that for
the time moments preceding $t_{\alpha}$ and following
$T_{\alpha}$, the particle  ${\alpha}$ does not contribute to the
process.

Let us suppose now that a strong magnetic field is applied for
beam guiding through the generation area.
Thus, the problem appears one-dimensional (components $v_x$ and
$v_y$ are suppressed).
Averaging the right-hand side of (\ref{vp_5}) over the particle
positions inside the beam, points of particle entrance to the
resonator $r_{\perp 0 \alpha}$ and time of particle entrance to
the resonator $t_\alpha$ one can obtain:
\begin{eqnarray}
& & e^{-i(\vec{k}_{\perp} \vec{r}_{\perp} + \phi_{1z}(z)-\omega
t)} 4 \pi \vec{e} \left( \frac{1}{c^2} \frac{\partial
\vec{j}(\vec{r},t)}{\partial t} + \vec{\nabla} \rho (\vec{r},t)
\right) \nonumber \\
& & = -\frac{4 \pi i \omega \rho \, \vartheta_1 \, u(t) \, e}{c^2}
\frac{1}{S} \int d^2 \vec{r}_{\perp 0} \frac{1}{T} \int_0^t e^{-i(
\phi_{1}(\vec{r},\vec{r}_{\perp},t,t_0) + \vec{k}_{\perp}
\vec{r}_{\perp  0} -\omega t)} dt_0 \nonumber \\
& & = -\frac{4 \pi i \omega \rho\,  \vartheta_1 \, u(t) \, e}{c^2}
\langle\langle e^{-i( \phi_{1}(\vec{r},\vec{r}_{\perp},t,t_0) +
\vec{k}_{\perp} \vec{r}_{\perp  0} -\omega t)} dt_0
\rangle\rangle,  \label{vp_6}
\end{eqnarray}
where $\rho$ is the electron beam density, $u(t)$ is the mean
electron beam velocity, which depends on time due to energy
losses, $\vartheta_1=\sqrt{1 - \frac{\omega^2}{\beta^2 k_1^2
c^2}}$, $\beta^2=1-\frac{1}{\gamma^2}$,
$\langle\langle~~\rangle\rangle$ indicates averaging over the
transversal coordinate of the point of particle entrance to the
resonator $r_{\perp 0 \alpha}$ and time of particle entrance to
the resonator~$t_\alpha$.

According to \cite{vfel_Batrakov+Sytova}, the averaging procedure in
(\ref{vp_6}) can be simplified, when consider that random phases,
appearing due to random transversal coordinate and time of
entrance, presents in (\ref{vp_6}) as differences.
Therefore, double integration over $d^2 \vec{r}_{\perp 0} \, d
t_0$ can be replaced by single integration
\cite{vfel_Batrakov+Sytova}.

The system (\ref{vp_4}) in this case converts to:
\begin{eqnarray}
& & 2i {k}_{1z}(z) \frac{\partial A_1}{\partial z} +i
\frac{\partial {k}_{1z}(z)}{\partial z}
 A_1  - (k_{\perp}^2+k_{1z}^2({z}) ) A_1  \nonumber  \\
& & + \frac{\omega^2}{c^2} \varepsilon_0(\omega,z) A_1 + i
\frac{1}{c^2} \frac{\partial \omega^2
\varepsilon_0(\omega,z)}{\partial \omega} \frac{\partial
A_1}{\partial t} + \frac{\omega^2}{c^2}
\varepsilon_{-\tau}(\omega,z) A_2  \nonumber \\
& & + i \frac{1}{c^2} \frac{\partial \omega^2
\varepsilon_{-\tau}(\omega,z)}{\partial \omega}
\frac{\partial A_2}{\partial t}= i \frac{2 \omega}{c^2} J_{1} (k_{1z}(z)), \\
& & 2i {k}_{2z}(z) \frac{\partial A_2}{\partial z} +i
\frac{\partial {k}_{2z}(z)}{\partial z}
 A_2  - (k_{\perp}^2+k_{2z}^2({z}) ) A_2  \nonumber \\
& & + \frac{\omega^2}{c^2} \varepsilon_0(\omega,z) A_2 + i
\frac{1}{c^2} \frac{\partial \omega^2
\varepsilon_0(\omega,z)}{\partial \omega} \frac{\partial
A_2}{\partial t} + \frac{\omega^2}{c^2}
\varepsilon_{\tau}(\omega,z) A_1  \nonumber \\
& & + i \frac{1}{c^2} \frac{\partial \omega^2
\varepsilon_{\tau}(\omega,z)}{\partial \omega} \frac{\partial
A_1}{\partial t} = i \frac{2 \omega}{c^2} J_{2} (k_{2z}(z)) ,
\nonumber
\label{vp_sys2}
\end{eqnarray}
where the currents $J_1$, $J_2$ are determined by the expression
\begin{eqnarray}
J_m=2 \pi j \vartheta_m \int_0^{2 \pi} ~ \frac{2 \pi-p}{8
\pi^2}(e^{-i \phi_m(t,z,p)}+ e^{-i \phi_m(t,z,-p)})~dp, ~~ m=1,2\nonumber\\
\label{vp_current}
\end{eqnarray}
\[
\vartheta_m = \sqrt{1 - \frac{\omega^2}{\beta^2 k_m^2 c^2}}~,~
\beta^2=1-\frac{1}{\gamma^2}~,
\]
$j=en_0 v$ is the current density, $A_1 \equiv A_{\tau=0}$, $A_2
\equiv A_{\tau}$, $\vec{k}_1 = \vec{k}_{\tau=0}$, $\vec{k}_2 =
\vec{k}_1 +\vec{\tau}$.
The expressions for $J_1$ and $k_1$ independent on $z$ was
obtained in \cite{vfel_Batrakov+Sytova}.

When more than two waves participate in the diffraction process, the
system (\ref{vp_sys2}) should be supplemented with equations for
waves $A_m$, which are similar to those for $A_1$ and $A_2$.

Now we can find the equation for the phase. From (\ref{vp_phi1}),
(\ref{vp_phi2}) follows that
\begin{eqnarray}
\frac{d^2 \phi_m}{dz^2} + \frac{1}{v}\frac{d v}{dz}\frac{d
\phi_m}{dz} =  \frac{d k_{m}}{d z}+ \frac{k_{m}}{v^2} \frac{d^2
z}{dt^2}, \label{vp_phase}
\end{eqnarray}
Let us introduce a new function $C(z)$ as follows:
\begin{eqnarray}
& &\frac{d\phi_m}{dz} = C_m(z) e^{-\int_0^z \frac{1}{v}
\frac{dv}{dz^{\prime}} d z^{\prime}} = \frac{v_0}{v(z)}
C_m(z),~~\\
& &\phi_m(z)=\phi_m (0)+\int_0^z \frac{v_0}{ v(z^{\prime})}
C_m(z^{\prime}) d z^{\prime} \nonumber \label{vp_cz}
\end{eqnarray}

Therefore,
\begin{eqnarray}
\frac{d C_m(z)}{dz} = \frac{v(z)}{v_0} \left( \frac{d k_m}{d z}+
\frac{k_m}{v^2} \frac{d^2 z}{dt^2} \right) . \label{vp_cz1}
\end{eqnarray}

In a one-dimensional case,  equation (\ref{vp_sys11}) can be
written as:
\begin{eqnarray}
\frac{d^2 z_\alpha}{dt^2}= \frac{e \vartheta}{m
\gamma(z_{\alpha},t,p)} \textrm{Re} E(z_{\alpha},t),
\end{eqnarray}
therefore,
\begin{equation}
\frac{d C_m(z)}{dz} = \frac{v(z)}{v_0} \frac{d k_m}{d z}+
\frac{k_m}{v_0 v(z)} \frac{e \vartheta_m}{m \gamma^3(z,t(z),p)} Re
\{ A_m (z,t(z))  e^{i \phi_m(z,t(z),p)} \}, \label{vp_cz2}
\end{equation}

\[
\frac{d \phi_m (t,z,p)}{dz}|_{z=0} = k_{mz} - \frac{\omega}{v},~
\phi_m (t,z,p)|_{z=0} = p,
\]
\[
A_1 |_{z=L} = E_1^0,~A_2 |_{z=L} = E_2^0,~
\]
\[
A_m |_{t=0} = 0, ~ m=1,2,
\]
\[
t>0,~z\in [0,L],~ p \in [-2 \pi, 2 \pi], ~L ~\textrm{is the length
of the photonic crystal}.
\]

These equations should be supplied with the equations for
$\gamma(z,p)$.
It is well-known that
\begin{eqnarray}
mc^2 \frac{d \gamma}{dt} = e \vec{v} \vec{E}. \label{vp_gamma}
\end{eqnarray}
Therefore,
\begin{eqnarray}
\frac{d \gamma(z,t(z),p)}{dz} = \sum_l \frac{e \vartheta_l}{ m
c^2} \textrm{Re} \{ \sum_l A_l (z,t(z)) e^{i \phi_l (z,t(z),p)}
\}. \label{vp_gamma1}
\end{eqnarray}

The above obtained equations (\ref{vp_sys2}),
(\ref{vp_cz}), (\ref{vp_cz2}), (\ref{vp_gamma1}) enable describing
the generation process in a FEL with varied parameters of diffraction
grating (photonic crystal).
The analysis of the system (\ref{vp_cz2}) can be simplified by
replacement of the $\gamma(z,t(z),p)$ with its averaged by the
initial phase value
\[
<\gamma (z,t(z))>=\frac{1}{2 \pi} \int_0^{2 \pi} \gamma(z,t(z),p)
\, d p.
\]
Note that the law of parameters change can be both smooth and
stair-step.

\begin{figure}[tbp]
\epsfxsize = 10 cm \centerline{\epsfbox{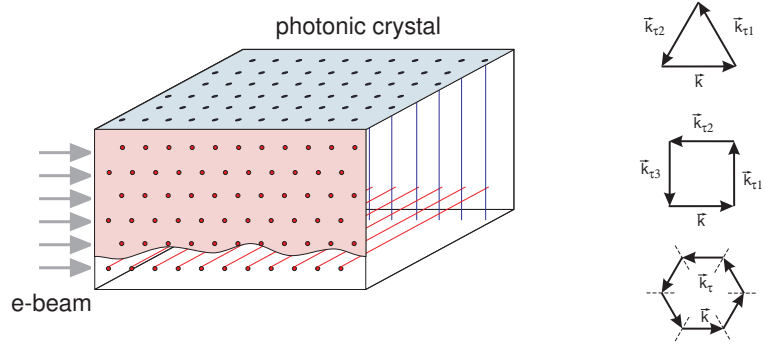}} \caption{An
example of photonic crystal with the thread arrangement providing
multi-wave volume distributed feedback. Threads are arranged to
couple several waves (three, four, six and so on), which appear
due to diffraction in such a structure, in both the vertical and
horizontal planes. The electronic beam takes the whole volume of
photonic crystal.} \label{vp_volume}
\end{figure}

Using photonic crystals provide the development of different VFEL
arrangements (see Figure \ref{vp_volume}).

It should be noted that, for example, in the FEL (TWT, BWO)
resonator with changing in space parameters of grating (photonic 
crystal), the electromagnetic wave with a spatial
period depending on $z$  is formed.
This means that the dynamical undulator with a period depending on
$z$ appears along the whole resonator length, i.e., a tapering
dynamical wiggler becomes settled.
It is well known that a tapering wiggler can significantly increase
the efficiency of the undulator FEL.
The dynamical wiggler with varied period, which is proposed, can
be used for development of double-cascaded FEL with parameters
changing in space.
The efficiency of such a system can be significantly higher that of
a conventional system.
Moreover, the period of a dynamical wiggler can be made much shorter
than that available for wigglers using static magnetic fields.
%

It should also be noted that the compression of the radiation pulse in such a system is possible because the phase
velocity of the electromagnetic wave depends on time.

Thus, the equations providing the description of the generation process in FEL with
varied parameters of diffraction grating (photonic crystal) are
obtained.
It is shown that applying diffraction gratings (photonic crystal)
with the variable period one can significantly increase radiation
output.
It is mentioned that diffraction gratings (photonic crystal) can
be used for creation of the dynamical wiggler with variable period
in the system.
This makes possible to develop double-cascaded FEL with variable
parameters changing, which efficiency can be significantly higher
that of conventional system.

In conclusion, it should be noted that  in the photonic crystals
built from  metal (or dielectric) threads, waves are effectively
diffracted even when their lengths are much smaller than the
distance between the threads. That is why it is possible to
generate radiation in the terahertz range using  diffraction
gratings with a millimeter period alone.

Above, we have studied the features of generation for the sources
with transverse dimensions much greater than the radiation
wavelength. In the case when the finite dimensions of a resonator
should be taken into account, it is necessary to follow the rules
given in
\cite{vfel_VFELreview,rins_66,lanl_13,lanl_22}.

\subsection{Radiative instability of a relativistic electron beam
moving in a finite-dimensional photonic (or natural) crystal}

According to the analysis given in \cite{lanl_22},
%
%
%
%
%
the first and most important step in describing the generation
process in VFELs (FELs and so on) is the analysis of the problem
of the electron beam instability in the resonator. The theoretical
study of the instability of electron beams moving in natural and
artificial (photonic) crystals was carried out above for the ideal
case of an infinite medium (see the review in
\cite{vfel_VFELreview} and
\cite{berk_133,vfel_Batrakov+Sytova,rins_66,lanl_13,hyb_FirstLasing}).
The question arising in this regard is how the finite dimensions
of the photonic crystal placed inside the resonator affect the law
of electron beam instability. It is known, for example, that the
discrete structure of the modes in waveguides and resonators is
crucial for effective generation in the microwave range
\cite{vfel_Leavitt3a,vfel_Leavitt3b,L_2,L_4,L_5}.

In present section the radiative instability of a beam moving in a
photonic crystal is studied. The dispersion equation describing
instability in this case is obtained. It is shown that the law
$\Gamma\sim\rho ^{1/\left( {s + 3} \right)}$ is also valid and
caused by the mixing of the electromagnetic field modes in the
finite volume due to the periodic disturbance from the photonic
crystal.

The system of equations describing generation of induced radiation
in photonic (and natural) crystals  can be obtained similarly to
those in section \ref{vp_sec:2} (equations
(\ref{vp_sys0})-(\ref{vp_1})).

In the general case, the susceptibility of the photonic crystal
reads
 $\hat {\chi} \left( {\vec {r}} \right) = \sum\limits_{i} {\hat {\chi
}_{cell} \left( {\vec {r} - \vec {r}_{i}}  \right)} ,$ where $\hat
{\chi }_{cell} \left( {\vec {r} - \vec {r}_{i}}  \right)$ is the
susceptibility of the crystal unit cell. The susceptibility of an
infinite perfect crystal $\hat {\chi} \left( {\vec {r}} \right)$
can be expanded into the Fourier series as follows: $\hat {\chi}
\left( {\vec {r}} \right) = \sum\limits_{\vec {\tau} } {\hat
{\chi} _{\vec {\tau} } e^{i\vec {\tau} \vec {r}}} ,$ where $\vec
{\tau} $ is the reciprocal lattice vector of the crystal.

To be more specific, let us consider in details a practically
important case when a photonic crystal is placed inside a smooth
 waveguide of rectangular cross-section.

The eigenfunctions and the eigenvalues of such a waveguide are
well-studied \cite{WIND_Jackson-a,L_15}.

Suppose the $z$-axis to be directed along the waveguide axis. Make
the Fourier transform of (\ref{vp_sys01}) over time and
longitudinal coordinate $z$ . Expanding thus obtained equation for
the field $\vec {E}\left( {\vec {r}_{ \bot}  ,k_{z} ,\omega}
\right)$ over a full set of vector eigenfunctions of a rectangular
waveguide $\vec {Y}^{\lambda }_{mn} \left( {\vec {r}_{ \bot}
,k_{z}} \right)$ (where $m,n = 1,2,3....$, while $\lambda $
describes the type of the wave \cite{66},  one can obtain for the
field $\vec {E}$ the equality

\begin{equation}
\label{eq7} \vec {E}\left( {\vec {r}_{ \bot}  ,k_{z} ,\omega}
\right) = \sum\limits_{mn\lambda}  {C^{\lambda} _{mn} \left(
{k_{z} ,\omega} \right)\vec {Y}^{\lambda} _{mn}}  \left( {\vec
{r}_{ \bot}  ,k_{z}} \right).
\end{equation}

As a result, the following equation can be written

\begin{equation}
\label{eq8}
\begin{array}{l}
 \left[ {\left( {k_{z} ^{2} + \kappa _{mn\lambda}  ^{2}} \right) -
\frac{{\omega ^{2}}}{{c^{2}}}} \right]C^{\lambda} _{mn} \left(
{k_{z}
,\omega}  \right) - \\
 - \frac{{\omega ^{2}}}{{c^{2}}}\,\frac{{1}}{{2\pi
}}\sum\limits_{m'n'\lambda '} {\int {\vec {Y}^{\lambda ^{\ast}
}_{mn} \left( {\vec {r}_{ \bot}  ,k_{z}}  \right)\hat {\chi}
\left( {\vec {r}} \right)\vec {Y}^{\lambda '}_{m'n'} \left( {\vec
{r}_{ \bot}  ,k'_{z}}  \right)e^{ - i\left( {k_{z} - k'_{z}}
\right)z}} \,} d^{2}r_{ \bot}  C_{m'n'}^{\lambda
'} \left( {k'_{z} ,\omega}  \right)dk'_{z} dz = \nonumber \\
 = \frac{{4\pi i\omega} }{{c^{2}}}\int {\vec {Y}^{\lambda ^{\ast} }_{mn}
\left( {\vec {r}_{ \bot}  ,k_{z}}  \right)\left\{ {\vec {j}\left(
{\vec {r}_{ \bot}  ,z,\omega}  \right) + \frac{{c^{2}}}{{\omega
^{2}}}\vec {\nabla }\left( {\vec {\nabla} \vec {j}\left( {\vec
{r}_{ \bot}  ,z,\omega}
\right)} \right)} \right\}e^{ - ik_{z} z}} d^{2}r_{ \bot}  dz\nonumber \\
 \end{array}
\end{equation}
\noindent where $\kappa _{mn\lambda}  ^{2} = k_{xm\lambda} ^{2} +
k_{yn\lambda} ^{2} $.

The beam current and density appearing on the right-hand side of
(\ref{eq8}) are complicated functions of the field $\vec {E}$. To
study the problem of the system instability, it is sufficient to
consider the system in the approximation linear  over
perturbation, i.e., one can expand the expressions for $\vec {j}$
and $\rho $ over the field amplitude $\vec {E}$ and abridge
oneself with the linear approximation.

As a result, a closed system of equations comes out. For further
consideration, one should obtain the expressions for the
corrections $\delta \vec {j}$ and $\delta  \rho $ due to beam
perturbation by the field. Considering the Fourier transforms of
the current density and the beam charge $\vec {j}\left( {\vec
{k},\omega} \right)$ and $\rho \left( {\vec {k},\omega} \right)$,
 one can obtain from (\ref{eq2}) that

\begin{equation}
\delta \vec {j}\left( {\vec {k},\omega}  \right) =
e\sum\limits_{\alpha = 1}^{N} {e^{ - i\vec {k}\vec {r}_{\alpha
_{0}} } } \left\{ {\delta \vec {v}_{\alpha}  \left( {\omega - \vec
{k}\vec {u}_{\alpha} }  \right) + \vec {u}_{\alpha}  \frac{{\vec
{k}\delta \vec {v}_{\alpha}  \left( {\omega - \vec {k}\vec
{u}_{\alpha} }  \right)}}{{\omega - \vec {k}\vec {u}_{\alpha} } }}
\right\},
\end{equation}

\noindent where $\vec {r}_{\alpha _{0}}  $ is the original
coordinate of the electron, $\vec {u}_{\alpha}  $ is the
unperturbed velocity of the electron.

For simplicity, let us consider a cold beam, for which $\vec
{u}_{\alpha}  \approx \vec {u}$, where $\vec {u}$ is the mean
velocity of the beam. The general case of a hot beam is obtained
by averaging $\delta \vec {j}\left({\vec {k},\omega}  \right)$
over the velocity $\vec {u}_{\alpha}$ distribution in the beam.

According to (\ref{vp_sys1}), the velocity $\delta \vec
{v}_{\alpha} $ is determined by the field $\vec {E}\left( {\vec
{r}_{\alpha} ,\omega}  \right)$ taken at the electron location
point  $\vec {r}_{\alpha}$. The Fourier transform of the field
$\vec {E}\left( {\vec {r}_{\alpha}  ,\omega}  \right)$ has a form
\[
\vec {E}\left( {\vec {r}_{\alpha}  ,\omega}  \right) =
\frac{{1}}{{\left( {2\pi}  \right)^{3}}}\int {\vec {E}\left( {\vec
{k}',\omega} \right)e^{i\vec {k}'\vec {r}_{\alpha} } d^{3}k'}.
\]
As a result, the formula for $\delta \vec {j}\left( {\vec
{k},\omega} \right)$ includes the sum $\sum\limits_{\alpha}  {e^{
- i\left( {\vec {k} - \vec {k}'} \right)\,\vec {r}_{\alpha} } } $
over the particle distribution in the beam. Suppose that the
electrons in an unperturbed beam are uniformly distributed over
the area occupied by the beam. Therefore
\[
\sum\limits_{\alpha}  {e^{ - i\left( {\vec {k} - \vec {k}'}
\right)\,\vec {r}_{\alpha} } } = \left( {2\pi}  \right)^{3}\rho
_{0} \,\delta \left( {\vec {k} - \vec {k}'} \right),
\]
\noindent where $\rho _{0} $ is the beam density (the number of
electrons per 1 cm$^{3}$).

As a result, the following expression for $\delta \vec {j}\left(
{\vec {k},\omega} \right)$ can be obtained
\cite{berk_135,berk_144}:

\begin{equation}
\label{eq9} \delta \vec {j}\left( {\vec {k},\omega}  \right) =
\frac{{i\vec {u}e^{2}\rho \left( {k^{2} - \frac{{\omega
^{2}}}{{c^{2}}}} \right)}}{{\left( {\omega - \vec {k}\vec {u}}
\right)^{2}m\gamma \omega} }\vec {u}\vec {E}\left( {\vec
{k},\omega}  \right).
\end{equation}

Using the continuity equation, one immediately obtains the
expression for $\rho \left( {\vec {k},\omega}  \right)$.
Expression (\ref{eq9}), the inverse Fourier transform of $\vec
{E}\left( {\vec {k},\omega}  \right)$, and the expansion
(\ref{eq7}) enable writing the system of equations (\ref{eq8}) as
follows:
\begin{equation}
\label{eq10}
\begin{array}{l}
 \left[ {\left( {k_{z} ^{2} + \kappa _{mn\lambda}  ^{2}} \right) -
\frac{{\omega ^{2}}}{{c^{2}}}} \right]C^{\lambda} _{mn} \left(
{k_{z}
,\omega}  \right) - \\
 - \frac{{\omega ^{2}}}{{c^{2}}}\,\frac{{1}}{{2\pi
}}\sum\limits_{m'n'\lambda '} {\int {\vec {Y}^{\lambda ^{\ast}
}_{mn} \left( {\vec {r}_{ \bot}  ,k_{z}}  \right)\hat {\chi}
\left( {\vec {r}} \right)\vec {Y}^{\lambda '}_{m'n'} \left( {\vec
{r}_{ \bot}  ,k'_{z}}  \right)e^{ - i\left( {k_{z} - k'_{z}}
\right)z}} \,} d^{2}r_{ \bot}  C_{m'n'}^{\lambda
'} \left( {k'_{z} ,\omega}  \right)dk'_{z} dz = \\
 = - \frac{{\omega _{L}^{2} \left( {k_{mn}^{2} c^{2} - \omega ^{2}}
\right)}}{{\gamma c^{4}\left( {\omega - \vec {k}_{mn} \vec {u}}
\right)^{2}}}\left\{ {\frac{{1}}{{2\pi} }\left| {\int {\vec
{u}\,\vec {Y}^{\lambda} _{mn} \left( {\vec {k}_{ \bot}  ,k_{z}}
\right)d^{2}k_{ \bot }} }  \right|^{2}} \right\}C^{\lambda} _{mn}
\left( {k_{z} ,\omega}
\right), \\
 \end{array}
\end{equation}

\noindent where $\vec {Y}^{\lambda} _{mn} \left( {\vec {k}_{ \bot}
,k_{z}}  \right) = \int {e^{ - i\vec {k}_{ \bot}  \vec {r}_{ \bot}
} \vec {Y}_{mn}^{\lambda} } \left( {\vec {r}_{ \bot}  ,k_{z}}
\right)d^{2}r_{ \bot}  $.

Note that within the limit where the transverse dimensions of a
photonic crystal tend to infinity, the expression between the
braces takes the form $\left( {\vec {e}\vec {u}} \right)^{2}$,
where $\vec{e}$  is the unit polarization vector of the wave
emitted by the beam \cite{berk_135,berk_144}.

Now let us consider the integrals on the left-hand side of
equation (\ref{eq10}). Note that according to
\cite{66,WIND_Jackson-a,L_15}, the eigenfunctions $\vec
{Y}_{mn}^{\lambda} \left({\vec {r}_{ \bot} ,k_{z}}  \right)$ of a
rectangular waveguide  include the combinations of sines and
cosines of the form $sin\frac{{\pi m}}{{a}}x,cos\frac{{\pi
m}}{{a}}x$ ($sin\frac{{\pi n}}{{b}}y,cos\frac{{\pi n}}{{b}}y$),
i.e., in fact, the combinations $e^{i\frac{{\pi
m}}{{a}}x},e^{i\frac{{\pi n}}{{b}}y}$. Hence, the left-hand side
of the equation includes the integrals of the type

\[
I = \int {e^{ - i\frac{{\pi m}}{{a}}x}} \sum\limits_{i} {\hat
{\chi} _{cell} } \left( {x - x_{i} ,y - y_{i} ,z - z_{i}}
\right)e^{i\frac{{\pi m'}}{{a}}x}dx.
\]

The substitution of variables $x - x_{i} = \eta $ gives the sums
of the form

\[
S_{x} = \sum\limits_{i} {e^{ - i\frac{{\pi} }{{a}}\left( {m - m'}
\right)x_{i}} }
\]

\noindent where $x_{i} = d_{x} f_{1} $, where $d_{x} $ is the
period of the photonic crystal along the $x$-axis, $f_{1} =
1,2,...N_{x} $, where $N_{x} $ is the number of cells along the
$x$-axis.

The above-mentioned sum

\begin{equation}
\label{eq11} S_{x} = \sum\limits_{i} {e^{ - i\frac{{\pi}
}{{a}}\left( {m - m'} \right)x_{i}} } = e^{i\frac{{\pi}
}{{2a}}\left( {m - m'} \right)\left( {N_{x} - 1} \right)d_{x}}
\frac{{sin  \frac{{\pi \left( {m - m'} \right)d_{x} N_{x} }}{{2a}}
} }{{sin\frac{{\pi \left( {m - m'} \right)d_{x}} }{{2a}}}} .
\end{equation}

If $m - m' = 0$, then $S_{x} = N_{x} $.

 Let us now discuss what this sum is equal to when $m - m' = 1$. In the
 numerator $d_{x} N_{x} = a$,
 hence, the nominator {is equal to} 1 ($sin\frac{{\pi} }{{2}} = 1$), while in the denominator $sin\frac{{\pi
d_{x}} }{{2a}} \approx \frac{{\pi} }{{2N_{x}} }$. As a result, the
relation $\frac{{S_{x} \left( {m - m' = 1} \right)}}{{S_{x} \left(
{m - m' = 0} \right)}} = \frac{{2}}{{\pi} } \approx 0.6$.

With growing difference $m - m'$, the contribution to the sum of
the next terms diminishes until the following equality is
fulfilled
\begin{equation}
\label{eq12} \frac{{\pi \left( {m - m'} \right)d_{x}} }{{2a}} =
\pi {\rm P},
\end{equation}

\noindent where ${\rm P} = \pm 1, \pm 2...$ In this case the sum
$S_{x} = N_{x} $.

The similar reasoning is valid for summation along the $y$-axis.

It follows from the aforesaid that if the equalities like
(\ref{eq11}), (\ref{eq12}) are fulfilled, that is, the equalities
$k_{xm} - k'_{xm'} = \tau _{x} $ are fulfilled (i.e., $k'_{xm'} =
k_{xm} - \tau _{x} $), where $\tau _{x} = \frac{{2\pi} }{{d_{x}}
}F$ is the $x$-component of the reciprocal lattice vector of the
photonic crystal,  $F = 0, \pm 1, \pm 2...$ and $k_{yn} - k'_{yn'}
= \tau _{y} $ (i.е., $k'_{yn'} = k_{yn} - \tau _{y} $), where
$\tau _{y} = \frac{{2\pi} }{{d_{y}} }F'$ is the $y$-component of
the reciprocal lattice vector of the photonic crystal,  $F' = 0,
\pm 1, \pm2...$), then the major contribution to the sums comes
from the amplitudes $C_{m'n'}^{\lambda '} \left( {k'_{z} ,\omega}
\right) \equiv C^{\lambda '}\left( {\vec {k}_{ \bot mn} - \vec
{\tau} _{ \bot}  ,k_{z} - \tau _{z} ,\omega}  \right) = C^{\lambda
'}\left( {\vec {k}_{mn} - \vec {\tau} ,\omega}  \right)$.

In describing the system we shall further consider only those
modes that satisfy the equalities of the type (\ref{eq11}),
(\ref{eq12}). As stated above, the contribution of other modes is
suppressed.

As a result, one can rewrite the system of equations (\ref{eq10})
as

\begin{equation}
\label{eq13}
\begin{array}{l}
 \left( {\vec {k}_{mn}^{2} - \frac{{\omega ^{2}}}{{c^{2}}}}
\right)C^{\lambda} \left( {\vec {k}_{mn} ,\omega}  \right) -
\frac{{\omega ^{2}}}{{c^{2}}}\sum\limits_{\lambda '\tau}  {\chi
_{mn}^{\lambda \lambda '} \left( {\vec {\tau} } \right)}
C^{\lambda '}\left( {\vec {k}_{mn} - \vec
{\tau} ,\omega}  \right) = \\
 - \frac{{\omega _{L}^{2} \left( {k_{mn}^{2} c^{2} - \omega ^{2}}
\right)}}{{\gamma c^{4}\left( {\omega - \vec {k}_{mn} \vec {u}}
\right)^{2}}}\left\{ {\frac{{1}}{{2\pi} }\left| {\int {\vec
{u}\,\vec {Y}^{\lambda} _{mn} \left( {\vec {k}_{ \bot}  ,k_{z}}
\right)d^{2}k_{ \bot }} }  \right|^{2}} \right\}C^{\lambda} \left(
{\vec {k}_{mn} ,\omega}
\right), \\
 \end{array}
\end{equation}

\noindent i.e.,

\begin{equation}
\label{eq14}
\begin{array}{l}
 \left( {\vec {k}_{mn}^{2} - \frac{{\omega ^{2}}}{{c^{2}}}\left( {1 + \chi
_{mn}^{\lambda \lambda}  \left( {0} \right) - \frac{{\omega
_{L}^{2} \left( {k_{mn}^{2} c^{2} - \omega ^{2}} \right)}}{{\omega
^{2}\gamma c^{2}\left( {\omega - \vec {k}_{mn} \vec {u}}
\right)^{2}}}\left\{ {\frac{{1}}{{2\pi }}\left| {\int {\vec
{u}\,\vec {Y}^{\lambda} _{mn} \left( {\vec {k}_{ \bot} ,k_{z}}
\right)d^{2}k_{ \bot} } }  \right|^{2}} \right\}} \right)}
\right)C^{\lambda} \left( {\vec {k}_{mn} ,\omega}  \right) \\
 - \frac{{\omega ^{2}}}{{c^{2}}}\sum\limits_{\lambda '\tau}  {\chi
_{mn}^{\lambda \lambda '} \left( {\vec {\tau} } \right)}
C^{\lambda '}\left(
{\vec {k} - \vec {\tau} ,\omega}  \right) = 0 \\
 \end{array}
\end{equation}

\noindent where $\chi _{mn}^{\lambda \lambda '} \left( {\tau}
\right) = \frac{{1}}{{d_{z}} }\int {\vec {Y}^{\lambda \ast} _{mn}
\left( {\vec {r}_{ \bot}  ,k_{z}}  \right)\hat {\chi} \left( {\vec
{r}_{ \bot}  ,\tau _{z}} \right)} \vec {Y}_{m'n'}^{\lambda '}
\left( {\vec {r}_{ \bot}  ,k_{z} - \tau _{z}}  \right)d^{2}r_{
\bot}  $, $\hat {\chi} \left( {\vec {r}_{ \bot} ,\tau _{z}}
\right) = \sum\limits_{x_{i} ,y_{i}}  {\int {\hat {\chi }_{cell}
\left( {x - x_{i} ,y - y_{i} ,\zeta}  \right)}}  \,e^{ - i\tau
_{z} \zeta} d\zeta $, $m'$ and  $n'$ are found by the conditions
like (\ref{eq12}), $\omega _{L} $ is the Langmuir frequency,
$\omega _{_{L}} ^{2} = \frac{{4\pi e^{2}\rho _{0} }}{{m}}$.

This system of equations coincides in form with that describing
the instability of a beam passing through an infinite crystal
\cite{berk_135,berk_144}. The difference between them is that the
coefficients appearing in these equations are defined differently
and that in the case of an infinite crystal, the wave vectors
$\vec {k}_{mn} $ have a continuous spectrum of eigenvalues rather
than a discrete one.

These equations enable one to define the dependence $k\left(
{\omega} \right)$, thus defining the expressions for the waves
propagating in the crystal. By matching the incident wave packet
and the set of waves propagating inside the photonic crystal using
the boundary conditions, one can obtain the explicit expression
describing the solution of the considered equations.

The result obtained is formally analogous to that given in
\cite{berk_139}.

According to (\ref{eq14}), the expression between the square
brackets acts as the dielectric permittivity $\varepsilon $ of the
crystal under the conditions when diffraction can be neglected:
\[\varepsilon _{0} = n^{2} = 1 + \chi _{mn}^{\lambda \lambda}
\left( {0} \right) - \frac{{\omega _{L}^{2} \left( {k_{mn}^{2}
c^{2} - \omega ^{2}} \right)}}{{\omega ^{2}\gamma c^{2}\left(
{\omega - \vec {k}_{mn} \vec {u}} \right)^{2}}}\left\{
{\frac{{1}}{{2\pi} }\left| {\int {\vec {u}\,\vec {Y}^{\lambda}
_{mn} \left( {\vec {k}_{ \bot}  ,k_{z}} \right)d^{2}k_{ \bot} } }
\right|^{2}} \right\},
\]
 $n $  is the refractive index.

As is seen, in this case the contribution to the refractive index
comes
 not only from the scattering of waves by the unit cell of the crystal
 lattice, but also from the scattering of waves by the beam electrons
 (the term proportional to $\omega _{L}^{2} $): the photonic crystal
 penetrated by a beam of electrons is a medium that can be described
  by a ceratin refractive index $n$ (or the dielectric permittivity
  $\varepsilon _{0} $).

According to (\ref{eq14}), the beam contribution increases when
$\omega \to \vec {k}\vec {u}$.

Since this system of equations is homogeneous, its solvability
condition is the vanishing of the system determinant.

In the beginning, let us assume that the diffraction conditions
are not fulfilled. Then the amplitudes of diffracted waves are
small. In this case the sum over $\tau $ can be dropped, and the
conditions for the occurrence of the wave in the system is
obtained by the requirement that the expression between the square
brackets equal zero.

This expression can be written in the form (the velocity $\vec {u}
|| oz$)

\[
\left( {\omega - k_{z} u} \right)^{2}\left( {k_{mn}^{2} -
\frac{{\omega ^{2}}}{{c^{2}}}n_{0}^{2}}  \right) = - \frac{{\omega
_{L}^{2} \left( {k_{mn}^{2} c^{2} - \omega ^{2}} \right)}}{{\gamma
c^{4}}}\left\{ {\frac{{1}}{{2\pi} }\left| {\int {\vec {u}\,\vec
{Y}^{\lambda} _{mn} \left( {\vec {k}_{ \bot}  ,k_{z}}
\right)d^{2}k_{ \bot} } }  \right|^{2}} \right\},
\]

\noindent where  $n_{0} $ is the refractive index of the photonic
crystal in the absence of the beam $\varepsilon _{0} = n_{0}^{2} =
1 + \chi _{mn}^{\lambda \lambda}  \left( {0} \right)$,

\noindent i.e.,

\begin{equation}
\label{eq15} \left( {k_{z}^{2} - \left( {\frac{{\omega
^{2}}}{{c^{2}}}n_{0}^{2} - \kappa_{mn}^{2}}  \right)}
\right)\left( {\omega - k_{z} u} \right)^{2} = - \frac{{\omega
_{L}^{2} \left( {k_{mn}^{2} c^{2} - \omega ^{2}} \right)}}{{\gamma
c^{4}}}\left\{ {\frac{{1}}{{2\pi} }\left| {\int {\vec {u}\,\vec
{Y}^{\lambda} _{mn} \left( {\vec {k}_{ \bot}  ,k_{z}}
\right)d^{2}k_{ \bot} } }  \right|^{2}} \right\}
\end{equation}

Since the nonlinearity is insignificant, let us consider as the
zero
 approximation the spectrum of the waves of equation (\ref{eq15}) with
  zero right-hand side.

Let us concern with the case when $\omega - k_{z} u \to 0$ (i.e.,
the Cherenkov radiation condition can be fulfilled) and $\left(
{k_{z}^{2} - \left( {\frac{{\omega ^{2}}}{{c^{2}}}n_{0}^{2} -
\kappa _{mn}^{2}}  \right)} \right) \to 0$, i.e, the
electromagnetic wave can propagate in a photonic crystal without
the beam. With zero right-hand side the equation reads

\begin{equation}
\label{eq16} \left( {k_{z}^{2} - \left( {\frac{{\omega
^{2}}}{{c^{2}}}n_{0}^{2} - \kappa _{mn}^{2}}  \right)} \right) =
0, \quad \left( {\omega - k_{z} u} \right) = 0
\end{equation}

As a consequence, in this case the roots of the equation are
\begin{equation}
\label{eq17} k_{1z} = \frac{{\omega} }{{c}}\sqrt {n_{0}^{2} -
\frac{{\kappa _{mn}^{2} c^{2}}}{{\omega ^{2}}}} , \quad k'_{1z} =
- k_{1z} , \quad k_{2z} = \frac{{\omega} }{{u}}.
\end{equation}

Since  $k_{2z} = \frac{{\omega} }{{u}}
> 0$ in view of the Cherenkov condition, we are concerned with the propagation of the
wave with $k_{1z} > 0$ in the photonic crystal. In this case in
the equation for $k_{z} $, one can take $\left( {k_{z} - k_{1z}}
\right) \left( {k_{z} + k_{1z}}  \right) \approx 2k_{1z} \left(
{k_{z} - k_{1z}} \right)$ and rewrite equation (\ref{eq15}) as
follows:

\begin{equation}
\label{eq18} \left( {k_{z} - k_{1z}}  \right)\left( {k_{z} -
k_{2z}}  \right)^{2} = - \frac{{\omega _{L}^{2} \omega ^{2}\left(
{n_{0}^{2} - 1} \right)}}{{2k_{1z} u^{2}\gamma c^{4}}}\left\{
{\frac{{1}}{{2\pi} }\left| {\int {\vec {u}\,\vec {Y}^{\lambda}
_{mn} \left( {\vec {k}_{ \bot}  ,k_{z}}  \right)d^{2}k_{ \bot }} }
\right|^{2}} \right\}
\end{equation}

\noindent i.e.,

\begin{equation}
\label{eq19} \left( {k_{z} - k_{1z}}  \right)\left( {k_{z} -
k_{2z}}  \right)^{2} = - A
\end{equation}

\noindent

where $A$ is real and $A > 0$ (as for the occurrence of the
Cherenkov effect, it is necessary that $n_{0}^{2} > 1$). We have
obtained the  cubic equation for  $k_{z} $. Let us consider the
case when the roots $k_{1z} $ and  $k_{2z} $ coincide $k_{1z} =
k_{2z} $. It is possible when the particle velocity satisfies the
condition

\begin{equation}
\label{eq20} u = \frac{{c}}{{\sqrt {n_{0}^{2} - \frac{{\kappa
_{mn}^{2} c^{2}}}{{\omega ^{2}}}}} }.
\end{equation}

\noindent Introduction of  $\xi = k - k_{1z} $ gives for $k_{1z} =
k_{2z} $

\begin{equation}
\label{eq21} \xi ^{3} = - A.
\end{equation}

\noindent The solution of equation (\ref{eq21}) gives three roots
$\xi _{1} = - \sqrt[{3}]{{A}}$,
 $\xi
_{2,3} = \frac{{1}}{{2}}\left( {1 \pm i\sqrt {3}}  \right)
\sqrt[{3}]{{A}}$.

As a consequence, the state corresponding to the root $\xi _{2} =
\frac{{1}}{{2}}\left( {1 + i\sqrt {3}}  \right)\sqrt[{3}]{{A}}$
grows with growing $z$, which indicates the presence of
instability in a beam \cite{L_21}. In this case $Im\,k_{z} =
Im\,\xi _{2}\sim\sqrt[{3}]{{\rho} }$.

Note here that the photonic crystal built from metallic threads
has the refractive index $n_{0} < 1$ for a wave with the electric
polarizability parallel to the threads, i.e., in this case the
Cherenkov instability of the beam does not exist
\cite{vfel_VFELreview} (but if the electric vector of the wave is
orthogonal to the metallic threads, the refractive index is $n_{0}
> 1$ , so for such a wave the Cherenkov instability exists
\cite{lanl_23}).

It should be pointed out, however, that, unlike an infinite
photonic crystal, the field in the crystal placed into the
waveguide has a mode character, and so the presence of $\kappa
_{mn}^{2} $ in the denominator of equation (\ref{eq20}) results in
reduction  of the radicand in (\ref{eq20}) to the magnitude
smaller than unity even when $n_{0}^{2}
> 1$. Hence, $u > c$, which is impossible.
Consequently, the radiative instability of the above type in the
waveguide can arise under the condition $n_{0}^{2} - \frac{{\kappa
_{mn}^{2} c^{2}}}{{\omega ^{2}}} >1$ rather than $n_{0}^{2}>1$.

Suppose now that in the photonic crystal the conditions can be
realized under which the wave amplitude $C_{mn} \left(
{\vec{k}_{mn} + \vec {\tau} } \right)$ is comparable with the
amplitude $C_{mn} \left( {\vec {k}_{mn}} \right)$. By analogy with
the standard diffraction theory for an infinite crystal
\cite{lanl_23,nim06_James}, in the case under consideration, when
$\chi < < 1$, it is sufficient that only the equations for these
amplitudes remain in (\ref{eq14}).

To be specific, let us further consider a photonic crystal formed
by parallel threads. Also assume that they are parallel to the
waveguide boundary $\left( {y,z} \right)$.

Analysis of diffraction of  a $\lambda $-type wave with the
electric vector in the plane $\left( {y,z} \right)$ (a TM-wave)
gives

\begin{equation}
\label{eq22} \left[ {k_{mn}^{2} - \frac{{\omega
^{2}}}{{c^{2}}}\varepsilon} \right]C^{\lambda} \left( {\vec
{k}_{mn} ,\omega}  \right) - \frac{{\omega ^{2}}}{{c^{2}}}\chi
_{mn}^{\lambda \lambda}  \left( { - \vec {\tau} }
\right)C^{\lambda} \left( {\vec {k}_{mn} + \vec {\tau} ,\omega}
\right) = 0
\end{equation}

\[
\left[ {\left( {\vec {k}_{mn} + \vec {\tau} } \right) -
\frac{{\omega ^{2}}}{{c^{2}}}\varepsilon _{0}}  \right]C^{\lambda}
\left( {\vec {k}_{mn} + \vec {\tau} ,\omega}  \right) -
\frac{{\omega ^{2}}}{{c^{2}}}\chi _{mn}^{\lambda \lambda}  \left(
{\vec {\tau} } \right)C^{\lambda} \left( {\vec {k}_{mn} ,\omega}
\right) = 0.
\]

Since the term containing $\left( {\omega - \left( {\vec {k} +
\vec {\tau} } \right)\vec {u}} \right)^{ - 1}$ is small when
$\left( {\omega - \vec {k}\vec {u}} \right)$ vanishes, in the
second equation it is dropped.

The dispersion equation defining the relation between $k_{z} $ and
$\omega $ is obtained by equating to zero the determinant of the
system (\ref{eq22}) and has a form:

\begin{equation}
\label{eq23}
\begin{array}{l}
 \left[ {\left( {k_{mn}^{2} - \frac{{\omega ^{2}}}{{c^{2}}}\varepsilon _{0}
} \right)\left( {\left( {\vec {k}_{mn} + \vec {\tau} } \right)^{2}
- \frac{{\omega ^{2}}}{{c^{2}}}\varepsilon _{0}}  \right) -
\frac{{\omega ^{4}}}{{c^{4}}}\chi _{\tau}  \chi _{ - \tau} }
\right]\left( {\omega -
k_{z} u} \right)^{2} = \\
 - \frac{{\omega _{L}^{2}} }{{\gamma c^{4}}}\left\{ {\frac{{1}}{{2\pi
}}\left| {\int {\vec {u}\,\vec {Y}^{\lambda} _{mn} \left( {\vec
{k}_{ \bot} ,k_{z}}  \right)d^{2}k_{ \bot} } }  \right|^{2}}
\right\}\left( {k_{mn}^{2} c^{2} - \omega ^{2}} \right)\left(
{\left( {\vec {k}_{mn} + \vec {\tau} } \right)^{2} - \frac{{\omega
^{2}}}{{c^{2}}}\varepsilon _{0}}  \right).
 \end{array}
\end{equation}

Because the right-hand side of the equation is small, one can
again seek the solution near the points where the right-hand side
is zero that corresponds the condition of occurrence of the
Cherenkov radiation and excitation of the wave which can propagate
in the waveguide:

\begin{equation}
\label{eq24}
\begin{array}{l}
 \left( {k_{z}^{2} - \left( {\frac{{\omega ^{2}}}{{c^{2}}}\varepsilon _{0} -
\kappa _{mn}^{2}}  \right)} \right)\left( {\left( {k_{z} + \tau}
\right)^{2} - \left( {\frac{{\omega ^{2}}}{{c^{2}}}\varepsilon
_{0} - \left( {\vec {\kappa} _{mn} + \vec {\tau} _{ \bot} }
\right)^{2}} \right)} \right)
- \frac{{\omega ^{4}}}{{c^{4}}}\chi _{\tau}  \chi _{ - \tau}  = 0 \\
 \left( {k_{z} - \frac{{\omega} }{{u}}} \right)^{2} = 0 \\
 \end{array}
\end{equation}

The roots of equations are sought near the conditions $k_{mn}^{2}
\approx \left( {\vec {k}_{mn} + \vec {\tau} } \right)$,

\begin{equation}
\label{eq25} k_{z} = k_{z0} + \xi , \quad k_{z}^{2} = k_{z0}^{2} +
2k_{z0} \xi + \xi ^{2}, \quad k_{z0}^{2} = \frac{{\omega
^{2}}}{{c^{2}}}\varepsilon _{0} - \kappa _{mn}^{2} ; \quad k_{z0}
= \frac{{\omega} }{{c}}\sqrt {\varepsilon _{0} - \frac{{\kappa
_{mn}^{2} c^{2}}}{{\omega ^{2}}}}
\end{equation}

\[
\left( {k_{z} + \tau _{z}}  \right)^{2} = \left[ {\left( {k_{0z} +
\tau _{z} } \right) + \xi}  \right]^{2} = \left( {k_{0z} + \tau
_{z}}  \right)^{2} + 2\left( {k_{0z} + \tau _{z}}  \right)\xi +
\xi ^{2}
\]

Hence,

\begin{equation}
\label{eq26}
\begin{array}{l}
 \left( {k_{0z} + \tau _{z}}  \right)^{2} + \left( {\vec {\kappa} _{mn} +
\vec {\tau} _{ \bot} }  \right)^{2} + 2\left( {k_{0z} + \tau _{z}}
\right)
+ 2\left( {k_{0z} + \tau _{z}}  \right)\xi + \xi ^{2} = \\
 \left( {\vec {k}_{mn} + \vec {\tau} } \right)^{2} + 2\left( {k_{0z} + \tau
_{z}}  \right)\xi + \xi ^{2} = k_{0mn}^{2} + 2\vec {k}_{0mn} \vec
{\tau} + \tau ^{2} + 2\left( {k_{0z} + \tau _{z}}  \right)\xi +
\xi ^{2}.
 \end{array}
\end{equation}

And one can get

\[
2k_{0z} \xi \left( {2\left( {k_{0z} + \tau _{z}}  \right)\xi +
\left( {2\vec {k}_{0mn} \vec {\tau}  + \tau ^{2}} \right)} \right)
- \frac{{\omega ^{4}}}{{c^{4}}}\chi _{\tau}  \chi _{ - \tau}  = 0
\]

\begin{equation}
\label{eq27} 4k_{0z} \left( {k_{0z} + \tau _{z}}  \right)\xi ^{2}
+ 2k_{0z} \left( {2\vec {k}_{0mn} \vec {\tau}  + \tau ^{2}}
\right)\xi - \frac{{\omega ^{4}}}{{c^{4}}}\chi _{\tau}  \chi _{ -
\tau}  = 0
\end{equation}

\[
\xi ^{2} + \frac{{\left( {2\vec {k}_{0mn} \vec {\tau}  + \tau
^{2}} \right)}}{{\left( {k_{0z} + \tau _{z}}  \right)}}\xi -
\frac{{\omega ^{4}}}{{c^{4}}}\frac{{\chi _{\tau}  \chi _{ - \tau}
} }{{4k_{0z} \left( {k_{0z} + \tau _{z}}  \right)}} = 0
\]

\[
\xi _{1,2} = - \frac{{\left( {2\vec {k}_{0} \vec {\tau}  + \tau
^{2}} \right)}}{{4\left( {k_{0z} + \tau _{z}}  \right)}} \pm \sqrt
{\left( {\frac{{2\vec {k}_{0} \vec {\tau}  + \tau ^{2}}}{{4\left(
{k_{0z} + \tau _{z}}  \right)}}} \right)^{2} + \frac{{\omega
^{4}}}{{c^{4}}}\frac{{\chi _{\tau}  \chi _{ - \tau} } }{{4k_{0z}
\left( {k_{0z} + \tau _{z}} \right)}}}
\]

If $\left( {k_{0z} + \tau _{z}}  \right) = - \left| {k_{0z} + \tau
_{z}} \right|$, the root can cross the zero point. At the same
time, the second equation should hold
\[ \omega - k_{z} u = \omega
- k_{0z} u - \xi u = 0. \]

Consequently,

\[
\xi = \frac{{\omega - k_{0z} u}}{{u}} = \frac{{\omega} }{{u}} -
k_{0z} = \frac{{\omega} }{{u}} - \frac{{\omega} }{{c}}\sqrt
{\varepsilon _{0} - \frac{{\kappa _{mn}^{2} c^{2}}}{{\omega
^{2}}}}.
\]

If $\varepsilon _{0} < 1$, then $\xi = \frac{{\omega} }{{u}}\left(
{1 - \beta \sqrt {\varepsilon _{0} - \frac{{\kappa _{mn}^{2}
c^{2}}}{{\omega ^{2}}}}}  \right) > 0$, $\xi = \frac{{\omega}
}{{u}} - k_{0z} $

Let the roots  $\xi _{1} $ and $\xi _{2} $ coincide ($\xi _{1} =
\xi _{2} $). This is possible at point

\[
\frac{{2\vec {k}_{0} \vec {\tau}  + \tau ^{2}}}{{4\left( {k_{0z} +
\tau _{z} } \right)}} = \pm \frac{{\omega
^{2}}}{{c^{2}}}\frac{{\sqrt {\chi _{\tau} \chi _{ - \tau} } }
}{{\sqrt {4k_{0z} \left| {k_{0z} + \tau _{z}}  \right|} }},
\]

\noindent here $k_{0z} + \tau _{z} < 0$.

The roots coincide when the following equality is fulfilled

 \[\frac{{\omega} }{{u}} - k_{0z} = \mp \frac{{\omega
^{2}}}{{c^{2}}}\frac{{\sqrt {\chi _{\tau}  \chi _{ - \tau} } }
}{{\sqrt {4k_{0z} \left| {k_{0z} + \tau _{z}}  \right|}} },
\]
 i.e.,
 \[\frac{{\omega} }{{u}}
= k_{0z} \mp \frac{{\omega ^{2}}}{{c^{2}}}\frac{{\sqrt {\chi
_{\tau}  \chi _{ - \tau} } } }{{\sqrt {4k_{0z} \left| {k_{0z} +
\tau _{z}}  \right|}} } \quad\mbox{and}\quad k_{0z} =
\frac{{\omega} }{{c}}\sqrt {\varepsilon _{0} - \frac{{\kappa
_{mn}^{2} c^{2}}}{{\omega ^{2}}}}
\]

Let $\varepsilon _{0} < 1$, then $\frac{{\omega} }{{u}} > k_{0z} $
(since $u < c$),  the situation for the solution $\frac{{\omega}
}{{u}} = k_{0z} - \frac{{\omega ^{2}}}{{c^{2}}}\frac{{\sqrt {\chi
_{\tau}  \chi _{ - \tau} } } }{{\sqrt {4k_{0z} \left| {k_{0z} +
\tau _{z}}  \right|}} }$ gets complicated and the
Vavilov-Cherenkov condition is not fulfilled.

Now let us consider the solution $\frac{{\omega} }{{u}} = k_{0z} +
\frac{{\omega ^{2}}}{{c^{2}}}\frac{{\sqrt {\chi _{\tau}  \chi _{ -
\tau} } } }{{\sqrt {4k_{0z} \left| {k_{0z} + \tau _{z}}  \right|}}
}$. At $\tau _{z} < 0$ the difference $k_{0z} + \tau _{z} $ can be
reduced so that the sum on the right would appear to become equal
to  $\frac{{\omega} }{{u}}$, and so  one could
 obtain 4 coinciding roots.

Interestingly enough, for backward diffraction, which is a typical
case of frequently used one-dimensional generators with a
corrugated metal waveguide (the traveling-wave tube, the
backward-wave tube), such a coincidence of roots is impossible.

Indeed, let the roots $\xi _{1} $ and $\xi _{2} $ coincide. In
this case for the backward Bragg diffraction
 $\left| {\tau _{z}}  \right| \approx
2k_{0z} $, $\tau _{z} < 0$. Then by substituting the expressions
for  $k_{0z} = \frac{{\omega} }{{c}}\sqrt {\varepsilon _{0} -
\frac{{\kappa _{mn}^{2} c^{2}}}{{\omega ^{2}}}} $ and $\varepsilon
_{0} = n_{0}^{2} = 1 + \chi _{mn}^{\lambda \lambda}  \left( {0}
\right)$ and retaining the first-order infinitesimal terms, the
relation $\frac{{\omega} }{{u}} \approx k_{0z} + \frac{{\omega
^{2}}}{{c^{2}}}\frac{{\left| {\chi _{\tau} } \right|}}{{2k_{0z}}
}$ can be reduced to the form $\frac{{\omega} }{{u}} \approx
\frac{{\omega }}{{c}}\left( {1 - \frac{{\left| {\chi
_{mn}^{\lambda \lambda}  \left( {0} \right)} \right|}}{{2}} -
\frac{{\kappa _{mn}^{2} c^{2}}}{{2\omega ^{2}}} + \frac{{\omega}
}{{c}}\frac{{\left| {\chi _{\tau} }  \right|}}{{2}}} \right) <
\frac{{\omega} }{{u}}$, i.e., the equality does not hold and the
four-fold degeneracy is impossible. Only ordinary three-fold
degeneration is possible.

However, if $\varepsilon_{0}>1$ and is appreciably large,  then in
a one-dimensional case, the four-fold degeneracy of roots is also
possible in a finite photonic crystal\footnote{The authors are
grateful to K. Batrakov, who drew our attention to the fact that
for an infinite crystal with $\varepsilon_{0}>1$, the intersection
of roots is possible in a one-dimensional case.}.

Thus, the left-hand side of equation (\ref{eq23}) has  four roots
$\xi_{1}$, $\xi_{2}$, and a double degenerated root $\xi_{3}$.
Hence, equation (\ref{eq23}) can be written as follows:

\[
\left( {\xi - \xi _{1}}  \right)\left( {\xi - \xi _{2}}  \right)
\left( {\xi - \xi _{3}}  \right)^{2} = B.
\]

If the roots coincide ($\xi _{1} = \xi _{2} = \xi _{3} $), one
obtains
 $\left( {\xi - \xi _{1}}  \right)^{4} = B,$, i.e., $\xi - \xi _{1} =
\sqrt[{4}]{{B}}$.

The fourth root of  $B$ has imaginary solutions depending on the
beam density as
 $Imk_{z}\sim\rho _{0} ^{1/4}$ (the parameter
$B \sim \omega _{L}^{2} $, i.e., $B \sim \rho _{0} $,  see the
right-hand side of (\ref{eq23})). This increment is larger than
the one we obtained for the case of the three-fold degeneracy.

The analysis shows that with increasing number of diffracted
waves, the law established in \cite{berk_104,berk_133,L_7} is
valid: the instability increment appears to be proportional to
$\rho ^{\frac{{1}}{{s + 3}}}$, where $s$ is the number of waves
emerging through diffraction. As a result,  the abrupt decrease in
the threshold generation current also remains in this case (the
threshold generation current $j_{th}\sim\frac{{1}}{{\left( {kL}
\right)^{3}\left( {k\chi _{\tau}  L} \right)^{2s}}}$, where $L$ is
the length of the interaction area).

It is interesting that according to \cite{rins_66}, for a photonic
crystal made from metallic threads, the coefficients $\chi \left(
{\tau}  \right)$, defining the threshold current and the growth of
the beam instability, are practically independent on $\tau $ up to
the terahertz range of frequencies because the diameter of the
thread can easily be made much smaller than the wavelength.
That is why photonic crystals with the period of about 1 mm can
 be used for lasing in terahertz range at high harmonics (for
example,  photonic crystal with 3 mm period provides the frequency
of the tenth harmonic of about 1 terahertz ($\lambda $=300
micron).

The analysis of laser generation in VFEL with a photonic crystal
 when the beam moves in an undulator (electromagnetic wave)
located in a finite crystal, made similarly to the above analysis,
shows that in this case the dispersion equation and the law of
instability also have the same form as in the case of an infinite
crystal. The procedure for going from  the dispersion equations
describing instability in the infinite case (\ref{eq14}) is
similar to that discussed earlier in this paper. It consists in
replacing the continuous $\vec {k}$ by the quantified values of
$\vec {k}_{mn} $  and redefining the coefficients appearing in
equations like (\ref{eq14}).

It is important to emphasize the general character of the rules
found in this paper for obtaining dispersion equations that
describe the radiative instability of the electron beam in a
finite photonic crystal. In particular, they are valid for
describing the processes of instability of an electromagnetic wave
in finite nonlinear photonic crystals.



\section{Hybrid systems with virtual cathode for high power microwaves generation (\cite{lanl_24})}
\label{hyb_sec:1}

The interest for high power microwave (HPM) sources has emerged in
recent years due to revealing new applications and offering novel
approaches to existing ones.

Vacuum electronic sources, which convert the kinetic energy from
an electron beam into the electromagnetic field energy, are a natural
choice for generating HPM.

The high current density electron beam, once generated, propagates
through an interaction region, which converts the beam's kinetic
energy to HPM.
It is the particular nature of the interaction that distinguishes
various classes of sources.

 High power HPM
sources generating high electromagnetic power density require the
high power densities in the electron beams, where space-charge
effects are essential.

When the magnitude of the current $I_b$ of an electron beam
injected into a drift tube exceeds the space-charge-limiting
current $I_{limit}$, an oscillating virtual cathode (VC) is formed
\cite{hyb_RuhadzeUFN,hyb_Selemir.rev1}.

According to \cite{hyb_RuhadzeUFN}, the following formula gives a good
approximation for the space-charge-limiting current:
\begin{equation}\label{hyb_I_limit}
    I_{limit}(kA)=\frac{m c^3}{e} G (\gamma^{2/3}-1)^{3/2}
\end{equation}
 where $m$ and $e$ are the mass and the charge of an electron, $c$ is the speed of light, $\gamma$ is the Lorentz factor of the electron beam
 and $G$ depends on the geometry \cite{hyb_RuhadzeUFN,hyb_NeskolkoVC}.
For example, for an annular electron beam in a cylindrical drift
tube $G$ reads as follows \cite{hyb_NeskolkoVC}:
\begin{equation}\label{hyb_G}
    G=\frac{1}{(\frac{r_b-r_b^{in}}{r_b}+2 \ln \frac{R}{r_b}) (1- sech \frac{\mu_1 L}{2
    R})},
\end{equation}
where $r_b$ and $r_b^{in}$ are the outer and inner radius of the
electron beam, $R$ and $L$ are the radius and length of the
cylindrical drift tube and $\mu_1$ is the first root of the Bessel
function $J_0 (\mu)=0$.

\begin{figure}[h]
\epsfxsize = 10 cm \centerline{\epsfbox{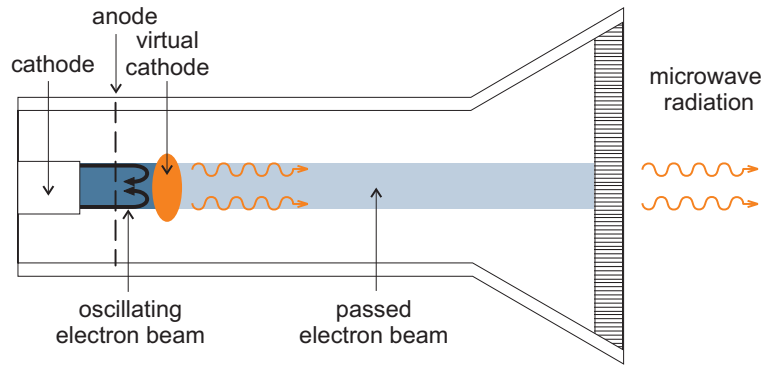}} \caption{An
oscillator with virtual cathode} \label{hyb_fig1}
\end{figure}

{When} the oscillating virtual cathode is formed, two types of
electrons exist: those oscillating in the vircator potential well
and passing through the vircator area (see Figure \ref{hyb_fig1}).

For electrons oscillating in the area ''cathode-anode-virtual
cathode'',  two radiation mechanisms provide a radio-frequency signal
\cite{hyb_Giri}:

1. one radiation mechanism originates from the oscillations of the
reflected electrons about the anode foil (electron oscillations in
the potential well ''cathode-anode-virtual cathode'').
A microwave signal is generated at  a {frequency of roughly}
$\frac{c}{2 d_d}$, where $d_d$ is the anode-to-cathode spacing.

2. The other radiation mechanism is the oscillation of the virtual
cathode at a frequency near the plasma frequency $\omega_p$ of the
space charge density that is formed. That is
\begin{equation}
\omega_p=\sqrt {\frac{4 \pi n_e e^2}{m}}
 \label{omegap}
\end{equation}
where $n_e$ is the number density of the electrons in the space
charge configuration (in the plane of the
anode grid) \cite{hyb_Giri}.

The essence of the above radiation mechanisms is bremsstrahlung
radiation ensuing from electron deceleration.

It should be emphasized that bremsstrahlung radiation from
electrons oscillating in an electron beam with a virtual cathode
is accompanied by transition radiation, which is originated by
electron velocity rather than acceleration.
Use of a photonic crystal enables one to construct several types
of hybrid systems with a virtual cathode, which could radiate due
to different radiation mechanisms (bremsstrahlung and diffraction
(transition) radiation) with different frequencies.

In vircator systems, a grid cathode and anode (or anodes) are
commonly used
\cite{hyb_NeskolkoVC,hyb_Selemir_patent,hyb_87Selemir,hyb_Selemir_patent2}.
The electron beam oscillates, making electrons periodically cross
the grid  anodes and cathode (see, for example
\cite{hyb_87Selemir}).
It is transition radiation that occurs when electrons pass through
a border between two media with different indices of refraction.
It is worth noting that periodical excitement of transition
radiation from electrons oscillating in vircator is similar to
diffracted radiation from a charged particle in a periodic
structure.
As a result, in a system with oscillating virtual cathode, the
vircator radiation, which  is actually the electron beam
bremsstrahlung, is accompanied by radiation excited by the
additional mechanism due to transition (diffraction) radiation
from oscillating current passing through the grid anodes
(cathode).

Let us turn to that part of the beam, which passes through the
virtual cathode area.
Recall that the oscillation of the virtual cathode can produce a
highly modulated electron beam, and, as a result, the energy from
the bunched transmitted beam can be recovered using slow-wave
structures \cite{hyb_SelemirTWT}.

\begin{figure}[h]
\epsfxsize = 10 cm \centerline{\epsfbox{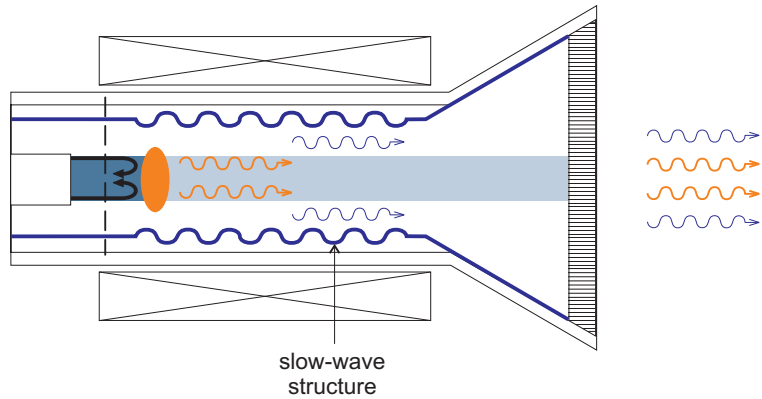}}
\caption{Hybrid system ''vircator + travelling wave tube''
\cite{hyb_SelemirTWT}} \label{hyb_fig2}
\end{figure}

Interaction of the electron beam with the slow-wave structure in,
for instance, a conventional TWT poses special challenges:
interaction is sufficiently effective only for electrons moving at
the distance $\delta$ from the slow-wave structure surface
\begin{equation}\label{hyb_delta}
\delta \le \frac{{\lambda \beta \gamma} }{{4\pi} },
\end{equation}
$\delta $ is the so-called  beam impact parameter, $\lambda $ is
the radiation wavelength, $\beta = v /c$ , $v$ is the electron
beam velocity, $\gamma $ is the electron Lorentz-factor.
For example, for electrons with the energy of 250 keV ($\beta=0.74$
and $\gamma=1.49$) and the radiation wavelength $\lambda= 10 mm $
(frequency 30 GHz), the impact parameter $\delta \approx 0.9 mm$.
It means that for efficient radiation generation by the annular
electron beam with the thickness $\Delta \leq \delta$ only a part of
the beam would contribute to radiation.

As we showed in
\cite{vfel_VFELreview,vfel_grid-t,vfel_new-grid-experiment,tu2_grid-ex,rins_66,
hyb_grid-FEL06-ex,hyb_grid-FEL06-th,hyb_grid-FEL07-ex},
 this challenge
can be overcome by applying a photonic crystal, formed by metal
grids (grid traveling wave tube (grid TWT), grid volume free
electron laser(grid VFEL)).

What is more, in accordance with
\cite{hyb_27Selemir,hyb_28Selemir,hyb_29Selemir}, the  application
of metal inserts (meshes, grids and so on) inside a resonator
enables increasing the electron beam limit current.

Therefore, in the grid TWT (grid VFEL), the  presence of the metal
grid (photonic crystal) serves both for forming the resonator,
where interaction of the beam and radiation occurs, and for
potential balancing that makes it possible to increase the beam
vacuum limit current.

And for the grid TWT (grid VFEL) with the supercritical current,
the electron beam executes compound motion exciting two radiation
mechanisms contributing to radiation: bremsstrahlung of
oscillating electrons and diffraction (transition) radiation from
downstream electrons interacting with the periodic grid
structure(photonic crystal).

This means that the hybrid system "vircator + grid TWT (grid
VFEL)" arise by analogy with the system described in
\cite{hyb_SelemirTWT}, where several vircators could appear due to
the presence of several anode grids (see also
\cite{hyb_NeskolkoVC}).
But in contrast to the system \cite{hyb_NeskolkoVC}, the hybrid system
"vircator + grid TWT (grid VFEL)" uses periodically placed grids
with either constant \cite{tu2_grid-ex,hyb_grid-FEL06-ex,hyb_grid-FEL06-th} or
variable period \cite{hyb_grid-FEL07-ex}.

The frequency of diffracted radiation excited by an electron beam
in a periodic structure with the period $d$ is determined by the
condition
\begin{equation}
\omega - \vec{k} n({k}){\vec{v}}= \vec{\tau} \vec{v} ~,
\label{hyb_eq:3-2}
\end{equation}
where $\vec{v}$ is the electron beam velocity, $\vec{\tau}$ is the
reciprocal lattice vector ($|\vec{\tau}|=\frac{2 \pi p}{d}$),
$n(k)$ is the refraction index of a periodic structure, $p$ is an
integer number ($p=1,2,3,...$).

When the electron beam velocity $\vec{v}$ is parallel to the
reciprocal lattice vector $\vec{\tau}$, (\ref{hyb_eq:3-2}) reads
\cite{rins_66}
\begin{equation}\label{hyb_frequency}
    \omega=\frac{2 \pi p \cdot v}{d(1-\beta n(\omega,k) \cos \theta)}
\end{equation}

\begin{figure}[h]
\epsfxsize = 15 cm \centerline{\epsfbox{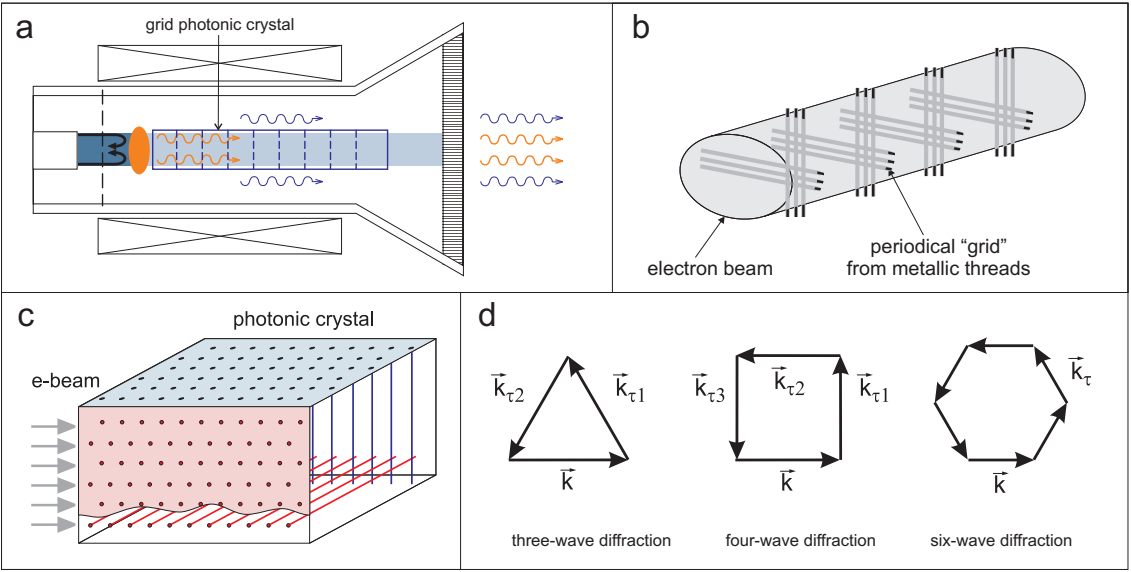}} \caption{Grid
TWT (grid VFEL) and photonic crystal arrangement} \label{hyb_fig3}
\end{figure}

Essentially, a photonic crystal in the grid TWT (grid VFEL) is
transparent to radiation as well as to an electron beam (see
Figure \ref{hyb_fig3}).
Moreover, several diffracted waves could exist in a photonic
crystal (see Figure \ref{hyb_fig3})d, which  makes it possible to
introduce a feedback in such a system at the frequency of
diffracted radiation and, hence, to couple several hybrid
''vircator + grid TWT (VFEL)'' generators making a phase-locked
source, in which diffracted waves from one photonic crystal (grid
resonator) excites oscillations in the neighbor resonators.

\begin{figure}[h]
\epsfxsize = 10 cm \centerline{\epsfbox{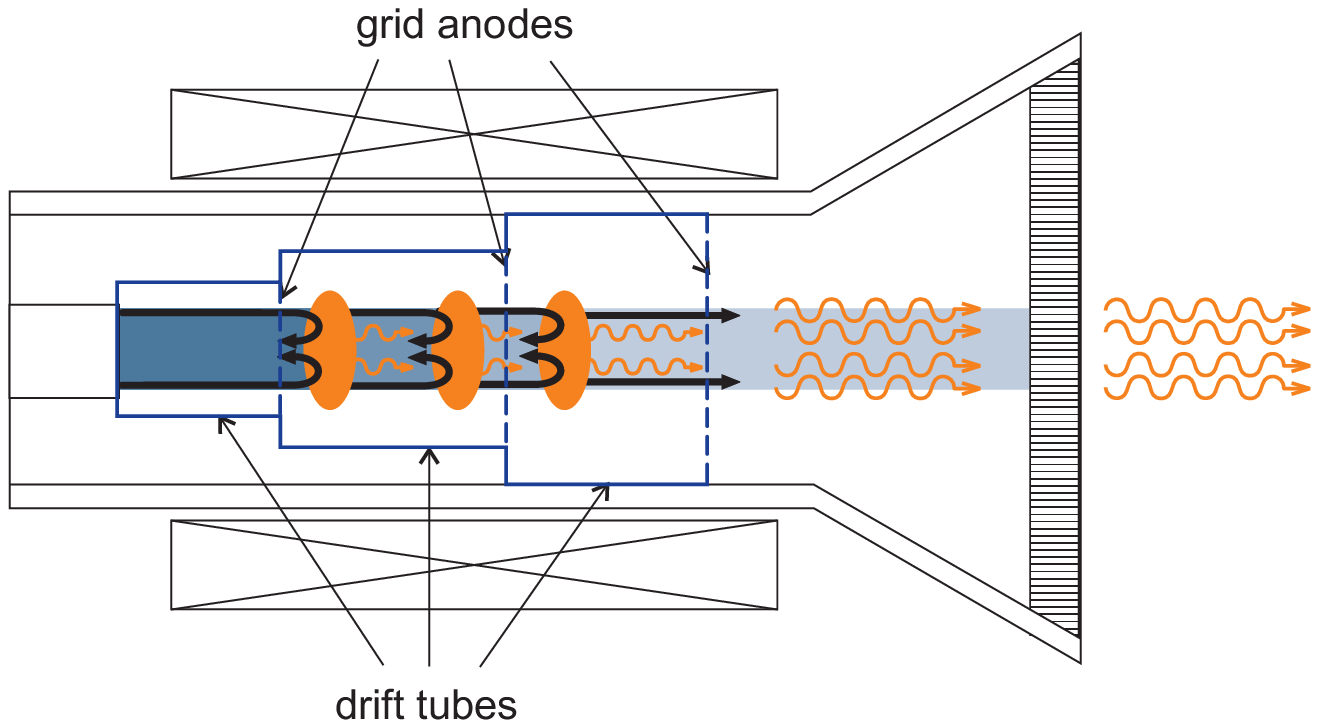}}
\caption{Several vircators \cite{hyb_NeskolkoVC}} \label{hyb_fig4}
\end{figure}

The proposed grid systems drastically differ from the system described in
\cite{hyb_NeskolkoVC}, where several grids serve only for forming
several vircators (see Figure \ref{hyb_fig4}).

Of course, radiation from a hybrid generator ''vircator + grid TWT
(VFEL)'' can be excited by several electron beams similar to
a phase-locked array.

\begin{figure}[h]
\epsfxsize = 10 cm \centerline{\epsfbox{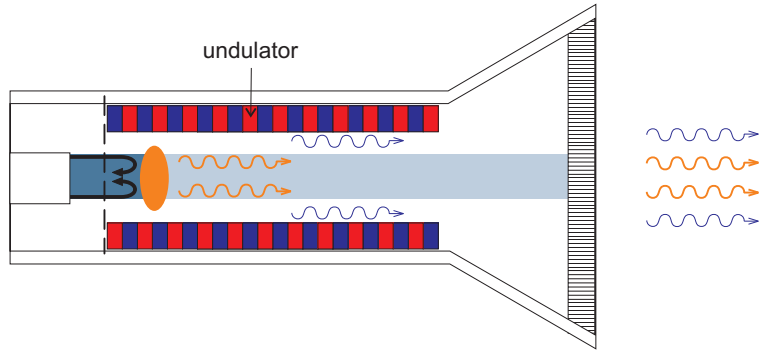}}
\caption{Hybrid system ''vircator + undulator FEL''}
\label{hyb_fig5}
\end{figure}

The bunched electron beam, which has passed through the virtual
cathode area, can also be  used for excitation of free electron
laser (ubitron) (see Figure \ref{hyb_fig5}) oscillation contributing
to the radiation power.

Thus, the use of a photonic crystal enables one to construct
several types of hybrid systems with virtual cathode, which could
radiate due to different radiation mechanisms (bremsstrahlung and
diffraction (transition) radiation) with different frequencies.
 The photonic crystal makes it possible to create
a phase-locked array of generators.
%



\section{Volume Free Electron Laser
(VFEL) as a new trend in development of high-power tunable
radiation sources: review of theoretical background and
experiments in millimeter range} \label{nova_sec1}


To conclude, let us mark the most important results obtained in
investigation of the VFEL. Use of non one-dimensional distributed
feedback in vacuum electronic devices enables frequency tuning in
a wide range and removes limits for available output power. It
also solves a problem of development of highly stable
mode-selective overmoded resonators, which open up new avenues for
extending the operating frequencies of all classes of microwave
vacuum electronic devices.

New advances in different areas require the development of
tunable, wide-band, high-power sources of coherent electromagnetic
radiation in gigahertz, terahertz and higher frequency ranges.
Conventional electron vacuum devices have restricted the possibility
of frequency tuning (usually it does not exceed 5-10\%) for the
certain carrier frequency at certain e-beam energy. They also have
power limits because high levels of power density inside the
system cause sparkovers and damage of mirrors.

Volume free electron laser (VFEL)
\cite{berk_104,vfel_VFELreview,hyb_FirstLasing} was proposed as a new type
of a tunable high-power source of electromagnetic radiation.

The most essential feature of the FEL and other types of generators is
a feedback, which is formed by a system of mirrors, or distributed
feedback based on diffraction in a spatially periodic medium, when
wave vectors of transmitted and reflected waves are colinear. The
distinction of volume FEL (VFEL) is non-one-dimensional multi-wave
volume distributed feedback (VDFB). Use of a non one-dimensional
distributed feedback in vacuum electronic devices gives the
possibility of frequency tuning in a wide range and removes limits
for available output power. It also solves a problem of
development of highly stable mode-selective overmoded resonators,
which open up new avenues for extending the operating frequencies
of all classes of microwave vacuum electronic devices.

It is well known that each radiative system is defined by its
eigenmodes and by the so-called dispersion equation, which in the
case of small perturbations (linear regime) describes possible
types of waves in a system and relation between frequency and wave
number of the system eigenmodes. Thorough analysis of FEL dispersion
equation \cite{nova_1_bvg} shows that: \\
1. dispersion equation for FEL in collective regime coincides with
that for
conventional traveling wave tube amplifier (TWTA) \cite{nova_2_bvg}; \\
2. FEL gain (increment of electron beam instability) is
proportional to $\rho ^{1/3}$, where $\rho$ is the electron beam
density.

But the law of instability of the electron beam can  essentially
change at passing through a spatially periodic medium. This fact
was first indicated in  \cite{berk_133}. Dispersion equations were
obtained and investigated for conditions of multi-wave
diffraction. It was shown that there is a new law of electron beam
instability at the points of diffraction roots degeneration. The
amplification and generation gain of the electromagnetic wave are
sharply changed at these points.

This result is also valid for an electron beam, which moves in
vacuum close to the surface of a spatially periodic medium
\cite{berk_104} (or in a vacuum slot made inside a periodic
medium).

First lasing of the VFEL was reported at FEL 2001
\cite{hyb_FirstLasing}.

\subsection{Volume FEL distinctive features}

The advantages of VFEL are exhibited in a wide spectral range from
microwaves to X-rays
\cite{berk_139,vfel_VFELreview,nova_bar4}. Frequency
tuning, possibility to use  wide electron beams (several
e-beams) and reduction of threshold current density necessary for
the start of generation provided by VFEL, make it a basis for
development of more compact, high-power and tunable radiation
sources than conventional electron vacuum devices could let.

Benefits given by VFEL:

1. Volume FEL provides frequency tuning by rotation of diffraction
grating;

2. Use of multi-wave diffraction reduces generation threshold and
the size of the generation zone. The starting current $j$ depends on
interaction length $L$ as \cite{berk_139}: $j_{start}\sim
1/\{(kL)^{3}(k\chi _{\tau }L)^{2s}\}$, $s$ is the number of
surplus waves appearing due to diffraction (for example, in the case
of two-wave Bragg diffraction $s=1$, for three-wave diffraction
$s=2$, and so on).

3. Wide electron beams and diffraction gratings of large volumes
can be used in VFEL. Two or three-dimensional diffraction gratings
allow one to distribute the interaction over a large volume and to overcome
power restrictions in a resonator. Volume distributed feedback
provides mode discrimination in a VFEL resonator.

4. VFEL can simultaneously generate radiation at several
frequencies.


\begin{thebibliography}{199}
\bibitem{1} M.L. Goldberger and R.M. Watson \emph{Collision Theory} (Wiley, New York, 1984).
\bibitem{Nuclear_optics} V.G. Baryshevsky  \emph{Nuclear Optics of Polarized Media} (Energoatomizdat, Moscow, 1995) [in Russian].
\bibitem{132} Chang Shih-Lin,\emph{ Multiple Diffraction of X-Rays in Crystals} (Springer-Verlag Berlin Heidelberg New-York Tokyo, 1984).
\bibitem{147} V.G. Baryshevsky, Pis'ma Zh. Tekh. Fiz.\textbf{2} (1976) 112-114; V.G. Baryshevsky, Zh.Exp.Teor Fiz. \textbf{70} (1976) 430-434[Sov. Phys. JETP].
\bibitem{201} V.G. Baryshevsky \emph{Nuclear Optics of Polarized Media} (Energoatomizdat, Moscow, 1995) [in Russian].
\bibitem{66} P.M. Morse, H. Feshbach \emph{Methods of Theoretical Physics} (Mc Graw Hill, New York, 1953), 
\bibitem{67} V.G. Baryshevsky, Zh.Exper.Teor Fiz. \textbf{67} (1974) 1651-1659 [Sov. Phys. JETP \textbf{40} (1975) 821].
\bibitem{63} V.G.Baryshevsky, I.D. Feranchuk, J. Physiq. \textbf{44} (1983) 913-922.
\bibitem{171} V.G. Baryshevsky, V.V. Tikhomirov,  Sov. Phys.-Uspekhi \textbf{32}, 11, (1989) 1013-1032 [Uspekhi Fiz. Nauk \textbf{159} (1989) 529-565].

\bibitem{chan_17} V.G. Baryshevsky, I.Ya. Dubovskaya, Phys. Status Solidi \textbf{B 82} (1977) 403.
\bibitem{chan_44} V.G. Baryshevsky, A.O. Grubich, I.Ya. Dubovskaya, Phys. Status Solidi  \textbf{B 88} (1978) 351.
\bibitem{chan_45} V.G. Baryshevsky, A.O. Grubich, I.Ya. Dubovskaya, Phys. Status Solidi  \textbf{ B 99} (1980) 205.
\bibitem{chan_67} V.G. Baryshevsky, A.O. Grubich, I.Ya. Dubovskaya,  Phys. Lett.  \textbf{A 77} (1980) 61.

\bibitem{absan_3} M.L. Ter-Mikaelian \emph{High Energy Electromagnetic Processes in Condensed Media} (Interscience Tracts in Physics and Astronomy, Vol 29, Willey, New York, 1972).
\bibitem{nim06_James} R.W. James  \emph{The Optical Principles of the Diffraction of X-rays }(Ox Bow Press, 1982)


\bibitem{WIND_Jackson-a}  J.D. Jackson \emph{Classical Electrodynamics}, 3rd ed. (Wiley,
1998)



\bibitem{berk_2} V.G. Baryshevsky \emph{Channeling, Radiation and Reactions in Crystals at High Energy} (Bel. State Univers., Minsk, 1982)

\bibitem{berk_10} L.M. Frank \emph{Vavilov-Cherenkov Radiation. Theoretical Questions} (Nauka, Moscow, 1988)

\bibitem{berk_13} V.G. Baryshevsky and I.Ya. Dubovskaya, Dokl. Akad. Nauk SSSR \textbf{231} (1976) 1336.

\bibitem{berk_19} V.G. Baryshevsky, I.Ya. Dubovskaya,  J. Phys. Solid State Phys.  \textbf{ C 16}, 19 (1983) 3663-3672.

\bibitem{berk_27} V.G. Baryshevsky, O.T. Gradovsky and  I.Ya. Dubovskaya, Izv. Akad. Nauk Bel.SSR Ser. Fiz-math., \textbf{6} (1987) 77.
\bibitem{berk_28} O.T. Gradovsky, Phys. Lett. \textbf{A 126} (1988) 291.
\bibitem{berk_129} V. G. Baryshevsky and I. Ya. Dubovskaya, Journal of Physics: Condensed Matter \textbf{3}, n 14 (1991) 2421-2430.

\bibitem{berk_29} A.O. Grubich and O.M. Lugovskaya, Izv. Akad. Nauk Bel.SSR, Ser.  Fiz-math., \textbf{1} (1991) 61.

\bibitem{berk_104} V.G. Baryshevsky, Dokl. Akad. Nauk SSSR   \textbf{299}, 6, (1988) 1363-1365.
\bibitem{berk_105} V.G. Baryshevsky in : \emph{Some Problems of Modern Physics to 80-th Anniversary of I.M. Frank }(Nauka, Moscow, 1989) 156.
\bibitem{berk_107} V.G. Baryshevsky and I.Ya. Dubovskaya, Phys. Status Solidi [in Russian] \textbf{19} (1977) 597.
\bibitem{berk_108} A.V. Andreev,  Sov. Phys. Usp. \textbf{28} (1985) 70Ц84. 
\bibitem{berk_109} A.A. Andriyanchik, I.Ya. Dubovskaya and A.N. Kaminsky, J. Phys.  \textbf{C 3} (1991) 5579.
\bibitem{andr} A.A. Andriyanchik, V.G. Baryshevsky and A.N. Kaminsky, Pis'ma v zhur. Tekh. Fiziki \textbf{17} (1991) 53.
\bibitem{andr1} A.A. Andriyanchik, V.G. Baryshevsky and A.N. Kaminsky, Physica Status Solidi (b) \textbf{184}, 2, (1994) 543-551.


\bibitem{berk_112} A.A. Andriyanchik, V.G. Baryshevsky and A.N. Kaminsky, Nucl. Instr. Methods \textbf{B 83}, 4, (1993) 482-494.
\bibitem{berk_114} V.G. Baryshevsky, Izv. Akad. Nauk Bel.SSR, Ser. Fiz-math.  \textbf{3} (1980) 117-122.
\bibitem{berk_115} V.G. Baryshevsky in : \emph{Abstr. 16-th Conf. on the Physics of Interaction of Charged Prticles with Crystals} (Moscow, 1986) 53.
\bibitem{berk_116} V.G. Baryshevsky and I.V. Polikarpov,  Izv. Akad. Nauk Bel.SSR, ser. fiz-mat.  \textbf{2} (1988) 86.
\bibitem{berk_117} V.G. Baryshevsky and I.V. Polikarpov, Zh. Eksp. Teor. Fiz. \textbf{94} (1988) 109.
\bibitem{berk_118} V.G. Baryshevsky and I.V. Polikarpov, Izv. Bel. Univers. Ser. 1 \textbf{1} (1989) 8.
\bibitem{berk_119} V.G. Baryshevsky and I.V. Polikarpov, Phys. Lett. \textbf{A 190} (1989) 205.
\bibitem{berk_120} V.G. Baryshevsky and I.V. Polikarpov,  Izv. Akad. Nauk Bel.SSR, ser. fiz-mat.  \textbf{2} (1989) 81.
\bibitem{berk_121} I.V. Polikarpov and V.V. Skadorov, Phys. Status Solidi  \textbf{B 143} (1987) 11.
\bibitem{berk_122} I.V. Polikarpov and V.V. Skadorov, Izv. Akad. Nauk Bel.SSR, Ser. Fiz-math.  \textbf{6} (1978) 95.
\bibitem{berk_123} I.V. Polikarpov and V.V. Skadorov, Izv. Akad. Nauk Bel.SSR, Ser. Fiz-math.  \textbf{3} (1988) 83.
\bibitem{berk_126} I.Ya. Dubovskaya and Truong Ba Ha, Izv. Bel. Univers. Ser. 1 \textbf{1} (1988) 11.
\bibitem{berk_127} A.O. Grubich and O.M. Lugovskaya, Izv. Akad. Nauk Bel.SSR Ser. Fiz-energ. \textbf{3} (1991) 61.
\bibitem{berk_128} A.A. Andriyanchik and A.I. Kaminsky, Izv. Bel. Univers. Ser. 1 \textbf{2} (1988) 54.
\bibitem{berk_130} V.G. Baryshevsky and I.D. Feranchuk in: \emph{Dokl. 11-th Conf. on the Physics of Interaction of Charged Prticles with Crystals} (Moscow, 1982) 208.
\bibitem{berk_131} V.G. Baryshevsky and I.D. Feranchuk, Dokl. Akad. Nauk Bel.SSR  \textbf{27} (1983) 995.
\bibitem{berk_132} V.G. Baryshevsky and I.D. Feranchuk, Dokl. Akad. Nauk Bel.SSR  \textbf{28} (1984) 336.
\bibitem{berk_133} V.G. Baryshevsky and I.D. Feranchuk, Phys. Lett.  \textbf{A 102} (1984) 141.
\bibitem{berk_134} V.G. Baryshevsky and I.D. Feranchuk, Izv. Akad. Nauk Bel.SSR, Ser. Fiz-math.  \textbf{2} (1985) 79.
\bibitem{berk_135} V.G. Baryshevsky, I.Ya. Dubovskaya and I.D. Feranchuk, Izv. Akad. Nauk Bel.SSR, Ser. Fiz-math.  \textbf{1} (1988) 92-97.
\bibitem{berk_136} V.G. Baryshevsky, K.G. Batrakov and I.Ya. Dubovskaya in : \emph{Dokl. 19-th Conf. on the Physics of Interaction of Charged Prticles with Crystals} (Moscow, 1990) 105.
\bibitem{berk_137} K.G. Batrakov and I.Ya. Dubovskaya, Izv. Akad. Nauk Bel.SSR, Ser. Fiz-math.   \textbf{5} (1990) 82.
\bibitem{berk_138} V.G. Baryshevsky, K.G. Batrakov and I.Ya. Dubovskaya, Izv. Akad. Nauk Bel.SSR Ser. Fiz-energ. \textbf{1} (1991) 53.
\bibitem{berk_139} V.G. Baryshevsky, K.G. Batrakov and I.Ya. Dubovskaya, J. Phys. Appl. Phys. \textbf{D 24} (1991) 1250.
\bibitem{berk_140}  V.G. Baryshevsky, I.Ya. Dubovskaya and A.V. Zege in : \emph{Dokl. 19-th Conf. on the Physics of  Interaction of Charged Prticles with Crystals} (Moscow, 1990) 102.
\bibitem{berk_141} V.G. Baryshevsky, I.Ya. Dubovskaya and A.V. Zege, Nucl. Instr. Methods \textbf{A 135} (1990) 368.
\bibitem{berk_142} V.G. Baryshevsky, I.Ya. Dubovskaya and A.V. Zege, Izv. Akad. Nauk Bel.SSR Ser. Fiz-energ., \textbf{3} (1990) 49.
\bibitem{berk_143} V.G. Baryshevsky, I.Ya. Dubovskaya and A.V. Zege, Phys. Lett. \textbf{A 149} (1990) 30-34.
\bibitem{berk_144} V.G. Baryshevsky and I.Ya. Dubovskaya, Izv. Akad. Nauk Bel.SSR Ser. Fiz-energ., \textbf{1} (1990) 30-36.
\bibitem{berk_145} A. Fridman, A. Gover, G. Kurizki, et. al, Rev. Mod. Phys. \textbf{60} (1988) 471.
\bibitem{berk_146} G. Kurizki in: \emph{Relativistic Channeling}, Eds. R.A. Carrigan and S.A. Ellison (M.Y. Plenum Press, 1987) 505.
\bibitem{berk_147} S.A. Bogaez and J.B. Ketterson, J. Appl. Phys. \textbf{60} (1986) 177.
\bibitem{berk_148} M. Strauss, P. Amend, N. Rostoker and A. Ron, Appl. Phys. Lett. \textbf{52} (1988) 866.
\bibitem{berk_149} M. Strauss and N. Rostoker, Phys. Rev. \textbf{A 39} (1989) 579.
\bibitem{berk_150} M.A. Piestrup and P.E. Finman, IEEE J. Of Quantum Electronics \textbf{QE-19} (1983) 357.
\bibitem{berk_151} M.A. Piestrup, IEEE J. Of Quantum Electronics \textbf{QE-24} (1988) 591.
\bibitem{berk_152} M.A. Yariv, Appl. Phys. \textbf{24} (1974) 105.

\bibitem{vfel_TWT1} R. Kompfner, Wireless World  \textbf{52} (1946) 369.

\bibitem{vfel_TWT2} R. Pierce,  Proc. of the IRE  \textbf{35}, 2, (1947) 111.

\bibitem{vfel_SP} S.J. Smith and E.M. Purcell, Phys. Rev. \textbf{92}
(1953) 1069.
\bibitem{vfel_Salisbury1} W.W. Salisbury, US Patent 2,634,372 (1953).
\bibitem{vfel_Salisbury2} W.W. Salisbury, J.Opt. Soc. Am. \textbf{60}, 10, (1970) 1279-1284.

\bibitem{vfel_Walsh1} G. Doucas, J.H. Mulvey, M.Omori, J.Walsh and
M.F.Kimmit, Phys.Rev.Lett. \textbf{69} (1992)  1761.
\bibitem{vfel_Walsh2} John E. Walsh US Patent 5,790,585 (1996).

\bibitem{vfel_Bolotovskii} B. M. Bolotovskii and G. V. Voskresenskii, Usp. Fiz.
Nauk. \textbf{88} (1966) 209. (Sov. Phys. Usp. \textbf{9}
(1966) 73).

\bibitem{vfel_orotron1} F.S. Rusin and G.D. Bogomolov, JETP Lett. \textbf{4} (1966) 160.
\bibitem{vfel_orotron2} F.S. Rusin and G.D. Bogomolov, (USSR inventors certificate no.195557 (1967));
\bibitem{vfel_orotron3} F.S. Rusin and G.D. Bogomolov, Proc. IEEE \textbf{57} (1969) 720.

\bibitem{vfel_ledatron} K.Mizuno, S.Ono and Y. Shibata, IEEE Trans. Electron Devices \textbf{ED-20} (1973) 749 .

\bibitem{vfel_Leavitt1} R. P. Leavitt et al., IEEE Jour. Quant. Electr. \textbf{QE-17},
no.8, (1981) 1333.

\bibitem{vfel_Leavitt2} D. E. Wortman et al., IEEE Jour. Quant. Electr. \textbf{QE-17}, 8,
(1981) 1341.

\bibitem{vfel_Leavitt3a} D. E. Wortman and R. P. Leavitt,
 \emph{Infrared and Millimeter Waves: Coherent
Sources and Applications}, Part II', Vol. 7, chapter 7, 321-375
(editted by K.J. Button,  Academic Press, New York, 1983);
\bibitem{vfel_Leavitt3b} D. E. Wortman, C.A. Morrison and R. P. Leavitt, US Patent 4,545,056 (1985).

\bibitem{vfel_PXRbook}   V.G. Baryshevsky, I.D. Feranchuk, A.P. Ulyanenkov, \textit{Parametric X-Ray Radiation in Crystals:
Theory, Experiment and Applications} (Series: Springer Tracts in
Modern Physics, Vol. 213  2005).

\bibitem{vfel_Kurizki} G.Kurizki, M.Strauss, I.Oreg and N.Rostoker, Phys.Rev. \textbf{A 35} (1987) 3427 .

\bibitem{vfel_VFELreview} V.G.Baryshevsky, Nucl. Instr. Methods \textbf{A 445} (2000) 281-283; LANL e-print archive physics$/$9806039.
\bibitem{vfel_FEL2002} V.G.Baryshevsky, K.G. Batrakov, A.A. Gurinovich et al., Nucl. Instr. Methods \textbf{A 507} (2003) 137.

\bibitem{vfel_grid-t} V.G.Baryshevsky, A.A. Gurinovich, LANL e-print arXiv: physics$/$0409107.


\bibitem{vfel_PC1} A. L. Pokrovsky and A. L. Efros, Phys. Rev. \textbf{B 65} (2002) 045110 .

\bibitem{vfel_PC2} A. L. Pokrovsky, Phys. Rev. \textbf{B  69} (2004) 195108 .

\bibitem{vfel_PC3}  E. I. Smirnova, C. Chen, M. A.  Shapiro et al., J. Appl.Phys. \textbf{91}, 3 (2002) 960. 

\bibitem{vfel_PC4}  E. I. Smirnova and C. Chen, M. A., J. Appl.Phys. \textbf{93}, 10, (2003) 5859 .

\bibitem{vfel_G0} N.S.Ginzburg, A.S.Sergeev, N.Yu.Peskov et al., Nucl. Instr. Methods \textbf{A 375} (1996) 202 .

\bibitem{vfel_G4} Iv.V. Konoplev, A.W. Cross, W. He, A.D.R. Phelps, K. Ronald, G. R. M. Robb, C. G. Whyte, N. S. Ginzburg, N. Yu. Peskov and A. S. Sergeev,  Nucl. Instr. Methods \textbf{A 445} (2000) 236.


\bibitem{vfel_G6} A.W. Cross, W. He, I.V. Konoplev, A.D. R. Phelps,
K. Ronald, G. R. M. Robb, C. G. Whyte, N. S. Ginzburg, N. Yu. Peskov and A. S. Sergeev,  Nucl. Instr. Methods  \textbf{A 475} (2001) 164.


\bibitem{vfel_G9} I.V. Konoplev, A.D.R. Phelps, A.W. Cross, K. Ronald, P.
McGrane, W. He, C.G. Whyte, N.S. Ginzburg, N.Yu. Peskov, A.S. Sergeev and M. Thumm,   Nucl. Instr. Methods \textbf{A 528} (2004) 101.


\bibitem{vfel_G10} N.S. Ginzburg, N. Yu. Peskov,A. S. Sergeev, I. V. Konoplev, K. Ronald, A. D. R. Phelps and A. W. Cross,  Nucl. Instr. Methods \textbf{A 528} (2004)  78.


\bibitem{vfel_Born} M.Born, E.Wolf, \emph{Principles of optics: Electromagnetic Theory of Propagation, Interference and Diffraction of Light } (Pergamon Press, 1965)

\bibitem{vfel_Nikolsky} V.V. Nikolsky, \emph{Electrodynamics and propagation of radio-wave} (Nauka, 1978).

\bibitem{vfel_Henl} H\"{o}nl H., Maue A., Wespfahl K., \emph{Theorie der Beugung} (Berlin:Springer-Verlag, 1961).

\bibitem{vfel_Batrakov+Sytova} K.G.Batrakov and S.N.Sytova, Computational Mathematics and Mathematical Physics \textbf{45}, 4, (2005) 666.

\bibitem{vfel_new-grid-experiment} V.G.~Baryshevsky, N.A.~Belous, V.A.~Evdokimov, A.A.~Gurinovich, A.S.~Lobko, P.V.~Molchanov, P.F.~Sofronov, V.I.~Stolyarsky, LANL e-print arXiv: physics$/$0605122.



\bibitem{laser_1} T.C. Marshall, \emph{Free-Electron Lasers} (Macmillan Publishing Company, London, 1985).

\bibitem{laser_8}  V.G.Baryshevsky, K.G.Batrakov, I.Ya.Dubovskaya, Nucl. Instr. Methods  \textbf{A 358}  (1995) 493-496.

\bibitem{laser_13} N. S. Yerokhin, M. V. Kuznetsov, S. S. Moiseev, et al. \emph{Nonequilibrium and Resonance Processes in Plasma Radiophysics} ( Nauka, Moscow (1982)) [in Russian].



\bibitem{nim03_nonrel} V.G.Baryshevsky, K.G.Batrakov, V.I.Stolyarsky in : \emph{Proceedings of 21 FEL Conference} (1999)  37-38.


\bibitem{nova_1_bvg} A.Gover, Z. Livni,   Optics Commun.  \textbf{26} (1978) 375.
\bibitem{nova_2_bvg} J.R.Pierci,  \emph{Travelling wave tubes} (Van Nostrand, Princeton, 1950);

\bibitem{nova_bar4} V.G.Baryshevsky, K.G.Batrakov, I.Ya. Dubovskaya, V.A.Karpovich, V.M.Rodionova,  Nucl. Instr. Methods \textbf{A 393} (1997) 71-75.


\bibitem{tu2_Granatstein} V.L.Granatstein, R.K.Parker  and C.M.Armstrong,  Proceedings of the IEEE  \textbf{87}, 5, (1999) 702-716.

\bibitem{tu2_grid-ex} V.G.~Baryshevsky, K.G.~Batrakov, N.A.~Belous, A.A.~Gurinovich, A.S.~Lobko, P.V.~Molchanov, P.F.~Sofronov,
V.I.~Stolyarsky, LANL e-print archive: physics/0409125.




\bibitem{tu3_Takagi}  S.Takagi, Acta Crystall. \textbf{15} (1962) 1311.



\bibitem{vp_Kuraev} M.P.Batura, A.A.Kuraev, A.K.Sinitzyn, \emph{Simulation and optimization of powerful microwave devices}  (Minsk, 2006) [in Russian]

\bibitem{vp_undulatorFEL} T.J.Orzechovsky, B.R.Anderson, J.C.Clark et.al., Phys.Rev.Lett. \textbf{57} (1986) 2172.



\bibitem{nim95_2} A. Yariv and P. Yeh, \emph{Optical Waves in Crystals} (Wiley, 1984).

\bibitem{rins_66} V.G. Baryshevsky, A.A. Gurinovich, Nucl. Instr. and Methods in Physics Research   \textbf{B 252}, 1, (2006) 92-101.

\bibitem{vesti_8} V.G. Baryshevsky, I.Ya Dubovskaya in : \emph{Abstarcts of reports at 14th All-Union Conference on the Physics of Interaction of Charged Particles with Crystals}, Izd-vo Mosk. Univ., Moscow, 1984, p. 50.
\bibitem{vesti_9} V.G. Baryshevsky, I.Ya Dubovskaya in: \emph{Proceedings of the 15th All-Union Conference on the Physics of Interaction of Charged Particles with Crystals}, Izd-vo Mosk. Univ., Moscow, 1986, pp. 60-62.

\bibitem{vesti_12} A.M. Fedorchenko, I. Ya. Kotzarenko, \emph{Absolute and Convective instability in Plasma and Solids} (Moscow, 1981) [in Russian].
\bibitem{vesti_13} V.I. Fadeeva, I.M. Terentiev, \emph{Value Tables of the Probability Integral of the Complex Argument} (Moscow, 1954) [in Russian].


\bibitem{dan_8} V.G. Baryshevsky, I.D. Feranchuk,  Pis'ma Zh. Tekh. Fiz. \textbf{10}, 19, (1984) 1157-1159.


\bibitem{para_1} V.G. Baryshevsky, Doklady Akad. Nauk BSSR \textbf{15}  (1971) 306. 
\bibitem{para_2} V.G. Baryshevsky, I.D. Feranchuk, Zh. Eksp. Teor. Fiz. \textbf{61} (1971) 944 [Sov. Phys. JETP \textbf{34}(1972) 50; Errata, ibid \textbf{64} (1973) 760 [\textbf{37} (1973) 605].
\bibitem{para_4} V.G. Baryshevsky, A.O. Grubich, Le Tien Hai, Zh. Eksp. Teor. Fiz. \textbf{94} (1988) 51 [Sov. Phys. JETP \textbf{67}(1988) 895].
\bibitem{para_5} V.G. Baryshevsky, I.D. Feranchuk, Phys. Lett \textbf{A 57} (1976) 183. 
\bibitem{para_6b} V.G. Baryshevsky, I.D. Feranchuk, Nucl. Instr. Methods \textbf{A 228} (1985) 490.
\bibitem{para_7a} A. Caticha,  Phys. Rev. \textbf{A 40},8  (1989) 4322-4329.
\bibitem{para_7b} A. Caticha,  Phys. Rev. \textbf{B 45}, 17 (1992) 9541-9550.
\bibitem{para_8a} H.Nitta, Phys. Lett. \textbf{A 158}, 5(1991) 270-274.
\bibitem{para_8b} H.Nitta, Phys. Rev. \textbf{B 45}, 14(1992) 7621-7626.
\bibitem{para_9a} R.B. Fiorito, D.W. Rule, M.A. Piestrup, Li Qiang, A.H. Ho, X.K.
Maruyama,
NIM \textbf{B79}, 1-4, (1993) pp. 758-761
\bibitem{para_9b} R. B. Fiorito, D. W. Rule, M. A. Piestrup, X. K. Maruyama, R. M. Silzer, D. M. Skopik, and A. V. Shchagin
Phys. Rev. \textbf{E 51} (1995) R2759.
\bibitem{para_10} V.G. Baryshevsky et al., Nucl. Instr. Methods \textbf{A 249} (1993) 304.
\bibitem{para_11} Yu.N. Adishchev et al., Nucl. Instr. Methods \textbf{B 44}, 2, (1989) 130-136.
\bibitem{para_12} {J. Freudenberger,} V. B. Gavrikov, M. Galemann, H. Genz, L. Groening1, V. L. Morokhovskii, V. V. Morokhovskii,
U. Nething, A. Richter, J. P. F. Sellschop, and N. F. Shul'ga,
Phys. Rev. Lett.  \textbf{74}, 13, (1995) 2487Ц2490.




\bibitem{vesti92_5} V.G. Baryshevsky, K.G. Batrakov, I.Ya Dubovskaya in :
\emph{Proceedings of the  All-Union Conference on the Physics of
Interaction of Charged Particles with Crystals}, Izd-vo Mosk.
Univ., Moscow, 1990.

\bibitem{vesti92_11} V.G. Baryshevsky, K.G. Batrakov, I.Ya Dubovskaya  in:
\emph{Abstracts of the  All-Union Conference on the Physics of
Interaction of Charged Particles with Crystals}, Izd-vo Mosk.
Univ., Moscow, 1990 p. 88.


\bibitem{lanl_1}  V.G. Baryshevsky, I.Ya. Dubovskaya,  \emph{Diffraction phenomena in spontaneous and
stimulated radiation by relativistic particles in crystals} (Review) (1991) Technical Report Lawrence Berkeley Lab., CA
(United States) DOI 10.2172/5808050
\bibitem{lanl_2} M.L. Ter-Mikaelian, Usp. Fiz. Nauk \textbf{171}, n. 6, (2001) 597  [Sov. Phys. Uspekhi \textbf{44} (2001) 571-596 ].
\bibitem{lanl_3} Ulrik I. Uggerh{\o}j, Rev. Mod. Phys. \textbf{77}, 4, (2005) 1131-1171 .
\bibitem{lanl_4} R. Yabuki, H. Nitta, T. Ikeda and Y. H. Ohtsuki, Phys. Rev.   \textbf{B 63}, 17, (2001) 174112.
\bibitem{lanl_5} T. Ikeda, Y. Matsuda, H. Nitta, Y. H. Ohtsuki, Nucl. Instr. Methods \textbf{B 115}, 1-4, (1996) 380.

\bibitem{lanl_6} Y. Matsuda, T. Ikeda, H. Nitta, H. Minowa, Y. H. Ohtsuki, Nucl. Instr. Methods  \textbf{B 115}, 1-4, (1996) 396-400.


\bibitem{lanl_7} K.B. Korotchenko, Yu.L. Pivovarov, T.A. Tukhfatullin, Nucl. Instr. Methods \textbf{B 266}, 17, (2008) 3753-3757.

\bibitem{lanl_7a} V. G. Baryshevsky, Nucl. Instr. Methods  \textbf{B 122}, 1, (1997) 13-18.

\bibitem{lanl_8} V.G. Baryshevsky, Vesti AN BSSR, Ser. phiz-math nauk, \textbf{1} (1992) 31-37.
%

\bibitem{lanl_9} A. Kostyuk, A. V. Korol, A. V. Solov'yov and W. Greiner, Journal of Physics B \textbf{43}, 15, (2010) 151001.

\bibitem{lanl_10} V. G. Baryshevsky and K. G. Batrakov, Nucl. Instrum. Methods \textbf{A 483} (2002) 531Ц533.
\bibitem{lanl_11} V.G. Baryshevsky, K.G. Batrakov, LANL e-print arXiv:physics/0209028v1 [physics.optics].

\bibitem{lanl_12} V. G. Baryshevsky, K. G. Batrakov and I. Ya. Dubovskaya Nucl. Instr. Methods \textbf{A 375}, 1-3,(1996) 292-294.



\bibitem{lanl_13} V.G. Baryshevsky and A.A. Gurinovich, Proc. FEL'06, Berlin, August 2006, TUPPH013, p.335,
{http://www.JACoW.org}; LANL e-print arXiv:physics/0608068v1[physics.acc-ph].

\bibitem{lanl_14} D.A. Baklanov, I.E. Vnukov, V.K. Grishin,  Yu.V. Zhandarmov, A.N. Ermakov, G.P. Pokhil, R.A. Shatokhin,
  Preprint MGU є 2008-1/837, 14p.
\bibitem{lanl_14a}"Charged and Neutral Particles Channeling Phenomena  - Channeling 2008", Proceedings of the 51st Workshop of the INFN  Eloisatron (Project S.B. Dabagov and L. Palumbo, Eds., World Scientific, 2010) (The Science and Culture series - Physics, Series Ed. A. Zichichi)
\bibitem{lanl_15} W. Wagner, B. Azadegan, H. B\"{u}ttig, J. Pawelke, M. Sobiella, L. Sh. Grigoryan,
in: "Charged and Neutral Particles Channeling Phenomena  - Channeling 2008", Proceedings of the 51st Workshop of the INFN
 Eloisatron (Project S.B. Dabagov and L. Palumbo, Eds., World Scientific, 2010) (The Science and Culture series - Physics, Series Ed. A. Zichichi) pp. 378-407.


\bibitem{lanl_16} A.V. Korol, A.V. Solov'yov and W. Greiner, Journal of Physics \textbf{G 24}, 5 (1998) L45-L53.

\bibitem{lanl_17} A.V. Korol, A.V. Solov'yov and W. Greiner, International Journal of Modern Physics \textbf{E 8}, 1 (1999) 49-100.
    %
\bibitem{lanl_18} V. T. Baranov, S. Bellucci, V. M. Biryukov, G. I. Britvich and C.~ Balasubramanian, et al.,  JETP Lett. \textbf{82}, 9 (2005) 562-564.  

\bibitem{lanl_19} V.G. Baryshevsky, Dok. Akad. Nauk BSSR \textbf{31}, 12 (1987) 1090-1092.
\bibitem{lanl_20} A. R. Mkrtchyan, H. A. Aslanyan, A. H. Mkrtchian, R. H. Gasparian, A. Kh. Mkhitarian, G. M. Garibian, R. H. Avakian, S. P. Taroyan, A. E. Avetisyan, H. S. Kisogyan, K. R. Dallakyan, V. A. Gyurdzhyan, S. S. Danagulyan, G. M. Elbakyan and, Ye. M. Boyakhchyan, Solid State Commun.  \textbf{79},  4 (1991) 287-288.
\bibitem{lanl_21} A. R. Mkrtchyan,  A. H. Mkrtchian, H. A. Aslanyan, et al, Izv. Akad. Nauk. Arm. SSR, Fizika \textbf{47} (2005) 282.
\bibitem{lanl_21a} P.N. Zhukova, M.S. Ladnykh, A. H. Mkrtchian,  A. R. Mkrtchyan, N.N. Nasonov,
Pis'ma Zh. Tekh. Fiz. \textbf{36}, 21 (2010) 29-37.

\bibitem{lanl_22} V.G. Baryshevsky, A.A. Gurinovich LANL e-print arXiv: physics.acc-ph/1011.2983 
\bibitem{lanl_23a} V.G. Baryshevsky, K.G. Batrakov, V.A. Evdokimov, A.A. Gurinovich, A.S. Lobko, P.V. Molchanov, P.F. Safronov, V.I. Stolyarsky, Nuclear Instruments and Methods  \textbf{B 252}, 1, (2006) 86-91 .
\bibitem{lanl_23} V.G. Baryshevsky, E.A. Gurnevich The possibility
of Cherenkov radiation generation in a photonic crystal formed by
parallel metallic threads, in: \emph{Proceedings of 2010
International Kharkov Symposium on Physics and Engineering of
Microwaves, Millimeter and Submillimeter Waves (MSMW)}, 21-26 June
2010. DOI: 10.1109/MSMW.2010.554 6019, p. 1-3.
\bibitem{lanl_24} V.G. Baryshevsky, A.A. Gurinovich,  LANL e-print arXiv:0903.0300v1 [physics.acc-ph]


\bibitem{hyb_RuhadzeUFN} L.S. Bogdankevich and A.A. Rukhadze, Sov. Phys.Usp. \textbf{14}, 2, (1971) 163-189.

\bibitem{hyb_Selemir.rev1} A.E. Dubinov, V.D. Selemir, Radiotekhnika i electronika. \textbf{47} (2002) 645.

\bibitem{hyb_NeskolkoVC} A.E. Dubinov, I.A.Efimova, Technical Physics \textbf{46}, 6, (2001) 723-728.
\bibitem{hyb_Selemir_patent} A.E. Dubinov et al., patent RU 2221306 C2

\bibitem{hyb_87Selemir} A.E. Dubinov et al., Izv. VUZov, Fizika \textbf{6}
67 (1999)

\bibitem{hyb_Giri} Clayborne D. Taylor, D.V.Giri, \emph{High-Power Microwave systems and effects} (Taylor and Francis, 1994).


\bibitem{hyb_Selemir_patent2} V.D. Selemir, A.E. Dubinov, B.G Ptitsyn, K.S. Shilin, patent RU 2260870 C1


\bibitem{hyb_SelemirTWT} V. D. Selemir, A. E. Dubinov, E. E. Dubinov, I. V. Konovalov and A. V. Tikhonov,
Technical Physics Lett. \textbf{27}, 7,  583-585 (2001).


\bibitem{hyb_grid-FEL06-ex}  Baryshevsky V.G., Belous N.A., Gurinovich A.A. et al, in: \emph{Proceedings of the 28th
Intern. Free Electron Laser Conference FEL2006} (2006) 331-334.

\bibitem{hyb_grid-FEL06-th} Baryshevsky V.G., Gurinovich A.A. in:  \emph{Proceedings of
the 28th Intern. Free Electron Laser Conference FEL2006} (2006) 335-338.

\bibitem{hyb_grid-FEL07-ex}  V.G.Baryshevsky, N.A.Belous, A.A.Gurinovich et al, in:  \emph{Proceedings of the 29th Intern. Free Electron Laser Conference FEL2007 }(2007) TUPPH012.

\bibitem{hyb_27Selemir} D.D. Ryutov, Pis'ma Zh. Tekh. Fiziki  \textbf{1} (1975) 581.

\bibitem{hyb_28Selemir} R.A. Richardson, J. Denavit, M. S. Di Capua, and P. W. Rambo  J.Appl.Phys. \textbf{69} 6261-6272 (1991).
\bibitem{hyb_29Selemir} R.J. Adler, B.Sabol, G.F. Kiuttu, IEEE Trans. \textbf{NS-30}, 4,  3198-31200 (1983)

\bibitem{hyb_FirstLasing} V. Baryshevsky, K. Batrakov, A. Gurinovich, I. Ilienko, A. Lobko, V. Moroz,
P. Sofronov, V. Stolyarsky Nucl. Instr. Methods  \textbf{A 483} (2002) 21-23.%

\bibitem{hyb_patent} V.G.Baryshevsky et al., Eurasian Patent 004665 B1




\bibitem{L_2} D.I.Trubeckov, A.E.Hramov, \emph{Lectures on microwave electronics for
physicists }(Moscow, FIZMATLIT, 2004) [in Russian].

\bibitem{L_4} R. A. Silin,\emph{ Periodic Waveguides} (Phasis, Moscow, 2002) [in Russian].

\bibitem{L_5} R.A.Silin, V.P.Sazonov, \emph{Slow-wave structures} (Soviet Radio, Moscow, 1966) [in
Russian]; R.A. Silin and V.P. Sazonov, \emph{Slow Wave Structures}
National Lending Library for Science and Technology, Boston SPA,
Eng (1971).

\bibitem{L_7} V.G. Baryshevsky, Doklady Akademy of Science of Belarus SSR, v.31, n.12 (1987) 1089.

\bibitem{L_14}  L.D. Landau, E.M. Lifshitz, \emph{The Classical Theory of Fields} in: L.D. Landau, E.M. Lifshitz \emph{Course of Theoretical Physics} Vol. 2 (Pergamon Press, 4ed.,
1975).

\bibitem{L_15} L.D.Landau, E.M.Lifshitz,  L.P.Pitaevskii,
\emph{Electrodynamics of Continuous Media} in: L.D. Landau, E.M. Lifshitz \emph{Course of Theoretical Physics} Vol. 8 (Butterworth Heinemann, 2ed., 1984).

\bibitem{L_21} L.P.Pitaevskii, E.M.Lifshitz, \emph{Physical Kinetics} in: L.D. Landau, E.M. Lifshitz \emph{Course of Theoretical Physics} Vol. 10 (Butterworth Heinemann, 1981).

\bibitem{L_23} Z. G., Pinsker, \emph{Dynamical scattering of X-rays in crystals}
 (Springer-Verlag Berlin, Heidelberg, New York, 1978).



\end{thebibliography}
\end{document}